\documentclass[twocolumn,prd,amsmath,amssymb,showpacs,nofootinbib,%
  superscriptaddress]{revtex4}

\usepackage{graphicx}
\graphicspath{{figures/}}

\usepackage{tikz}
\usetikzlibrary{patterns}
\usetikzlibrary{plotmarks}

\usepackage{longtable}
\usepackage{hyperref}

\newcommand{\Z}{\ensuremath{Z}}
\newcommand{\A}{\ensuremath{A}}
\newcommand{\grammage}{\ensuremath{\chi}}
\newcommand{\grammageunits}{\ensuremath{{\rm g~cm}^{-2}}}
\newcommand{\density}{\ensuremath{\varrho}}
\newcommand{\densityunits}{\ensuremath{{\rm g~cm}^{-3}}}

\newcommand{\E}{\ensuremath{E}}
\newcommand{\Ek}{\ensuremath{E_{\rm k}}}
\newcommand{\Ekn}{\ensuremath{\mathcal{E}_{\rm k}}}
\newcommand{\Eknunits}{GeV/\A}
\newcommand{\R}{\ensuremath{R}}
\newcommand{\Runits}{GV}
\newcommand{\p}{\ensuremath{p}}
\newcommand{\punits}{GeV/c}

\newcommand{\flux}{\ensuremath{F}}
\newcommand{\fluxunits}{\ensuremath{{\rm m}^{-2}~{\rm s}^{-1}~{\rm GeV}^{-1}}}
\newcommand{\intensity}{\ensuremath{J}}
\newcommand{\intensityunits}%
    {\ensuremath{{\rm m}^{-2}~{\rm s}^{-1}~{\rm GeV}^{-1}~{\rm sr}^{-1}}}
\newcommand{\intensityunitsn}%
    {\ensuremath{{\rm m}^{-2}~{\rm s}^{-1}~({\rm GeV}/\A)^{-1}~{\rm sr}^{-1}}}
\newcommand{\anisotropy}{\ensuremath{\delta}}

\newcommand{\thetamsp}{\ensuremath{\theta_{\rm MS}^{\rm plane}}}
\newcommand{\thetamss}{\ensuremath{\theta_{\rm MS}^{\rm space}}}

\newcommand{\aeff}{\ensuremath{A_{\rm eff}}}
\newcommand{\aeffnorm}{\ensuremath{\aeff^{\perp}}}
\newcommand{\accept}{\ensuremath{G}}
\newcommand{\fov}{{\rm Fo\kern-1ptV}}
\newcommand{\psf}{{\rm PSF}}

\newcommand{\obstime}{\ensuremath{T_{\rm obs}}}
\newcommand{\exposure}{\ensuremath{\mathcal{E}}}
\newcommand{\fexposure}{\ensuremath{\mathcal{E}_{\rm f}}}
\newcommand{\edisp}{\ensuremath{D_E}}
\newcommand{\eres}{\ensuremath{\frac{\sigma_E}{E}}}
\newcommand{\eresl}{\ensuremath{\sigma_E/E}}
\newcommand{\Fermi}{{\it Fermi}}

\newcommand{\cheren}{Cherenkov}
\newcommand{\bremss}{bremsstrahlung}

\newcommand{\tabrowspace}{\noalign{\vspace{1pt}}}

\newcommand{\smallurl}[1]{{\scriptsize\url{#1}}}

\begin{document}

\title{Space-Based Cosmic-Ray and Gamma-Ray Detectors: a Review}

\author{Luca Baldini}
\email{luca.baldini@pi.infn.it}
\homepage[Visit: ]{http://www.df.unipi.it/~baldini/index.html}
\affiliation{Universit\`a di Pisa and INFN-Sezione di Pisa}

\date{\today}

\begin{abstract}
Prepared for the 2014 ISAPP summer school, this review is focused on
space-borne and balloon-borne cosmic-ray and gamma-ray detectors.
It is meant to introduce the fundamental concepts necessary to understand the
instrument performance metrics, how they tie to the design choices and how they
can be effectively used in sensitivity studies.
While the write-up does not aim at being complete or exhaustive, it is largely
self-contained in that related topics such as the basic physical processes
governing the interaction of radiation with matter and the near-Earth
environment are briefly reviewed.
\end{abstract}


\maketitle


\section{Prelude}%
\label{sec:prelude}

This write-up was prepared for the ISAPP School \emph{``Multi-wavelength and 
  multi-messenger investigation of the visible and dark Universe} held in
Belgirate between 21 and 30 July 2014%
\footnote{The program of the school is available at
\url{https://agenda.infn.it/conferenceDisplay.py?confId=6968}.}.
As such the reader should not expect it to be accurate and/or terse as an
actual scientific publication. It is largely idiosyncratic and guaranteed to
contain errors, instead.

That said, this review might be expanded and/or improved in the future,
depending on whether people will find it useful or not. If you find a mistake
or you have suggestions, do not hesitate to inform the author---though you
should not expect him to be working on this project with a significant duty
cycle, nor necessarily proactive. (Note, however, that I am committed to fix
right away any instance where the work of individuals or research groups has
arguably been misrepresented.)

\subsection{Overview}

The main focus of this review is introducing the basic instrument performance
metrics, with emphasis on space-born%
\footnote{Among space detectors we shall focus exclusively on those in low-Earth
  orbit, i.e., orbiting around the Earth at an altitude of a few hundred~km.}
and balloon-borne cosmic-ray and gamma-ray detectors. While, in general,
instrument response functions can only be predicted accurately by means of
detailed Monte Carlo simulations, the main features can be typically understood
(say within a factor of two) by means of simple, back-of-envelope calculations
based on the micro-physics of the particle interaction in the detector. Such an
approach is effective in highlighting the connection between the underlying
design choices and the actual instrument sensitivity.

To give some context to the discussion, the current body of knowledge on
cosmic and gamma rays is briefly reviewed in historical prospective, and the
basics of interactions between radiation and matter, along with the main
characteristics of the near-Earth environment, are concisely covered.
A honest attempt has been made to provide a complete and up-to date list of
references to the relevant sources in the literature.

The organization of the material is not necessarily the most natural one, based
on the consideration that, if we were to write it down in strict logical order,
deriving each fact from the prime principles, it is unlikely that the average
reader would get past page 2.
Instead, we frequently switch back and forth and use concepts that, strictly
speaking, are only introduced later in the flow of the discussion. We put some
effort, though, in trying to provide explicit cross references between the
different bits in a sensible way.

In the attempt to make it useful and---more importantly---\emph{usable}, all the
material contained in this review is copylefted under the GPL General Public
License and available at
\url{https://bitbucket.org/lbaldini/crdetectors}.
This includes: the \LaTeX\ source code, the tables, high-resolution versions of
all the figures and the code (along with the necessary auxiliary data files)
used to generate them.

\section{Introduction}%
\label{sec:intro}

The discovery of cosmic rays (CR) is customarily attributed to Viktor Franz
Hess, who first demonstrated in a series of pioneering balloon flights performed
in 1911 and 1912 that the ionizing radiation in the atmosphere of the Earth was
of extraterrestrial origin%
\footnote{Sure enough putting an electroscope on a balloon is more germane
  to the subject of this review than immersing it under water. Nonetheless
  it is appropriate to mention here the (largely uncredited) work of Domenico
  Pacini around the same time. The reader is referred
  to~\cite{2011arXiv1103.4392D} and references therein for a historical account
  of the early developments of cosmic-ray studies.}.
More than 100~years after their discoveries, cosmic rays have been extensively
studied, both with space-based detectors and with ground observatories.
For reference, the cosmic-ray database described in~\cite{CRDB}, which we have
extensively used in the preparation of this review, contains some $\sim 1100$
data sets from $\sim 40$ different experiments at energies below
$\sim 10^{15}$~eV.

\begin{figure}[!htb]
  \includegraphics[width=\linewidth]{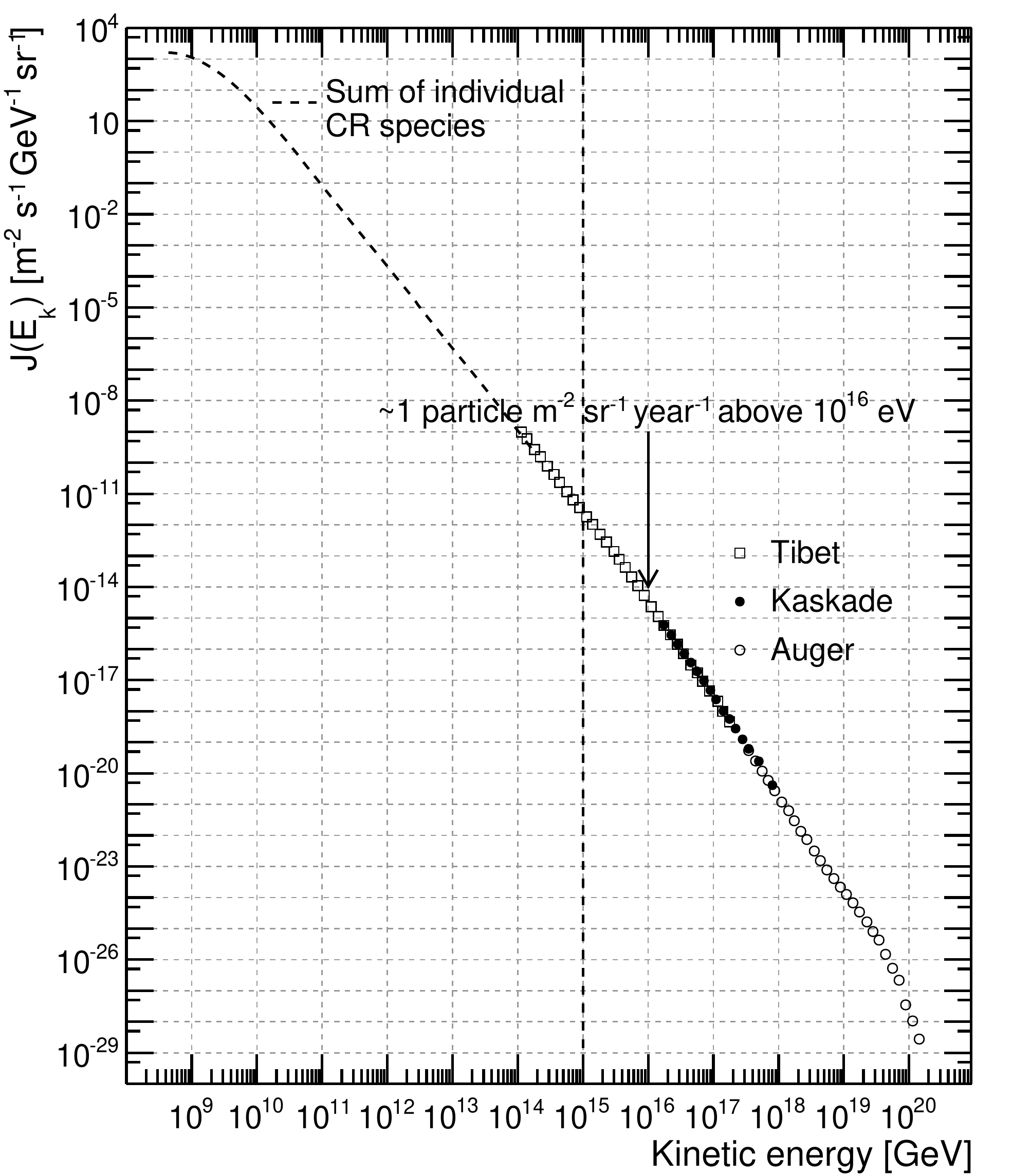}
  \caption{All-particle energy spectrum of primary cosmic rays, compiled from
    some of the most recent available measurements. The dashed line extending
    from $\sim 10^{9}$ to $\sim 10^{14}$~eV has been obtained by summing up
    the energy spectra of the most abundant CR species (p, He, C, O and Fe),
    measured by several different space- and balloon-borne experiments.
    The vertical dashed line marks the maximum energy ($\sim 1$~PeV) that can
    be practically measured by a CR detector in orbit.}
  \label{fig:cr_big_picture}
\end{figure}

Figure~\ref{fig:cr_big_picture} provides a largely incomplete, and yet
impressive, account of the tremendous body of knowledge that we have
accumulated along the way. The dashed line extending from $\sim 10^{9}$ to
$\sim 10^{14}$~eV is a weighted sum of the energy spectra of the most abundant
CR species (p, He, C, O and Fe), as measured by several different space- and
balloon-borne experiments. The reader is referred to~\cite{PDG} and references
therein for a more comprehensive compilation of the available all-particle
ultra-high energy cosmic-ray (UHECR) spectral measurements.

The immediate overall picture is that of a seemingly featureless power law
extending over some $12$~orders of magnitude in energy and $33$ orders of
magnitude in flux. As a matter of fact, cosmic-ray spectra are customarily
weighted by some power of the energy (typically ranging from $2$ to $3$) in
order to make relatively subtle intrinsic features more prominent---which is
a fairly sensible thing to do, as those features are connected with the
underlying physics. If we did so in this case, we would see the so-called
\emph{knee} emerging at a few $10^{15}$~eV, a second knee at $\sim 10^{18}$~eV,
the \emph{ankle} even higher energies and a spectral steepening, possibly
to be interpreted as the GZK cutoff, around $10^{20}$~eV.

Here we shall take a slightly different perspective and start by noting that
$10^{16}$~eV is roughly the energy where the integral all-particle cosmic-ray
spectrum amounts to $\sim 1$~m$^{-2}$~sr$^{-1}$~year$^{-1}$. If we forget for a
second the steradians (we shall see in section~\ref{sec:planar_detector}
that this essentially gives an extra factor of $\pi$, which is largely
irrelevant for the purpose of our argument) this means that
\emph{far away from the Earth there is about $1$ particle per year above
  $10^{16}$~eV crossing  any $1$~m$^2$ plane surface}.
Now, $1$~m$^2$ is a large figure by any space-experiment standard---and
$1$~year is not short either. This basic consideration naturally set the
PeV scale as the upper limit to the energies than can be practically studied
in space. It is fortunate that, though the atmosphere of the Earth is opaque
to primary particles above $\sim 10$~eV, ultra-high-energy cosmic rays
can be effectively studied from the ground through a variety of
experimental techniques involving the detection of their secondary
products---most notably the water-\cheren\ and florescence techniques, both
exploited by the Pierre Auger Observatory~\cite{2004NIMPA.523...50A},
and the extensive air shower arrays \emph{a la}
KASKADE~\cite{2003NIMPA.513..490A} or ARGO-YBJ~\cite{2012NIMPA.661S..50A}.
This is also true for high-energy gamma rays (above $\sim 100$~GeV), with at
least three major advanced imaging \cheren\ telescopes operating at the
time of writing (H.E.S.S., MAGIC, VERITAS) and the HAWC water-\cheren\
gamma-ray observatory currently under construction.
We note, in passing, that the dichotomy between space-borne (or balloon-borne)
and ground-based experiments has an important implication in that, while the
former provide measurements of separate cosmic-ray species, for the latter it
is much harder to infer the chemical composition. Important as it is, the
synergy between ground- and space-base observatories will not be discussed
further in this write-up.

At the other extreme of the energy spectrum (say below a few~GeV), primary
cosmic rays are heavily influenced by the heliospheric environment and
reprocessed by the atmosphere and the magnetic field of the Earth. While the
study of very low-energy primaries and of the geomagnetically trapped radiation
is by no means less interesting than that of their higher-energy fellows, an
informed discussion would require a large additional body of background
information and is therefore outside the scope of the paper.

In broad terms, this review is focused on the seven energy decades between 
$\sim 100$~MeV and $1$~PeV---and particularly on the experimental techniques
that have been (and are being) exploited to study cosmic rays and gamma rays in
this energy range. The subject is well defined and homogeneous enough---both
in terms of science topics and as far as experimental techniques are
concerned---that it can be usefully discussed in a unified fashion.

\section{Basic formalism and notation}%
\label{sec:notation}

A significant part of cosmic-ray physics is about \emph{energy spectra}.
That said, you should be wary when you happen to hear a cosmic-ray physicist
pronouncing the word \emph{spectrum}: it might indicate all sort of things.
On the $x$-axis you might find the particle (total or kinetic) energy, the
energy per nucleon, the momentum or the rigidity. On the $y$-axis you might
find a differential or integral flux or intensity, possibly multiplied by a
power of the variable plotted on the $x$-axis. We shall not refrain ourselves
from using the term spectrum in this somewhat loose sense, and the reader is
advised that sometimes it's pretty easy to get confused---so always pay
attention to the axis labels!

\begin{figure*}[!htb]
  \includegraphics[width=0.5\linewidth]{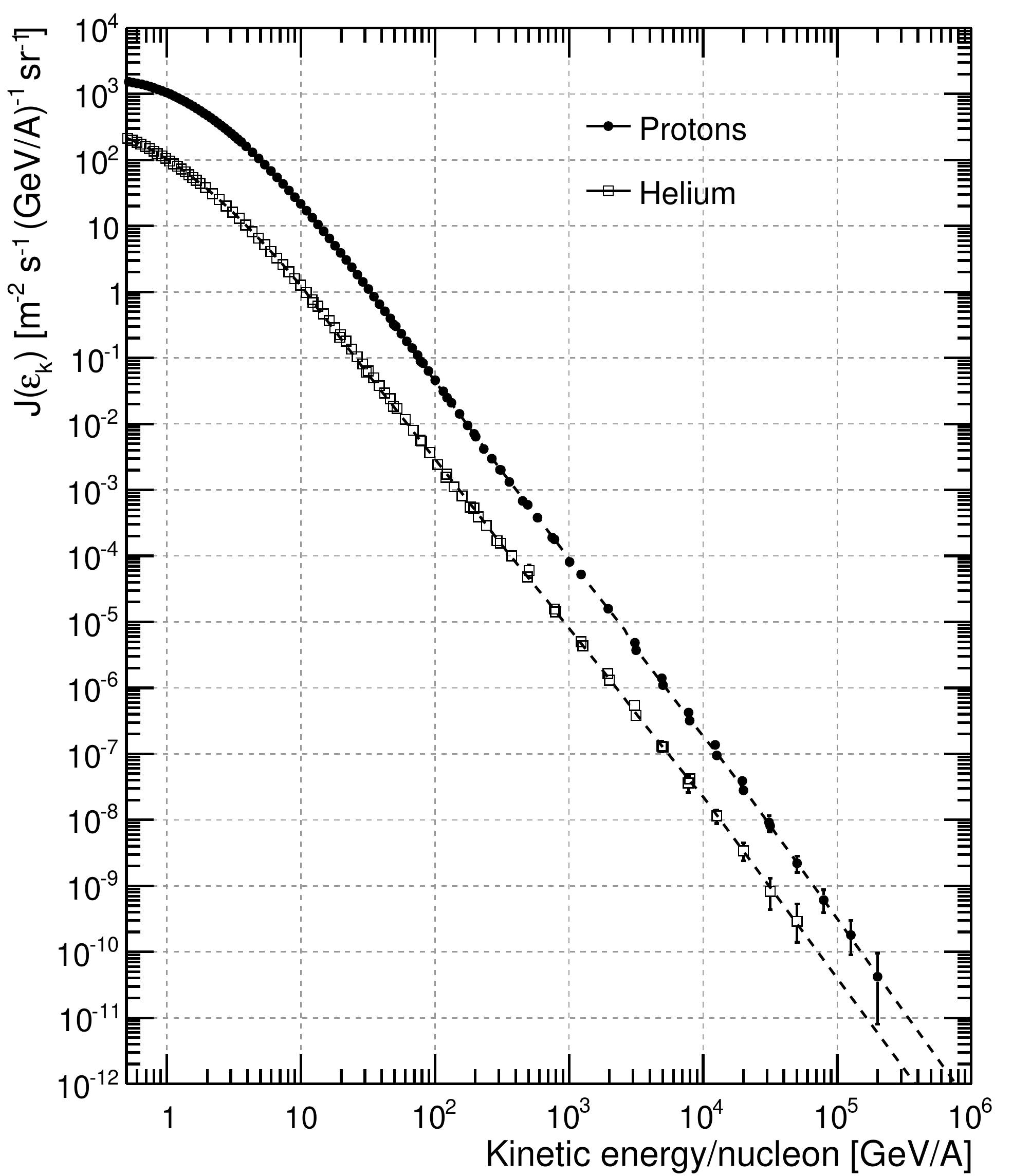}%
  \includegraphics[width=0.5\linewidth]{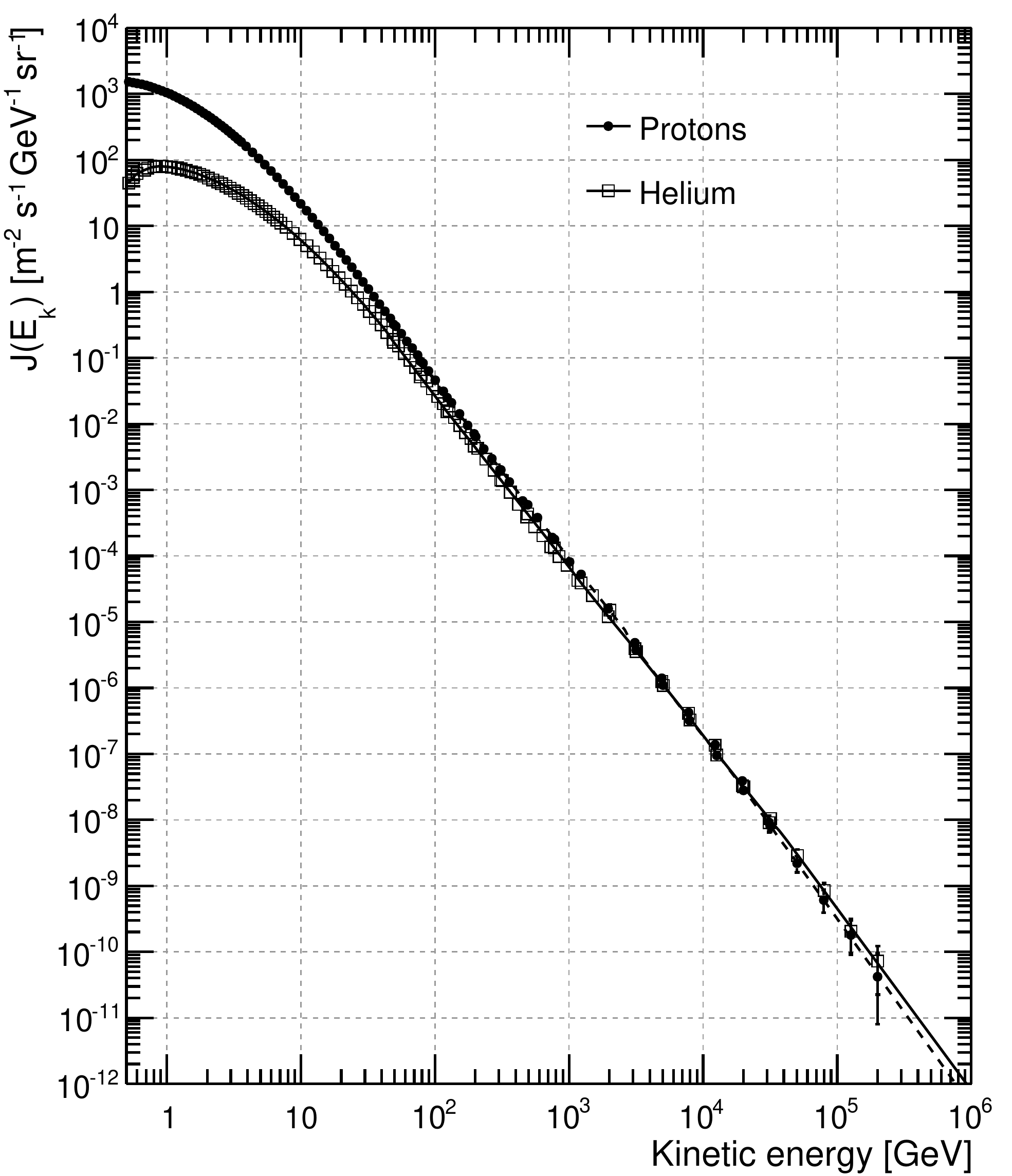}
  \caption{Differential intensities, plotted as a function of kinetic energy
    per nucleon (left) and total kinetic energy (right) for the two most
    abundant CR species: protons and He nuclei. While the He intensity is a
    factor of $\sim 10$ smaller than that of protons at, say,
    $\sim 1$~TeV/nucleon, the two are comparable, at $\sim 1$~TeV, when binned
    in total kinetic energy. See next section for more details about the data
    points and the dashed lines.}
  \label{fig:cr_spectra_p_he}
\end{figure*}

As a mere illustration, figure~\ref{fig:cr_spectra_p_he} shows how the proton
and helium spectra look different, relative to each other, depending on whether
the differential intensity is plotted as a function of the kinetic energy per
nucleon or the total kinetic energy. We note, in passing, that the first
choice is the one customarily adopted in this energy range and, clearly, the
one that we have in mind when we make statements such as
``\emph{protons account for $\sim 90\%$ of cosmic-rays}''. That said,
the second representation is not necessarily less meaningful---e.g., when
dealing with the energy measurement in a calorimetric experiment.
We'll come back to this shortly.

In this section we introduce some basic terminology and notation that
we shall use in the following.

\begin{table}[htb!]
  \begin{tabular*}{\linewidth}{l @{\extracolsep{\fill}} lll}
    \hline
    Symbol & Description & Units\\
    \hline
    \hline
    \Z & Atomic number & --\\
    \A & Mass number & --\\
    \density & Density & [\densityunits]\\
    \grammage & Grammage & [\grammageunits]\\
    \hline
    \E & Energy & [GeV]\\
    \Ek & Kinetic energy & [GeV]\\
    \Ekn & Kinetic energy/nucleon & [\Eknunits]\\
    \p & Momentum & [\punits]\\
    \R & Rigidity & [\Runits]\\
    \hline
    \flux & Differential flux & [\fluxunits]\\
    \intensity & Differential intensity & [\intensityunits]\\
    & & [\intensityunitsn]\\
    \anisotropy & Dipole anisotropy & --\\
    \hline
  \end{tabular*}
  \caption{Summary table of some useful quantities, along with the
    corresponding symbols and measurement units we shall use throughout this
    write-up.}
  \label{tab:basic_notation}
\end{table}

\subsection{Fluxes and Intensities}%
\label{sec:fluxes_intensites}

The term \emph{(differential) flux} indicates the number of particles per unit
time and energy crossing the unit vector area toward a given direction in the
sky and is customarily measured in \fluxunits. We shall indicate differential
fluxes with \flux\ throughout this manuscript.

The concept of differential flux essentially applies to \emph{point source}
studies, where the incoming particles all arrive from the same direction.
On the other hand, an isotropic flux of charged particles or photons is more
conveniently characterized by its \emph{intensity} \intensity\ (number of
particles per unit time, energy, area and solid angle) which is typically
measured in \intensityunits.

The distinction between differential fluxes and intensities is connected with
the dispersive nature of gamma-ray astronomy---in gamma rays you observe
a different patch of the sky at any time, while cosmic rays are approximately
the same in all directions. As we shall see in section~\ref{sec:irfs}, the
consequences are far reaching in terms of describing the instrumental
sensitivity. We shall try and stick to this nomenclature religiously throughout
the manuscript.

Sometimes it is handy to work with quantities related to the number of events
detected \emph{above} a given energy---\emph{integral} fluxes and intensities
are useful concepts that will be widely used in the following.

Depending on the situation, differential and integral fluxes and intensities
can be expressed as a function of energy, energy per nucleon, momentum and
rigidity%
\footnote{As a rule of thumb remember that, if a differential quantity is
  plotted as a function of a given variable $x$ (be it energy, momentum,
  rigidity or whatever), it is generally understood that the derivative is
  taken with respect to $x$.}%
. Trivial as this might seem, there is a few subtleties involved in
the conversion between different representation that we shall discuss in the
next section.

\subsection{Energy, momentum and all that}%
\label{sec:energy_momentum}

The total energy \E, kinetic energy \Ek\ and momentum \p\ of a particle or
nucleus are related to each other (through the rest mass $m$) by the well
known relativistic formul\ae
\begin{align}
  \p &= mc\beta\gamma\\
  \E^2 &= m^2 c^4 + \p^2 c^2 \\
  \Ek &= \E - mc^2,
\end{align}
that allow to switch from one variable to another when needed. When dealing
with ultra-relativistic particles (i.e., when $E \gg mc^2$, which is not
uncommon at all, in this context) energy, kinetic energy and momentum are
really the same thing---modulo the speed of light $c$---and one does not need
to bother about the differences. But for, e.g., protons and heavier nuclei
below $\sim 10$~GeV the spectra do look different depending on the variable
they are binned in (see figure~\ref{fig:cr_spectra_p_he}).

Since we are at it, here is a few other relativistic formul\ae\ that we shall
occasionally use in the following---they express $\beta$, $\gamma$ and
$\beta\gamma$ as a function of momentum, total energy and kinetic energy:
\begin{align}
  \beta & =
  \frac{pc}{\sqrt{m^2c^4 + \p^2c^2}} = 
  \sqrt{1 - \frac{m^2c^4}{{\E}^2}} =
  \frac{\sqrt{\Ek^2 + 2\Ek mc^2}}{\Ek + mc^2}\\
  \gamma & = 
  \sqrt{1 + \frac{\p^2c^2}{m^2c^4}} =
  \frac{\E}{mc^2} = 
  1 + \frac{\Ek}{mc^2}\\
  \beta\gamma &=
  \frac{\p}{mc} =
  \sqrt{\frac{{\E}^2}{m^2c^4} - 1} =
  \sqrt{\frac{{\Ek}^2 + 2\Ek mc^2}{m^2c^4}}.
\end{align}

The rigidity $R$ is essentially the ratio between the momentum and the
charge \Z\ (measured in units of the electron charge $e$) of a
particle\footnote{Don't get confused by the extra factors: a proton ($Z = 1$)
with a momentum $\p = 1$~GeV/c has rigidity of 1~GV.}:
\begin{align}
  \R = \frac{\p c}{\Z e}
\end{align}
Since we shall deal with magnetic fields---and it is easy to realize that
particles with the same rigidity behave the same way in a magnetic field---this
is a useful concept.

Finally, the CR spectra for He and heavier nuclei are more conveniently
expressed as a function of the kinetic energy per nucleon
\begin{align}
  \Ekn = \frac{\Ek}{\A}.
\end{align}
The kinetic energy per nucleon is a useful concept because, from the standpoint
of the hadronic interactions, a nucleus with mass number \A\ and kinetic energy
\Ek\ behaves, to a large extent, as the \emph{superposition} of \A\ nucleons
with kinetic energy $\Ek/\A$. (As a matter of fact, this is equally true---and
relevant---in the interstellar medium, in the Earth atmosphere and in the
calorimeter of a particle-physics detector.) It goes without saying that 
the kinetic energy per nucleon is conserved in spallation processes, and that
it's also the basic quantity determining whether a nucleus is relativistic or
not (which is the reason why differential intensities for different charged
species look more similar to each other in shape when binned in \Ekn).

Before we move on, we note that translating a differential intensity expressed
in kinetic energy per nucleon into the same intensity expressed in (total)
kinetic energy (or vice-versa) is slightly more complicated than scaling the
$x$-axis by a factor \A---one has also to scale the $y$-axis by a factor
$1/\A$, as by \emph{differential} we really mean \emph{differential in whatever
variable we have on the $x$-axis}. If the original differential intensity is
reasonably well described by a power-law with a spectral index $\Gamma$, a
scale factor \A\ on the $x$-axis is equivalent to a factor $\A^\Gamma$ on the
$y$-axis and the whole thing is effectively equivalent to multiplying the data
points by factor $\A^{\Gamma -1}$:
\begin{align}
  \intensity(\Ek) \approx \A^{\Gamma -1} \intensity(\Ekn).
\end{align}
Is is not by chance that the two He spectra in figure~\ref{fig:cr_spectra_p_he}
differ by a factor of $\sim 10$ in the high-energy regime, as
$4^{1.75} \approx 10$.

\subsection{Other observables}%
\label{sec:other_observables}

Charged-particle energy spectra (i.e. differential intensities) are not only
interesting \emph{per se}, but also in relation to each other. This aspect
customarily goes under the name of cosmic-ray \emph{chemical composition},
which is a crucial piece of information, as it probes the diffusion processes
in the Galaxy and provides an indirect measurement of the material traversed
by cosmic rays in their random walk from the source to the observer.
The isotopical composition, accessible by magnetic spectrometers, is an
interesting sub-chapter of this topic (more on this in section~\ref{sec:tof}).

On a related note, CR arrival directions generally bear no real memory of the
source---except, possibly, for the highest energies (see, e.g.,
\cite{2008arXiv0808.0417L}), which are not really of interest, in this context.
Nonetheless the anisotropy on large angular scales is another interesting
observable, especially for the leptons (since they rapidly loose energy due to
radiation they probe the nearby galactic space). In the somewhat na\"ive
scenario where the intensity is dominated by a single (nearby) source, one
expects a dipole anisotropy by the Firck's law. More generally, in a multipole
analysis one could expect detectable signatures on medium to large scales due
to the stochastic nature of the sources.

Anisotropies at all angular scales are also interesting in gamma rays, both in
connection with the study of the galactic diffuse emission and the contribution
of source populations to the extra-galactic diffuse emission.

As gamma rays \emph{do} point back to their sources, the position and
morphology of point/extended sources are yet other relevant observables.
We shall come back to this later in the write-up.

We end the list mentioning the exciting perspective of measuring gamma-ray
polarization in the pair production regime.

\subsection{The grammage}

When a particle traverses a homogeneous slab of a material---say some
compressible gas at a given pressure---the average number of scattering centers
it encounters is proportional to the product between the density \density\
and the thickness $l$ of the slab. Doubling one and reducing the other by $1/2$
does not really change anything: the quantity characterizing the amount of
material traversed is the \emph{grammage}
\begin{align}
  \grammage = \density l,
\end{align}
which is measured in g~cm$^{-2}$ in the cgs system. This quantity is also
referred to as the \emph{mass per unit area}, as
\begin{align}
  \density = \frac{m}{V} = \frac{m}{lS},
\end{align}
and therefore
\begin{align}
  \grammage = \frac{m}{S}.
\end{align}

There are many situations in which the grammage is a useful concept, e.g., in
characterizing the integrated column density that cosmic rays traverse
in their random walk from the source to the observer or the residual 
atmosphere at the typical balloon floating altitude. In both cases it is
really the product of the density times the path length (or, more precisely,
the integral of the density along the path, as the density is not constant)
that matters%
\footnote{Interestingly enough, in this case the enormous differences in 
  the basic length scales play out in such a way that the answer is pretty much
  the same ($\sim 5$~\grammageunits) in both situations.
  More on this in section~\ref{sec:atmospheric_grammage}}.

The grammage is not the only quantity measured in g~cm$^{-2}$ we shall encounter
in the following. The radiation and interaction length of materials---as we
shall see in sections~\ref{sec:em_showers} and \ref{sec:had_showers} these are
the typical length scales over which electromagnetic and hadronic showers
develop---are customarily measured in the same units, as they tend to be
smaller for denser material. In this case the mass scaling law is only
approximate and measuring these quantities in \grammageunits\ underlines the
intrinsic differences. One can then recover the actual length scale (in~cm)
just dividing by the density. We shall see in section~\ref{sec:earth_limb} how
comparing the grammage and the radiation length in a variable-density
environment (the atmosphere of the Earth) turns out to be handy.

\section{Galactic cosmic rays}%
\label{sec:cr}

In order to give some context for the following discussion, in this section we
briefly summarize some of the basic facts about galactic cosmic rays.

\subsection{The spectrum of galactic cosmic rays}%
\label{sec:gcr_spectra}

\begin{figure}[!htb]
  \includegraphics[width=\linewidth]{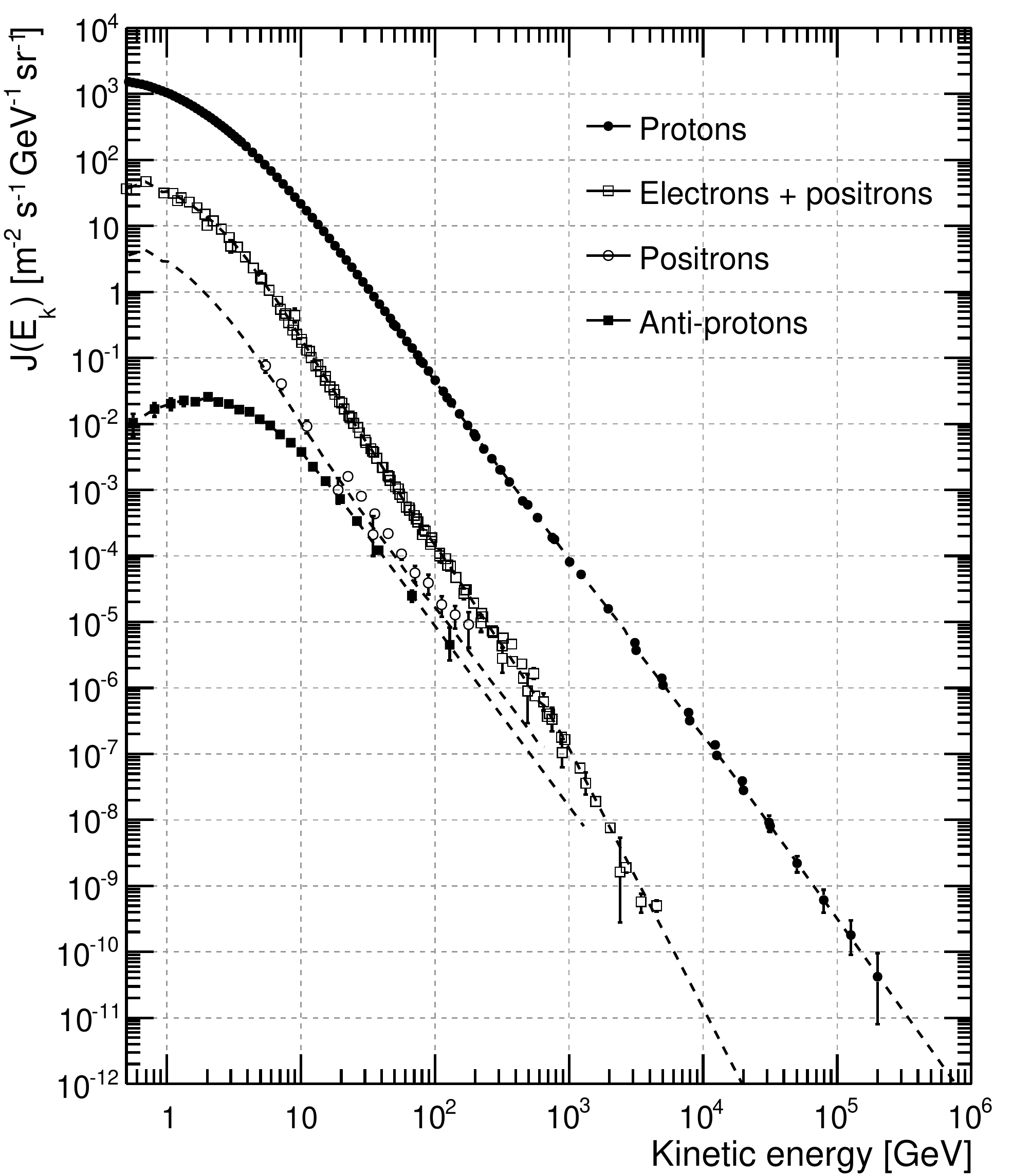}
  \caption{Spectra of the singly-charged components of the cosmic radiation.
    The data points are taken from~\cite{CRDB} and correspond to
    references~\cite{2009BRASP..73..564P, 2011ApJ...728..122Y, 2011Sci...332...69A, 2008Natur.456..362C, 2008PhRvL.101z1104A, 2010PhRvD..82i2004A, 2011PhRvL.106t1101A, 1996A&A...307..981R, 2012PhRvL.108a1103A, 1998ApJ...498..779B, 2013JETPL..96..621A}. For each species, the dashed line represents the weighted
  average of all the recent available measurements, and will be used in the
  following as the baseline for sensitivity estimates. For completeness,
  the model for the positron spectrum has been obtained by combining the
  $(e^{+} + e^{-})$ spectrum with the positron fraction measured by
  AMS-02~\cite{2013PhRvL.110n1102A}.}
  \label{fig:cr_spectra_single_charge}
\end{figure}

The energy spectrum of individual CR species has now been measured by
space- and balloon-borne detectors over some~7 decades in energy---at least
for the more abundant ones---and is largely dominated by protons, as
shown in figure~\ref{fig:cr_spectra_single_charge}.
Among the other singly-charged species, electrons amount to some
$10^{-2}$--$10^{-3}$ (depending on the energy) of the proton flux, and
positrons and antiprotons are even less abundant, the latter being
some $10^{-4}$ of the proton flux. We shall come back to these numbers in
section~\ref{sec:event_selection} when discussing the challenges one has to
face in separating the different species.

\begin{figure}[!htb]
  \includegraphics[width=\linewidth]{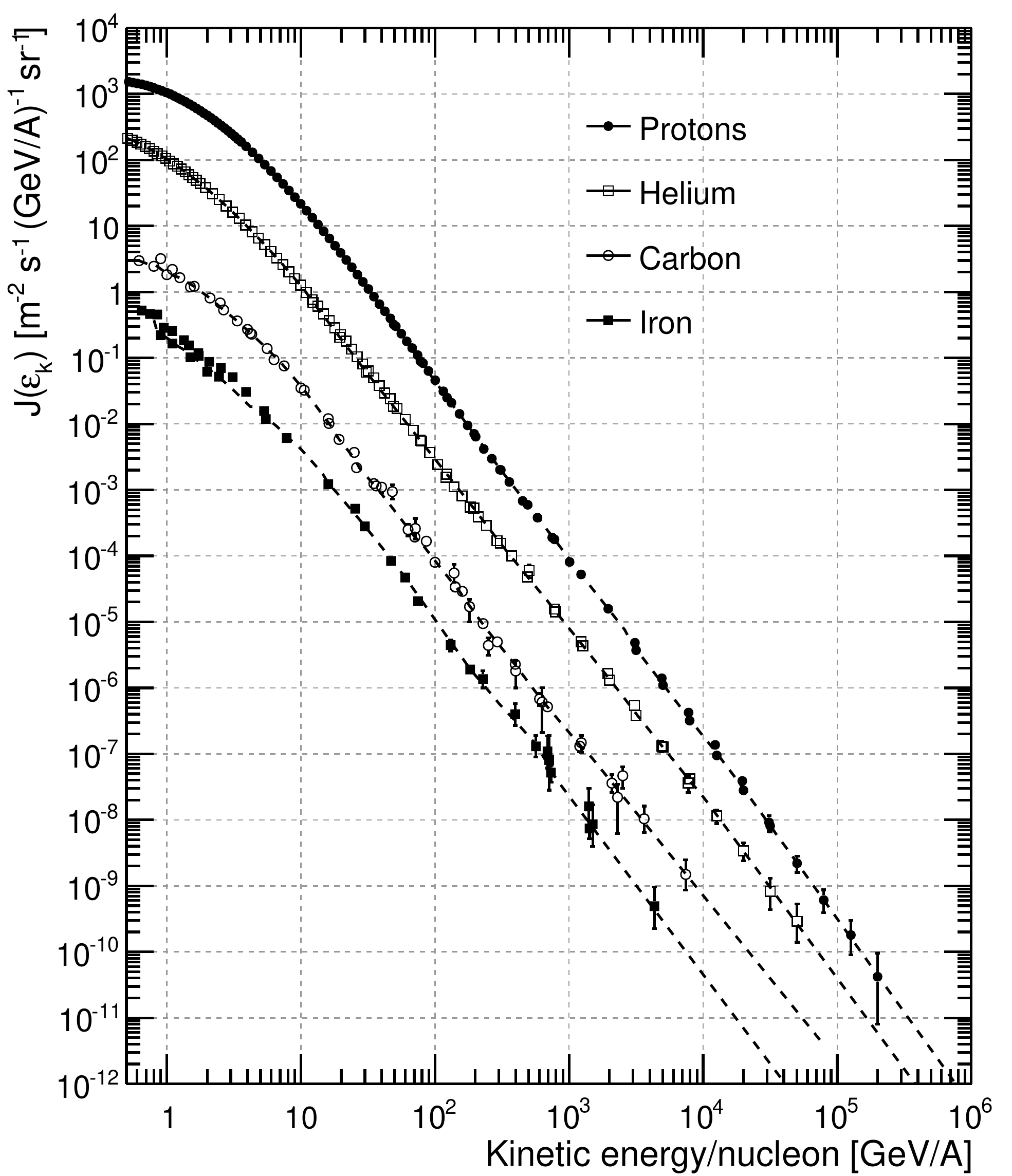}
  \caption{Spectra of some of the more abundant cosmic-ray species with $z > 1$,
    compared with the proton spectrum shown in
    figure~\ref{fig:cr_spectra_single_charge}.
    The data points are taken from~\cite{CRDB} and correspond to
    references~\cite{2009BRASP..73..564P, 2011ApJ...728..122Y, 2011Sci...332...69A, 2009ApJ...707..593A, 2011ApJ...742...14O, 1994ApJ...429..736B, 1980ApJ...239..712S, 1990A&A...233...96E, 2008ApJ...678..262A, 2005ApJ...628L..41D, 1981ApJ...248..847M, 1981ApJ...246.1014Y}.
    For each species, the dashed line represents the weighted
    average of all the recent available measurements, and will be used in the
    following as the baseline for sensitivity estimates.}
  \label{fig:cr_spectra_metals}
\end{figure}

As it turns out, cosmic rays include all sort of nuclei. Helium nuclei,
amounting to some $10\%$ of the protons, constitute the second more abundant
component, and carbon and oxygen are also relatively abundant, as shown in
figure~\ref{fig:cr_spectra_metals}.

For completeness, the dashed lines in
figures~\ref{fig:cr_spectra_single_charge} and \ref{fig:cr_spectra_metals}
represent weighted averages of all the recent available measurements for
each CR species, and we shall use them in the rest of this review for
sensitivity estimates. We shall be fairly liberal, within reason, in terms of
extrapolating differential and integral spectra at energies where there are
not yet measurements available.

\begin{table}[htb!]
  \begin{tabular}{p{0.3\linewidth}p{0.3\linewidth}p{0.3\linewidth}}
    \hline
    Z & Element & Relative\\
    & & abundance\\
    \hline
    \hline
    1 & H & 540 \\
    2 & He & 26\\
    3--5 & Li--Be & 0.4\\
    6--8 & C--O & 2.20\\
    9--10 & F--Ne & 0.3\\
    11--12 & Na--Mg & 0.22\\
    13--14 & Al--Si & 0.19\\
    15--16 & P--S & 0.03\\
    17--18 & Cl--Ar & 0.01\\
    19--20 & K--Ca & 0.02\\
    21--25 & Sc--Mn & 0.05\\
    26--28 & Fe--Ni & 0.12\\
    \hline
  \end{tabular}
  \caption{Relative abundances of several groups of cosmic-ray species at
    $10.6$~\Eknunits\ (adapted from~\cite{PDG}), normalized to that of oxygen
    ($3.29 \times 10^{-2}$~\intensityunitsn.}
  \label{tab:cr_abundances}
\end{table}


\subsection{The cosmic-ray gamma-ray connection}

Though it is not very common to see cosmic-ray and gamma-ray differential
intensities overlaid on the same plot, cosmic rays and gamma rays are tightly
tied to each other.
The vast majority of celestial gamma rays in the GeV energy range are produced
by interactions of cosmic rays with the interstellar medium and with galactic
magnetic and radiation fields. The study of this galactic \emph{diffuse}
emission provides a prospective on the diffusion of cosmic rays in the
galaxy complementary to direct measurements---as a matter of fact, it is
the realization that cosmic-ray interactions were bound to produce gamma rays
that provided one of the earliest stimuli to the development of gamma-ray
astronomy.
On a slightly different note, since gamma rays do point back to their sources,
they provide a direct view on the likely sources of cosmic rays---supernova
remnnants (SNR).
Finally, as we shall see in the following, the two fields of investigation
share much in terms of experimental techniques.

\begin{figure}[!htb]
  \includegraphics[width=\linewidth]{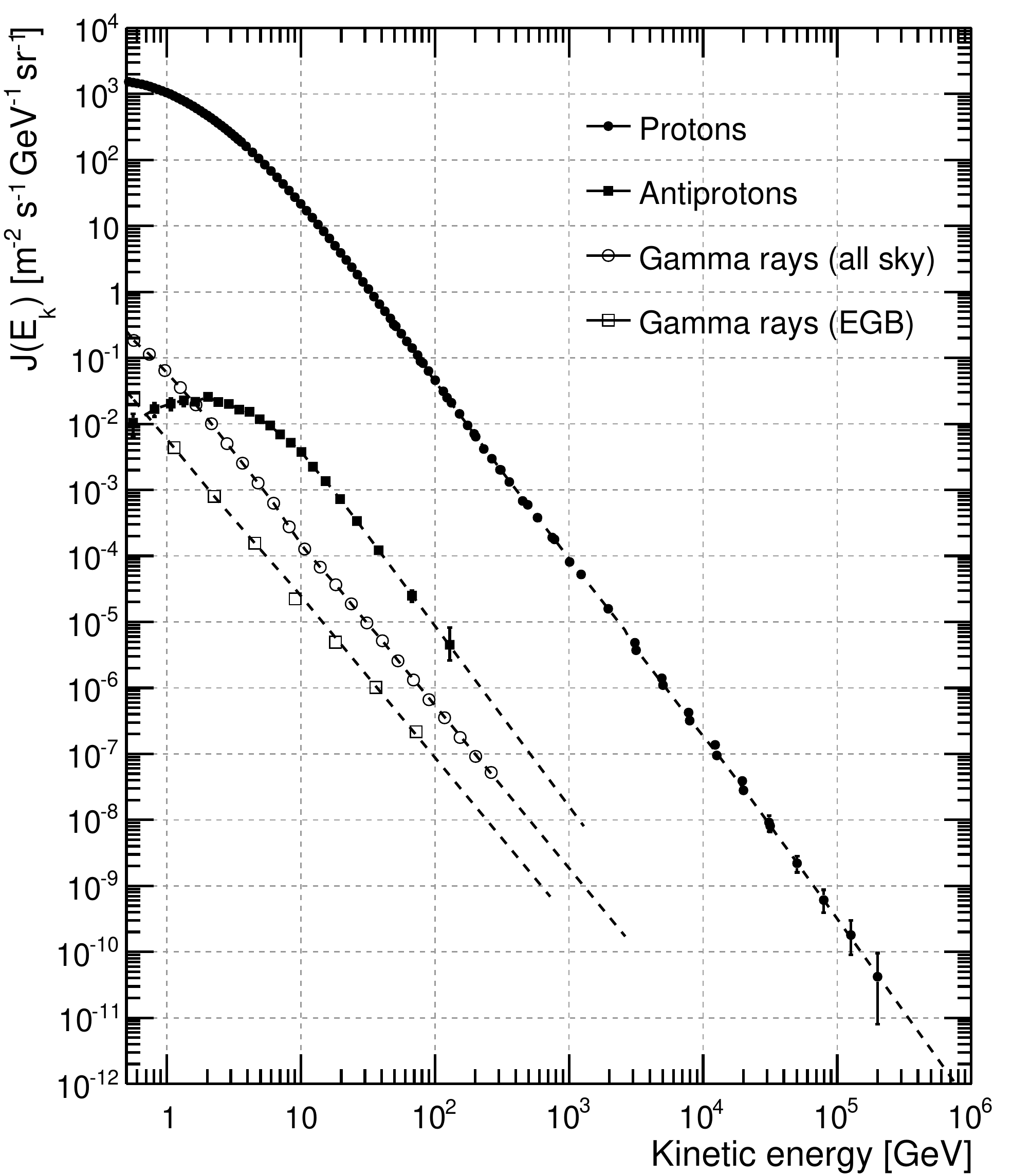}
  \caption{Energy spectrum of the all-sky gamma-ray intensity and of
    the extra-galactic gamma-ray background~\cite{2010PhRvL.104j1101A},
    compared with the most and least abundant singly charged CR species
    (i.e., protons and antiprotons). Cosmic-ray data are taken from~\cite{CRDB}
    and correspond to references~\cite{2009BRASP..73..564P, 2011ApJ...728..122Y, 2011Sci...332...69A, 2009ApJ...707..593A, 2011ApJ...742...14O, 1994ApJ...429..736B, 1980ApJ...239..712S, 1990A&A...233...96E, 2008ApJ...678..262A, 2005ApJ...628L..41D, 1981ApJ...248..847M, 1981ApJ...246.1014Y, 2013JETPL..96..621A}.}
  \label{fig:cr_spectra_gamma}
\end{figure}

Figure~\ref{fig:cr_spectra_gamma} shows the all-sky gamma-ray differential
intensity measured by the \Fermi-LAT, and makes it fairly obvious that celestial
gamma-rays constitute a tiny fraction of the cosmic radiation. Above $1$~GeV
the entire gamma-ray sky intensity is more than an order of magnitude weaker
than that of the rarest singly-charged species of the cosmic
radiation---antiprotons---and five to six orders of magnitude less abundant
than that of cosmic-ray protons.
It goes without saying that the relative paucity of gamma-ray fluxes
exacerbates the need for large instruments relative to charged species in the
same energy range.

The difficulties in separating gamma rays out of the bulk of the charged
cosmic-ray component are somewhat mitigated by the fact that efficient
anticoincidence detectors can be realized to distinguish between neutral and
charged particles and, at least for the analysis of gamma-ray sources,
spatial---and sometimes temporal---signatures can be exploited. Nonetheless
the measurement of the faint isotropic gamma-ray background may require
a proton rejection factor as high as $10^6$. We shall discuss this in somewhat
more details in section~\ref{sec:event_selection}.

\subsection{The gamma-ray sky}%
\label{sec:gamma_ray_sky}

In broad terms, the gamma-ray sky can be roughly subdivided in three main
components: the galactic diffuse emission (DGE), point and extended sources,
and the isotropic gamma-ray background (IGRB), sometimes referred as the
extra-galactic background. In the next three subsections we shall briefly
introduce these basic components.

\subsubsection{Galactic diffuse gamma-ray emission}%
\label{sec:dge}

As mentioned in the previous section, the vast majority of celestial gamma rays
above 1~GeV are produced by the interaction of charged cosmic rays with the
interstellar gas and radiation fields, resulting in a structured diffuse
emission, brighter along the galactic plane---and especially toward the
galactic center. The general subjects of characterizing and modeling the
galactic diffuse gamma-ray emission are way beyond the scope of this write-up
and we refer the reader to~\cite{2012ApJ...750....3A} and references therein
for an in-depth description of the state of the art.

In this section we limit ourselves to introduce the model of the galactic
diffuse emission that the \Fermi-LAT collaboration makes available
as one of the analysis components for point-source analysis of public
\Fermi-LAT data (as we shall use it in the following for sensitivity studies)%
\footnote{The model is publicly available at
  \url{http://fermi.gsfc.nasa.gov/ssc/data/access/lat/BackgroundModels.html}
  in the form of a fits file containing intensity maps binned in galactic
  coordinates in 30 energy slices from $\sim 60$~MeV to $\sim 500$~GeV. For
  completeness, the model used here is that named \texttt{gll\_iem\_v05.fit},
  though it is worth emphasizing that, from the standpoint of basic sensitivity
  studies such as the ones we are concerned about, the differences between
  different releases of the LAT diffuse model are hardly relevant.}.

\begin{figure}[htb!]
  \includegraphics[width=\linewidth]{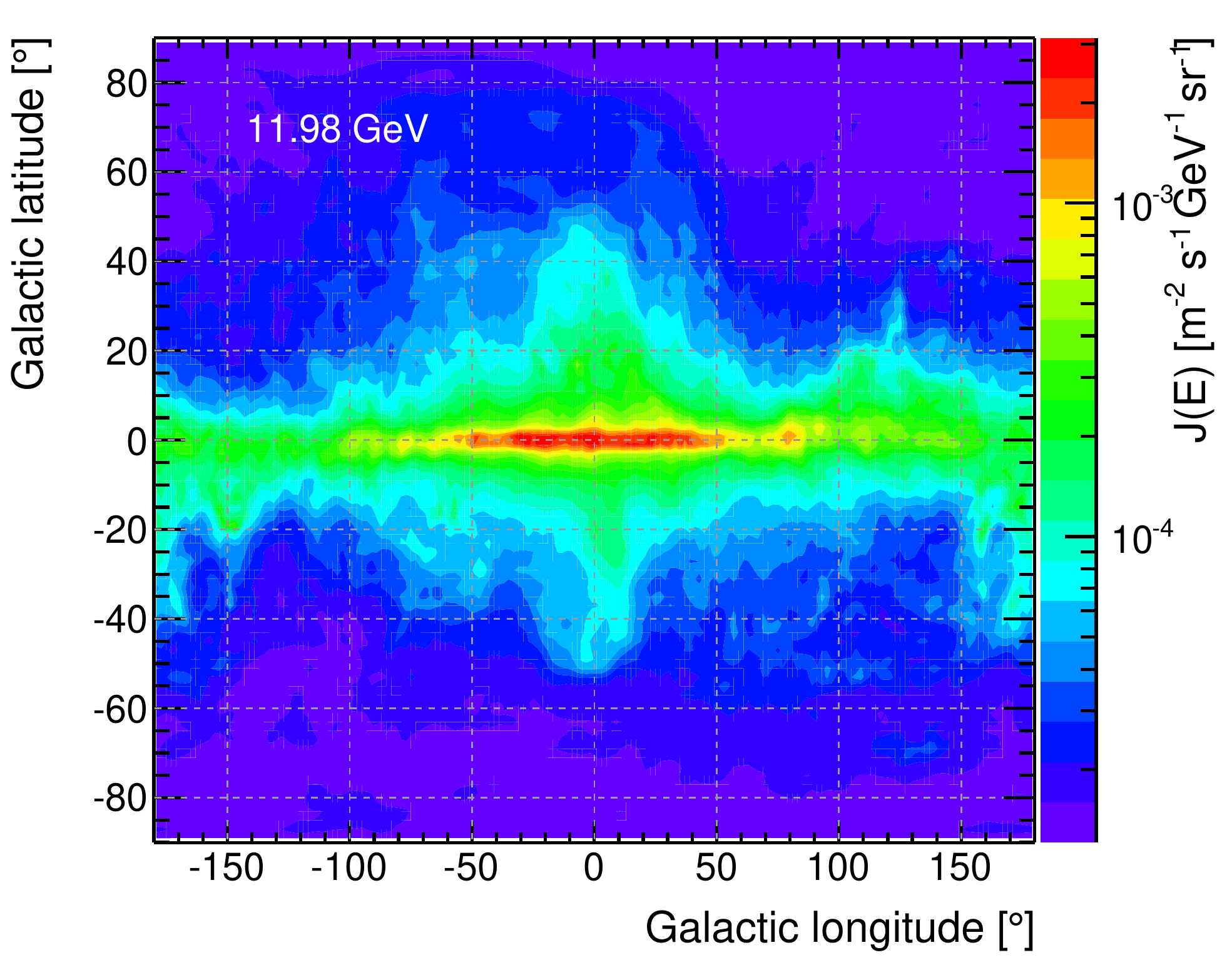}
  \caption{Map, in galactic coordinates, of the differential intensity of the
    LAT \texttt{gll\_iem\_v05.fit} model of the galactic diffuse emission in
    the energy slice centered at $\sim 12$~GeV.}
  \label{fig:gde_model}
\end{figure}

Figure~\ref{fig:gde_model} shows an example of the spatial morphology of the
\Fermi-LAT diffuse model in the energy slice centered at $\sim 12$~GeV%
\footnote{For completeness, the model has been re-binned from the original
$0.125^\circ$ grid into a coarser $2^\circ$ grid.}. The prominent emission from
the galactic plane, brightest toward the galactic center, is clearly visible.

\begin{figure}[htb!]
  \includegraphics[width=\linewidth]{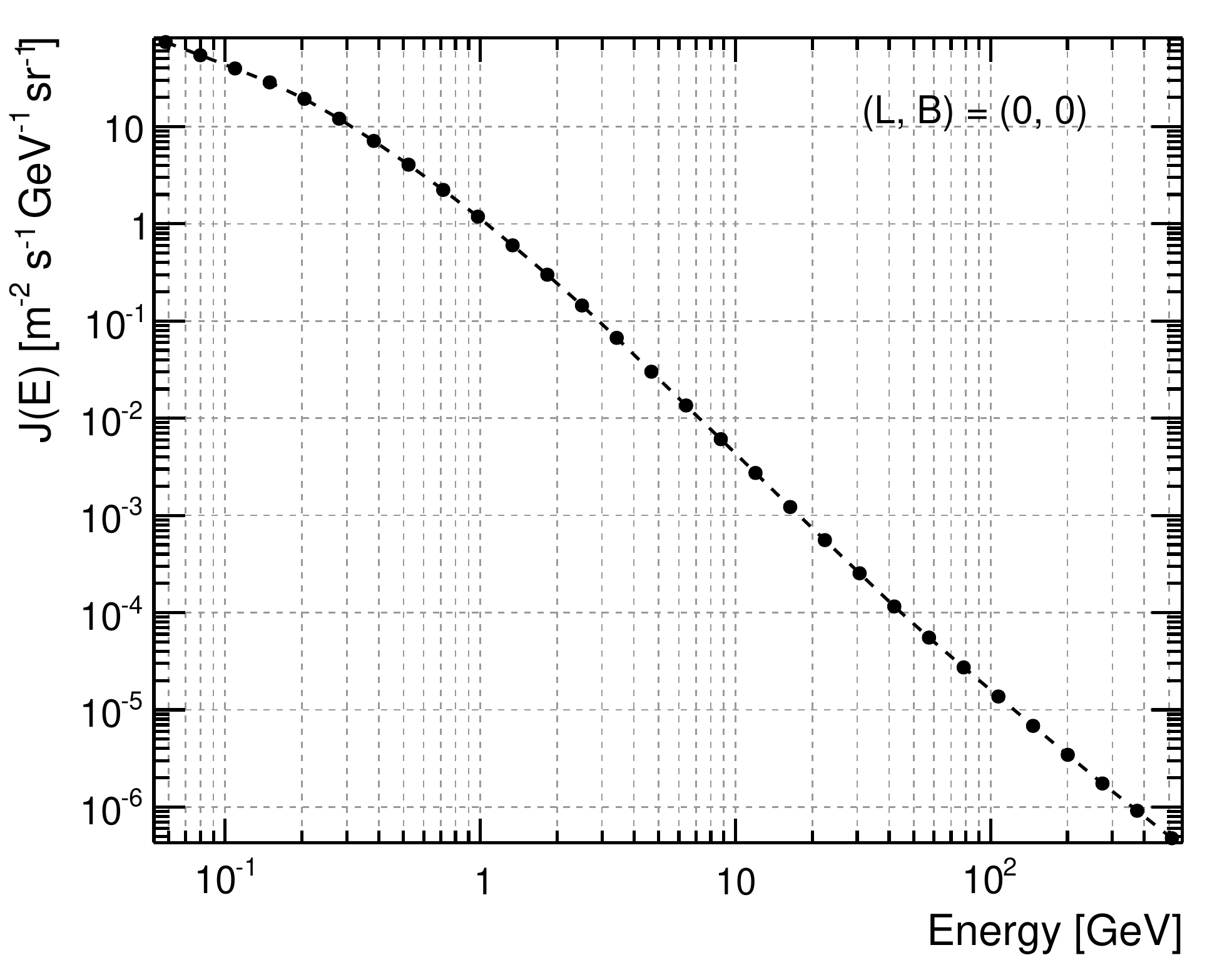}
  \caption{Differential intensity of the galactic diffuse emission around the
    galactic center for LAT \texttt{gll\_iem\_v05.fit} model.
    (Note that, this being a \emph{model}---though tuned to the LAT data---it
    comes with no error bars.)}
  \label{fig:gde_spectrum}
\end{figure}

Figure~\ref{fig:gde_spectrum} shows the differential intensity for the
$2^\circ \times 2^\circ$ square around the galactic center (the GDE spectrum
in other part of our Galaxy, normalization aside, is fairly similar.)

\subsection{Gamma-ray point sources}

Based on two years of sky survey, the second \Fermi-LAT source catalog
(2FGL~\cite{2012ApJS..199...31N}) is the deepest ever gamma-ray catalog
between 100~MeV and 100~GeV. It contains positional, variability and spectral
information for 1873 sources detected by the LAT in this energy range.

We refer the reader to the fellow LAT second pulsar
catalog~\cite{2013ApJS..208...17A}, the second catalog of active galactic
nuclei~\cite{2011ApJ...743..171A} and first catalog of sources above
10~GeV~\cite{2013ApJS..209...34A} for more information about specific source
populations.

\subsection{Gamma-ray isotropic background}

What remains after the galactic diffuse emission and the point and extended
sources have been subtracted from the gamma-ray all-sky intensity is generally
referred to as isotropic gamma-ray background,
or IGRB~\cite{2010PhRvL.104j1101A}. The IGRB is an observation-dependent
quantity, as its intensity depends on how many sources are resolved in a give
survey---and populations of faint sources below the detection threshold are
guaranteed to contribute to it. The sum of the IGRB and the extra-galactic
sources is, in some sense, a more fundamental quantity which is generally
referred to as extra-galactic background (EGB).

From the experimental point of view, the measurement of the IGRB is
challenging in that one has to discriminate an isotropic flux of gamma rays
against the much brighter charged-particle foreground without any spatial
or temporal signature to be exploited. We shall briefly come back to this
in section~\ref{sec:event_selection}.

\subsection{State of the art: modeling and measurements}

Beautiful as they are, figures~\ref{fig:cr_spectra_single_charge},
\ref{fig:cr_spectra_metals}, \ref{fig:cr_spectra_gamma}, \ref{fig:gde_model}
and \ref{fig:gde_spectrum} provide some of the most striking evidence for the
tremendous body of knowledge about cosmic and gamma rays accumulated over the
last century. On the theoretical side the progress has been no less
spectacular, and we refer the reader to~\cite{2013A&ARv..21...70B} for an
impressive, up-to-date and comprehensive review of the basic theoretical ideas
at the basis of the so-called supernova paradigm for the origin of galactic
cosmic rays.

In fact the discussion in the previous sections might very well give the
reader the warm and fuzzy feeling that there is little left to discover, while
nothing is farthest from truth. The aforementioned
review~\cite{2013A&ARv..21...70B} makes a honest effort at highlighting the
most important loose ends in our understanding of cosmic-ray production and
propagation, and \cite{2012ApJ...750....3A} provides a good summary of the
difficulties involved in a self-consistent modeling of the galactic diffuse
gamma-ray emission and the propagation of cosmic rays.

\begin{figure}[!htb]
  \includegraphics[width=\linewidth]{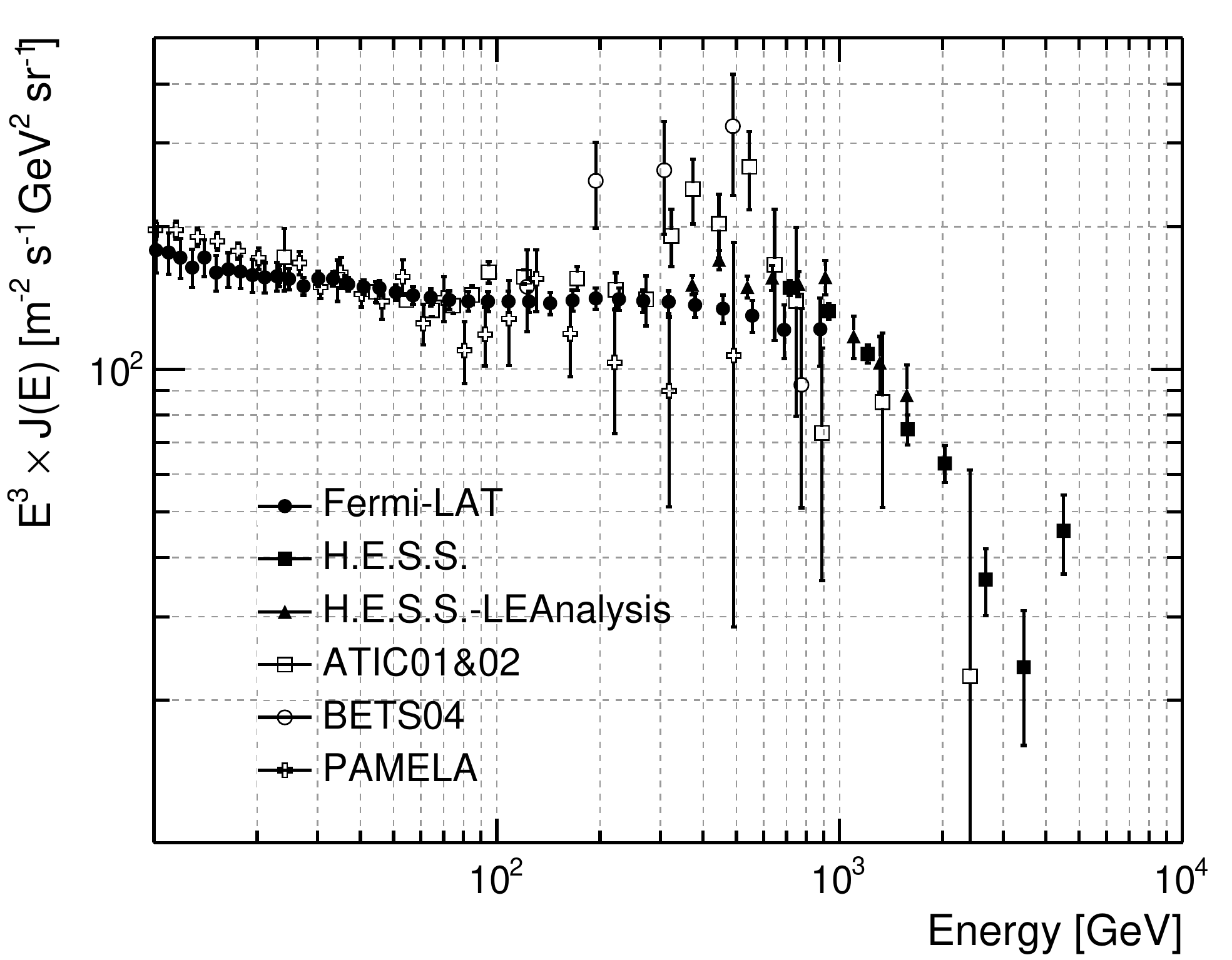}
  \caption{Compilation of some of the most recent measurements of the
    high-energy cosmic-ray $e^+ + e^-$ spectra. Data points are from~\cite{2010PhRvD..82i2004A, 2008PhRvL.101z1104A, 2009A&A...508..561A, 2008Natur.456..362C, 2008AdSpR..42.1670Y, 2011PhRvL.106t1101A}.}
  \label{fig:cr_allelectron_spectrum}
\end{figure}

More importantly, at least from the prospective of this review, there are
substantial pieces of observational evidence that are, to date, either missing
or controversial. The feature in cosmic ray electron spectrum reported by the
ATIC and BETS experiments in 2008, while not confirmed, at the time of writing,
by at least three experiments (\Fermi, PaMeLa and H.E.S.S) is a good example of
an unexpected finding that has stirred the interest of the community
triggering literally hundreds of follow-up papers (see
figure~\ref{fig:cr_allelectron_spectrum}).

More recently, the PaMeLa experiment has reported~\cite{2011Sci...332...69A} an
abrupt change of slope around $\sim 230$~GV in the proton and helium spectra,
whose extrapolation seem to nicely match the measurements at higher energies by
CREAM~\cite{2011ApJ...728..122Y} and ATIC~\cite{2009BRASP..73..564P}.
This tantalizing piece of evidence, if confirmed, would be of tremendous
interest, and is one of the notable examples of direct measurements where the
community is holding its breath for the results from the AMS-02 experiment
operating on the space station%
\footnote{Preliminary results from the AMS-02 experiment do not confirm
  the break reported by PaMeLa, but the issue is important enough that waiting
  for a refereed publication is in order before commenting further.}.

\begin{figure}[!htb]
  \includegraphics[width=\linewidth]{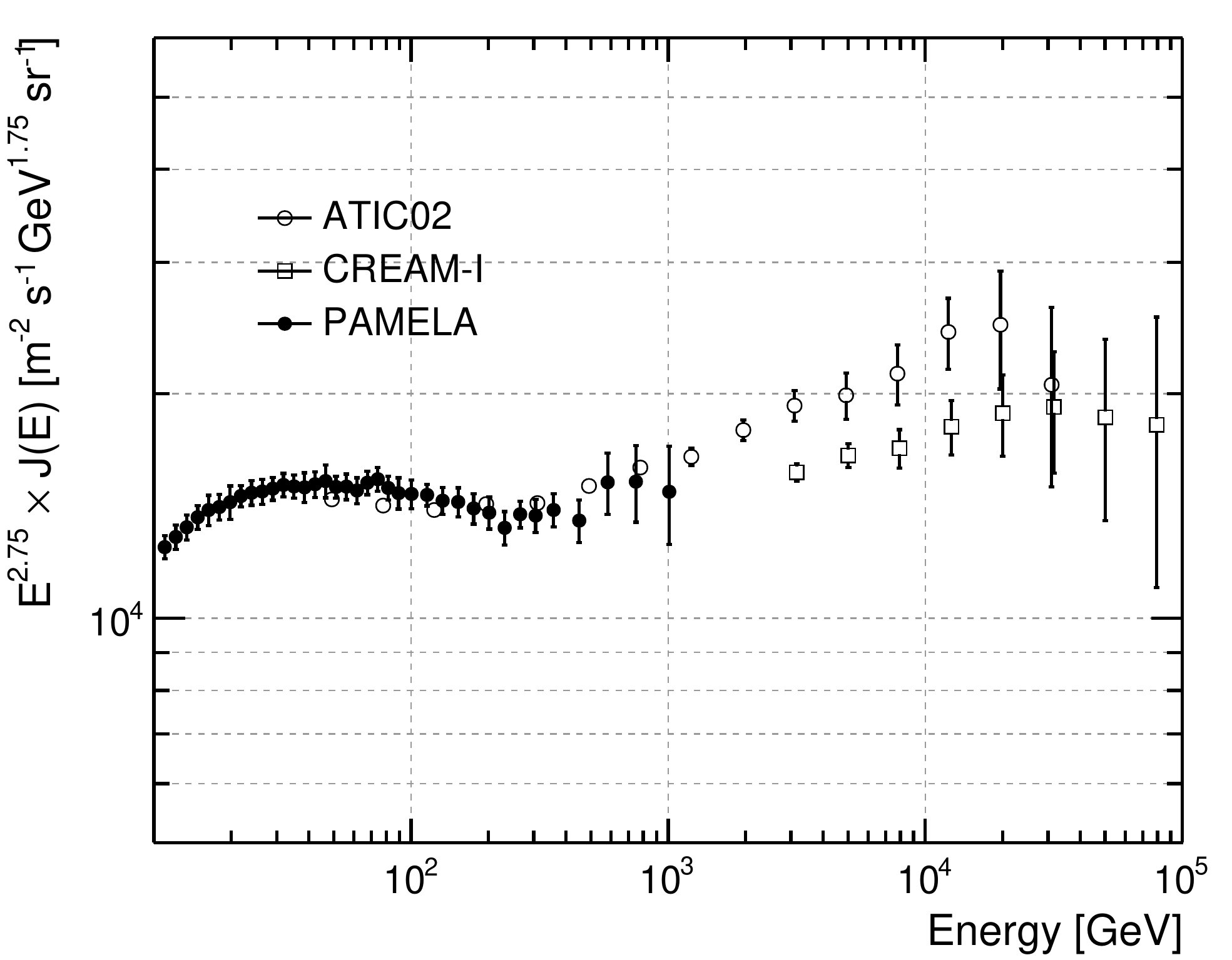}
  \caption{Compilation of the measurements of the high-energy proton spectrum
    by the PaMeLa~\cite{2011Sci...332...69A}, CREAM~\cite{2011ApJ...728..122Y}
    and ATIC~\cite{2009BRASP..73..564P} experiments.}
  \label{fig:cr_proton_spectrum}
\end{figure}

\section{Historical overview}%
\label{sec:history}

As we mentioned in the previous section, the discovery of cosmic rays is
customarily credited to Victor Hess for his balloon flights in the summer of
1912~\cite{HessDiscovery}. In fact, around the same time, several different
scientists were carrying out investigations on the penetrating radiation with
Wulf electroscopes, including Pacini~\cite{Pacini}, Gockel and Wulf himself.
By 1915 such instruments had been flown on balloons up to more than 8000~m,
measuring a level of radiation much larger than that recorded by Hess in his
first flight. The evidence for the extraterrestrial origin of the radiation was
compelling.

\begin{figure}[!htb]
  \includegraphics[width=\linewidth]{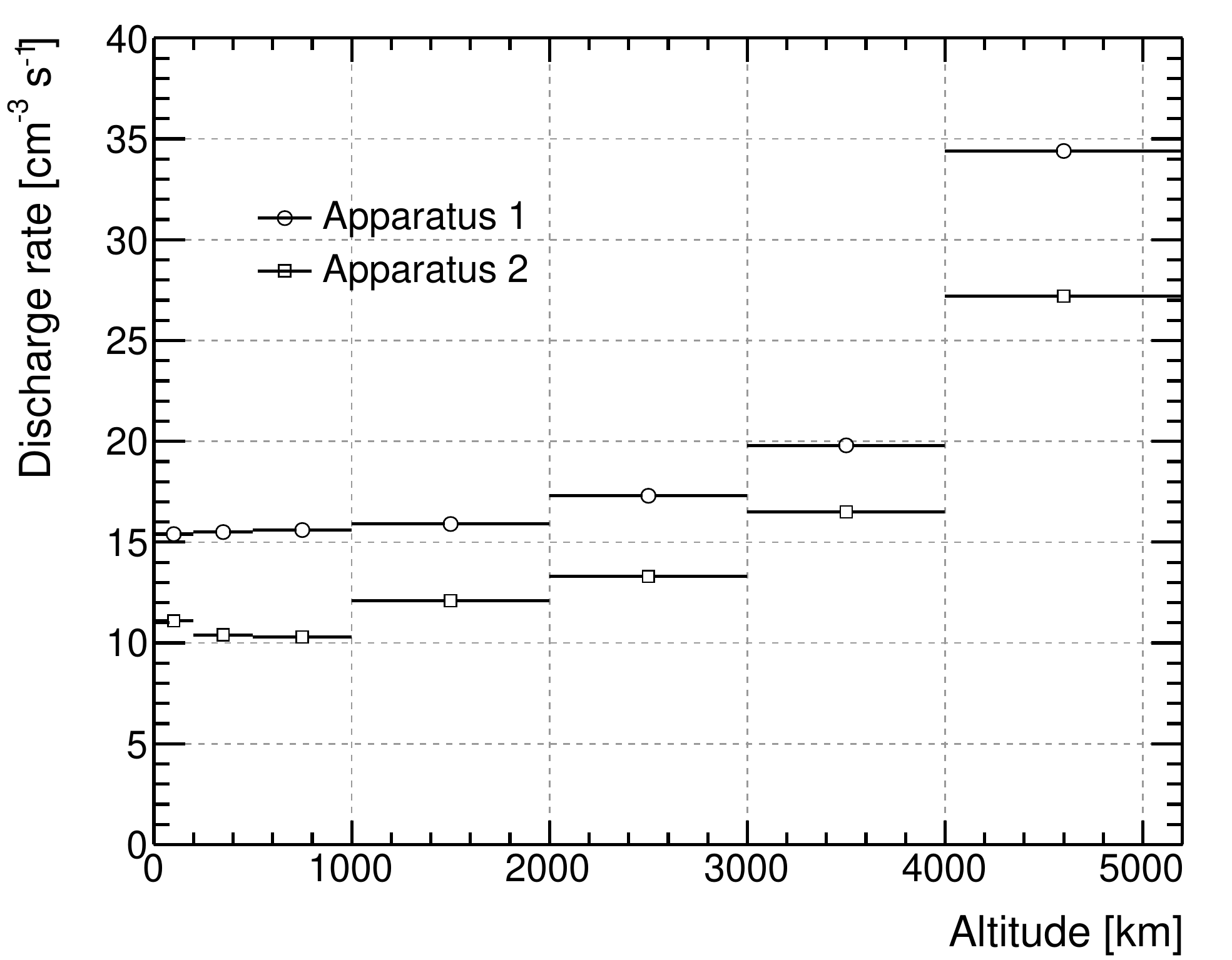}
  \caption{Original measurements from one of the ascents performed by
    Viktor Hess (adapted from~\cite{HessDiscovery}). The rate of discharge
    increasing with altitude implies that the ionizing radiation responsible
    for it is coming from the outer space. (On an un-related note, no plot is
    included in the original article, which is by no means less beautiful.)}
  \label{fig:hess_electroscope}
\end{figure}

\subsection{The early days}%
\label{sec:history_early}

A vibrant and enlightening (though not necessarily unbiased) account of the
first 50~years of research on cosmic rays is given by Bruno Rossi~\cite{Rossi}.
Interestingly enough, the question of the \emph{nature} of the cosmic radiation
did not get much attention until the end of the 1920s. The most striking
known feature of cosmic rays was their high penetrating power and, for the 
first 15 years after their discovery, scientists implicitly assumed that
they were gamma rays---the most penetrating radiation known at the time%
\footnote{While the term \emph{comic rays} was apparently coined by Millikan in
  the the 1920s, prior to 1930 the penetrating radiation was customarily
  referred to as \emph{ultragammastrahlung}, or ultra-gamma radiation, in the
  German literature.}.
This points to one of the most prominent difficulties that had to be faced
in the early studies of the cosmic radiation: the fact that little or nothing
was known about the physical interaction processes experienced by high-energy
photons and charged particles. The first satisfactory theory of
electromagnetic showers (see section~\ref{sec:em_showers}), due to Bethe and
Heitler, was published in 1934~\cite{1934RSPSA.146...83B}; before that, people
could only assume that high-energy gamma rays only interacted with matter via
Compton scattering, whose cross section has been shown to decrease with energy
by Dirac~\cite{1926RSPSA.111..405D} and Klein and
Nishina~\cite{1929ZPhy...52..853K}.

Robert Millikan formulated the first complete theory of comic rays, based on
all the measurements of attenuation in the atmosphere and in water available at
the time~\cite{1928SciAm.139..136M}. He proposed that the primary cosmic
radiation was composed of gamma rays of well-defined energies, produced in the
interstellar space by the the fusion of hydrogen atoms in heavier
elements---and that the charged particles observed in the Earth atmosphere were
electrons produced via Compton scattering.
Though we know, a posteriori, that this idea did not pass the test of time,
the original papers are still an interesting reading.

The begin of the post-electroscope era, around 1929, largely relies on a few
fundamental technical breakthroughs: the development of the Geiger-M\"uller
tubes, the first practical implementations of the coincidence technique,
introduced by Bothe and Kolh\"orster~\cite{1929ZPhy...56..751B} and refined by
Bruno Rossi~\cite{1930Natur.125..636R}, and the introduction of \emph{imaging}
devices such as the bubble chamber (and, later, the cloud chambers and the
stacks of photographic emulsions sensitive to single charged particles).

It was thanks to different clever arrangements of Geiger tubes in coincidence
and shielding materials that an incredible amount of new information about
cosmic rays was made available in the 1930s. It soon became clear that some of
the particles observed could pass through very noticeable amounts of material,
which casted serious doubts on the interpretation of the primary component as
consisting of photons.

At about the same time, physicists realized that the interaction between
radiation and matter was much more complicated that they had anticipated.
In 1933 Blackett and Occhialini~\cite{1933RSPSA.139..699B} published the
results of the first observations performed with a cloud chamber triggered by
Geiger-M\"uller tubes, clearly showing the copious production of secondary
radiation and the new phenomenon of the \emph{showers}.
This, in turn, forced scientists to focus the attention on
the \emph{genetic} relation between the primary and the secondary components
of the cosmic radiation, and to consider seriously the hypothesis that most of
the particles observed near the surface were actually produced in the
atmosphere.

The idea that the magnetic field of the Earth could be used to shed light
on the nature of the cosmic radiation, and establish unambiguously whether
primary cosmic rays were photons or charged particles occurred early on in the
1920s. It was clear that, in the second case, they would be somewhat channeled
along the field lines and one would expect a larger intensity at the magnetic
poles compared to the equator. Searches for this \emph{latitude effect} were
carried out as soon as 1927, but it is fair to say that in
1930, when Bruno Rossi started the first quantitative analysis of the
problem, evidence for an influence of the geomagnetic field on the intensity
of the cosmic radiation were far from being compelling (the latitude effect
was only established in the 1930s thanks to a monumental measurement campaign
led by A.~H.~Compton).
Building on top of the work by the Norwegian geophysicist Carl~St\"ormer,
Rossi set the stage for the ray-tracing techniques which are nowadays
customarily used to study the motion of charged particles in a magnetic field
(see section~\ref{sec:raytracing}).
He predicted that, if the primary cosmic rays were charged, and predominantly
of one sign (either positive or negative), one should observe an East-West flux
asymmetry, which would be maximal around the geomagnetic equator (see
section~\ref{sec:east_west_effect}).
In 1934 Rossi~\cite{1930PhRv...36..606R} and two other groups independently
measured this \emph{East-West effect}.
It was an incontrovertible evidence that primary cosmic rays are
charged---and, even more, the sign of the effect allowed to predict the
prevalent sign of their charge.
\emph{``The results of these experiments [\ldots] confirm the view,
  supported by the early experiments of the writer, that cosmic
  rays consist chiefly of a charged corpuscular radiation with a
  continuous energy spectrum extending to very great energies.
  Moreover the new results on the azimuthal effect show that the
  charge is predominantly positive.
  It is however possible that, in addition to the positive particles,
  a smaller amount of other kind of rays (negative particles,
  photons, neutrons) is contained in the cosmic radiation. In fact,
  some results are rather difficult to explain by supposing that the
  cosmic radiation consists merely of positive particles.''}
These brief except from~\cite{1930PhRv...36..606R}, written in 1934, still
constitutes a substantially correct description of our current understanding of
cosmic rays%
\footnote{One should be careful, however, in not trying and read
  \emph{too much} in these few sentences. At the time the relation between the
  primary and secondary components of the cosmic radiation was far from being
  completely understood and the correctness of the conclusions rest on
  the fact that the products of the interactions with the atmosphere
  retain much of the angular information of the primaries.}.

On a slightly different note, we should emphasize that we deliberately left
out from this short summary at least two fundamental items. The first is the
deep connection between cosmic rays and the early stages of development of
particle physics, with the positron~\cite{1933PhRv...43..491A},
muon~\cite{1937PhRv...51..884N,1937PhRv...52.1003S} and
pion~\cite{1947Natur.160..453L} all being discovered in the
cosmic radiation. The other is the discovery of extensive air showers
\cite{1939RvMP...11..288A}, which originated an independent (and incredibly
prolific) line of research---that of the study of ultra-high-energy cosmic
rays from the ground.

All this said, it is fair to say that between 1940 and 1950 a complete and
coherent picture of the phenomena connected with the cosmic radiation
emerged, with most of the primary cosmic rays being protons and nuclei of
heavier elements and most of the particles observed near to the surface being
secondary products of their interaction with the atmosphere. As we shall see in
a second, the development of stratospheric balloons and the beginning of the
space age allowed to customarily observe the primary radiation at the
top of the atmosphere in the following decades.

\subsection{The latter days}

The launch of the Sputnik~I artificial satellite by the Soviet Union on October
4, 1975 signals the begin of the space era. Within six months the Sputnik~II
soviet satellite and the Explorer~I and Explorer~III American satellites
were launched---all the three of them were equipped with Geiger-M\"uller
counters with the aim of mapping the comic-ray intensity beyond the altitudes
reachable by balloons (the Explorer~I and Explorer~III were placed into an
elliptical orbit reaching out to some 2500~km).
One of the most interesting discoveries was that of the Van Allen
belts~\cite{1959JGR....64.1683V}---regions around the Earth where low-energy
charged particles trapped in the geomagnetic field make the cosmic-ray
intensity several orders of magnitude more intense than that on the surface.

As it turns out, the radiation belts are not directly relevant for instruments
in low-Earth orbit (which is the main topic of this write-up), with the notable 
exception of the \emph{South Atlantic Anomaly} that we shall briefly introduce
in section~\ref{sec:saa}. We shall glance through the basics of geomagnetically
trapped radiation (which, in general, \emph{is} relevant for the low-Earth
orbit environment) in section~\ref{sec:trapped_radiation}.

\subsubsection{Charged cosmic rays}

It is fair to say that, among the first \emph{modern} instruments for charged
cosmic-ray measurements are the pioneering magnetic spectrometers flown on
balloons in the 1960s and 1970s for the study of the positron
\cite{1969ApJ...158..771F,1975ApJ...198..493D,1974PhRvL..33...34B} and the
antiproton~\cite{1979ICRC....1..330B,1984ApL....24...75G} components of the
cosmic radiation. They generally had limited (at least by any modern standard)
energy range and particle identification capabilities---typically provided by a
\cheren\ detector.

On a related note, it is somewhat amusing to note how back in the 1980s the
apparent increase of the positron fraction above $\sim 5$~GeV that these early
measurements seemed to indicate was actively discussed, and both pulsars and
dark matter annihilation were already proposed as viable candidates for its
origin~\cite{1989ApJ...342..807B}. And then the history repeated itself two
decades later (this time for real) with the measurement published by the
PaMeLa collaboration~\cite{2009Natur.458..607A}.

At about the same time emulsion
chambers~\cite{2012ApJ...760..146K,2006JPhCS..47...31C} and calorimetric
experiments~\cite{1975ApJ...197..219M,1977PhRvL..38.1368H,1984ApJ...278..881T}
were used for the measurement of the all-electron and the proton and nuclei
spectra.

WiZard~\cite{1990NCimB.105..191G} was the name of the magnetic spectrometer
concept selected in the late 1980s for the Astromag facility planned to operate
on the U.S. Space Station Freedom---which never saw the light as originally
conceived and later evolved into the International Space Station. As this
project was abandoned, the WIZARD collaboration~\cite{2003NuPhS.122...66S}
started a long and incredibly successful campaign of balloon-borne
experiments---including MASS89, MASS91, TS93, Caprice94, Caprice97 and
Caprice98---which finally winded up in the PaMeLa space-based magnetic
spectrometer, currently in operation.
HEAT and BESS (see section~\ref{sec:unconventional_detectors}) constitute
two additional notable examples of magnetic spectrometers flown around the same
time. Generally speaking, the 1990s signal a dramatic leap forward in the
particle identification capabilities of the instruments, with transition
radiation detectors and advanced \cheren\ detectors effectively exploited, and
modern imaging calorimeters providing shower-topology information in addition
to the basic energy measurement.

At this point we are straight into the present, with the AMS-02 magnetic
spectrometer operating on the International Space Station---preceded by the
Shuttle flight of the path-finder AMS-01 in 1998---and, on the calorimetric
side of the panorama, the \emph{fabulous four} balloon-borne detectors: ATIC,
CREAM, TIGER and TRACER. One a related note, the record-breaking 161 days of
exposure integrated by CREAM in its six flights signal the exciting perspectives
nowadays made available by the development for Long Duration (LD) circumpolar
balloon flights.

\subsubsection{Gamma rays}%
\label{sec:history_latter_gamma}

The Explorer XI satellite, launched in 1961, carried a gamma-ray detector
on board, consisting of a crystal scintillator and a \cheren\ counter,
surrounded by an anti-coincidence shield~\cite{1962PhRvL...8..106K}.
While the satellite could not be actively pointed and the photon direction was
loosely determined by the solid angle defined by the geometry of the telescope,
Explorer XI performed the first observation of the gamma-ray sky and
in 1962 the beginning of gamma-ray astronomy was announced to the world on
Scientific American:
\emph{``An ingenious telescope in a satellite has provided the
  first view of the Universe at the shortest wavelength of the electromagnetic
  spectrum. This historic glimpse is supplied by just 22 gamma
  rays.''}~\cite{1962SciAm.206e..52K}.

\begin{table}[htb!]
    \begin{tabular}{p{0.20\linewidth}p{0.28\linewidth}p{0.20\linewidth}%
        p{0.25\linewidth}}
      \hline
      Experiment & Energy range & Date & $\gamma$ candidates\\
      \hline
      \hline
      Explorer XI & $>50$~MeV & 1961 & 22\\
      OSO-3 & $>50$~MeV & 1967--1968 & 621\\
      SAS-2 & 20~MeV--1~GeV & 1972--1973 & 13,000\\
      COS-B & 30~MeV--3~GeV & 1975--1983 & 200,000\\
      EGRET & 30~MeV--10~GeV & 1991--1999 & 1,500,000\\
      \Fermi-LAT & 20~MeV--$>1$~TeV & 2008--?? & 500,000,000\\
      \hline
    \end{tabular}
    \caption{Total number of gamma-ray candidates collected by specific
      space-born detectors through their missions. Note that the energy
      ranges for the various experiments are purely indicative.}
    \label{tab:gamma_stat}
\end{table}

A similar detector concept was flown in 1967 on board the third Orbiting Space
Observatory (OSO-3). OSO-3 operated continuously for 16 months (at which
point the last spacecraft tape recorder failed), performing a complete sky
survey and recording 621 photons above 50 MeV. Most notably, the experiment
demonstrated that celestial gamma-rays are anisotropically
distributed---concentrated in the direction of the galactic plane and,
particularly, toward the galactic center.

In 1969 and 1970 the array of military satellites VELA, launched by the United
States to monitor possible nuclear experiments carried out by the Soviet Union,
serendipitously discovered the transient flashes of gamma radiation
that generally go under the name of gamma-ray bursts (GRB).

It is generally acknowledged that SAS-II~\cite{1972NucIM..98..557D}, launched
on November 1972%
\footnote{Unfortunately a failure of the low voltage power supply
stopped the data collection on June 1973.},
provided the first detailed information about the gamma ray sky and effectively
demonstrated the ultimate promise of gamma-ray astronomy, showing that the
galactic plane radiation was strongly correlated with the galactic structural
features---not event mentioning the first detection of gamma-ray point sources,
most notably the Vela and Crab pulsars.
SAS-II was the first satellite entirely devoted to gamma-ray astrophysics, with
a gamma-ray telescope on board composed by spark chambers interleaved with
tungsten conversion foils%
\footnote{The energy information was (loosely) derived by the multiple
  scattering, measured by means of the tracking detectors.}
, and an anti-coincidence system featuring a set of plastic scintillator tiles
and directional \cheren\ detectors placed below the spark chambers. With a peak
effective area of $\sim 120$~cm$^2$ and a PSF of the order of a few degrees,
the gamma-ray detector on board SAS-II is effectively one of the first
incarnations of the pair conversion telescope concept (see
section~\ref{sec:pair_conversion_telescopes}).

In 1975 the European Space Agency launched the COS-B~\cite{1975SSI.....1..245B}
satellite, that operated successfully for 6 years and 8 months---well beyond
the original goal of two years.
The gamma-ray telescope on board COS-B was conceived following the heritage
of that on SAS-II, with the crucial addition of a $4.7~X_0$ calorimeter to
improve the energy measurement. It was sensitive to photons between 30~MeV and
several~GeV over a field of view of almost 2~sr, with a peak effective area of
some 50~cm$^2$.
Among the key science results from the COS-B mission are the first catalog of
gamma-ray sources (including 25 entries) and a complete map of the disc of the
milky way.

With a weight of approximately 17~tons, the Compton Gamma Ray Observatory
(CGRO), launched by NASA in 1991, is possibly the heaviest scientific payload
ever flown in low-Earth orbit. The Energetic Gamma Ray Experiment Telescope
(EGRET~\cite{1988SSRv...49...69K}) on board CGRO, a pair conversion telescope
with far superior sensitivity than any of its predecessors, made the first
complete survey of the gamma-ray sky in the energy range between 30~MeV and
$\sim 10$~GeV (detecting 271 discrete sources) and is at the base of the
last-generation instruments exploiting the silicon-strip technology such as
AGILE and the \Fermi-LAT~\cite{2009ApJ...697.1071A}.

\begin{figure}[htb!]
  \includegraphics[width=\linewidth]{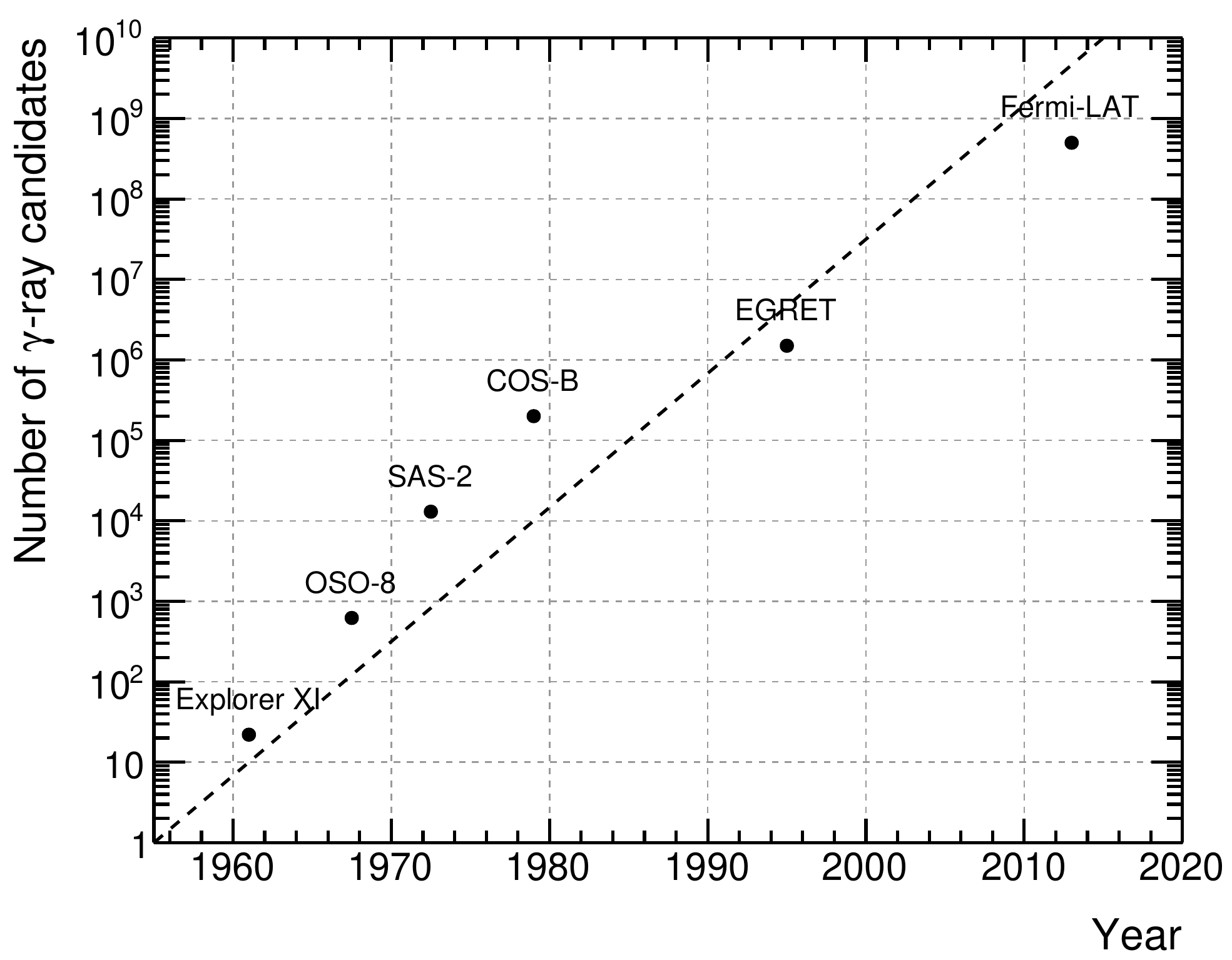}
  \caption{Total number of gamma-ray candidates collected by specific
    space-born detectors through their missions. The points on the $x$-axis
    are the average values between the mission start and stop times.}
  \label{fig:gamma_exp_stat}
\end{figure}

Table~\ref{tab:gamma_stat} summarizes the number of gamma-ray candidates
collected by the gamma-ray detectors listed in this section through the
duration of the corresponding mission. While the numbers surely give a sense
of the continuous advance in performance, there are good reasons (primarily
geometrical dimensions and weight) to presume that it is going to be hard to
keep up with this sort of Moore's law (see figure~\ref{fig:gamma_exp_stat})
over the next few years.

\subsection{Instrument concepts}

\begin{figure*}[htbp]
  \includegraphics[width=\textwidth]{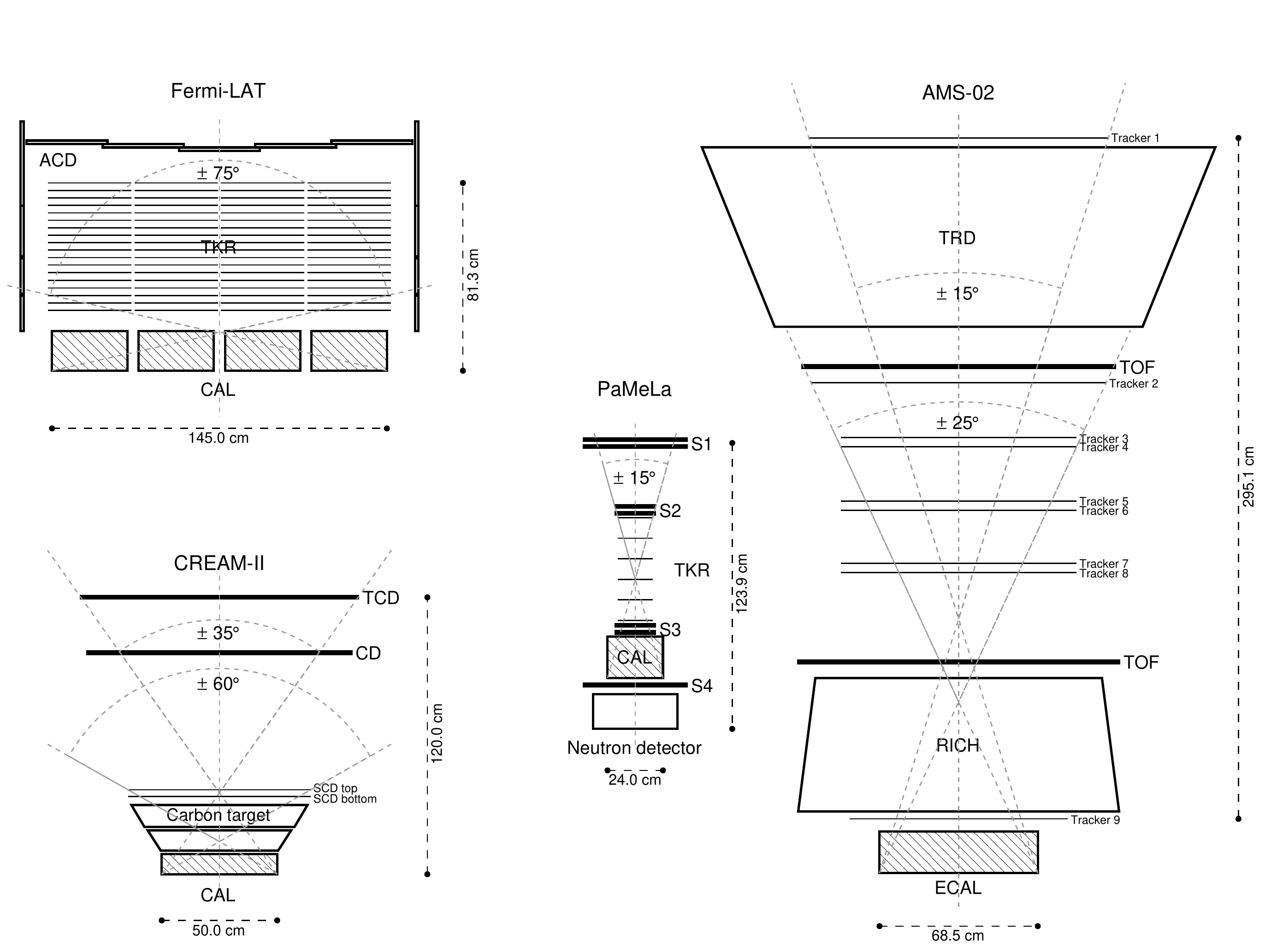}
  \caption{Schematic view of some of the most recent space- or balloon-borne
    cosmic-ray and gamma-ray detectors (all the dimensions are meant to be
    in scale). Only the detector subsystems are sketched (note that the magnets
    for PaMeLa and AMS-02 are not represented). For the \Fermi-LAT: the
    anti-coincidence detector (ACD), the tracker (TKR) and the calorimeter
    (CAL). For AMS-02: the transition radiation detector (TRD), the time of
    flight (TOF), the tracker layers, the electromagnetic calorimeter (ECAL)
    and the ring imaging \cheren\ detector (RICH). For CREAM-II: the
    timing charge detector (TCD), the \cheren\ detector (CD), the silicon
    charge detector (SCD) and the calorimeter (CAL). For PaMeLa: the time of
    flight scintillators (S1--S3), the tracker (TKR), the calorimeter (CAL)
    and the neutron detector. For each instrument the relevant combination
    of sub-detectors determining whether a given event is reconstructable
    (and hence the field of view) is also sketched. Trivial as it is, we also
    note that the calorimeters are made of different material and therefore
    the relative thickness does not necessarily reflect the depth in terms
    of radiation lengths (this information is included in
    table~\ref{tab:pres_fut_instruments}).}
  \label{fig:detectors}
\end{figure*}

\begin{table*}[htb!]
  \begin{tabular*}{\textwidth}{l @{\extracolsep{\fill}} ccccccr}
    \hline
    Experiment &
    \multicolumn{3}{c}{Peak~\accept~[m$^2$~sr]} &
    \obstime~[year]
    & \multicolumn{2}{c}{$\sigma E/E$} &
    References\\
    & $e^\pm$ & $\gamma$ & $p$/nuclei &  & $e^\pm$, $\gamma$ & $p$/nuclei & \\
    \hline
    \hline
    Agile & -- & 0.1 & -- & $>7$ &
    $\sim 100\%$ \@ 400~MeV & -- &
    \cite{2009AA...502..995T}\\

    AMS-02 & \multicolumn{3}{c}{0.05} & 20 &
    2\% @ 50 GeV & -- & 
    \cite{Brient:2013hsa,2013NuPhS.243...12T}\\

    ATIC & \multicolumn{3}{c}{0.24} & 0.15 &
    2\% \@ 150 GeV & 35\% &
    \cite{2005NIMPA.552..409G}\\

    CREAM & -- & -- & 0.43 & 0.5 &
    6\% \@ 200~GeV & 40\% &
    \cite{2011ApJ...728..122Y}\\

    Fermi &
    2.8 @ 50~GeV\footnote{For reference, the acceptance for
      $e^\pm$ @ 1 TeV is $\sim 0.9$~m$^2$~sr\cite{2010PhRvD..82i2004A}.} &
    2.0 @ 10~GeV & -- & 10 &
    5--15\% & -- &
    \cite{2012ApJS..203....4A,2010PhRvD..82i2004A} \\
    
    PaMeLa &
    0.00215 & -- & 0.00215 & 7 &
    5--10\% & -- &
    \cite{2007APh....27..296P}\\

    TRACER & -- & -- & 4.73\footnote{Before the selection cuts.}
    & 0.05 & -- & See~\cite{2011ApJ...742...14O} &  \cite{2011ApJ...742...14O}\\

    \hline
    \hline

    CALET &
    \multicolumn{3}{c}{0.12} & 5 &
    2\% @ 1~TeV & 40\% @ 1 TeV &
    \cite{2013NuPhS.239..199M}\\

    DAMPE &
    0.3 & 0.2 & 0.2 & 3 &
    1.5\% @ 800~GeV & 40\% @ 800~GeV &
    --\\

    Gamma-400%
    \footnote{\parbox[t]{\textwidth}{The high-energy angular resolution for
        gamma rays ($\sim 0.01^\circ$) is one of the salient aspects of the
        instrument design.\hfill\phantom{a}}}
    & \multicolumn{3}{c}{0.5} & 7 &
    1\% @ 10 GeV & -- &
    \cite{2013AIPC.1516..288G}\\

    Gamma-400 (CC%
    \footnote{\parbox{\textwidth}{Alternative design including the large-\fov\
        CALOCUBE calorimeter concept.\hfill\phantom{a}}}) &
    \multicolumn{2}{c}{3.4 @ 1 TeV} & 3.9 @ 1 TeV & 7 &
    2\% @ 1 TeV & 35\% @ 1 TeV\footnote{As good as 15\% when exploiting
    a dual readout.} &
    --\\

    HERD &
    \multicolumn{2}{c}{$>3$} & $>2$ & 10 &
    1\% @ 100 GeV & 20\% @ 1 TeV &
    \cite{2014arXiv1407.4866Z}\\
    \hline
  \end{tabular*}
  \caption{Summary table with the main characteristics of some of the
    cosmic-ray and gamma-ray instruments that have started operation after the
    year 2000. Note that, despite the effort that has been put in extracting
    the numbers from the available literature, it is likely that some of them
    will be slightly incorrect.
    Although the topic is is not discussed (yet) in this write-up,
    the bottom part of the table does include some of the most prominent
    instrument that are now in the design or implementation phase (the reader
    is advised that that corresponding figures are subject to change without
    notice).}
  \label{tab:pres_fut_instruments}
\end{table*}

The brief summary of the history of cosmic-ray measurements we have outlined
in the previous two sections is also an illustration of the basic detector
concepts and experimental techniques that have been exploited over the last
$\sim 50$~year. The dichotomy between magnetic spectrometers and calorimetric
experiments is a fundamental one and we shall elaborate a little bit more
on it in the next section. We postpone a (much) more in-depth discussion of
some related technical aspects to sections~\ref{sec:techniques} and
\ref{sec:irfs}. Schematic views of a few actual instruments, either recent or in
operation, are shown (in scale!) in figure~\ref{fig:detectors}.

\subsubsection{Spectrometers and calorimeters}

Most modern comic-ray detectors fall in either of the two categories:
magnetic spectrometers or calorimetric experiments. Strictly speaking all
the advanced magnetic spectrometers feature an electromagnetic calorimeter
for energy measurement, so the basic difference between the two is really the
presence of the magnet.

The key feature of magnetic spectrometers is their ability of distinguishing
the charge sign---e.g., separating electrons and positrons or protons and
antiprotons. In addition, they are typically equipped with additional
sub-detectors (\cheren\ detectors and/or transition radiation detectors) aimed
at particle identification, e.g., for isotopical composition studies (see
section~\ref{sec:tof}). This all comes at a cost, in that the magnet is a
passive element (and typically a heavy one) contributing to the mass budget and
limiting the field of view (not even mentioning the potential issue of the
background of secondary particles). Both effect conspire to make the acceptance
of this kind of instrument relatively smaller.

Calorimetric experiments, on the other hand, typically feature a larger
acceptance and energy reach---and are best suited for measuring, e.g., the
inclusive $e^+ + e^-$ spectrum or the proton and nuclei spectra up to the
highest energies---but cannot readily separate charges (see, however,
section~\ref{sec:geomeg_charge_sep} for yet another twist to the story).
Modern electromagnetic imaging calorimeters, be they homogeneous or sampling,
provide excellent electron/hadron discrimination and are typically instrumented
with some kind of external active layer for the measurement of the absolute
value of the charge (e.g., to distinguish between singly-charged particles and
heavier nuclei). Since flying an accelerator-type hadronic calorimeter in space
is impractical due to mass constraints (see section~\ref{sec:had_cal}),
a fashionable alternative to measure the energy for hadrons is that of 
exploiting a passive low-Z target (as done, e.g., in the CREAM and ATIC
detectors) to promote a nuclear interaction and then recover the energy from
the electromagnetic component of the shower.

\subsubsection{Pair-conversion telescopes}%
\label{sec:pair_conversion_telescopes}

In the basic scheme laid out in the previous section gamma-ray pair conversion
telescopes are essentially calorimetric experiments featuring a dedicated
tracker-converter stage in which foils of high-$Z$ materials are interleaved
with position sensitive detection planes. The basic detection principle is
easy: the conversion foils serve the purpose to promote the conversion of
high-energy (say above $\sim 20$~MeV) gamma rays into an electron-positron pair
which is in turn tracked to recover the original photon direction.
The pair is then absorbed into the calorimeter for the measurement of the
gamma-ray energy. Last but not least, pair conversion telescopes feature some
kind of anti-coincidence detector for the rejection of the charged-particle
background that, as we have seen, outnumbers the signal by several orders of
magnitude in typical low-Earth orbit.

As mentioned in section~\ref{sec:history_latter_gamma},
EGRET~\cite{1988SSRv...49...69K} on-board the CGRO mission, AGILE and the
\Fermi-LAT~\cite{2009ApJ...697.1071A} are prototypical examples of
pair-conversion telescopes.

\subsubsection{Unconventional (or just old-fashioned) implementations}%
\label{sec:unconventional_detectors}

While most of the detectors we shall consider in the following are built around
either a magnetic spectrometer \emph{a la}~AMS-02 or an electromagnetic
calorimeter, there exist less conventional implementations that deserve to be
briefly mentioned, here.

Among the magnetic spectrometers, BESS (see, e.g., \cite{2002AdSpR..30.1253Y})
is a notable example where the instruments features a thin superconducting
solenoid magnet enabling a large geometrical acceptance with a horizontally
cylindrical configuration (note that in this case the particles go
\emph{through} the magnet!). Several different versions of the instrument
underwent a long and very successful campaign of balloon flight for the
measurement (and monitoring in time) of the
antiproton~\cite{2008PhLB..670..103B} and proton and helium
spectra~\cite{2000ApJ...545.1135S}.

With its unrivaled imaging granularity, the emulsion chamber technique
played a prominent role in the early days of calorimetric experiments, as it
readily allowed to assemble large---and yet relatively simple---detectors.
In this case it is the data analysis process that is significantly more
difficult to scale up with the available statistics with respect to that of
modern \emph{digital} detectors. Reference~\cite{2012ApJ...760..146K}, e.g,
contains a somewhat detailed summary of more than 30~years of observations of
high-energy cosmic-ray electrons---in several balloon flights from 1968 to
2001---with a detector setup consisting of a stack of nuclear emulsion plates,
X-ray films, and lead/tungsten plates. Electromagnetic showers were detected by
a naked-eye scan of the X-ray films and energies were determined by counting
the number of shower tracks in each emulsion plate within a 100~$\mu$m wide
cone around the shower axis. At even higher energies, a nice account of the
emulsion chambers (e.g., JACEE~\cite{1986NIMPA.251..583B} and
RUNJOB~\cite{2005ApJ...628L..41D}) flown on balloons with the aim of measuring
the cosmic-ray chemical composition near the \emph{knee} is given
in~\cite{2006JPhCS..47...31C}.

\begin{figure}[!hbt]
  \includegraphics[width=\linewidth]{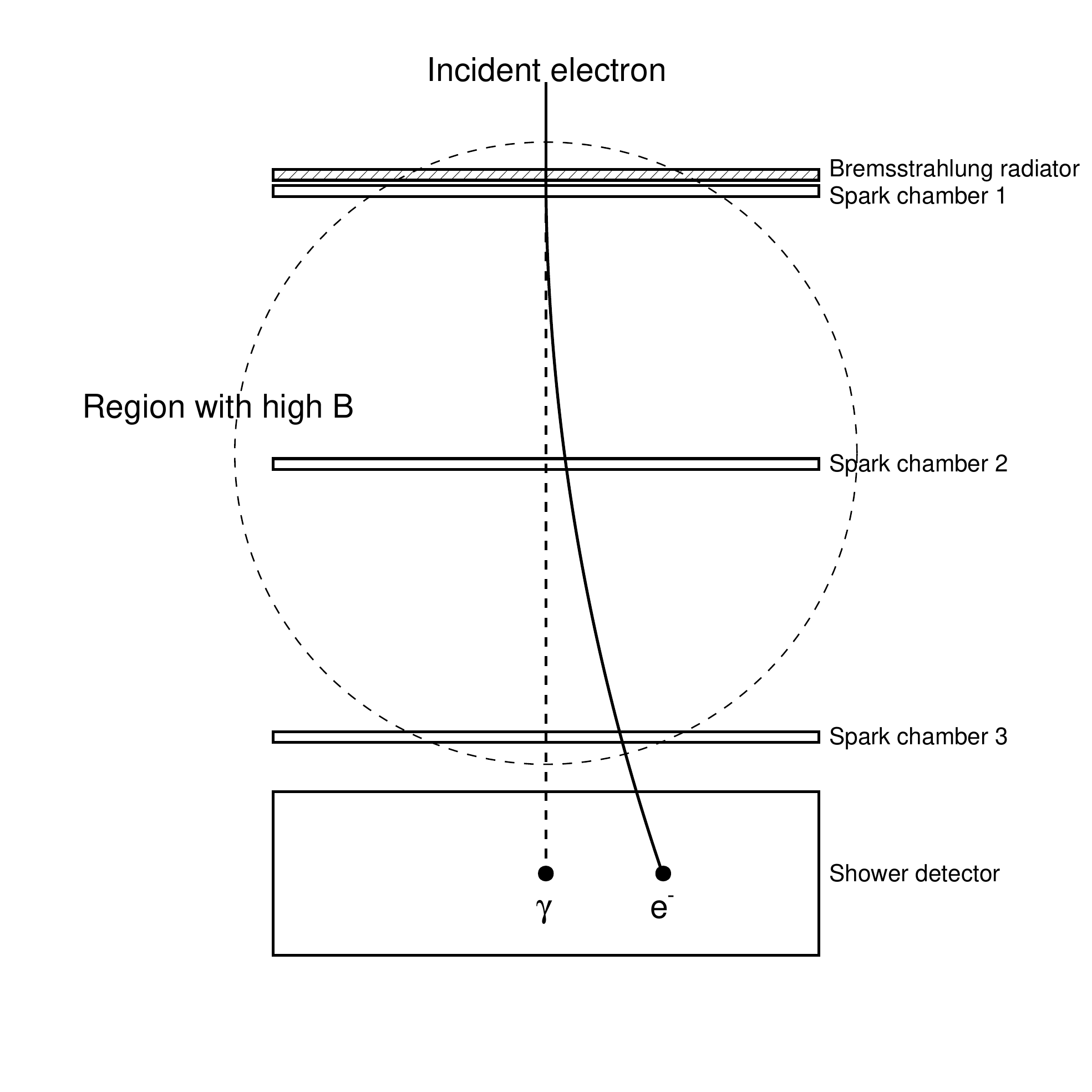}
  \caption{Schematic diagram of a magnetic spectrometer equipped for the
    \bremss-identification technique (adapted from~\cite{1974PhRvL..33...34B}).}
  \label{fig:bremss_id}
\end{figure}

We close this section by mentioning the \bremss-identification technique,
used, e.g., in ~\cite{1974PhRvL..33...34B}, which strikes the author as one of
the neatest and most clever attempts to overcome the limited particle
identification capabilities of the early detectors.
The basic idea, illustrated in figure~\ref{fig:bremss_id}, is that of using
a thin radiator and select the electrons and positrons producing
a \bremss\ photon---identified as an additional individual shower in the
\emph{calorimeter} (or really, in the \emph{shower detector}). Though the
overall fraction of useful signal events is somewhat reduced, this detector
concept provides a clear and distinct signature (that is very hard to mimic
for heavy particles) allowing proton rejection factors of the order of $10^5$.

\subsubsection{Different instrument concepts}

All the cosmic-ray detectors we have mentioned so far exploit nuclear
interactions to measure the energy of protons and heavier nuclei. In contrast
to calorimetric measurements, it is possible, at least in principle, to
measure the cosmic-ray charge and energy through their electromagnetic
interaction only. The main advantage is that this concept makes it possible
to realize instruments with very large geometrical aperture (of the order of
several m$^2$~sr before selection cuts), as only low-density material are
necessary. On the other hand, the main drawback is that the quality of the
energy measurement is largely non uniform across the energy range covered---and
poor-to-non existent in some significant portions of the phase space.

\begin{figure}[!hbt]
  \includegraphics[width=\linewidth]{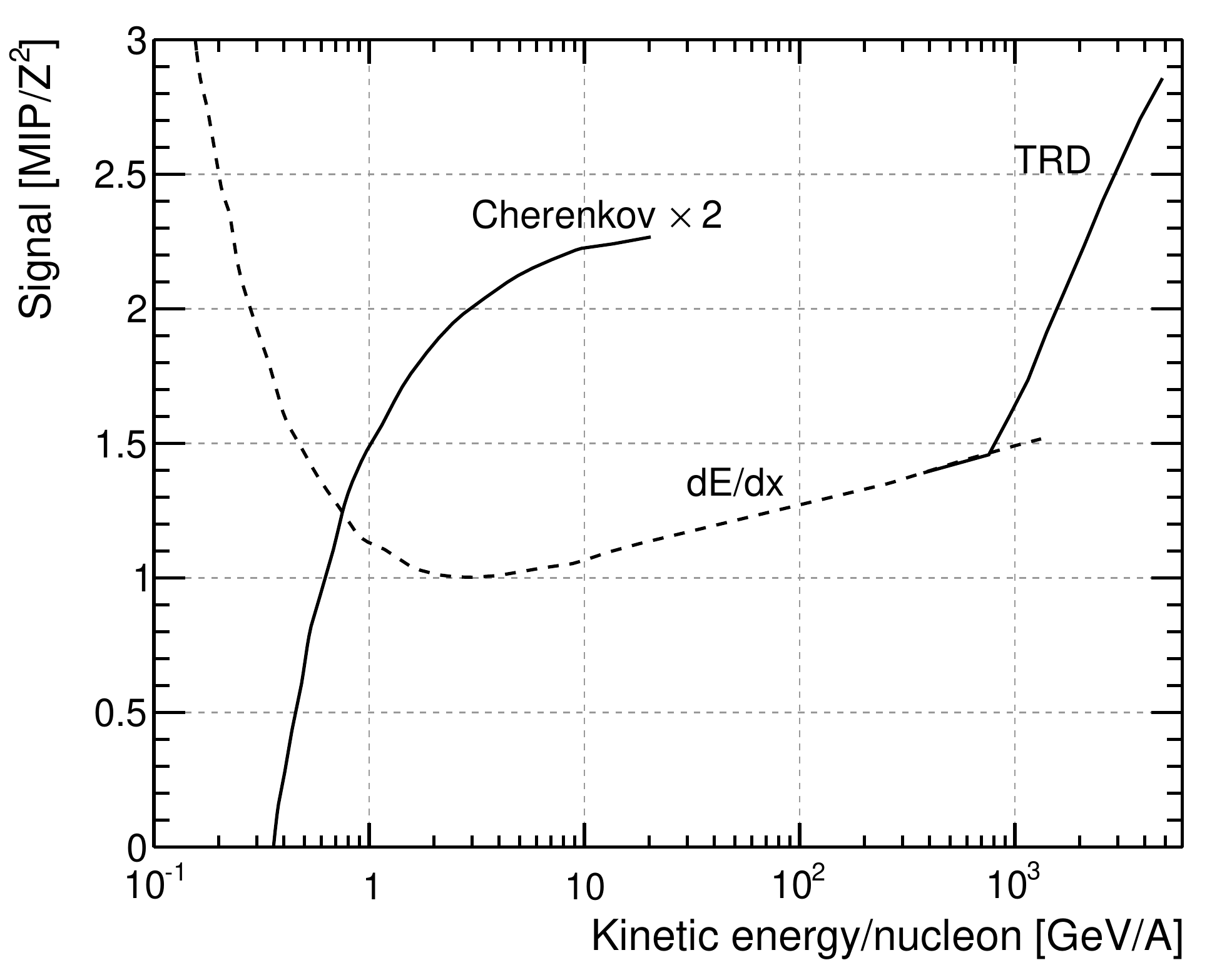}
  \caption{Response functions, normalized by $Z^2$ for the three basic
    TRACER sub-detectors (adapted from~\cite{2011ApJ...742...14O}).
    Much more details about the basic physical interaction processes and
    how these can be exploited in actual particle detectors can be found in
    sections~\ref{sec:interactions} and~\ref{sec:techniques}.}
  \label{fig:tracer_signal}
\end{figure}

The TRACER balloon-borne detector~\cite{2011ApJ...742...14O} is possibly the
most notable example of this type of instrument concept. By a clever combination
of scintillators, \cheren\ detectors, transition radiation detectors and a
$dE/dx$ array (see sections~\ref{sec:interactions} and~\ref{sec:techniques}
for more details), TRACER was able to measure the charge and the energy of
cosmic-ray nuclei with $Z > 5$ between $\sim 1$~GeV and a few~TeV.
Figure~\ref{fig:tracer_signal} shows the response functions, normalized by
$Z^2$, for the three basic TRACER sub-detectors: the charge is measured by the
$dE/dx$ array---with the \cheren\ detectors breaking the degeneracy between the
two parts of the curve on the two sides of the minimum---while the energy is
recovered from the \cheren\ detectors near the \cheren\ threshold, from the
$dE/dx$ array in the intermediate range (admittedly with limited resolution),
and from the TRD at very high energy.

On a completely different subject, we briefly mention the
\emph{Compton telescope} concept, which constitutes one of the possible
experimental approaches for studying gamma rays below $\sim 20$~MeV---where
Compton scattering is the interaction physical process with largest cross
section.
The basic idea is that of measuring both the (first) Compton interaction
point through the scattered electron and the direction of the scattered
photon by absorbing it. This, in turn, allows to kinematically constraint the
direction in the sky of the original gamma-ray in the so called
Compton cone. As it turns out, any practical implementation of this seemingly
simple concept is quite challenging, for many different reasons.
It is not by coincidence that the last Compton telescope flow in space was
COMPTEL~\cite{1993ApJS...86..657S} on-board the CGRO, totaling $\sim 50$~cm$^2$
on-axis effective area---a somewhat meager figure when considering that the
total weight of the instrument was of the order of 1~ton.

\section{The near-Earth environment}%
\label{sec:environment}

This section deals with the basic characteristics of the environment in
which balloon and/or satellites in low-Earth orbit operate. 
There are two main aspects of the problem, namely the atmosphere and the
geomagnetic field.

\subsection{The atmosphere of the Earth}%
\label{sec:earth_atmoshere}

The simplest possible model for the Earth's atmosphere is that of an isothermal
gas in hydrostatic equilibrium. Under the assumption that the magnitude of the
gravitational field of the Earth does not change significantly with the
altitude $h$%
\footnote{This is a reasonable assumption as long as the altitude $h$ is much
  smaller that the radius of the Earth, i.e. surely for the stratosphere, as
  we shall see in a second.}%
, the problem can be analytically integrated and the result is
that the pressure follows a simple barometric profile
\begin{align}\label{eq:atmospheric_pressure}
  p(h) = p_0 \exp\left(-\frac{h}{h_0}\right),
\end{align}
with a pressure at sea level $p_0 \sim 101.325$~kPa.
It goes without saying that, in this na\"ive model, the density follows the
very same scale law:
\begin{align}\label{eq:atmospheric_density}
  \density(h) = \density_0 \exp\left(-\frac{h}{h_0}\right),
\end{align}
with a density at sea level of $\density_0 \sim 1.24 \times 10^{-3}$~g~cm$^{-3}$.
One can actually show that the isothermal scale height at a given temperature
$T$ is related to the basic physical properties of the atmosphere by the simple
relation
\begin{align}\label{eq:isothermal_scale_height}
  h_0 = \frac{RT}{g\mu},
\end{align}
where $R$ is the ideal gas constant, $g$ is the gravitational acceleration
and $\mu$ is the average molecular mass. By plugging the numbers into
equation~\eqref{eq:isothermal_scale_height}, one obtains a scale height of
$h_0 \sim 7.9$~km at $270^\circ$K (dry air contains about 78\% of N$_2$,
21\% of O$_2$ and small amounts of other gases, for an average molecular
mass $\mu \sim 29$).

\begin{figure}[htb!]
  \includegraphics[width=\linewidth]{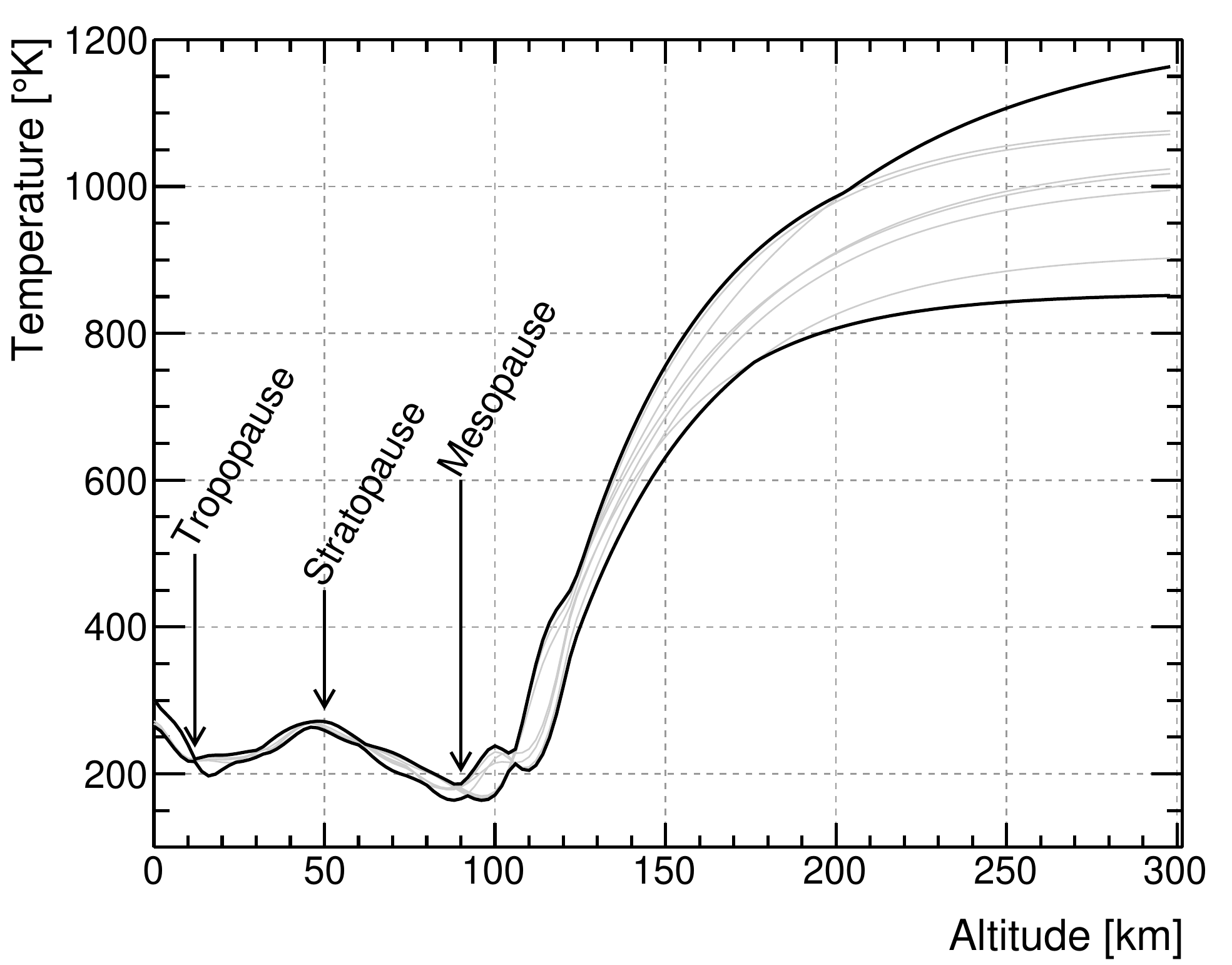}
  \caption{Illustrative atmospheric temperature profiles provided by the
    NRLMSISE-00 model~\cite{JGRA:JGRA16630}. The thin gray lines represent the
    model values for 8 reference points on the Earth's surface (4 on the
    equator and 4 at $60^\circ$ latitude), while the black thick lines represent
    the envelope of such models.
    Data are taken from~\url{http://ccmc.gsfc.nasa.gov/modelweb/models/nrlmsise00.php}.}
  \label{fig:atmospheric_temperature}
\end{figure}

Unfortunately life is not that easy, as the temperature profile of the Earth
atmosphere is a complicated function of the altitude, with its different
regimes defining the standard atmospheric layers: the \emph{troposphere}
($0$--$12$~km), the \emph{stratosphere} ($12$--$50$~km), the \emph{mesosphere}
($50$--$90$~km) and at even larger altitudes, the \emph{thermosphere} and the
\emph{exosphere}. Figure~\ref{fig:atmospheric_temperature} shows some
illustrative temperature profiles, as given by the
NRLMSISE-00~\cite{JGRA:JGRA16630} empirical atmospheric model, for different
locations on the surface of the Earth.
One of the most striking features of such profiles is the temperature increase
above $\sim 100$~km, which is essentially due to the absorption of ultraviolet
radiation from the Sun. Before the reader starts wondering whether a satellite
in low-Earth orbit (i.e. orbiting at a few hundred km above the Earth's surface)
is really immersed in a thermal bath at $\sim 1000^\circ$K, it is worth
stressing that this figure is merely a measure of the average molecular kinetic
energy. As a matter of fact, as we shall see in a moment, at these altitudes
the average density is so small that convection play essentially no role as
a heat exchange mechanism.
Discussing in details all the physical processes playing a role in the physics
of the atmosphere is obviously beyond the scope of this review.

\begin{figure}[htb!]
  \includegraphics[width=\linewidth]{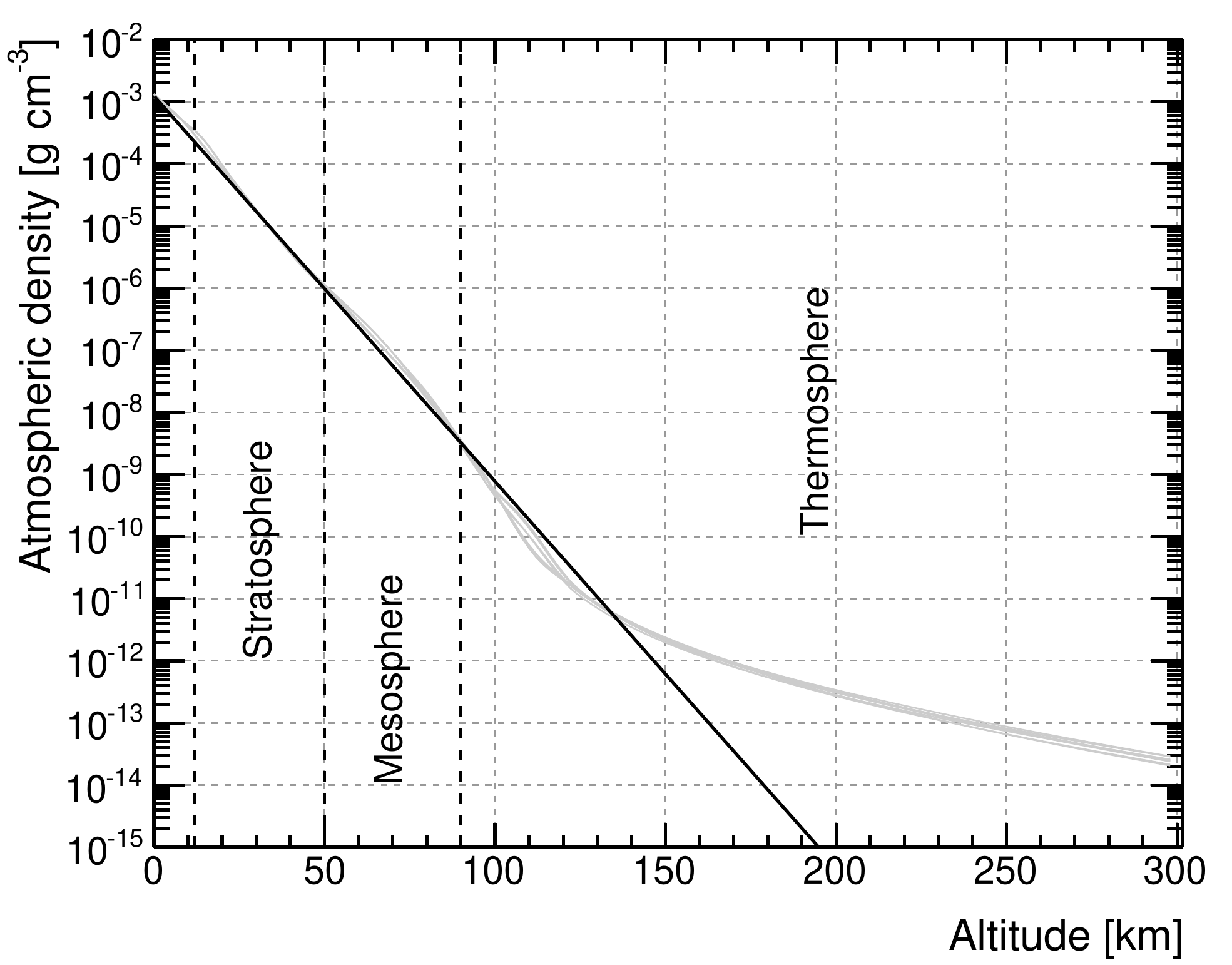}
  \caption{Illustrative atmospheric density profiles provided by the
    NRLMSISE-00 model. The thin gray lines represent the model values for
    8 reference points used for figure~\ref{fig:atmospheric_temperature},
    while the black line is the prediction of an isothermal
    model~\eqref{eq:atmospheric_density} with scale height $h_0 = 7$~km.}
  \label{fig:atmospheric_density}
\end{figure}

That all said, we note that for altitudes smaller than $\sim 100$~km the
temperature is, roughly speaking, approximately constant, so that our
simple-minded isothermal model~\eqref{eq:atmospheric_density} can be expected
to work reasonably well.
This is illustrated in figure~\ref{fig:atmospheric_density}, where the
density profiles from the same empirical model used in
figure~\ref{fig:atmospheric_temperature} are compared with an isothermal
model with scale height $h_0 = 7$~km.

\subsubsection{Typical balloon floating altitude}

The basic figures in the previous section can be used for a rough calculation
of the typical balloon floating altitude. We shall assume a volume $V_b$ at full
expansion of $10^6$~m$^3$, filled with He gas, and a mass of the payload
$m_p$ of 3 tons. At the same pressure and temperature the ratio between the
densities of He and air is approximately equal to ratio of the atomic weights:
\begin{align*}
  r_\density = \frac{\density_{\rm He}}{\density_{\rm Air}}
  \sim \frac{4}{29} \sim 0.14.
\end{align*}
The floating altitude is determined by Archimedes' principle---essentially
we have to equate the buoyant force to the weight of the payload:
\begin{align*}
  (1 - r_\density) V_b \density_0 \exp\left(-\frac{h_{\rm float}}{h_0}\right) = m_p,
\end{align*}
which is readily solved for $h_{\rm float}$:
\begin{align}
  h_{\rm float} = -h_0 \ln \left[ \frac{m_p}{(1 - r_\density) V_b \density_0}\right].
\end{align}

By plugging in the actual numbers one gets a value of $h_{\rm float} \sim 41$~km,
which is only slightly in excess of the typical floating altitude of
actual scientific balloons.

\subsubsection{The atmospheric grammage}%
\label{sec:atmospheric_grammage}

\begin{figure}[htb]
  \includegraphics[width=\linewidth]{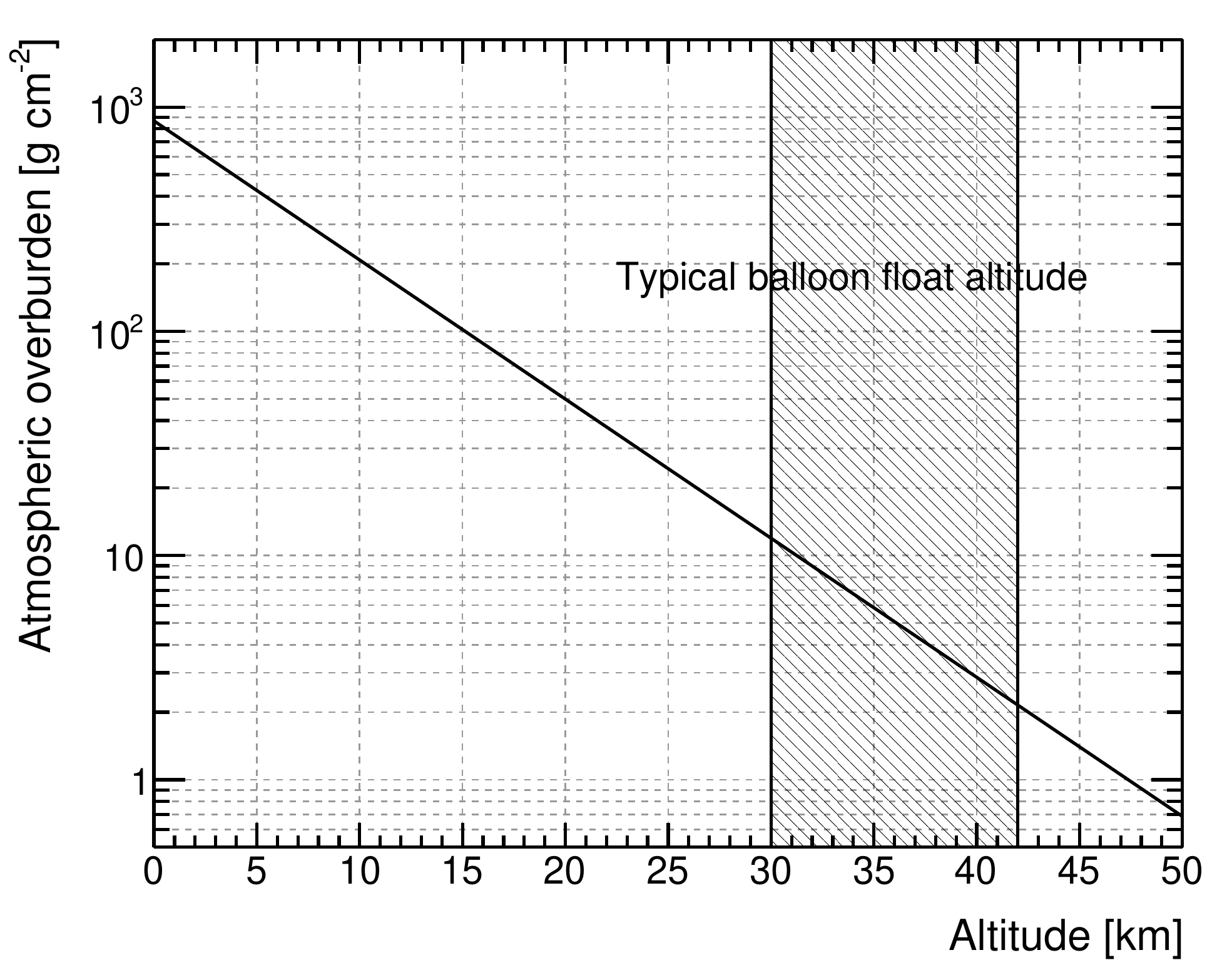}
  \caption{Atmospheric overburden as a function of the altitude calculated
    by integrating our baseline isothermal model. The shaded region
    indicates the typical float altitude of stratospheric balloons, where
    the effective overburden is of the order of a few g~cm$^{-2}$.}
  \label{fig:atmospheric_overburden}
\end{figure}

A relevant physical quantity related to the atmospheric density is the
integrated column density above a given altitude, sometimes called the 
\emph{atmospheric overburden}
\begin{align}\label{eq:atmospheric_overburden}
  \grammage(h) = \int_{h}^{\infty}\density(h')\;dh'.
\end{align}
If we limit ourselves to the stratosphere, the part of the integral above
$\sim 100$~km can be effectively neglected and we can use our isothermal
model~\eqref{eq:atmospheric_density}, which is readily integrated:
\begin{align}\label{eq:atmospheric_overburden_isothermal}
  \grammage(h) = \density_0 h_0 \exp\left(-\frac{h}{h_0}\right).
\end{align}
(The result of the integration is shown in
figure~\ref{fig:atmospheric_overburden}.)

First of all, the atmospheric overburden at sea level is
$\density_0 h_0 \sim 900$~g~cm$^{-2}$. In physical terms this means that the
integrated density profile is equivalent to a $7$~km long column with constant
density $\density = 1.24 \times 10^{-3}$~g~cm$^{-3}$. Just to put things in
context, this translates into $\sim 80$~cm, when re-scaled to the density of lead
($11.35$~g~cm$^{-3}$). In other words, when viewed as a gigantic calorimeter,
the Earth's atmosphere is effectively equivalent to $\sim 80$~cm of lead%
\footnote{This is actually not entirely true, as $7$~km of dry air at
  the see-level density correspond to some $25$ radiation lengths, while
  $80$~cm of lead correspond to about $140$~radiation lengths.
  But we haven't defined the concept of radiation length
  (see section~\ref{sec:radiation_length}), nor that of calorimeter
  (see section~\ref{sec:calorimetry}), yet, and the comparison is suggestive,
  anyway.}%
. This is the basic reason why primaries above $\sim 10$~eV do not reach the
Earth's surface.

For reference, the atmospheric overburden at $40$~km above the sea level
(which is the typical float height for balloons), is $\sim 3$--$4$~g~cm$^{-2}$.
This is an important number, as it is comparable with the propagation path
length of cosmic rays in the Galaxy, which means that balloon-borne experiments
have to correct the measured flux to recover the actual flux at the top of the
atmosphere.

\subsubsection{The Earth Limb}%
\label{sec:earth_limb}

So far we have convinced ourselves that the Earth's atmosphere is important
because it effectively prevents CR primaries from reaching the surface---and
constitute some target material even for stratospheric balloon experiments.
(Last but not least, as we know, the atmosphere makes the balloons float,
which, although kind of obvious, it is indeed relevant for this review.)

There is yet another twist to the story that it's worth mentioning---the fact
that the Earth's atmosphere, acting as a target for high-energy cosmic-ray
protons and nuclei, effectively constitutes the strongest high-energy gamma-ray
source in low-Earth orbit. The detailed modelization of the so called Earth
limb emission requires a fair number of inputs, but the main characteristics
can be understood on general grounds.

\begin{figure}[htb!]
  \includegraphics[width=\linewidth]{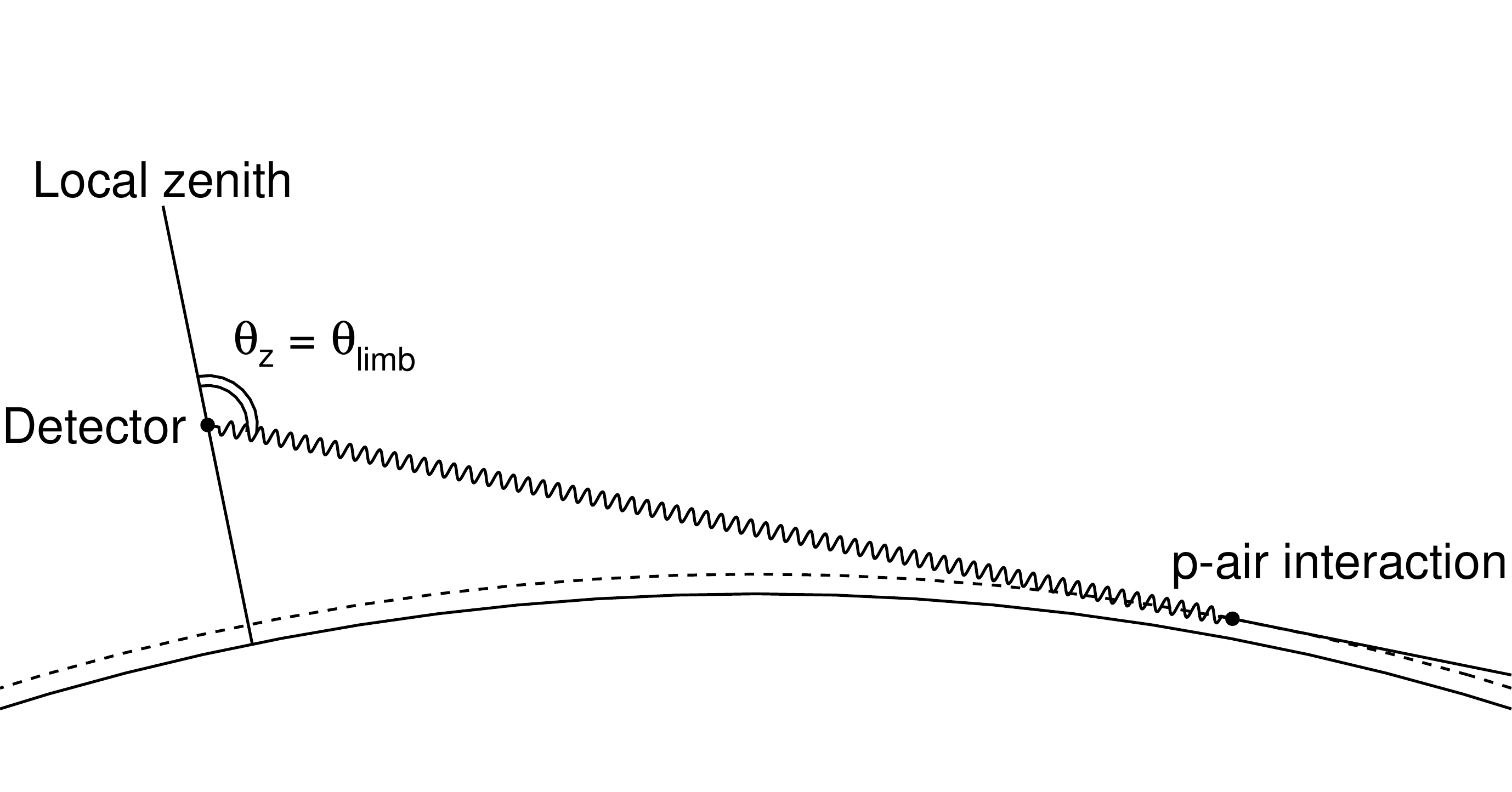}
  \caption{Sketch of the Earth limb gamma-ray emission mechanism: a primary
    cosmic-ray proton interacts at the top of the atmosphere (represented by
    the dashed line) and, among the final products, a photon (typically from
    the decay of a $\pi^0$) escapes the atmosphere in the forward direction.}
  \label{fig:earth_limb}
\end{figure}

\begin{figure}[htb]
  \includegraphics[width=\linewidth]{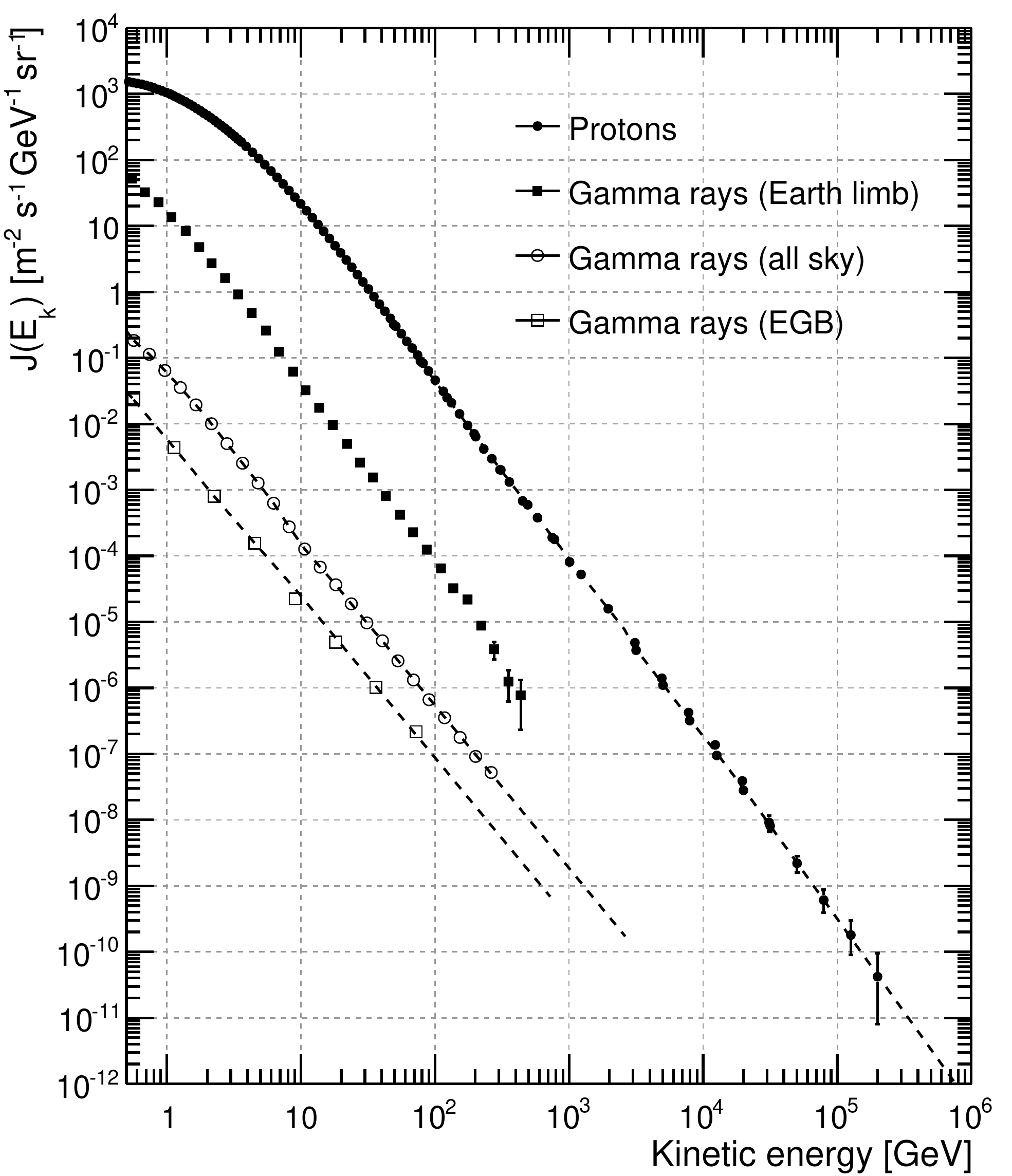}
  \caption{Intensity of the Earth limb measured by Fermi for the range of
    zenith angles $110^\circ < \theta_z < 114^\circ$ (adapted
    from~\cite{2009PhRvD..80l2004A}, notably with a different convention for
    how $\theta_z$ is measured). At the peak of its emission the limb
    outshines the average gamma-ray all-sky intensity by 1 to 2 orders of
    magnitude. It should also be noted that the high-energy spectral index
    follows that of primary protons, from which the limb emission originates.}
  \label{fig:cr_spectra_limb}
\end{figure}

The limb emission is mainly originating from cosmic-ray protons tangentially
interacting near the top of the atmosphere and producing gamma rays in the
forward direction, as sketched in figure~\ref{fig:earth_limb} (this is
especially true at high energy, where the secondaries are highly collimated due
to momentum conservation). The viewing angle at a given detector altitude
$h_{\rm det}$ reads
\begin{align}\label{eq:earth_limb_angle}
  \theta_{\rm limb} = 90^\circ +
  \arccos\left( \frac{R_E + h_{\rm atm}}{R_E + h_{\rm det}}\right),
\end{align}
where $h_{\rm atm}$ is the height of the top of atmosphere and $R_E$ the radius
of the Earth. (We shall see in a moment that the \emph{top of the atmosphere},
in this context, is really the maximum height at which a cosmic-ray protons
impinging tangentially encounters enough material to have a non-negligible
interaction probability.) For $h_{\rm atm} = 50$~km and $h_{\rm det} = 565$~km
(the latter being representative of the height of the \Fermi\ orbit), one gets
$\theta_{\rm limb} \sim 112^\circ$. At smaller zenith angles cosmic rays do
not interact while at larger angles the photons produced in the atmosphere
are readily absorbed, so that the limb emission is (more or less narrowly,
depending on the energy) peaked around $\theta_{\rm limb}$%
\footnote{For completeness: the azimuthal profile reflects the East-West effect
  (see section~\ref{sec:east_west_effect}), which vanishes at high energy.}.
As shown in figure~\ref{fig:cr_spectra_limb}, the gamma-ray emission from the
limb of the Earth typically outshines the average gamma-ray all-sky intensity
by 1 to 2 orders of magnitude (though it should be emphasized that it is only
covering a relatively limited solid angle).

We can actually move a little bit further in understanding the basics of the
limb emission before things get \emph{too} complicated.
Following~\cite{2009PhRvD..80l2004A} we shall take the $p^{14}$N total inelastic
cross section of $\sim 275$~mb as representative of the $p$-air cross section.
Given the number density $n$ of targets per unit volume (expressed in nitrogen
atoms per cm$^3$), this figure can be converted in a proton mean free path
through the well known relation%
\footnote{While \eqref{eq:mean_free_path} is derived in many textbooks, we note
  that it is effectively the only possibility simply on dimensional grounds and
  it scales as expected with the number density and the cross section.} 
\begin{align}\label{eq:mean_free_path}
  \lambda = \frac{1}{n\sigma}.
\end{align}
The number density and the actual density are related to each other by
\begin{align}
  \density = \frac{nA}{N_A},
\end{align}
where $A$ is the mass number of the target and $N_A$ the Avogadro number.
We can therefore rewrite \eqref{eq:mean_free_path} as
\begin{align}
  \lambda\density = \frac{A}{N_A\sigma}
\end{align}
and, by plugging in the actual numbers ($A \sim 14$ and $\sigma \sim 275$~mb),
we get a mean free path $\lambda_p \sim 85$~g~cm$^{-2}$. The other relevant
scale, here, is the radiation length of the air, which is $\sim 37$~g~cm$^{-2}$
(see table~\ref{tab:exp_radlen}). We can now define the concept of the
``top of the atmosphere'' introduced before somewhat more precisely: gamma-ray
production takes place with a reasonable efficiency when the integrated column
density is not negligible compared to mean inelastic free path $\lambda_p$, but
is significantly attenuated when the column density is comparable or larger than
the radiation length of the air---which leaves a relatively small angular window
left, where the column density is of the order of a few times 
$\sim 10$~g~cm$^{-2}$, for efficient production.

\begin{figure}[htb]
  \includegraphics[width=\linewidth]{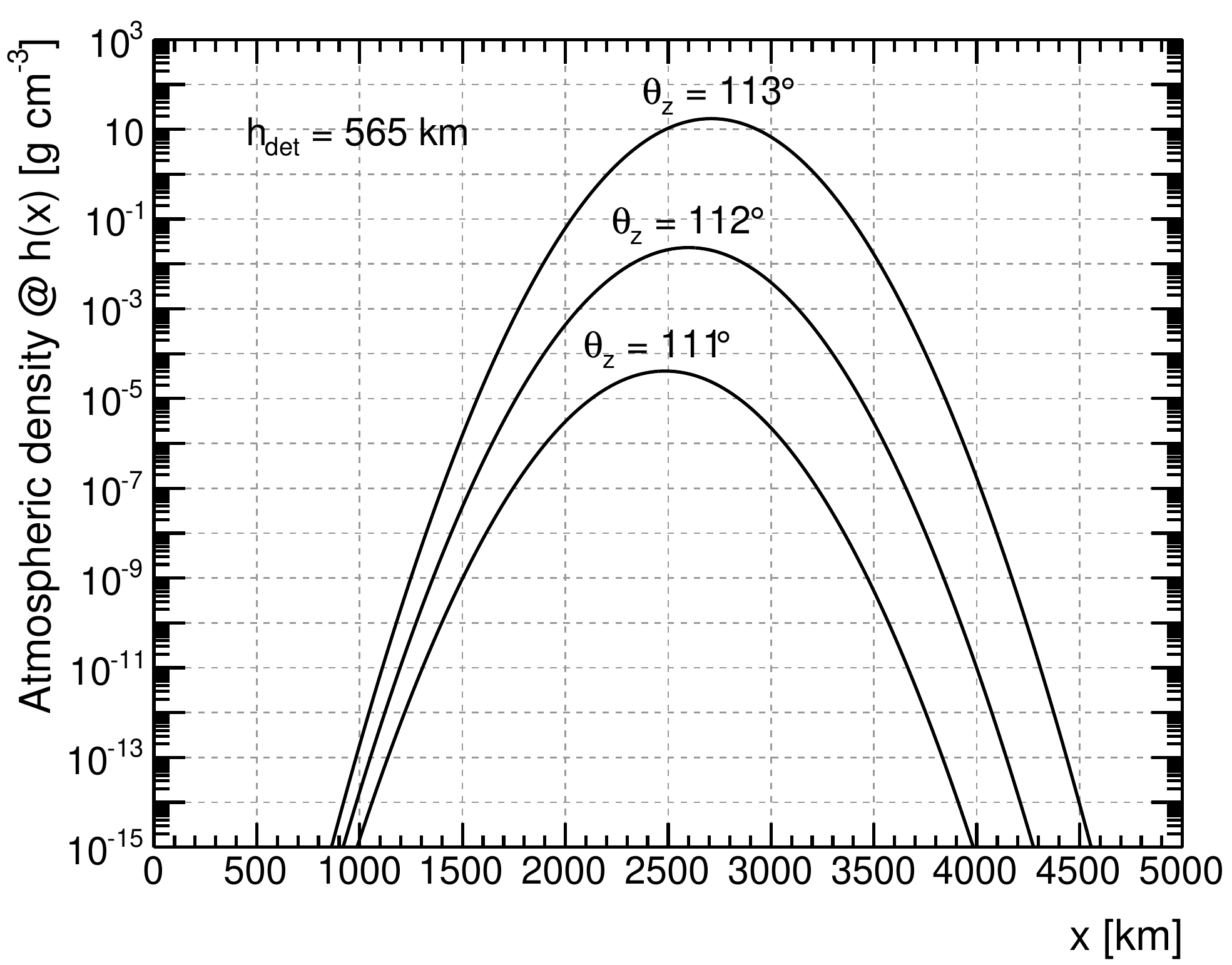}
  \caption{Profile of the atmospheric density traversed by a primary cosmic
    ray impinging on a detector at 565~km altitude for three (slightly)
    different zenith angles.}
  \label{fig:limb_col_profile}
\end{figure}

The integrated column density at a given zenith angle $\theta_z$ can be
readily calculated given an atmospheric model. The altitude at a distance
$x$ from the detector along the line of incidence of the primary cosmic ray
can be written (by the Carnot theorem) as
\begin{align*}
  h(x) = \sqrt{R_{\rm det}^2 + x^2 + 2xR_{\rm det}\cos(\theta_z)} - R_E,
\end{align*}
where $R_{\rm det} = R_E + h_{\rm det}$ (see figure~\ref{fig:earth_limb} for a
sketch of the geometry of the problem). Assuming a simple isothermal model such
as \eqref{eq:atmospheric_density} with a scale height of 7~km, one
has to calculate the integral
\begin{align}\label{eq:atm_col_density}
  \grammage(\theta_z) =
  \int_{0}^{\infty} \density_0\exp\left(-\frac{h(x)}{h_0}\right)
\end{align}
for a fixed zenith angle $\theta_z$.
The exponential in~\eqref{eq:atm_col_density} is such that the integrand
features a violent dependence on the zenith angle: as shown
in figure~\ref{fig:limb_col_profile} a difference of $1^\circ$ in $\theta_z$
translates into a difference of several orders of magnitude in the maximum
atmospheric density along the line of incidence of the primary particle.

\begin{figure}[htb]
  \includegraphics[width=\linewidth]{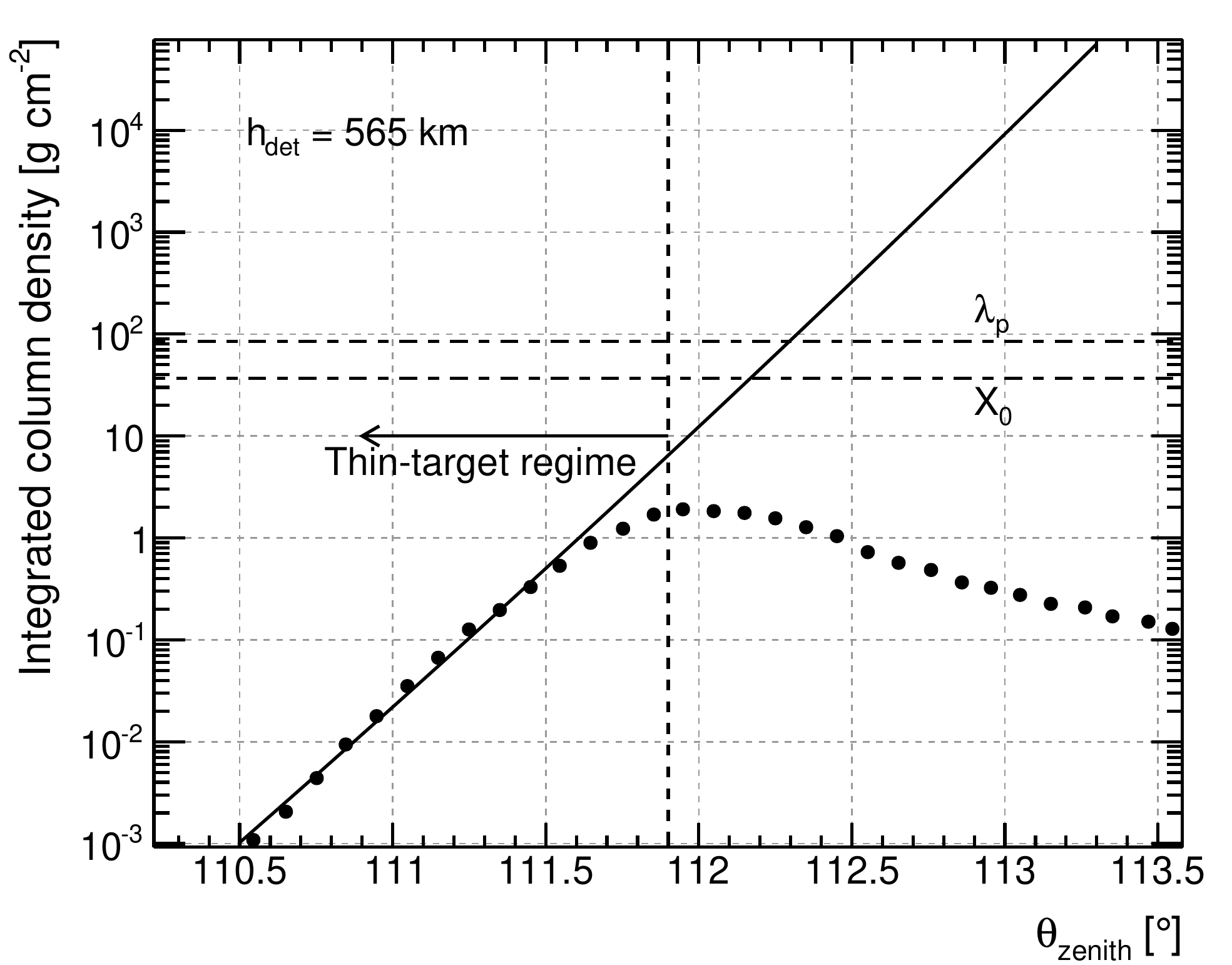}
  \caption{Integrated atmospheric column density, as a function of the
    zenith angle, for a detector in orbit at 565~km altitude (i.e.,
    \emph{a la}~\Fermi). The calculation is based on a simple isothermal model
    of the atmosphere with a scale height of 7~km. The points represent the
    zenithal profile of the limb emission, deconvolved for the PSF, measured
    by \Fermi\ above~3.6~GeV, adapted from~\cite{2009PhRvD..80l2004A}, and
    follows the integrated column density profile in the \emph{thin-target}
    regime. (When comparing with~\cite{2009PhRvD..80l2004A} keep in mind that
    we are using a different definition of the zenith angle.)}
  \label{fig:limb_col_density}
\end{figure}

Figure~\ref{fig:limb_col_density} shows the integrated column density as a
function of the zenith angle for our na\"ive isothermal atmospheric model.
When its value is less than $\sim 10$~g~cm$^{-2}$ (i.e., much smaller than
the radiation length of the air) we are in the so-call \emph{thin-target}
regime, where gamma-ray production involves a single interaction and the
gamma-ray spectrum is tracing that of primary cosmic
rays~\cite{2014PhRvL.112o1103A}. This makes the limb of the Earth an
excellent gamma-ray calibration source for instruments in low-Earth orbit
(see, e.g.,~\cite{2012ApJS..203....4A}).
For completeness, since the coefficient of in-elasticity of the process is of
the order of $k = 0.17$, gamma rays of, say, 10~GeV are produced on average by
protons of $\sim 60$~GeV.

\subsection{The Earth Magnetic Field}%
\label{sec:earth_field}

The fact that the Earth generates a magnetic field, and the magnitude of this
field is $\sim 0.5$~G, is a notion everybody has been taught in his/her
undergraduate education%
\footnote{Yes, I am kidding.}%
. As it turns out, enormous progress has been made, over the last century or
so, in mapping the basic properties of the geomagnetic field.

In the following of this section we shall use spherical coordinates
$(r, \theta, \phi)$ centered and aligned with the magnetic dipole generating
the lowest-order component of the geomagnetic field. In addition to the
magnetic co-latitude $\theta$, we shall also occasionally make use of the
magnetic latitude $\lambda = 90^\circ - \theta$, which is widely used in the
literature.

In source-free regions (i.e., above the Earth's surface), since
$\nabla \times {\mathbf B} = 0$, the static magnetic field can be expressed as
the negative gradient of a scalar potential $\psi$, just like the ordinary
electrostatic field. This potential, in turn, can be expanded in spherical
harmonics as
\begin{align}\label{eq:geomag_expansion}
  \psi(r, \theta, \phi) =&
  R_E \sum_{n=1}^{\infty}\left(\frac{R_E}{r}\right)^{n+1}\nonumber\\
  &\sum_{m=0}^{n}(g_n^m\cos m\phi + h_n^m\sin m\phi)P_n^m(\cos\theta).
\end{align}
(The reader is referred to \cite{Walt} for the exact definition of the
terms; we limit ourselves to note that, with the expansion written in this
form, $g_n^m$ and $h_n^m$ have the dimensions of a magnetic field.)
Modern \emph{professional} descriptions of the geomagnetic field, such as
the eleventh generation of the International geomagnetic Reference Field
(IGRF) contain the coefficients of such expansion up to order 13
\cite{2010GeoJI.183.1216F}.

\subsubsection{The ideal dipole field}

Many of the features of the geomagnetic field can be effectively illustrated
by truncating the expansion to the lowest-order (i.e., the dipole) term, with
$n = 1$ and $m = 0$ (which is the same as saying that the Earth's magnetic
field is, to a reasonable approximation, a dipole field):
\begin{align}
  \psi(r, \theta, \phi) = R_E \left(\frac{R_E}{r}\right)^{2} g_1^0 \cos\theta.
\end{align}
The two non-trivial components of the magnetic field (the component along
the $\phi$ versor is identically $0$ for symmetry reasons), read
\begin{align}\label{eq:dipole_field_components}
  B_r(r, \theta, \phi) =& -\frac{\partial\psi}{\partial r} =
  2 \left(\frac{R_E}{r}\right)^3 g_1^0 \cos\theta,\nonumber\\
  B_\theta(r, \theta, \phi) =& -\frac{1}{r}\frac{\partial\psi}{\partial\theta} =
  \left(\frac{R_E}{r}\right)^3 g_1^0 \sin\theta,
\end{align}
i.e., the field is purely radial at the poles and purely tangential at the
equator.
For completeness, the intensity of the dipole field at any given point is
given by
\begin{align}
  B(r, \theta, \phi) =&
  \sqrt{B_r^2(r, \theta, \phi) + B_\theta^2(r, \theta, \phi)} = \nonumber\\
  =& \left(\frac{R_E}{r}\right)^3 g_1^0 \sqrt{1 + 3\cos^2\theta}.
\end{align}
From the above expression it is easy to recognize that physically
$g_1^0$ represents the field intensity $B_0$ at the equator ($\theta = 90^\circ$)
on the Earth's surface ($r = R_E$). It is also easy to recognize that the
field intensity on the Earth's surface is minimum at the equator and twice
as large at the magnetic poles.

Even more important, it follows from~\eqref{eq:dipole_field_components} that
the equation for a geomagnetic field line, is
\begin{align}
  \frac{1}{r}\frac{{\rm d}r}{{\rm d}\theta} = \frac{B_r}{B_\theta} =
  \frac{2\cos\theta}{\sin\theta},
\end{align}
which is readily integrated to give the equation of the field lines:
\begin{align}
  r = R_EL\sin^2\theta.
\end{align}
(Mind when you do that you have logarithms on both sides that simplify.)
At this point the quantity $L$ in the equation above (which is customarily
referred to as the McIlwain~$L$ coordinate) is just a constant of integration,
but one with a pretty straightforward physical interpretation---it is the
distance, measured in units of Earth's radii, at which a given field line
crosses the magnetic equator ($\theta = 90^\circ$). For completeness, the
McIlwain~$L$ coordinate can be expressed in term of the geomagnetic coordinates
as
\begin{align}
  L = \frac{r}{R_E \sin^2\theta} = \frac{r}{R_E \cos^2\lambda},
\end{align}
which becomes
\begin{align}
  L = \frac{1}{\sin^2\theta} = \frac{1}{\cos^2\lambda}
\end{align}
on the surface of the Earth ($r = R_E$).
(Note that for an ideal dipole the field is azimuthally symmetric and
the McIlwain~$L$ coordinate does not depend on $\phi$.)

\begin{figure}[!htb]
  \includegraphics[width=\linewidth]{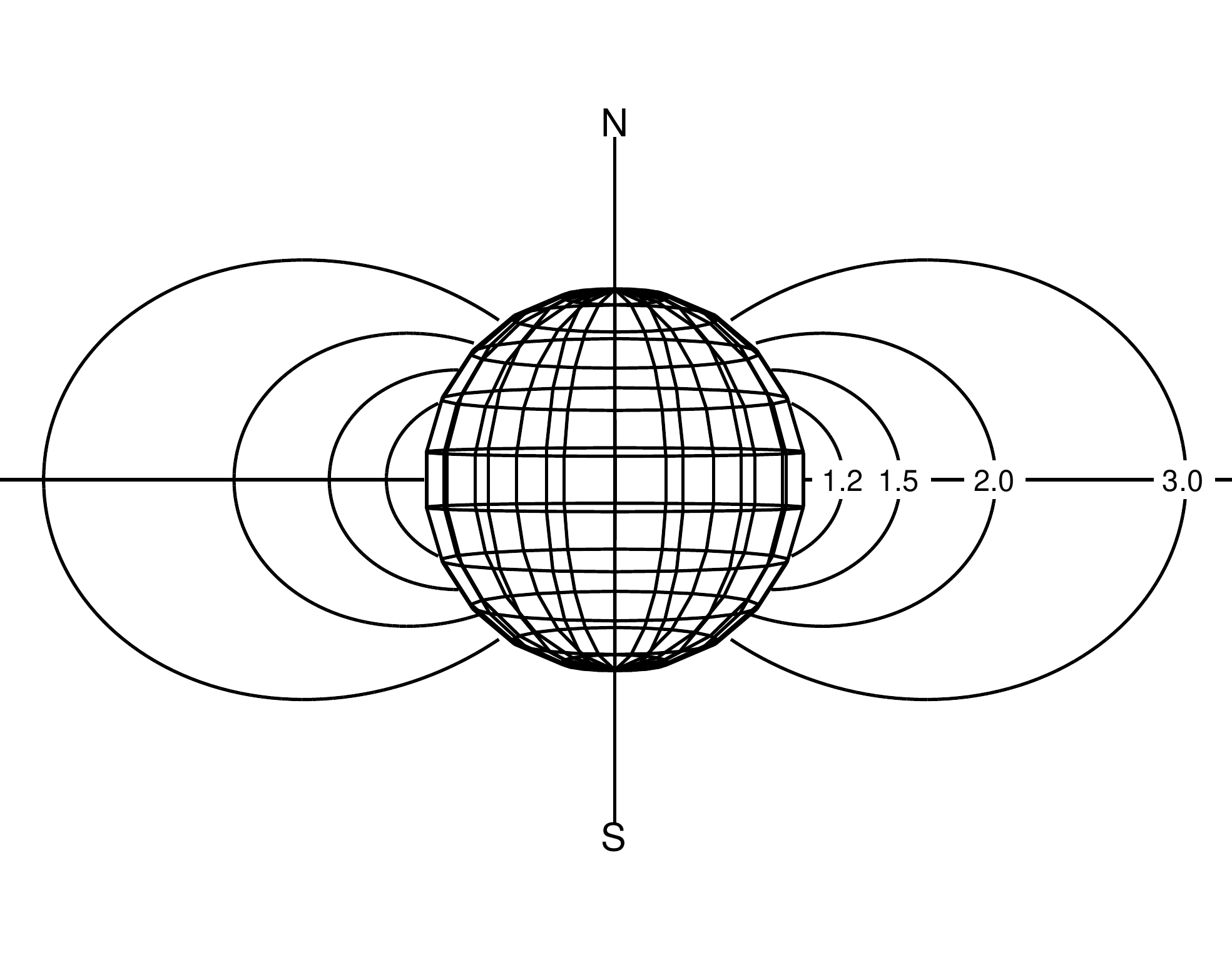}
  \caption{Basic sketch of the field lines (or L-shells) for a centered and
    aligned dipole field. The numbers indicate the McIlwain~$L$ coordinate
    (i.e., the distance from the dipole center to the intersection with the
    equator, measured in units of Earth's radii) for each shell.}
  \label{fig:dipole_field}
\end{figure}

Figure~\ref{fig:dipole_field} shows some illustrative geomagnetic field
line in our dipole approximation. Each of those line defines, by rotation
in the $\phi$ coordinate, a shell intersecting the equatorial plane at
a given distance from the dipole center. By definition all the points on such a
shell have the same McIlwain~$L$ coordinate (for instance, the points at
$L = 2$ are those on the magnetic field lines intersecting the equator at $2$
Earth radii).
As we shall briefly see in section~\ref{sec:trapped_radiation}, the
McIlwain~$L$ coordinate is convenient to describe the population of trapped
particles in the Earth's magnetosphere (see, however,
\cite{2010ASTRA...6....9P} for a succinct account of some of the subtleties
involved in the definition of the McIlwain~$L$ coordinate for non-dipolar field
geometries).

We note, in passing, that the field magnitude is not constant along the
field lines, so that, strictly speaking, different points on the same
$L$-shell are not magnetically equivalent. In fact the field
intensity decreases monotonically with the co-latitude $\theta$ along a field
line
\begin{align}
  B(\theta) = \frac{B_0}{L^3}\frac{\sqrt{1 + 3\cos^2\theta}}{\sin^6\theta}.
\end{align}
The $B$ coordinate defined by the above equation provides an obvious
parameterization of the geomagnetic environment within each given $L$ shell
and the McIlwain~$(B,~L)$ coordinates are widely used to describe the motion
of slow charged particles in the Earth's magnetic field.

\subsubsection{The actual geomagnetic field}

The actual geomagnetic field is not a perfectly aligned and centered dipole.
First of all the dipole axis is misaligned by $\sim 11^\circ$ with respect to
the spin axis of the Earth. In addition, the center of the dipole does not
coincide with the center of the Earth (the offset being of the order of 700~km).
Finally, asymmetries in the interior current system generating the magnetic
field produce higher-order terms in the expansion~\eqref{eq:geomag_expansion}.

\begin{figure}[!htb]
  \includegraphics[width=\linewidth]{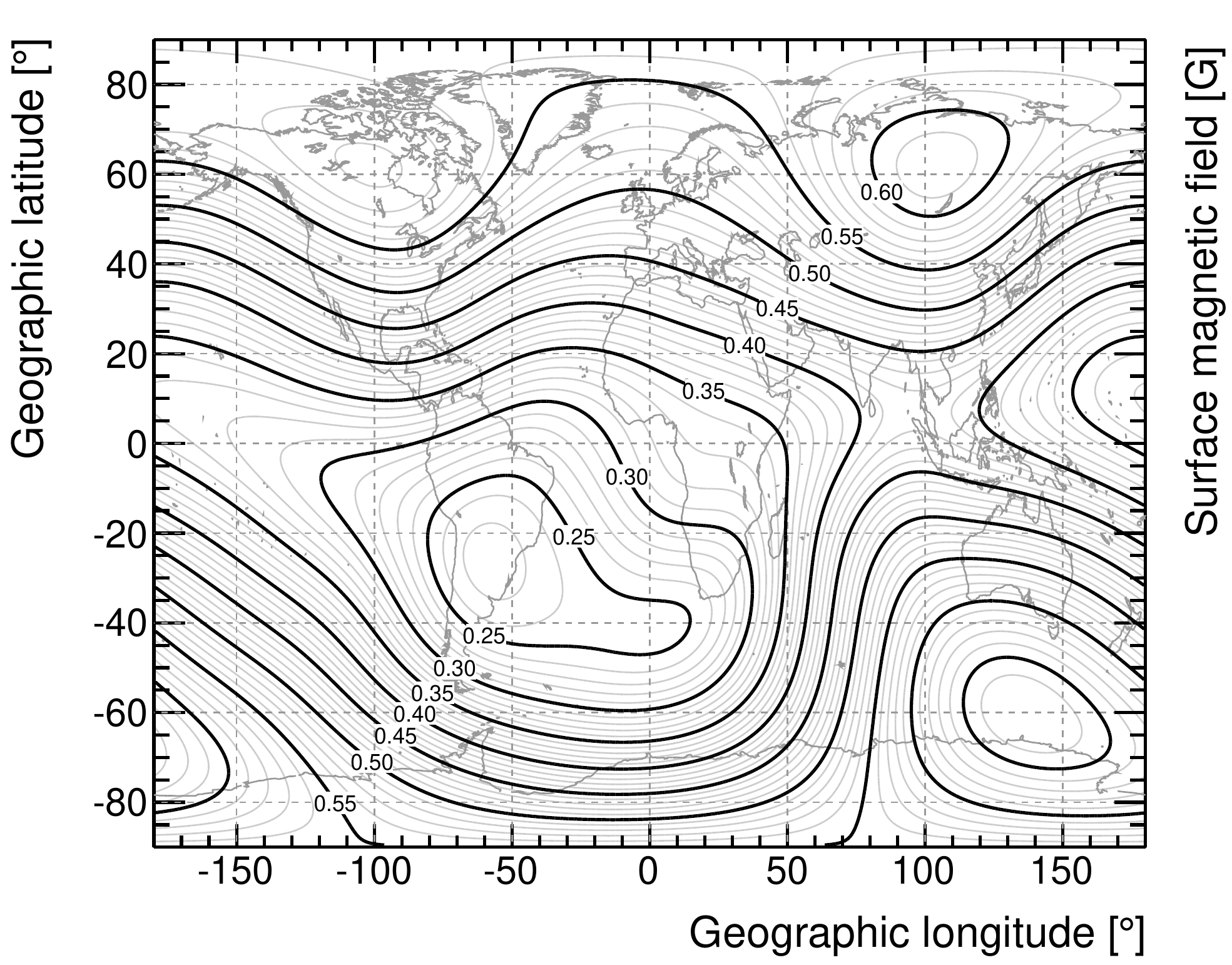}
  \caption{Iso-contours of the Earth's magnetic field, in G, at sea-level.
    The data have been taken from the IGRF-11 model. (For completeness,
    the reference time is January 1, 2013.)}
  \label{fig:sea_level_mag_field}
\end{figure}

Figure~\ref{fig:sea_level_mag_field} shows a typical map of the intensity of
the magnetic field at the Earth's surface. If our ideal dipole approximation
was a perfect description of reality, the iso-intensity lines would  be
parallel to the equator (i.e., there would be no $\phi$ dependence) and the
intensity at the poles would be exactly twice that at the equator.
Figure~\ref{fig:sea_level_mag_field} gives a fairly good idea of the departure
of the geomagnetic field from the simplistic dipole approximation.

\begin{figure}[htb]
  \includegraphics[width=\linewidth]{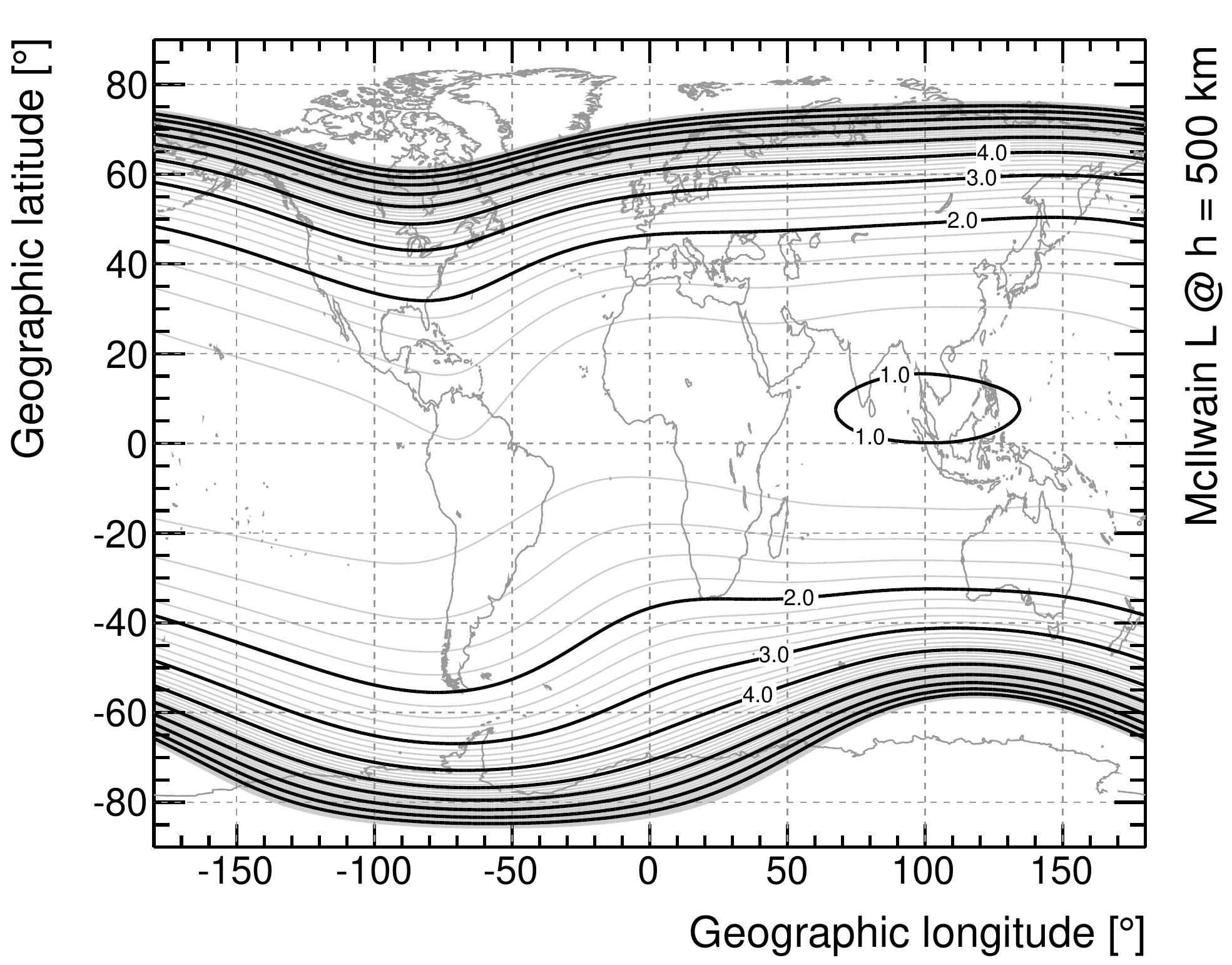}
  \caption{Iso-intensity line of the McIlwain~$L$ parameter at an altitude of
    $500$~km for the same field model shown in
    figure~\ref{fig:sea_level_mag_field}. (For completeness, the reference time
    is January 1, 2013.)}
  \label{fig:high_alt_mc_ilwain_L}
\end{figure}

That all said, the McIlwain parameterization of the geomagnetic environment is
a useful concept even in the real geomagnetic field.
Figure~\ref{fig:high_alt_mc_ilwain_L} shows a map of the McIlwain~$L$
iso-intentity lines at an altitude of $500$~km above the sea level
(i.e., in low-Earth orbit). Again, in the dipole approximation those
lines would be parallel to the equator.

\subsubsection{Ray-tracing techniques}%
\label{sec:raytracing}

\begin{figure}[htb]
  \includegraphics[width=\linewidth]{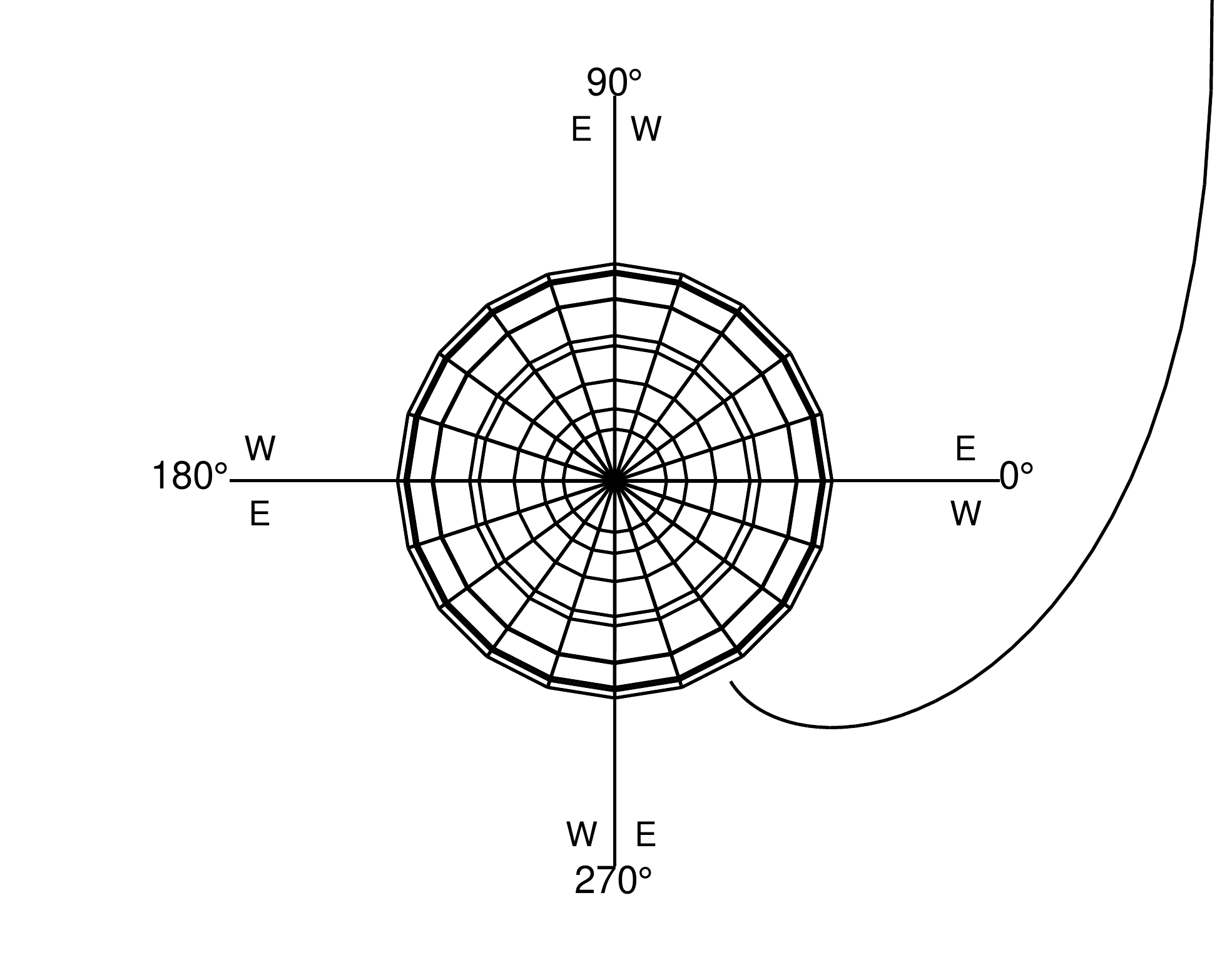}
  \caption{Numerical ray-tracing of an outgoing (i.e., traveling
    counter-clockwise) 12~GV antiproton in the geomagnetic field.
    The particle is generated at an altitude of 500~km over the magnetic
    equator, at a magnetic longitude of $-60^\circ$, and the initial momentum is
    directed along the local zenith. The trajectory is also representative of
    an in-going (i.e., traveling clockwise) 12~GV proton coming from infinity.}
  \label{fig:ray_tracing}
\end{figure}

Computer programs exists that, given a detailed model of the geomagnetic field
(e.g., the aforementioned IGRF model) allow to numerically solve the classical
equations of motion and provide reliable predictions for the actual
trajectories of charged particles, as illustrated in
figure~\ref{fig:ray_tracing}.
The reader is referred to~\cite{Walt} for an elementary introduction to
the motion of charged particles in the geomagnetic field and
to~\cite{2005AdSpR..36.2012S} for a brief overview of numerical ray-tracing
techniques.

Figure~\ref{fig:ray_tracing} deserves a somewhat detailed description, as we
shall encounter several similar instances in the rest of the section---and
since, when dealing with the motion of charged particle in a magnetic field,
it is easy to get confused with the signs.
The sphere represents the Earth viewed from the \emph{top}, with the geomagnetic
North pole at the center of the image. In this representation the geomagnetic
field is directed perpendicularly out of the page, so that positively charged
particles travel clockwise and negatively charged particles travel
counter-clockwise---if you're not persuaded try it yourself applying the
right-hand rule to the expression for the Lorentz force
\begin{align*}
  \mathbf{F} = q \mathbf{v} \times \mathbf{B}.
\end{align*}

In applications where one is interested in the properties of the cosmic-ray
particle populations at a given point in the magnetosphere (e.g., the point
where the detector is placed) it is not practical to simulate an isotropic
flux coming from large distances and select the particles that happen to arrive
in the vicinity of the region of interest---that would be way too inefficient.
It is customary, instead, to generate particles \emph{with the opposite charge}
at the location of the detector and propagate them to infinity.
When doing that, at relatively low energies, it does happen that part of the
trajectories intersect the surface of the Earth or end up deep in the
atmosphere. These trajectories are called \emph{forbidden trajectories} as
a primary cosmic ray cannot reach the Earth from large distances along any of
them. In the following we shall see several examples were the concept of
allowed and forbidden trajectory is important.

\subsubsection{The geomagnetic cutoff}%
\label{sec:geomagnetic_cutoff}

The geomagnetic field effectively acts like a shield for (relatively) low-energy
charged primary cosmic rays impinging on the Earth. The field being
strongest at the magnetic poles, one might naiv\"ely think that the shield
effect is stronger there than at the equator. In fact is quite the opposite,
as the field at the poles is radial and charged particles coming from the
zenith direction can travel unaffected along the field lines. At the equator,
on the other hand, the field is orthogonal to the zenith direction and the
shielding effect is maximum.

A widely used concept is that of the vertical rigidity cutoff, i.e. the minimum
rigidity that is required for a charged particle to reach a point above the
Earth surface, at a given altitude, from the direction of the local zenith.
Phrased in a different way, for any given geographic position and altitude,
particles below the vertical rigidity cutoff from the local zenith direction
are effectively shielded by the geomagnetic field.

\begin{figure}[htb]
  \includegraphics[width=\linewidth]{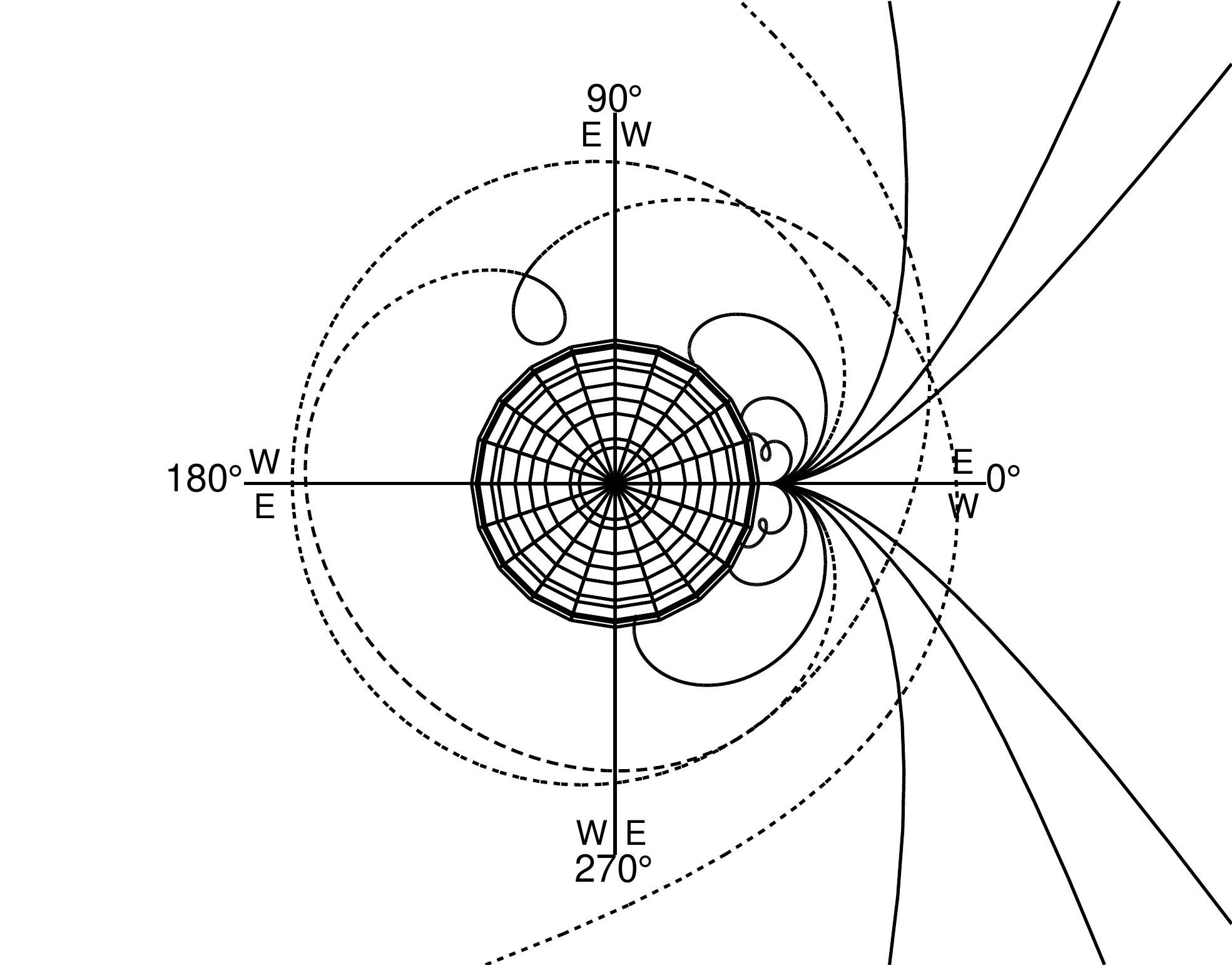}
  \caption{Illustration of the vertical rigidity cutoff numerical calculation.
    The solid lines corresponds to the trajectories of electrons
    and positrons with rigidity of $25$, $20$, $15$, $10$, $7.5$ and $5$~GV,
    starting from a point $500$~km above the equator at a magnetic longitude of
    $0^\circ$ (for completeness, positrons are coming in from the East direction
    and electrons are coming from the West). 
    The two dashed lines correspond to the rigidity values separating the
    trajectories reaching out to infinity (\emph{allowed}) and those
    intersecting the Earth's atmosphere (\emph{forbidden}). These values signal
    the vertical rigidity cutoff for electrons and positrons.}
  \label{fig:vertical_cutoff}
\end{figure}

Strictly speaking, the rigidity cutoff can only be calculated through a full
numerical particle tracing in a detailed model of the Earth's magnetic field.
As anticipated in the previous section, the basic strategy to calculate the
rigidity cutoff for, say, electrons at a given position is to simulate the
trajectories of positrons moving out from the very same position in the
direction of the local zenith for several different rigidity values. The
rigidity separating allowed and forbidden trajectories defines the vertical
rigidity cutoff, as illustrated in figure~\ref{fig:vertical_cutoff}.

\begin{figure}[htb]
  \includegraphics[width=\linewidth]{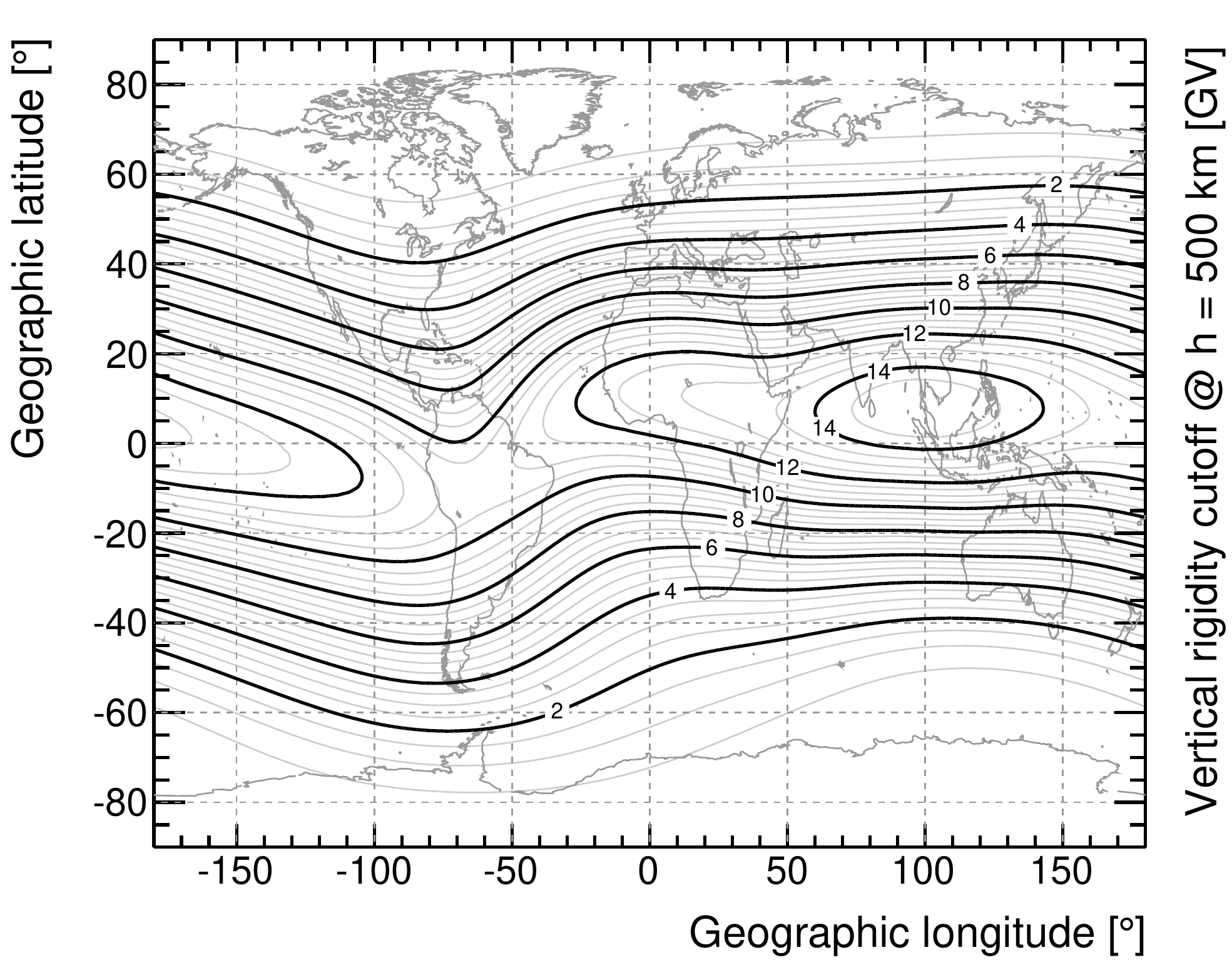}
  \caption{Iso-intensity lines of the vertical rigidity cutoff, in~GV, at an 
    altitude of $500$~km for the same field model shown in
    figure~\ref{fig:sea_level_mag_field}.}
  \label{fig:high_alt_rigidity_cutoff}
\end{figure}

As it turns out, however, in a dipolar geomagnetic field approximation the
equations of motion of a charged particle can be analytically
integrated~\cite{2005AdSpR..36.2012S} to get an explicit expression for the
vertical rigidity cutoff
\begin{align}
  R_v = \frac{R_0\sin^4\theta}{r^2} = \frac{R_0\cos^4\lambda}{r^2} =
  \frac{R_0}{L^2},
\end{align}
with $R_0 = 14.5$~GV.
Figure~\ref{fig:high_alt_rigidity_cutoff} shows the iso-intensity contours
of the vertical rigidity cutoff at an altitude of $500$~km. Near the
equator the typical cutoff value is of the order of $\sim 15$~GV, while
sub-GV values can be reached moving close to the magnetic poles.

We stress that this is not only relevant if one aims at measuring primary
cosmic rays down to the lowest possible energies, as in any event it has
practical implications on the level of backgrounds and the overall trigger
rates. We also note, in passing, that while balloon-borne experiments
operate at roughly constant McIlwain~$L$ (and rigidity cutoff), satellite
experiments orbiting at a fixed inclination span a range of L values
and therefore have variable background and trigger rate as a function of the
position in the orbit.

\subsubsection{The East-West effect}%
\label{sec:east_west_effect}

As mentioned in section~\ref{sec:history_early}, Bruno Rossi predicted early
in the 1930s that, if the primary cosmic rays carry predominantly one charge
sign (and we know that the vast majority of primary cosmic rays are positively
charged), one should observe an East-West flux asymmetry---which is maximal
around the geomagnetic equator.

\begin{figure}[htb]
  \includegraphics[width=\linewidth]{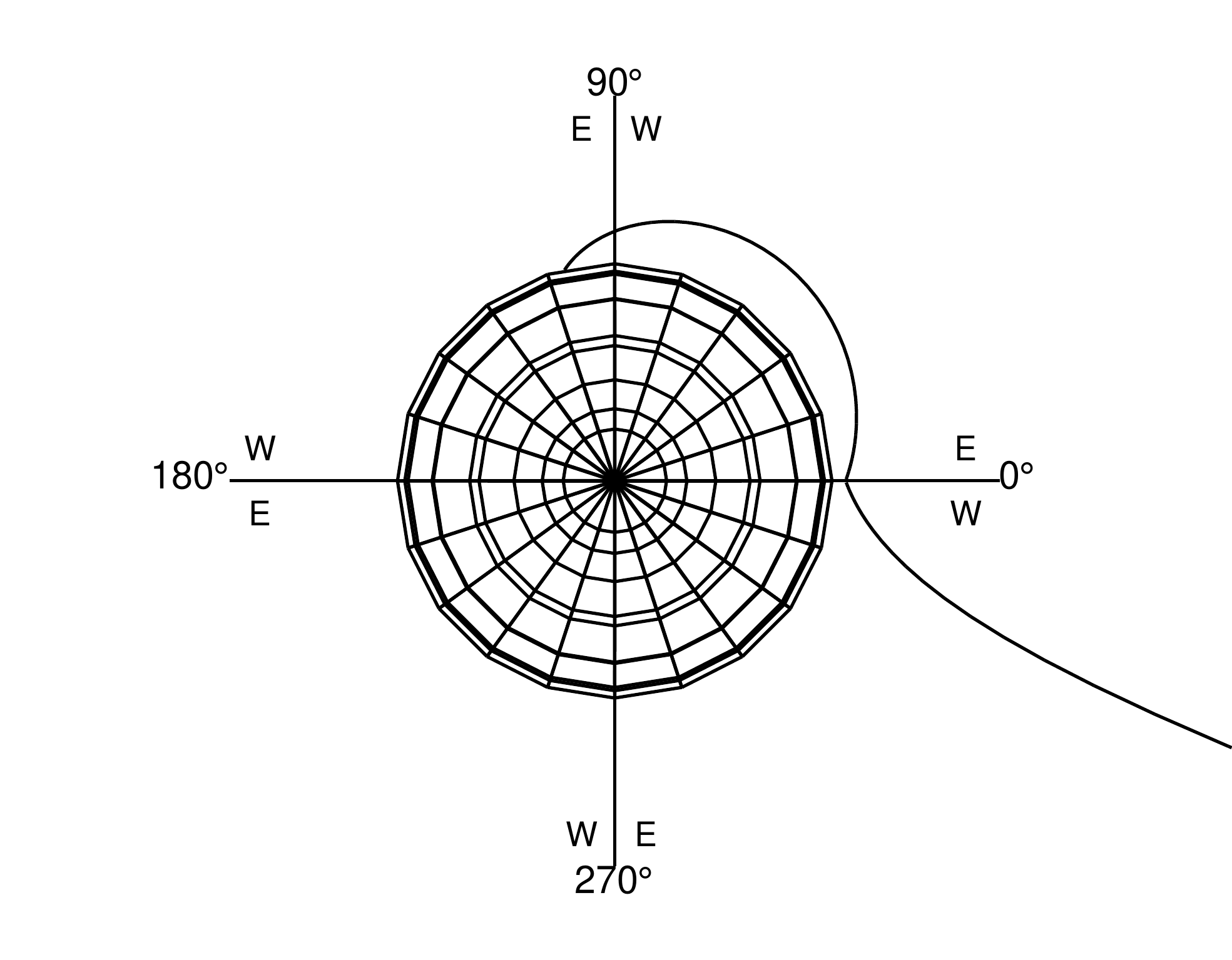}
  \caption{Illustration of the East-West effect. The figure shows the
    trajectories of two 30~GV protons reaching a point 500~km above the
    geomagnetic equator at $0^\circ$ geomagnetic longitude and a zenith angle
    of $70^\circ$. The trajectory coming from the East direction is forbidden
    (i.e., protons from this direction are effectively shielded), while
    that from the West direction is allowed.}
  \label{fig:east_west_effect}
\end{figure}

The effect is sketched in figure~\ref{fig:east_west_effect}, where the
trajectories of two 30~GV protons incoming from the East and the West directions
are shown. In the first case the trajectory intersects the Earth's surface
and is therefore forbidden---it cannot possibly represent a primary
proton coming from infinity. On the other hand the trajectory from the West is
allowed. As the rigidity increases the trajectories get more and more
straight and the magnitude of the effect tends to vanish, but there is a
relatively large energy range, between a few GV and a few tens of GV, where
the shadow of the Earth tends to suppress the flux of positively charged
particles coming from the East. Since the primary cosmic-ray spectrum is a
steep power law, this has a prominent effect of the integral flux.

We shall see in section~\ref{sec:geomeg_charge_sep} that the same kind of
effect can be used, at least to some extent, to distinguish the charge sign by
means of instrument that are not equipped with a magnet.

\subsubsection{Secondary radiation in low-Earth orbit}%
\label{sec:trapped_radiation}

Below the geomagnetic cutoff primary cosmic rays are reprocessed by the
combined effect of the geomagnetic field and the atmosphere of the Earth.
While this is not really the focus of this review, the subject is relevant
for all space-based instruments, as the sub-cutoff particle populations
can have a major impact on the trigger rate and constitute an important source
of background---especially for gamma-ray detectors.

\begin{figure}[htb]
  \includegraphics[width=\linewidth]{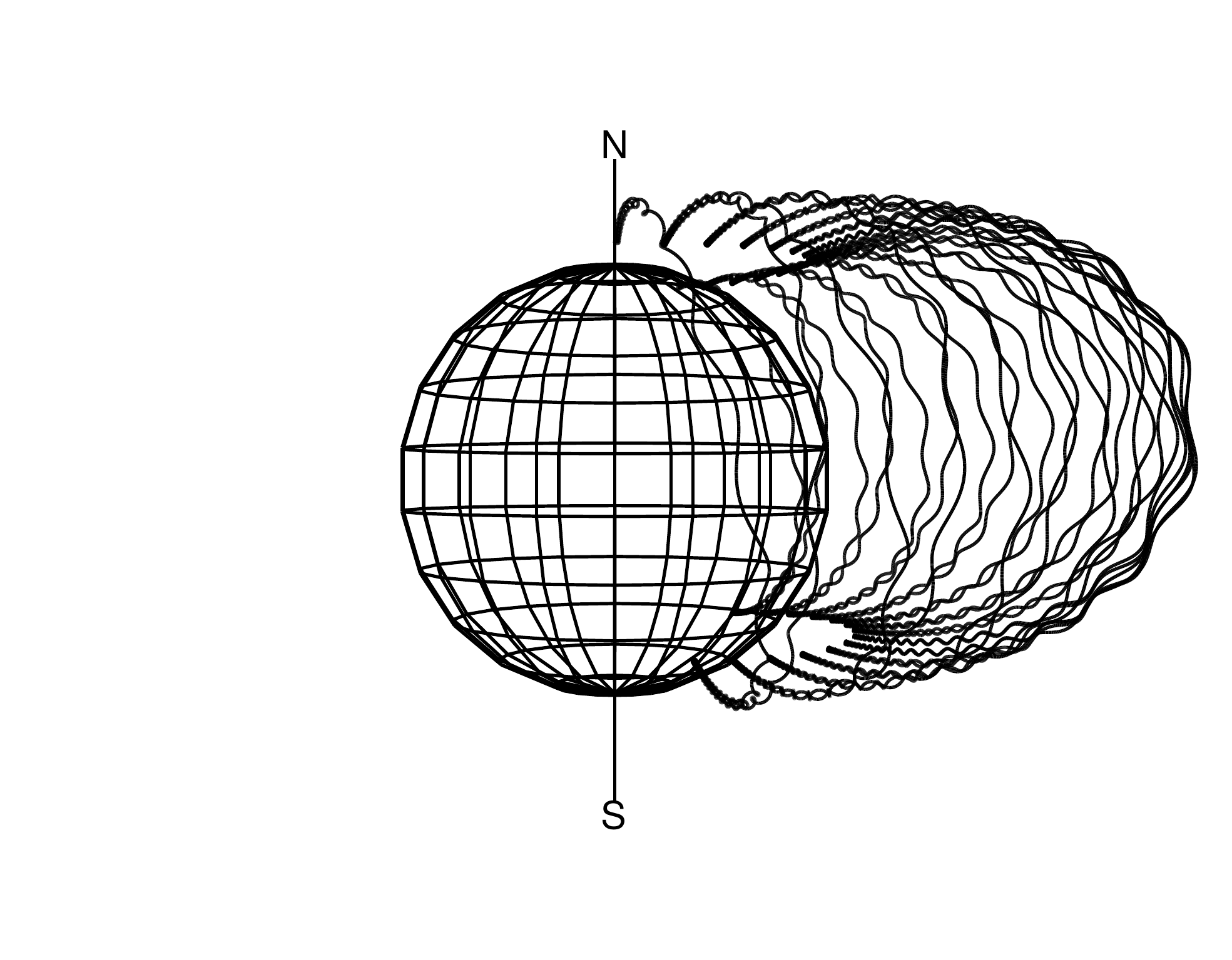}
  \caption{Trajectory of a $1$~GeV proton starting at an altitude
    of $2000$~km. The figure illustrates the complex motion of the low-energy
    trapped radiation, bouncing back and fourth multiple times between the
    magnetic mirror points.}
  \label{fig:trapped_particle}
\end{figure}

The basic theory of geomagnetically trapped radiation is thoroughly discussed
in~\cite{Walt} (and~\cite{2002APh....16..295L} is another good source of
information). Trapped particles are generated by the interaction of primary
cosmic rays with the atmosphere and generally undergo a characteristic
\emph{bouncing} motion around the geomagnetically field lines between two
mirror points (placed symmetrically with respect to the geomagnetic equator),
with a contestual \emph{drift} in the azimuthal direction, before being
re-absorbed in the atmosphere. The number of times a particle
bounces back and forth between the mirror points determines its trapping time,
with the radius of the Earth determining the minimum timescale
$R_E/c \sim 20$~ms. Figure~\ref{fig:trapped_particle} illustrates the bouncing
and drift motion characteristic of the trapped radiation. It should be noted
that it takes place on a fixed McIlwain~$L$ shell, which is the basic reason
why the McIlwain~$L$ coordinate is a useful concept from the standpoint of
studying the trapped radiation.

While a detailed description of all the salient characteristics of the trapped
radiation (in terms of geographical distribution, energy spectra and chemical
composition) is way beyond the scope of this paper, we do mention here a
striking feature---namely the fact that the positron ratio (i.e., the ratio
between the $e^+$ and the $e^+  + e^-$ intensity) is much higher than that
of primary cosmic rays. As a matter a fact sub-cutoff positrons can outnumber
electrons by a factor of $\sim 4$. This can be readily understood by noting that
trapped radiation is mostly injected in the magnetosphere by primaries
impinging tangentially on the atmosphere, as the lower-energy secondary
products of the interactions are beamed in the forward direction due to
momentum conservation and are immediately absorbed in the atmosphere if the
primary is impinging radially. Now: if the primary cosmic ray is impinging from
the west, positively-charged secondaries (e.g., positrons) are deflected away
from the atmosphere by the magnetic field, while negatively-charged ones
(electrons) are deflected toward the atmosphere and readily absorbed. In other
words primary cosmic rays from the West preferentially inject positrons
in the magnetosphere. For particles impinging from the East direction it's
exactly the opposite---but they are less numerous due to the East-West
effect.

\begin{figure}[htb]
  \includegraphics[width=\linewidth]{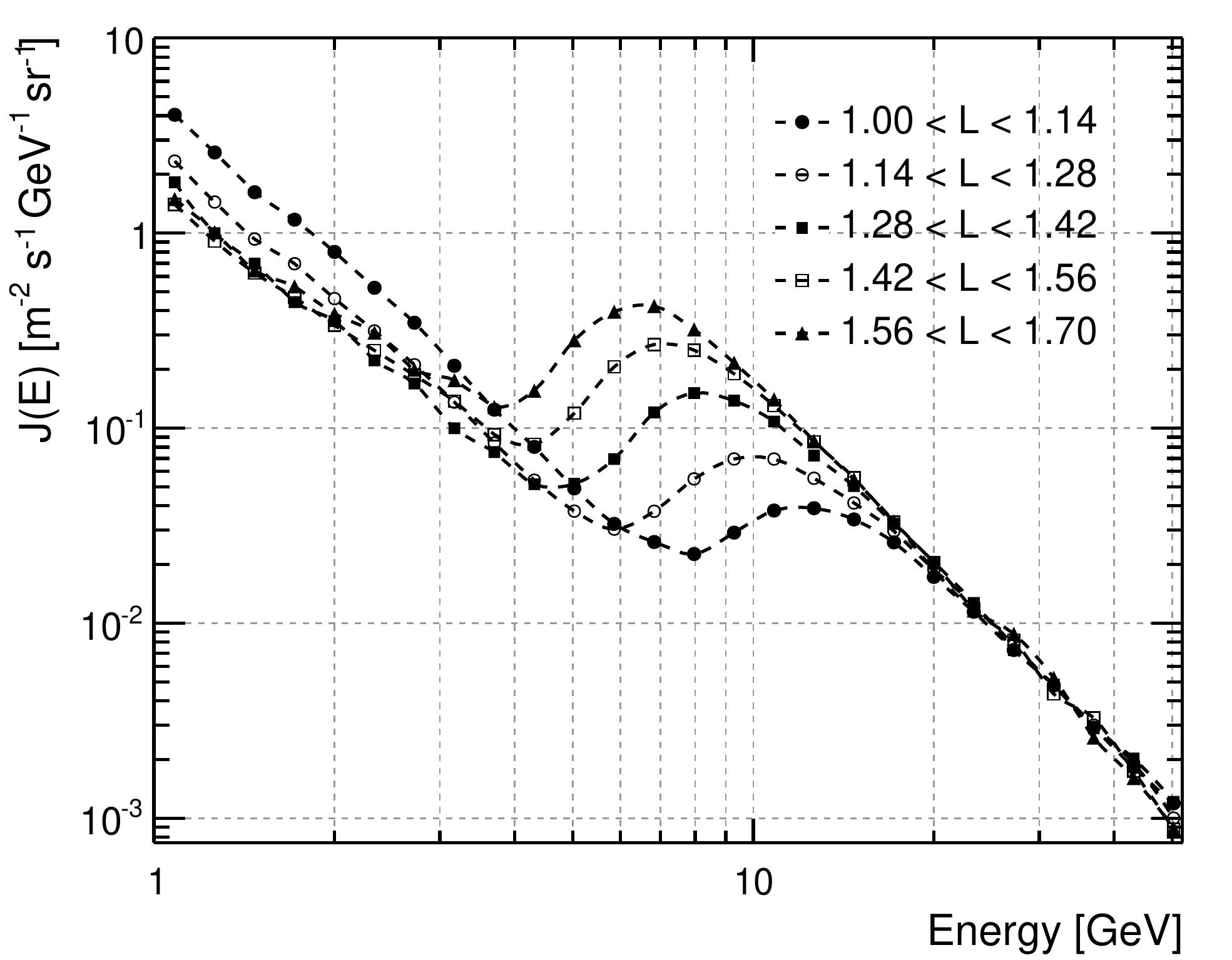}
  \caption{Inclusive cosmic-ray electron spectrum as measured by the
    \Fermi\ Large Area Telescope in bins of McIlwain~$L$ (adapted
    from~\cite{2010PhRvD..82i2004A}). For each of the series of data point
    the peak-like structure is given by the convolution of the primary
    power-law spectrum and the geomagnetic rigidity cutoff---averaged over
    a relatively wide range of $L$ values and zenith angles.
    Below the cutoff the spectrum is that of the trapped positron and
    electrons.}
  \label{fig:cre_broadband_spectra}
\end{figure}

Figure~\ref{fig:cre_broadband_spectra} shows the cosmic-ray electrons and
positron spectra around the cutoff (i.e., around the transition between the
trapped population and the primary cosmic rays) measured by the \Fermi\ Large
Area Telescope in several bins of McIlwain~$L$. We shall see in
section~\ref{sec:calib_energy_scale} how this can be exploited for an in-flight
calibration of the absolute energy scale.

\subsubsection{The South Atlantic Anomaly}%
\label{sec:saa}

The South Atlantic Anomaly (SAA) is the region where the radiation belts comes
closest to the Earth's surface---down to an altitude of $\sim 200$~km.
Interestingly enough, the radiation belts are, generally speaking, not 
terribly relevant for instruments in low-Earth orbit, as their intensity tends
to be maximum at much higher altitudes. The SAA, roughly located above
Brazil, is a notable exception. The level of radiation in the SAA can exceed
by orders of magnitude that in other regions at the same altitude---at the
level that space-based instruments may have to interrupt ordinary data
taking during the SAA passages (this is the case, e.g., for \Fermi, which
spends some $\sim 15\%$ of the time in the SAA).

\subsubsection{Separating charges using the geomagnetic field}%
\label{sec:geomeg_charge_sep}

\begin{figure}[htb]
  \includegraphics[width=\linewidth]{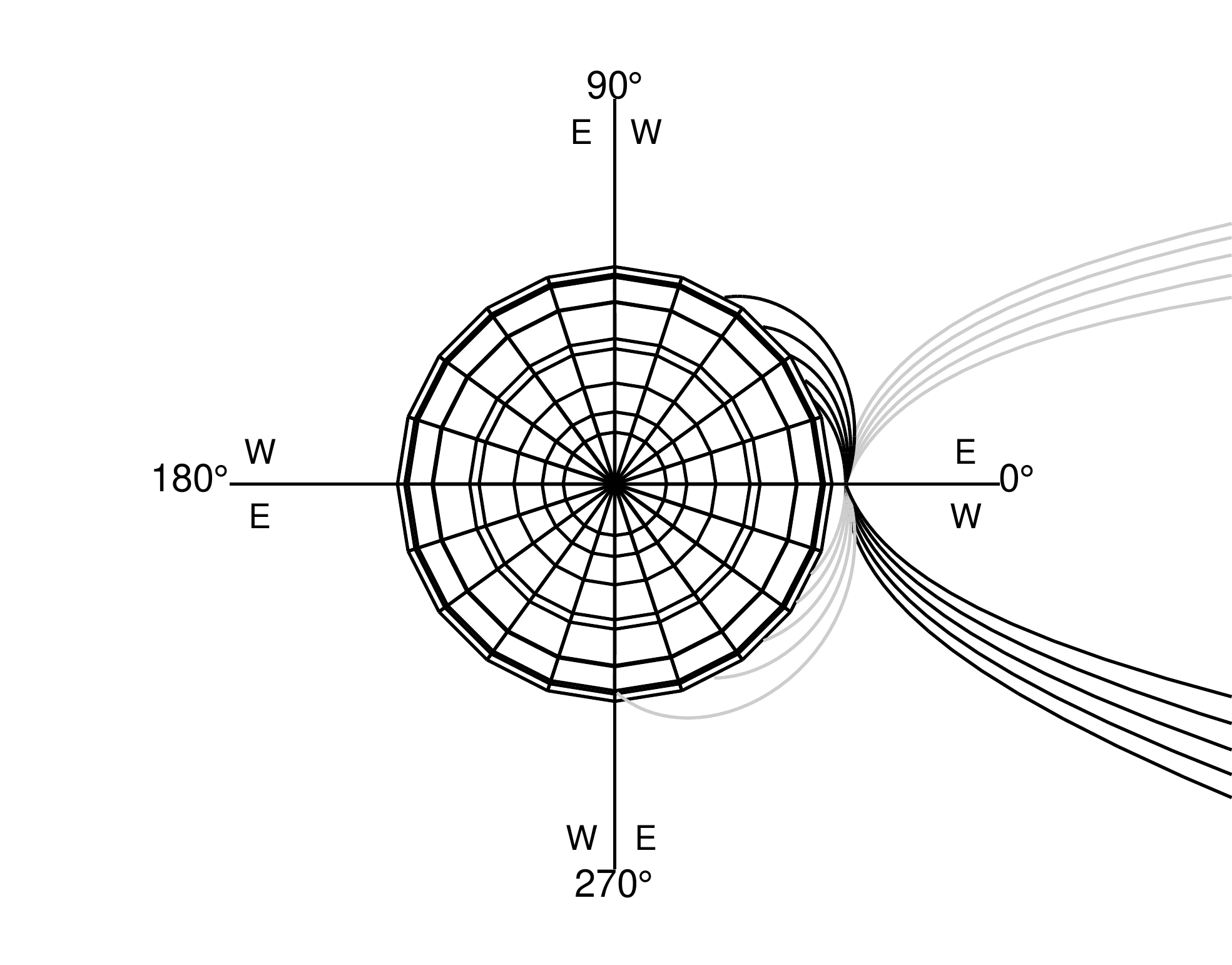}
  \caption{Sample trajectories of 25~GV in-going electrons (in gray) and
    positrons (in black) on the geomagnetic equatorial plane. For any given
    energy, position and altitude above the Earth's surface well defined
    regions can be defined in zenith/azimuth coordinates where only one charge
    sign is allowed.}
  \label{fig:geoelepos}
\end{figure}

The geomagnetic field can be used, under certain conditions, to distinguish
the charge sign without making use of a magnetic spectrometer.
As illustrated in figure~\ref{fig:geoelepos}, for any given energy, position
and altitude above the Earth's surface, one can define specific regions, in
zenith/azimuth coordinates, where only primaries with positive or negative
charge are allowed.
More specifically, there exists a region toward the East where only negative
charges (e.g., electrons) are allowed and a similar (but not exactly specular)
region toward the West where only positive charges (e.g., positrons) are
allowed. These regions become smaller and smaller as the particle energy
increases, up to the point where the angular resolution of the instrument
becomes the limiting factor---determining the upper limit of the energy range
where the measurement is practically feasible.

The basic idea dates back to the late 1960s~\cite{1965PhRvL..15..769D} and
has effectively been used to determine the positron fraction between 10 and
20~GeV using the difference in geomagnetic cutoff for positrons and
electrons from the East and West~\cite{1987ApJ...312..183M}.
More recently the \Fermi-LAT has used the same principle, exploiting a full
particle ray-tracing in the geomagnetic field, to measure the separate electron
and positron spectra between 20 and 200~GeV~\cite{2012PhRvL.108a1103A}.
Instruments planned for the near future are apparently planning to exploit the
same principle~\cite{2014AdSpR..53.1438R}.

We note, in passing, that this concept is more easily applied to separate
electrons and positrons, as opposed to, e.g., protons and antiprotons, because
one needs an accurate energy measurement to propagate the particle into the
geomagnetic field (and, as we shall see in section~\ref{sec:had_cal} that
hadronic calorimetry is a challenging business).

\section{Interaction of radiation with matter}%
\label{sec:interactions}

This section does not contain more information (and in fact it contains a lot
less information) that any of the excellent references floating
around~\cite{PDG,Leo,Knoll}---except maybe for the fact that we made an
effort to put this information in the specific context of the review.

\subsection{Charged particles: energy losses}

Energy losses of charged particles are customarily discussed separately for
heavy (e.g., protons, alpha particles and nuclei) and light (e.g. electrons)
particles---the reason behind that being that radiation losses are, generally
speaking, negligible for the the former. Given that, in principle, all
particles---be they heavy or light---suffer both ionization and radiation
losses, here we take a slightly different approach, focusing our discussion,
at the top level, on the distinction between these two types of losses.

\subsubsection{Ionization losses}

Relativistic charged heavy particles passing through matter loose energy by
ionization at an average rate that is reasonably well described by the Bethe
equation over most of energies we are interested:
\begin{align}\label{eq:stopping_power_full}
  -\left<\frac{dE}{dx}\right>_{\rm ion} \!\!\!\!\!\!
  & = 4\pi N_Ar_e^2m_ec^2 z^2 \frac{Z}{A}\frac{1}{\beta^2}\times
  \nonumber\\
  & \left[\frac{1}{2}\ln \frac{2m_ec^2\beta^2\gamma^2T_{\rm max}}{I^2}
    - \beta^2 -\frac{\delta(\beta\gamma)}{2} \right]
\end{align}
(here $z$, $\beta$ and $\gamma$ refer to the projectile, while $Z$ and $A$
refer to the target material; see \cite{PDG} for a detailed description of
all the other terms in the equation).

Complicated as this might seem, the main features are readily evident if one
rewrites it as
\begin{align}\label{eq:stopping_power}
  -\left<\frac{dE}{dx}\right>_{\rm ion} \!\!\!\!\!\!
  \propto z^2 \frac{Z}{A}\frac{1}{\beta^2}
  \left[ \frac{1}{2} \ln(C_0 \beta^2\gamma^2) - \beta^2 +
    \text{corrections} \right].
\end{align}
The energy loss per unit length is proportional to the $Z/A$ ratio of the
target material and the \emph{square} of the charge $z$ of the incident
particle. When the latter is \emph{slow} ($\beta \ll 1$) the energy losses
decrease as $1/\beta^2 \propto 1/E$ as $\beta$ increases.
In the ultra-relativistic regime ($\beta \approx 1$) the losses increase
as $\ln \beta^2\gamma^2$ (this generally goes under the name of
\emph{relativistic rise} or \emph{logarithmic raise}). These basic facts are
customarily used, as we shall briefly see in the following, for charge
measurement and/or particle identification.

We should note that, since ionization losses are fundamentally due to
interactions between the incoming particle and the atomic electrons of the
medium, there are significant kinematic differences between the two cases of
heavy and light projectiles. In addition to that, quantum-mechanical effects due
to the fact that the projectile and the target are identical particles come
into play in case of electrons. But we are neglecting a whole lot of details
anyway and won't push this any further.

\begin{figure}[htb!]
  \includegraphics[width=\linewidth]{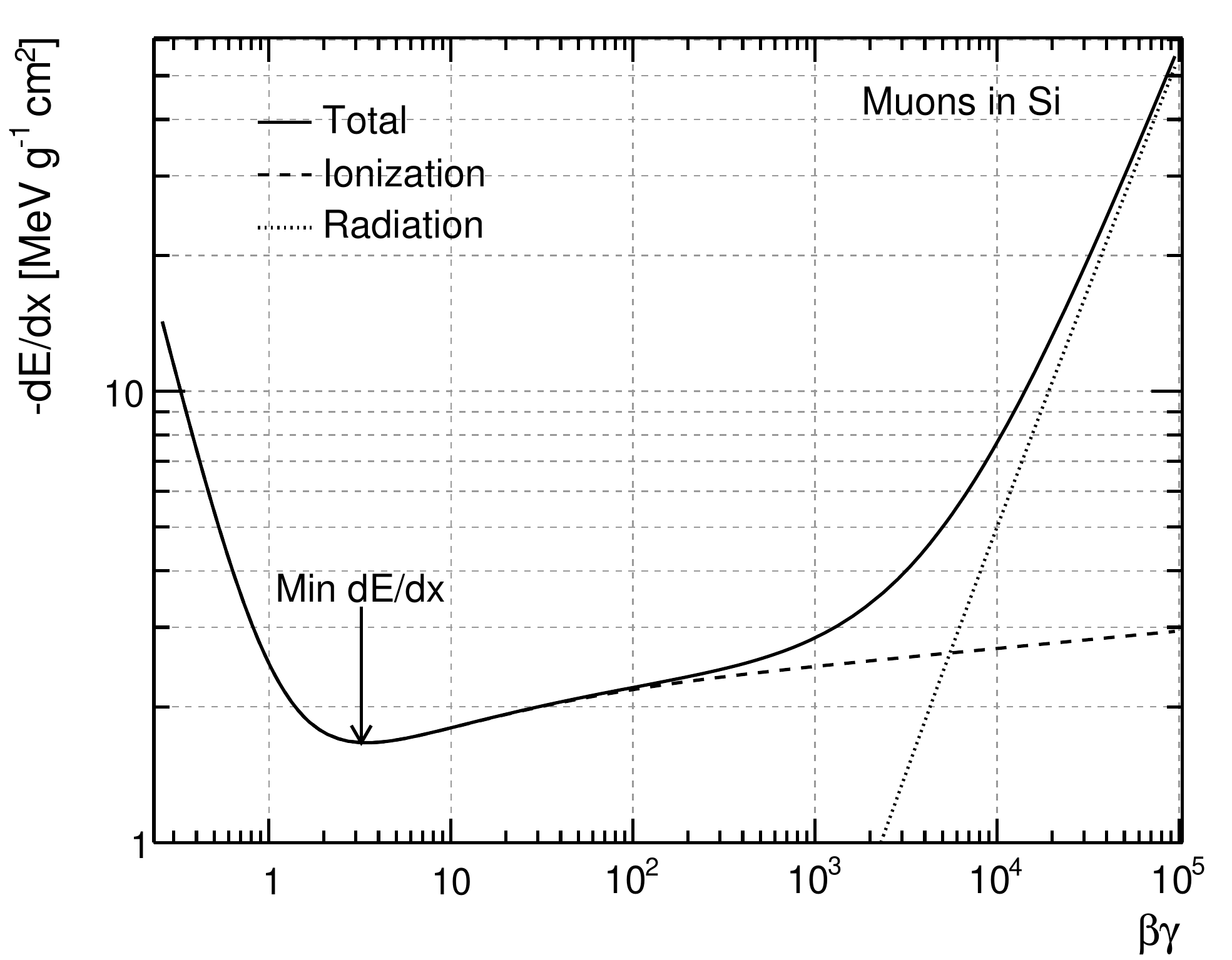}
  \caption{Average energy loss $dE/dx$, as a function of $\beta\gamma$ for
    muons in silicon. The data points are taken from
    \url{http://pdg.lbl.gov/2012/AtomicNuclearProperties/MUON_ELOSS_TABLES/muonloss_014.dat}.}
  \label{fig:stopping_power}
\end{figure}

When plotted as a function of $\beta\gamma$ (i.e., the momentum) of the
incident particle, the energy-loss curve has the typical shape shown in
figure~\ref{fig:stopping_power}, with a relatively broad minimum at
$\beta\gamma \sim 3$, where $dE/dx$ (normalized to the density of the target)
is of the order of 1--2~MeV~g$^{-1}$~cm$^{2}$.
A particle with the energy corresponding to this minimum ionization is
customarily called a \emph{minimum ionizing particle} (MIP).

\subsubsection{Radiation losses and critical energy}

At sufficiently high energy any charged particle radiates. Radiation losses
are nearly proportional to the particle energy
\begin{align}\label{eq:rad_losses}
  -\left<\frac{dE}{dx}\right>_{\rm rad} \!\!\!\!\!\! \propto E,
\end{align}
and, since ionization losses only grow logarithmically, the two loss rates,
when seen as function of energy, are bound to cross each other at some point,
as shown in figure~\ref{fig:stopping_power}. The energy $E_c$ at which this
happens is called \emph{critical energy}. Since the constant of proportionality
in equation~\eqref{eq:rad_losses} scales as $1/m^2$, for any given material
$E_c$ is different for different particles---and it is much higher for heavier
particles. For reference, the critical energy in silicon is $\sim 580$~GeV
for muons and $\sim 40$~MeV for electrons (and in the tens to hundreds of TeV
for protons).

\begin{figure}[htb!]
  \includegraphics[width=\linewidth]{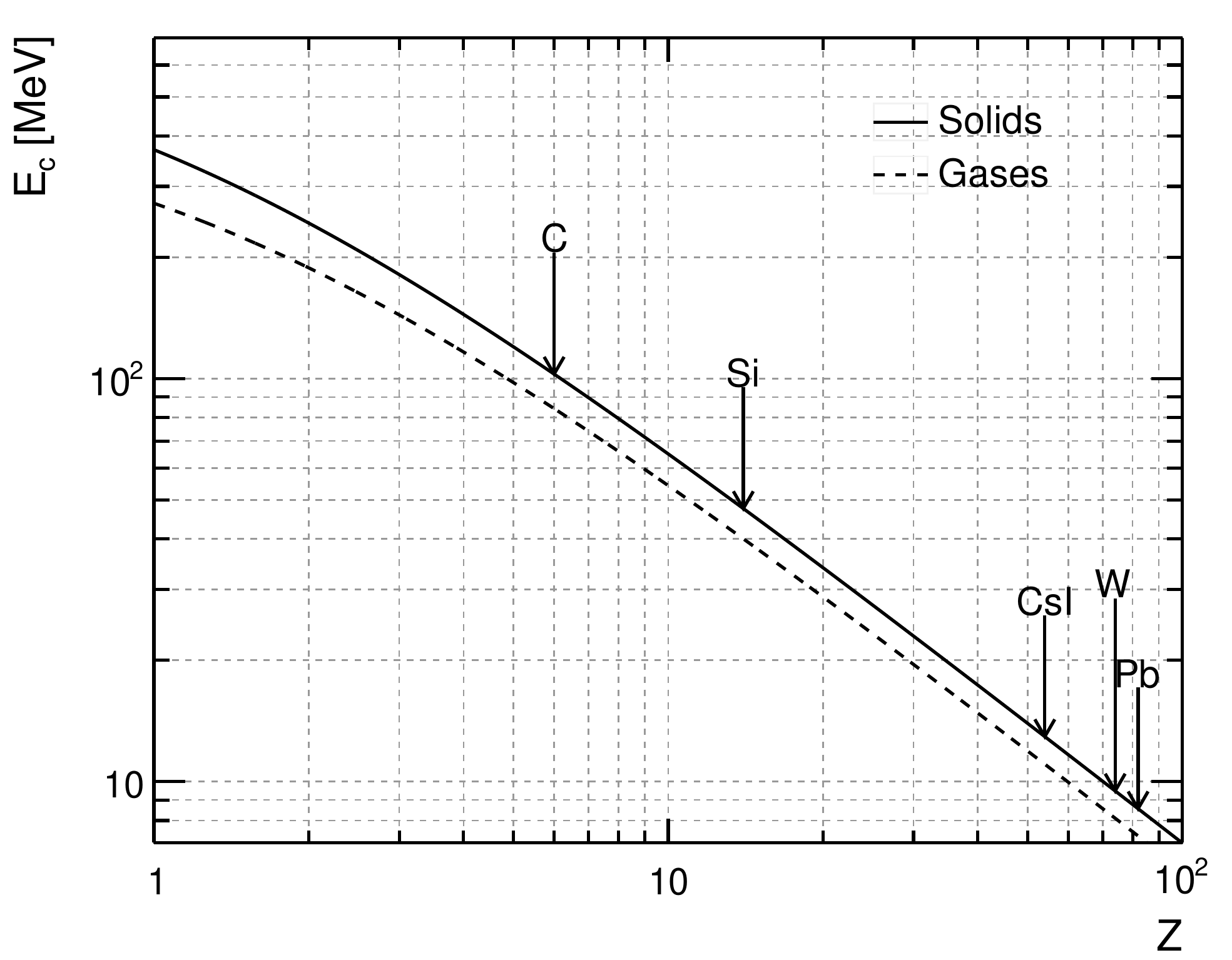}
  \caption{Empirical parameterization of the electron critical energy as a
    function of the atomic number from~\cite{PDG}.}
  \label{fig:critical_energy}
\end{figure}

To all practical purposes radiation losses are mainly relevant for electrons
(and positrons) and it is customary to refer to the critical energy for
electrons as \emph{the} critical energy%
\footnote{We note explicitly that, in a given material, the critical energies
  for electrons and positrons are, in general, slightly different.}%
. A popular empirical approximation for $E_c$ as a function of the atomic
number of the material (for solids) is given by
\begin{align}
  E_c = \frac{710}{Z + 0.92}\ {\rm MeV}
\end{align}
and plotted in figure~\ref{fig:critical_energy}.

\subsubsection{Radiation length}%
\label{sec:radiation_length}

Above the critical energy (i.e. in the regime where radiation losses dominate)
equation~\eqref{eq:rad_losses} can readily integrated in a homogeneous material,
yielding the electron energy as a function of the distance $x$ traversed:
\begin{align}
  E(x) = E_0 \exp\left(-\frac{x}{X_0}\right).
\end{align}
The quantity $X_0$, representing the typical length over which an electron
looses all but $1/e$ of its energy due to \bremss, is called
\emph{radiation length} and is characteristic of the material.
The radiation length is conveniently expressed in g~cm$^{-2}$, i.e.,
factoring out the density of the medium. A popular parameterization is given by
\begin{align}\label{eq:param_radlen}
  X_0 = \frac{716 A}{Z(Z + 1) \ln(287/\sqrt{Z})}~{\rm g\ cm^{-2}}
\end{align}
and table~\ref{tab:exp_radlen} shows numerical values for a few materials of
interest (see also figure~\ref{fig:rad_int_len}). 

\begin{table}[htb!]
  \begin{tabular}{p{0.23\linewidth}p{0.23\linewidth}p{0.23\linewidth}%
      p{0.23\linewidth}}
    \hline
    Material & $X_0$~[g~cm$^{-2}$] & $\density$ [g~cm$^{-3}$] & $X_0$ [cm]\\
    \hline
    \hline
    Pb & 6.37 & 11.350 & 0.561\\
    BGO & 7.97 & 7.130 & 1.12\\
    CsI & 8.39 & 4.510 & 1.86\\
    W & 6.76 & 19.3 & 0.350\\
    C (graphite) & 42.70 & 2.210 & 19.3\\
    Si & 21.82 & 2.329 & 9.37\\
    Air & 36.62 & $1.2 \times 10^{-3}$ & 30,500\\
    \hline
  \end{tabular}
  \caption{Tabulated values of the radiation length for some materials of
    interest.}
  \label{tab:exp_radlen}
\end{table}

The radiation length is the natural scale for all the most relevant
electromagnetic phenomena we shall deal with in the following: multiple
scattering, electron \bremss, pair production and electromagnetic
showers. We shall customarily indicate with the letter $t = x/X_0$ any
distance measured in units of $X_0$.

\subsection{Multiple Coulomb scattering}%
\label{sec:inter_mcs}

A charged particle traversing a medium undergoes multiple Coulomb scatterings
on the atomic nuclei of the material. In the so-called gaussian approximation
the root mean square of the deviation angle projected in any of the planes
containing the incoming particle direction can be parameterized as
\begin{align}\label{eq:theta_ms}
  \thetamsp =
  \frac{0.0136~{\rm GeV}}{\beta c p}z\sqrt{t} (1 + 0.038 \ln t)~\text{rad},
\end{align}
where $\beta c$, $p$ and $z$ are the velocity, momentum and charge number
of the incoming particle, and $t$ is the thickness of the traversed
material in units of radiation lengths.

The deflection space angle being the sum in quadrature of the independent
deflections on two orthogonal planes, it is simply given by
\begin{align}
  \thetamss = \sqrt{2}\thetamsp.
\end{align}

Whether one or the other is more relevant depends on the problem at hand
(e.g., typically the pointing accuracy of a detector is parameterized in terms
of the space angle, but in a magnetic spectrometer the important figure is
really the deflection angle in the bending plane). In the following we shall use
both and we shall try and make clear which one we are referring to.

\begin{figure}[htb!]
  \includegraphics[width=\linewidth]{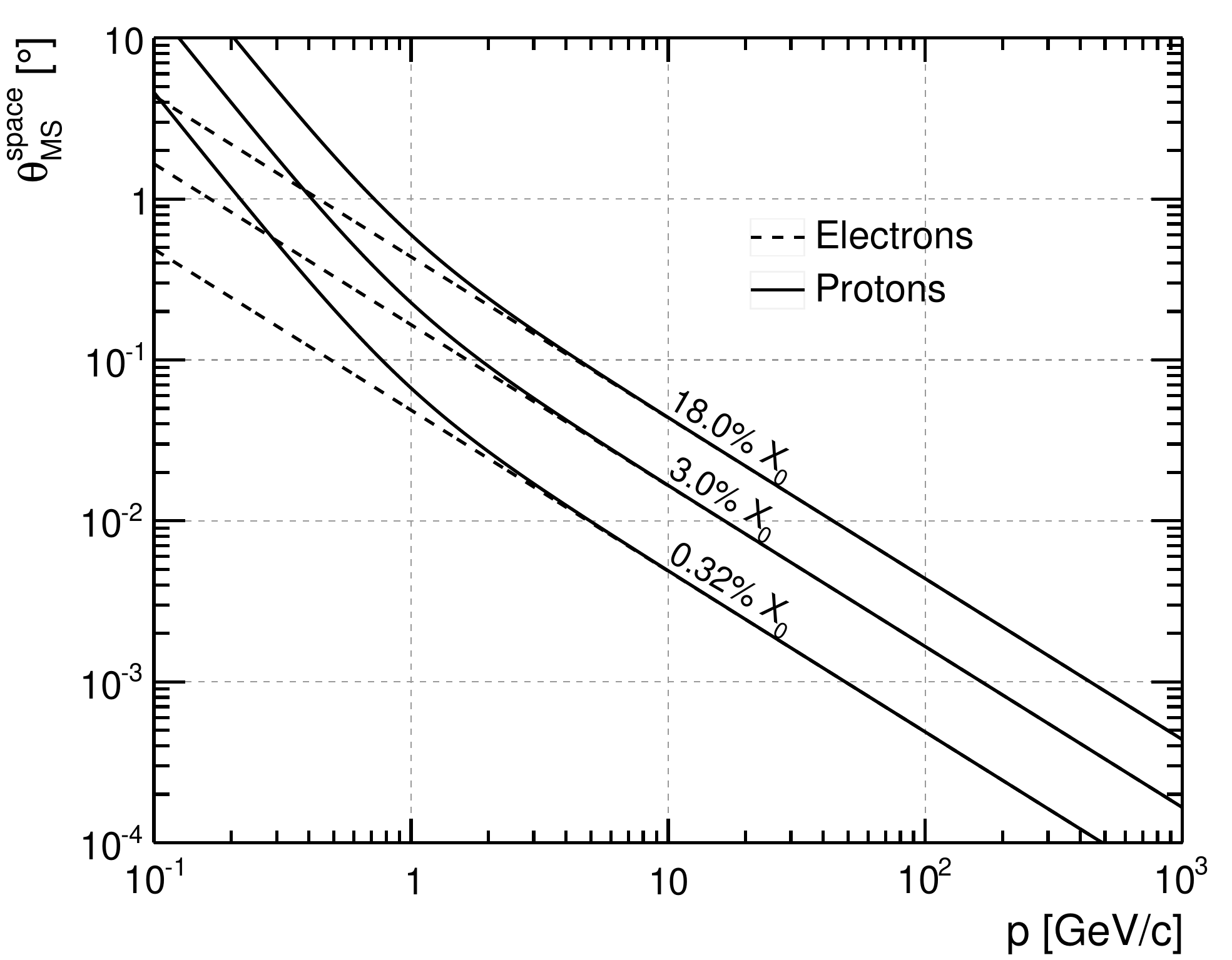}
  \caption{Multiple scattering angle, as a function of the momentum of the
    incoming particle, for three sample values of $t$, for electrons and
    protons. The three values of $t$ corresponds to $300~\mu$m of silicon
    (the thickness of the AMS-02 silicon detectors) and the \Fermi-LAT thin
    and thick radiators.}
  \label{fig:multiple_scattering}
\end{figure}

Figure~\ref{fig:multiple_scattering} shows the (space) multiple scattering
deflection angles for three sample values of $t$, for electrons and protons, as
a function of their momentum. In the ultra-relativistic regime
$\beta \approx 1$ (i.e. above a few MeV for electrons and above a few GeV for
protons) the curves are indistinguishable and fall as $1/p$. The difference
below 1~GeV has implication for the momentum resolution in a magnetic
spectrometer and we shall come back to it in section~\ref{sec:ams_mdr}.

\subsection{High-energy photons}%
\label{sec:inter_gammas}

High-energy photons (above a few tens of MeV) interact with matter mainly by
pair production in the field of atomic nuclei. As it turns out, the
processes of electron \bremss\ and pair production are intimately
related to each other, and the mean free path for pair production for a
high-energy photon is given by
\begin{align}\label{eq:lambda_pair}
  \lambda_{\rm pair} = \frac{9}{7}X_0
\end{align}
(i.e., it is $9/7$ of the scale length over which a high-energy electron
looses all but $1/e$ of its energy).

It should be noted that, while pair production is a destructive process
(the gamma-ray no longer exists afterwards), \bremss\ only degrades the
electron energy. As we shall see in a second, these two physical processes are
at the base of the development of electromagnetic showers.

\subsubsection{More on the pair production}%
\label{sec:more_pair_production}

The $e^+/e^-$ pair production by high-energy gamma rays is the basic physical
process used in pair conversion telescopes. It is pretty much conventional
wisdom that the kinematic of the process is closed and the tracks of the
electron and the positron can in principle be combined to recover \emph{exactly}
the original photon direction. It turns out that, even neglecting the finite
detector resolution, this is not quite true, and there are a few subtleties
involved that we briefly mention in this section.

The average opening angle of the electron-positron pair scales as
\begin{align}
  \theta_{\rm open} \propto \frac{m_ec^2}{E},
\end{align}
where $E$ is the photon energy. Depending on $E$ and on the layout of the
tracking detectors, this angle might be too small to be resolved, in which case
the two tracks effectively overlap (i.e., from an experimental standpoint, one
really sees one track).

At low enough energy, where the opening angle is relatively large and the tracks
are well separated, in order to recover the original photon direction one
should in principle combine the momenta of the two particles
\emph{covariantly} which brings up the question of the track energies.
The differential distribution in the fractional electron (or positron) energy
is relatively flat%
\footnote{The distribution is actually slightly peaked at $0$ and $1$, and more
and more so the highest the energy.}
so that asymmetric (energy-wise) pairs are quite common---if you think about,
this implies that the average energy of the particle in the pair with the
\emph{highest} energy is approximately $3/4$ of the original photon energy.
Now, depending on the experimental setup, one might measure the single track
energies with good or bad resolution---or just not measure them at all. As one
might imagine, this has profound implications on the maximum attainable angular
resolution for pair conversion telescopes.

There is one last ingredient to the mix that we haven't mentioned yet, namely
the nucleus recoil (remember, typically gamma rays pair-produce in the field
of the nucleus). In practical applications this is largely irrelevant, as one
is usually limited by the multiple scattering in the low-energy regime, as
we shall see in section~\ref{sec:lat_psf}. When this is not true the nucleus
recoil (which is essentially impossible to measure) becomes the ultimate
limiting factor to the accuracy with which one can reconstruct the photon
direction. It is shown in~\cite{2014APh....59...18H} that this \emph{kinematic}
limit $\theta_{\rm kl}$ is approximately given by
\begin{align}
  \theta_{\rm kl} \sim 5^\circ \left( \frac{10~{\rm MeV}}{E} \right)
\end{align}
and this figure is possibly relevant when designed gamma-ray detectors
optimized for the low-energy end of the pair production regime.

\subsection{Electromagnetic showers}%
\label{sec:em_showers}

As explained in the previous section, high-energy electrons and photons
produce in matter secondary photons by \bremss\ and electron-positron
pairs by pair production. These secondaries, in turn, can produce other
particles with progressively lower energy and start an
\emph{electromagnetic shower} (or \emph{cascade}).
The process continue until the average energy of the electron component falls
below the critical energy of the material---at which point the rest of the
energy is released via ionization.

\begin{figure}[!htb]
  \includegraphics[width=\linewidth]{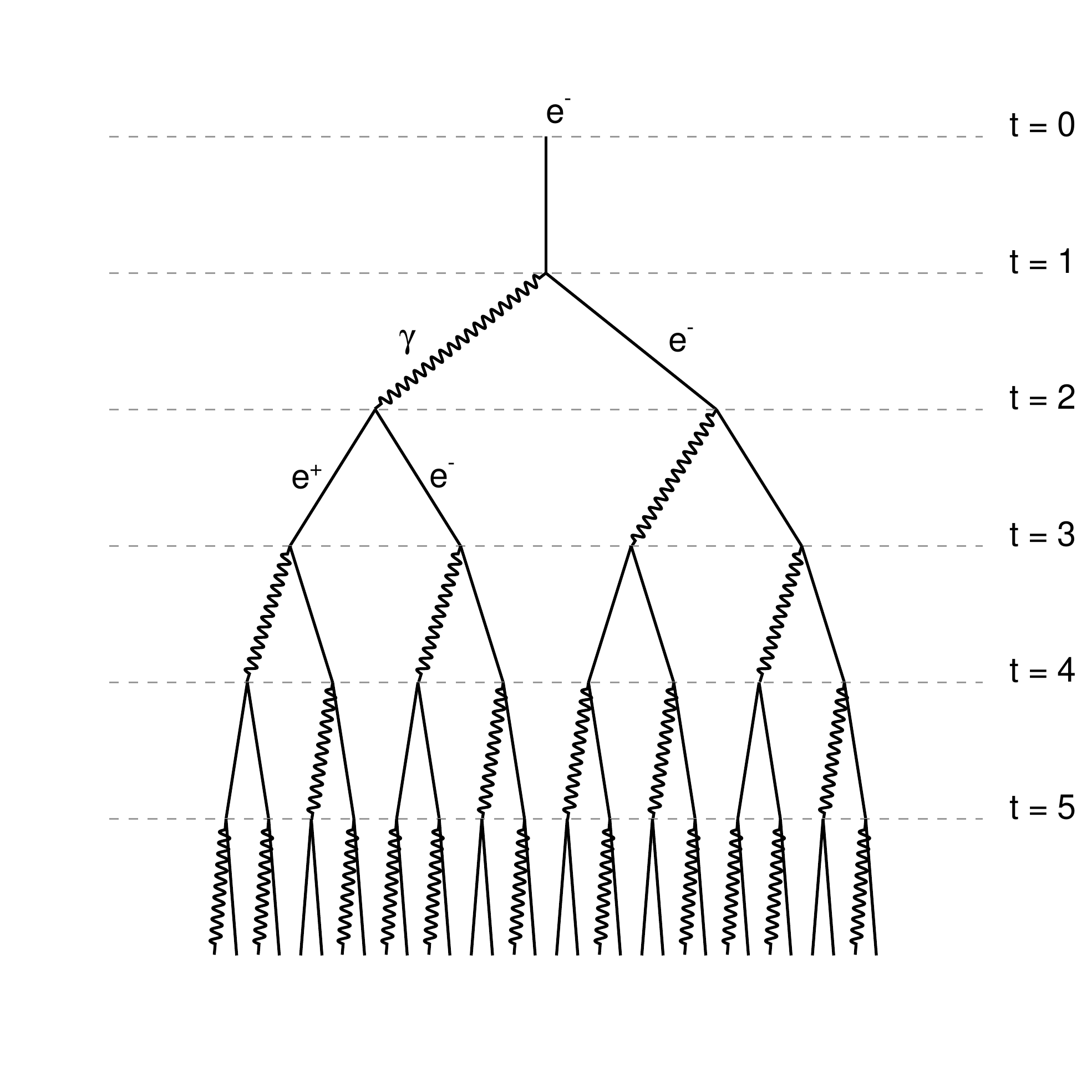}
  \caption{Sketch of the development of an electromagnetic shower in the 
    simplest possible toy model.}
  \label{fig:toy_em_shower}
\end{figure}

A simple toy model for an electron-initiated shower is schematically
represented in figure~\ref{fig:toy_em_shower}, where after $t$ radiation lengths
of material, the cascade has developed in $2^t$ particles (a mix of electrons,
positrons and photons) with average energy $E_0/2^t$. Rough as this model is
(for one thing we neglect the stocasticity of the process altogether, along
with the $9/7$ in~\eqref{eq:lambda_pair} and the fact that the electron
can---and does---radiate multiple photons of different energy per radiation
length) it is able to reproduce one of the main features of electromagnetic
showers, namely the fact that the position of the shower maximum scales
logarithmically with the energy. In fact the stop condition reads
\begin{align}
  \frac{E_0}{2^{t_{\rm max}}} \sim E_c,
\end{align}
and hence
\begin{align}
  t_{\rm max} \sim \ln \left(\frac{E_0}{E_c}\right).
\end{align}

At a slightly higher level of sophistication, the longitudinal profile of an
electromagnetic shower can be effectively described as
\begin{align}\label{eq:long_profile}
  \frac{dE}{dt} = E_0 b \frac{(bt)^{a-1} e^{-bt}}{\Gamma(a)},
\end{align}
where $a$ and $b$ are parameters related to the nature of the incident particle
(electron or photon) and to the characteristics of the medium ($b = 0.5$
is a reasonable approximation in many cases of practical interest).
The position of the shower maximum occurs at
\begin{align}\label{eq:shower_max}
  t_{\rm max} = \frac{(a - 1)}{b} \approx
  \ln \left(\frac{E_0}{E_c}\right) + t_0
\end{align}
where $t_0 = -0.5$ for electrons and $t_0 = 0.5$ for photons. For the sake
of clarity, the way one typically uses these relation is to plug $E_0$, $E_c$
and $t_0$ in \eqref{eq:shower_max} to find $a$ (assuming $b = 0.5$) and then
use~\eqref{eq:long_profile} to describe the longitudinal profile of the shower.

\begin{figure}[htb]
  \includegraphics[width=\linewidth]{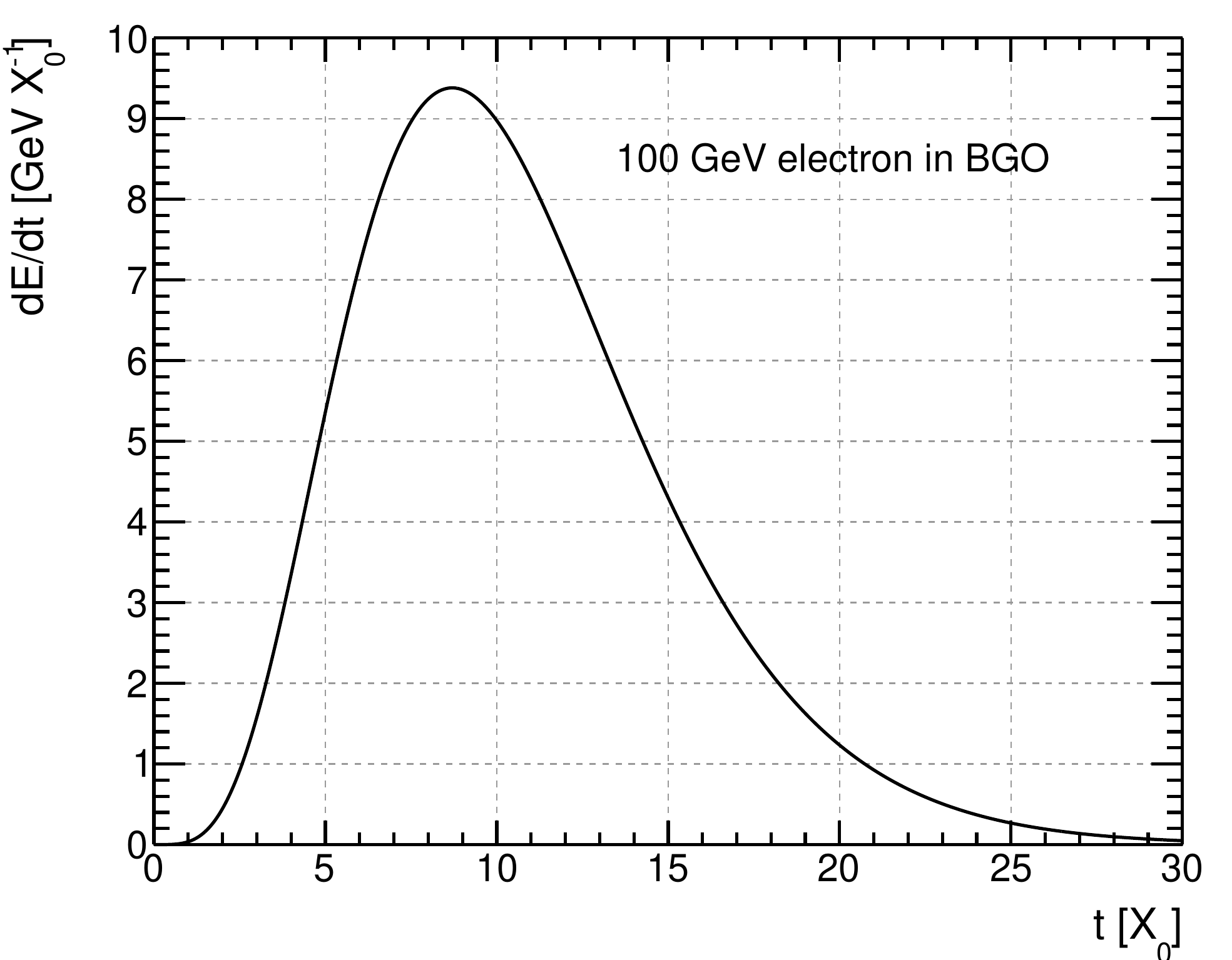}
  \caption{Average longitudinal profile of the shower generated by a
    $100$~GeV electron in a homogeneous slab of BGO.}
  \label{fig:shower_profile}
\end{figure}

Figure~\ref{fig:shower_profile} shows the average shower profile for
$100$~GeV in BGO. With a critical energy of $10.1$~MeV, the position of
the shower maximum is located, according to equation~\eqref{eq:shower_max},
at $8.7~X_0$.

We mention, in passing, that the transverse development of an electromagnetic
shower, mainly due to the multiple scattering of electrons and positrons away
from the shower axis, scales with the so-called the Moli\`ere radius,
that can be empirically parameterized as
\begin{align}
  R_M \approx \frac{21 X_0}{E_c~{[\rm MeV]}}~{\rm MeV}.
\end{align}

\subsection{Hadronic showers}%
\label{sec:had_showers}

While the development of electromagnetic showers is determined, as we have seen,
by two well-understood QED processes, the energy degradation of hadrons proceeds
through both strong and electromagnetic interactions in the medium.
The complexity of the hadronic and nuclear processes produce a multitude of
phenomena that make hadronic showers intrinsically more complicated than the
electromagnetic ones. In other words for hadronic showers there is no such a
thing as the toy model illustrated in figure~\ref{fig:toy_em_shower}.

\begin{figure}
  \includegraphics[width=\linewidth]{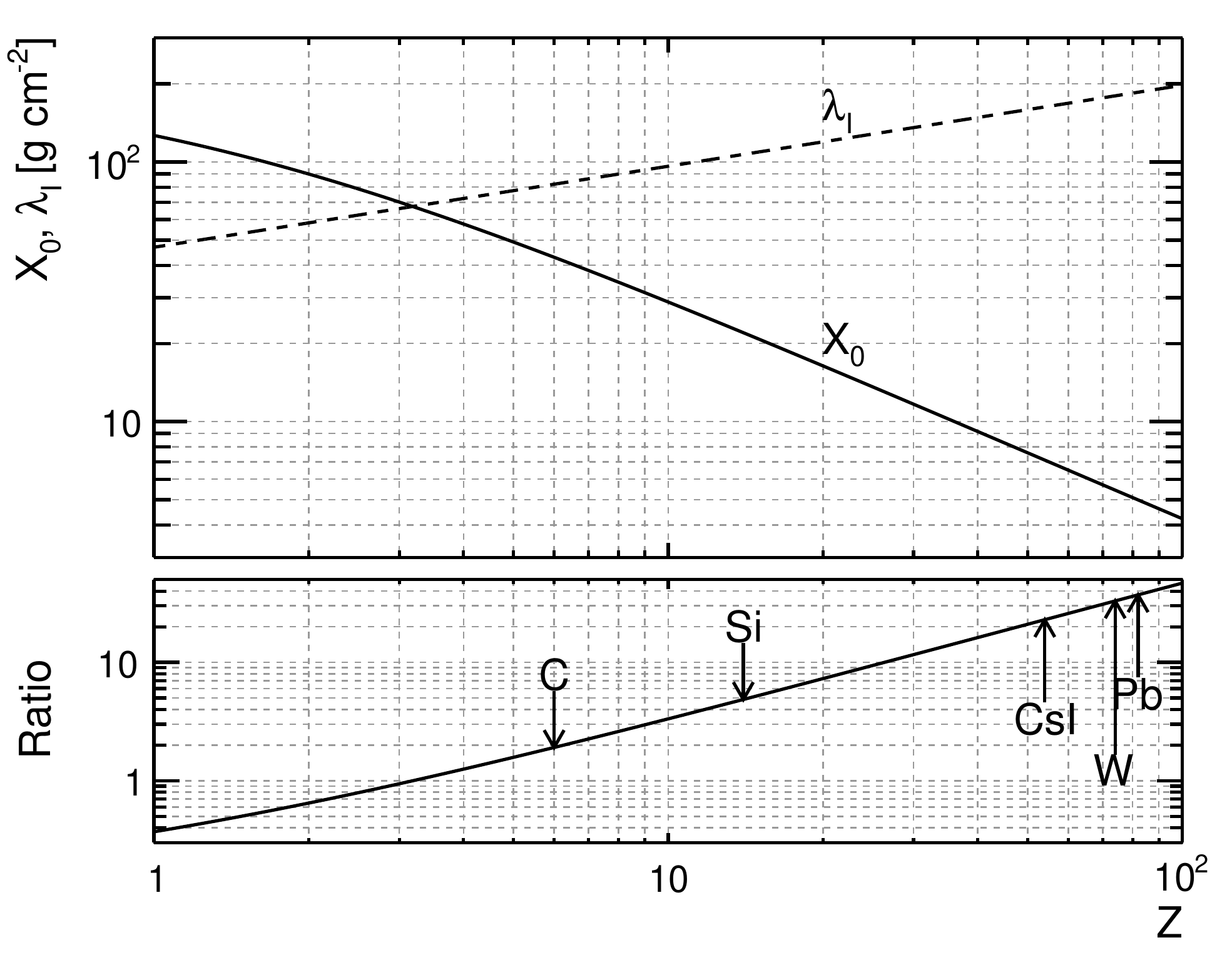}
  \caption{Approximate dependence of the radiation length $X_0$ and the
    nuclear interaction length $\lambda_I$ on the atomic number $Z$ of the
    material. The parameterizations used are those in equations
    \eqref{eq:param_radlen} and \eqref{eq:param_intlen}, with the additional
    assumption $A = 2Z$.}
  \label{fig:rad_int_len}
\end{figure}

As a matter of fact hadronic showers consist in general of two distinctly
different components: (i) an electromagnetic component due to $\pi^0$ and
$\eta$ generated in the absorption process and decaying into photons
(which in turn develop electromagnetic showers) before they have a chance to
undergo a new strong interaction; and (ii) a non-electromagnetic component,
which combines essentially everything else that takes place in the absorption
process.
Most importantly, the two components evolve with different length scales: the
radiation length $X_0$ for the first and the nuclear interaction length
$\lambda_I$ (which is at least on order of magnitude larger for $Z > 30$, as
shown in figure~\ref{fig:rad_int_len}) for the second. A useful
parameterization for $\lambda_I$ is
\begin{align}\label{eq:param_intlen}
  \lambda_I = 37.8 A^{0.312}~\text{g~cm}^{-2}
\end{align}
and the actual values for some relevant materials are shown in
table~\ref{tab:exp_intlen}.

\begin{table}[htb!]
  \begin{tabular}{p{0.23\linewidth}p{0.23\linewidth}p{0.23\linewidth}%
      p{0.23\linewidth}}
    \hline
    Material & $\lambda_I$~[g~cm$^{-2}$] & $\density$ [g~cm$^{-3}$] &
    $\lambda_I$ [cm]\\
    \hline
    \hline
    Pb & 199.6 & 11.350 & 17.6 \\
    BGO & 159.1 & 7.130 & 22.3 \\
    CsI & 171.5 & 4.510 & 38.0\\
    W & 191.9 & 19.3 & 9.94\\
    C (graphite) & 85.8 & 2.210 & 38.8\\
    Si & 108.4 & 2.329 & 46.5\\
    Air & 90.1 & $1.2 \times 10^{-3}$ & 75,000\\
    \hline
  \end{tabular}
  \caption{Tabulated values of the nuclear interaction length for some materials
    of interest.}
  \label{tab:exp_intlen}
\end{table}

The fractional non-electromagnetic component of a hadronic shower $F_h$ (as
opposed to electromagnetic component $F_e = 1 - F_h$) decreases with energy and
is generally~\cite{PDG} parameterized as
\begin{align}
  F_h(E) = \left( \frac{E}{E_0} \right)^{k-1},
\end{align}
where typical values are $E_0 \approx 1$~GeV and $k \approx 0.8$ (roughly
speaking, the fraction of energy the non-electromagnetic component accounts
for is of the order of $50\%$ at $100$~GeV and $30\%$ at $1$~TeV).

In broad terms, hadronic showers tend to start developing at relatively large
depths in the material and they are typically larger and more irregular when
compared with electromagnetic showers. As we shall see in the following, all
these differences are customarily used in modern space-based imaging
calorimeters for particle identification---particularly to discriminate
electrons and photons against the much larger proton background.

\subsection{\cheren\ radiation}%
\label{sec:cherenckov_rad}

\begin{figure}[htb]
  \includegraphics[width=\linewidth]{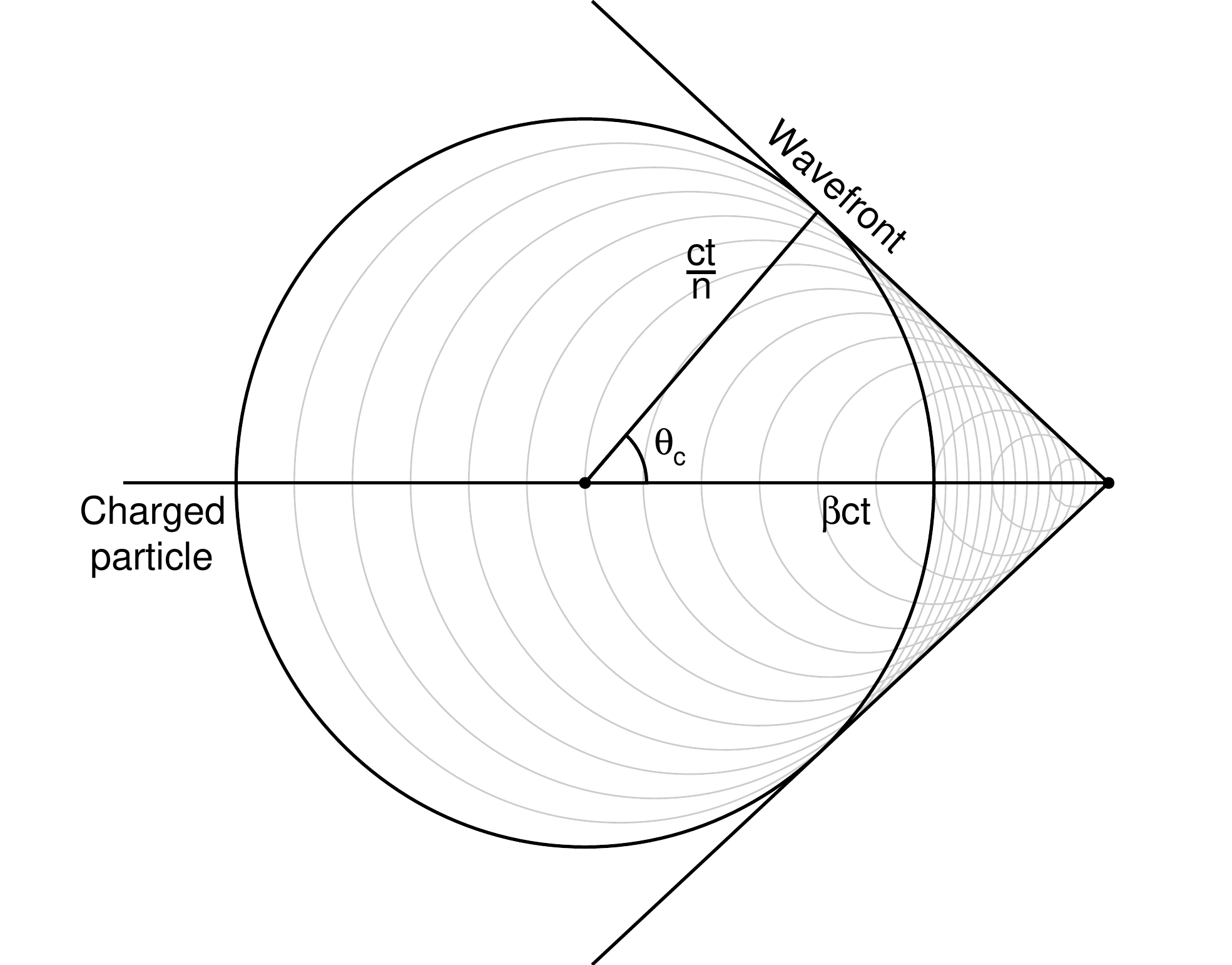}
  \caption{Sketch of the geometry relevant for the \cheren\ effect.}
  \label{fig:cherenkov_sketch}
\end{figure}

\cheren\ radiation is emitted when a charged particle moves in a medium at 
a speed greater than the speed of light \emph{in that medium}
\begin{align}
  \beta > \frac{1}{n}.
\end{align}
(Here $\beta$ refers to the incident particle and $n$ is the index of
refraction of the material.) In the ideal case of a non-dispersive medium, the
\cheren\ wave front form an acute angle with respect to the particle velocity
given by
\begin{align}
  \cos\theta_c = \frac{1}{n\beta}
\end{align}
as sketched in figure~\ref{fig:cherenkov_sketch}.

The (double differential) spectrum of the photons produced per unit path length
and wavelength is given by
\begin{align}\label{eq:cherenkov_loss}
  \frac{d^2N}{dxd\lambda} &= \frac{2\pi\alpha z^2}{\lambda^2}
  \left(1 -  \frac{1}{\beta^2 n^2(\lambda)}\right) = \nonumber\\
  &= \frac{2\pi\alpha z^2}{\lambda^2}
  \left(1 - \frac{1 + \beta^2\gamma^2}{\beta^2\gamma^2n^2(\lambda)}\right),
\end{align}
where $z$ is the charge of the projectile, and the index of refraction $n$
is evaluated at the generic photon wavelength $\lambda$.
Equation~\eqref{eq:cherenkov_loss} cannot be readily translated into a detector
signal as, in practice, one has to convolve it with the response of the
transducer and integrate over the photon wavelengths of interest.
That all said, plotting~\eqref{eq:cherenkov_loss} for a reference value of
the index of refraction is a useful illustrative exercise.

\begin{figure}[htb]
  \includegraphics[width=\linewidth]{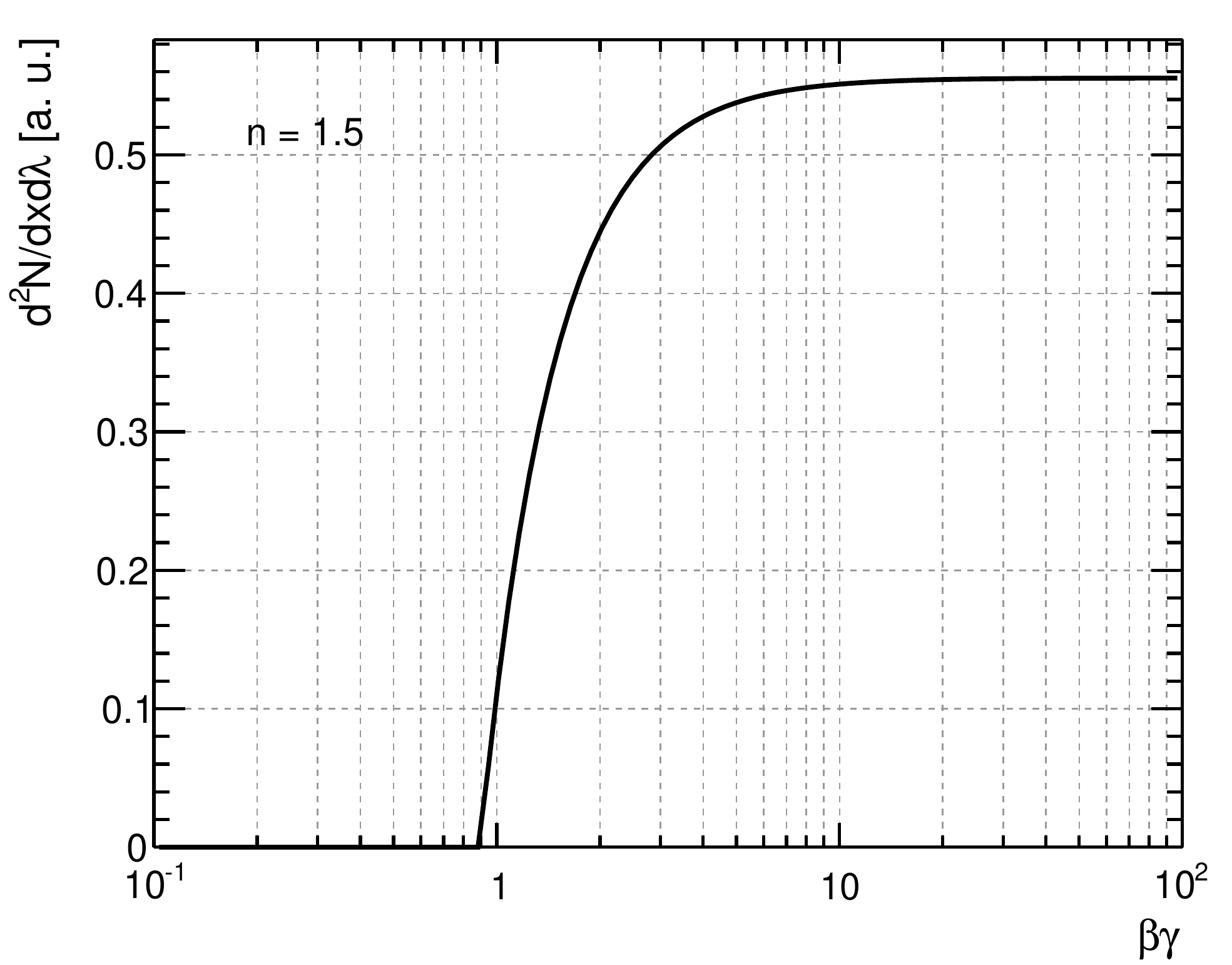}
  \caption{Number of \cheren\ photons emitted per unit path length
    and wavelength as a function of the $\beta\gamma$ of the projectile,
    for a reference index of refraction $n = 1.5$.}
  \label{fig:cherenkov_betagamma}
\end{figure}

While, strictly speaking, the \cheren\ radiation is generally not important in
terms of energy losses, we shall see in the following that its basic properties
(the existence of a threshold and the dependence of $\theta_c$ and
$d^2N/dxd\lambda$ on the velocity of the particle) are customarily exploited
in high-energy physics for particle identification and velocity measurements.
In particular figure~\ref{fig:cherenkov_betagamma} shows that the steep slope
near the threshold potentially allows to achieve a good velocity or momentum
resolution in that region (while the amount of radiation saturates at higher
energies).

\subsection{Transition radiation}%
\label{sec:transition_rad}

Transition radiation is emitted when a ultra-relativistic particle crosses the
interface between two media with different indices of refraction.
As we shall see in the following, in practical implementations the useful
(i.e., detectable) photons are in the x-ray band, which requires the
incidence particle to have a relativistic $\gamma$ factor of the order of
$10^3$.

For a single interface the fractional energy emitted into x-rays above a given
energy $\hbar\omega_0$ is given by
\begin{align}\label{eq:tr_yield}
  F(\hbar\omega > \hbar\omega_0) = 
  \frac{\alpha z^2}{\pi}\left[
    \left( \ln\frac{\gamma\omega_p}{\omega_0} - 1\right)^2 +
    \frac{\pi}{12}
    \right],
\end{align}
where $\omega_p$ is the plasma frequency of the radiation. If $\gamma$ is
big enough it effectively scales as
\begin{align}
  F(\hbar\omega > \hbar\omega_0) \propto z^2 \ln\gamma.
\end{align}
This form of~\eqref{eq:tr_yield} highlights the two main features of the 
transition radiation---namely: it scales as $z^2$ with the charge of the
particle and it grows with $\gamma$.

Since, in practical situations, $F$ is small, transition radiation detectors
typically exploit multiple boundary crosses to enhance the signal. In this case 
interference effects lead to a saturation effect at a value $\gamma$ that,
depending on the actual design, ranges from a few~$10^3$ to $\sim 10^5$.

\section{Experimental Techniques}%
\label{sec:techniques}

In this section we briefly review some of the most widely used experimental
techniques (mind this largely relies on the content of the previous section).

\subsection{Tracking detectors}

The simplest---and yet most widely used in practical
implementations---configuration for a tracking stage of a modern experiment is
a stack of parallel detection planes. The basic building block of such an
instrument is a position-sensitive detector, i.e., a device capable of
measuring the position of passage of the particle---in one or both dimensions.

At the top level, the two main figures of merit of a tracking detectors
are the tracking efficiency (i.e., the efficiency of correctly reconstructing
the track of the incoming particle) and the spatial and/or angular resolution.
At a more fundamental level, these performance figure derive largely
from the hit efficiency and resolution of the position-sensitive detector used,
though in practice there are many more considerations involved (e.g., we shall
see in sections~\ref{sec:lat_psf} and \ref{sec:ams_mdr} two examples of
interplay between the hit resolution and the multiple scattering in practical
tracking applications.)

Other relevant figures of merit of tracking systems are, in no particular order:
triggering capabilities, time response, dead time per event, radiation
tolerance, robustness, reliability, power consumption.

\subsubsection{Position-sensitive detectors}

There are literally too many position-sensitive detectors used in tracking
applications to list them here, and we refer the reader to~\cite{PDG} and
references therein for a comprehensive review.
Roughly speaking, most tracking detectors exploit the ionization produced by
charged particles as they traverse matter. In order for the electron-ion
(or electron-hole) pairs not to recombine immediately a suitable electric field
is necessary---and drifting under the effect of this electric field the charges
induce on the readout electrodes a signal that is generally amplified,
processed and digitized by some form of readout electronics.

At a microscopic level, the amount of primary free charge that is created as a
particle traverse a medium is determined by the effective ionization energy.
This determines the counting statistic of the process and, as a consequence,
the energy resolution (in applications where this is relevant). The average
ionization energy is smaller, e.g., for semiconductor detectors relative to
gaseous detectors, though it should be noted that in the latter case avalanche
processes can be exploited to multiply the primary ionization
\emph{in the detector medium}.

The drift velocity of the ionization toward the readout electrodes is another
fundamental characteristic of a position-sensitive detector, as it determines
the time profile of the charge signal at the input of the readout electronics.
In broad terms, gas detectors are generally slower than semiconductor
detectors.

The hit efficiency is defined as the efficiency of signaling the passage of a
(minimum ionizing) particle and usually runs not too much below $100\%$ within
the active area for a decent modern detector. (We note, in passing, that it
does not really make sense to quote an efficiency without quoting the
corresponding noise level, or the rate of spurious hits. Here we really mean
``$\sim 100\%$ hit efficiency at a reasonable noise level'' whatever than means
in any specific context.) Hit resolutions of $\sim 10~\mu$m are not uncommon
in modern silicon-strip trackers.

The positional measurement capabilities are usually achieved by segmenting
the readout electrode(s), e.g., in strips or pixels. We note, in passing,
that the number of readout channels in a pixel detector scales as the square
of the active surface (as opposed to strip detectors, for which the scaling is
linear), and therefore pixel detectors are not particularly fashionable for use
in space due to power constraints (there are, however, exception, see
e.g.~\cite{2007NIMPA.570..286P,2003ICRC....4.1857Z}).
The \emph{hit resolution} denotes the resolution on the point where the
particle hit the detector and it is obviously related to the level of
segmentation of the sensor. For a strip detector with pitch $p$ it is given by
\begin{align}
  \sigma_{\rm hit} = \frac{p}{\sqrt{12}}
\end{align}
(i.e., the standard deviation of a uniform continuous distribution) if a single
strip is hit and can be significantly better is more strips are hit and
the pulse-height information is available to baricenter the position.

\subsubsection{Silicon detectors}%
\label{sec:silicon_detectors}

\begin{figure}[htb!]
  \includegraphics[width=\linewidth]{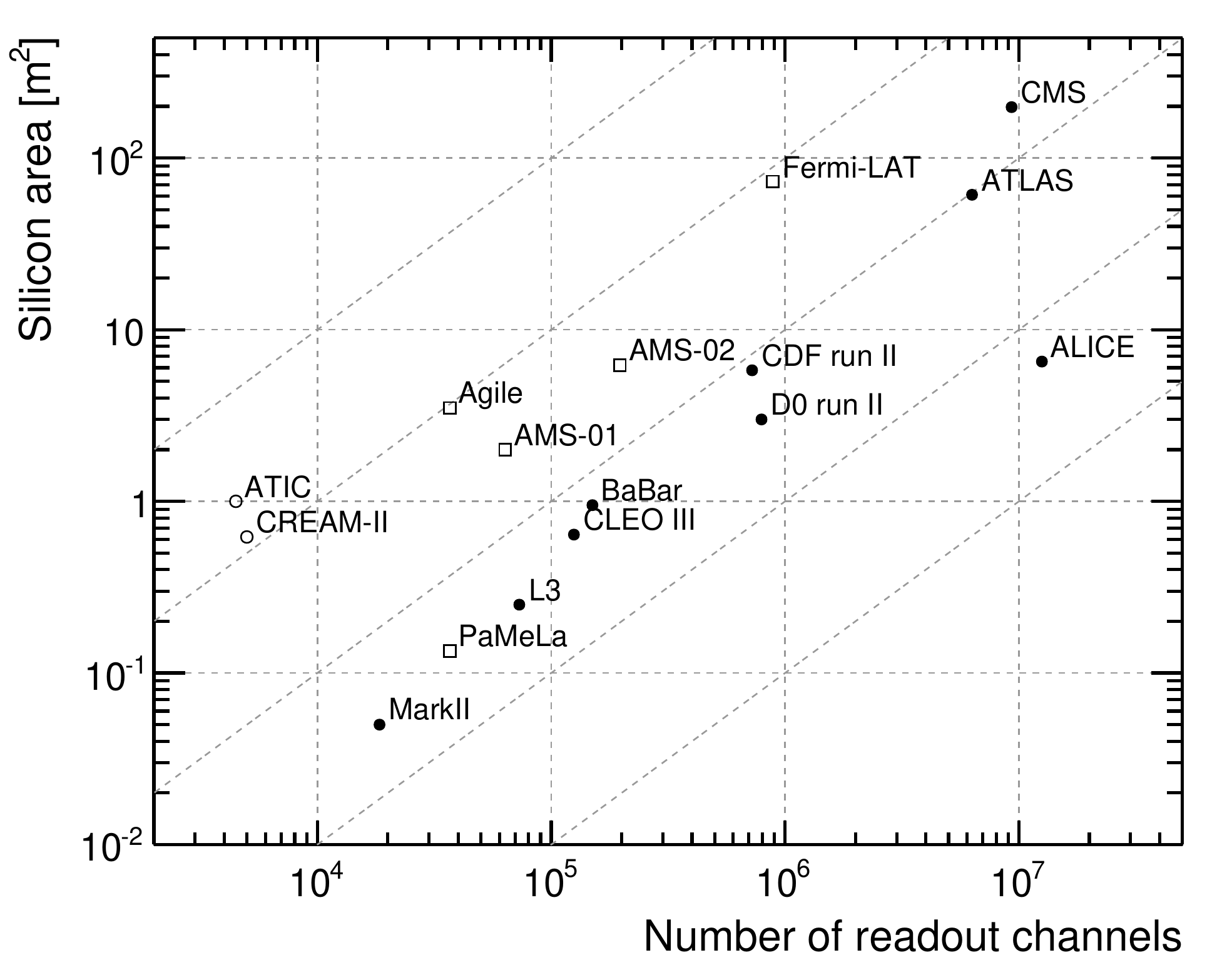}
  \caption{Silicon surface vs. number of readout channels for a compilation
    of tracking detectors operated at accelerators, in space and on
    balloons.}
  \label{fig:si_detector_size}
\end{figure}

Semiconductor detectors---and very especially silicon detectors---are widely
used in modern high-energy physics experiments (see
figure~\ref{fig:si_detector_size}), as integrated circuit technology allows
the realization on large scale of high-density electrode structures, with
excellent spatial resolution and very fast intrinsic response time.
All the last-generation space-based experiments where state-of the-art tracking
capabilities are needed exploit silicon (strip, mostly) detectors
(see, e.g., table~\ref{tab:exp_si_trackers}). In this section we shall
briefly review some of the basic related concepts and we refer the reader
to~\cite{2012NIMPA.666...25H} for a comprehensive review of silicon tracking
detectors in high-energy physics.

\begin{table}[htb!]
  \begin{tabular}{p{0.2\linewidth}p{0.18\linewidth}p{0.17\linewidth}%
      p{0.18\linewidth}p{0.17\linewidth}}
    \hline
    Experiment & Area [m$^2$] & Channels & Pitch [$\mu$m] & Power [W]\\
    \hline
    \hline
    AGILE & 3.5 & 36,864 & 242 & 15\\
    AMS-02 & 6.2 & 196,608 & 110/208 & 140\\
    \Fermi-LAT & 73.0 & 884,736 & 228 & 160\\
    PaMeLa & 0.134 & 36864 & 50/50 & 37 \\
    \hline
  \end{tabular}
  \caption{Main features of some of the silicon trackers built for space-based
    cosmic-ray and gamma-ray detectors. For the magnetic spectrometers the
    two figures for the strip pitch indicate the values for the
    bending/non-bending view, respectively.}
  \label{tab:exp_si_trackers}
\end{table}

Silicon detectors are essentially $p$-$n$ junction diodes operated at reverse
bias. The bias voltage has the primary purpose of increasing the depth of the
so-called depletion region, defining the active volume of the detector. Full
depletion can generally be achieved with 100--200~V bias over a typical
thickness of 300--400~$\mu$m.

One of the primary advantages of silicon detectors over other typologies of
sensors is the small (3.6~eV) average energy needed to create an electron-hole
pair---roughly an order of magnitude less than what is usually required in a
gaseous detector. For the same radiation energy this translates, as mentioned
in a the previous section, into a corresponding increase in terms of primary
charge carriers created. In addition they feature very fast intrinsic
response times (of the order of ns) and can be reliably produced with active
surfaces of tens (or hundreds) of m$^2$. They are self-triggering and
do not require consumables (e.g., gas), which makes them ideal candidates for
long-term operation in space.

\subsection{Magnetic Spectrometers}%
\label{sec:magnetic_spectrometers}

Momentum is typically measured by measuring the deflection of the
charged-particle trajectory in a magnetic field. There are actually two
slightly different concepts of \emph{magnetic spectrometers}: those measuring
the deflection angle and those measuring the sagitta. While this distinction
is partially artificial, the most notable difference is that the tracking
detectors are arranged in two arms \emph{outside} of the deflecting magnet in
the first case, while they are placed \emph{inside} the magnet in the latter.
Placing the tracking stage inside the magnet is key to make the spectrometer
compact and therefore this is typically the solution adopted in space.
We shall see, though, that AMS-02 in the permanent magnet configuration is
actually an interesting hybrid between the two concepts.

\begin{figure}[htb!]
  \includegraphics[width=\linewidth]{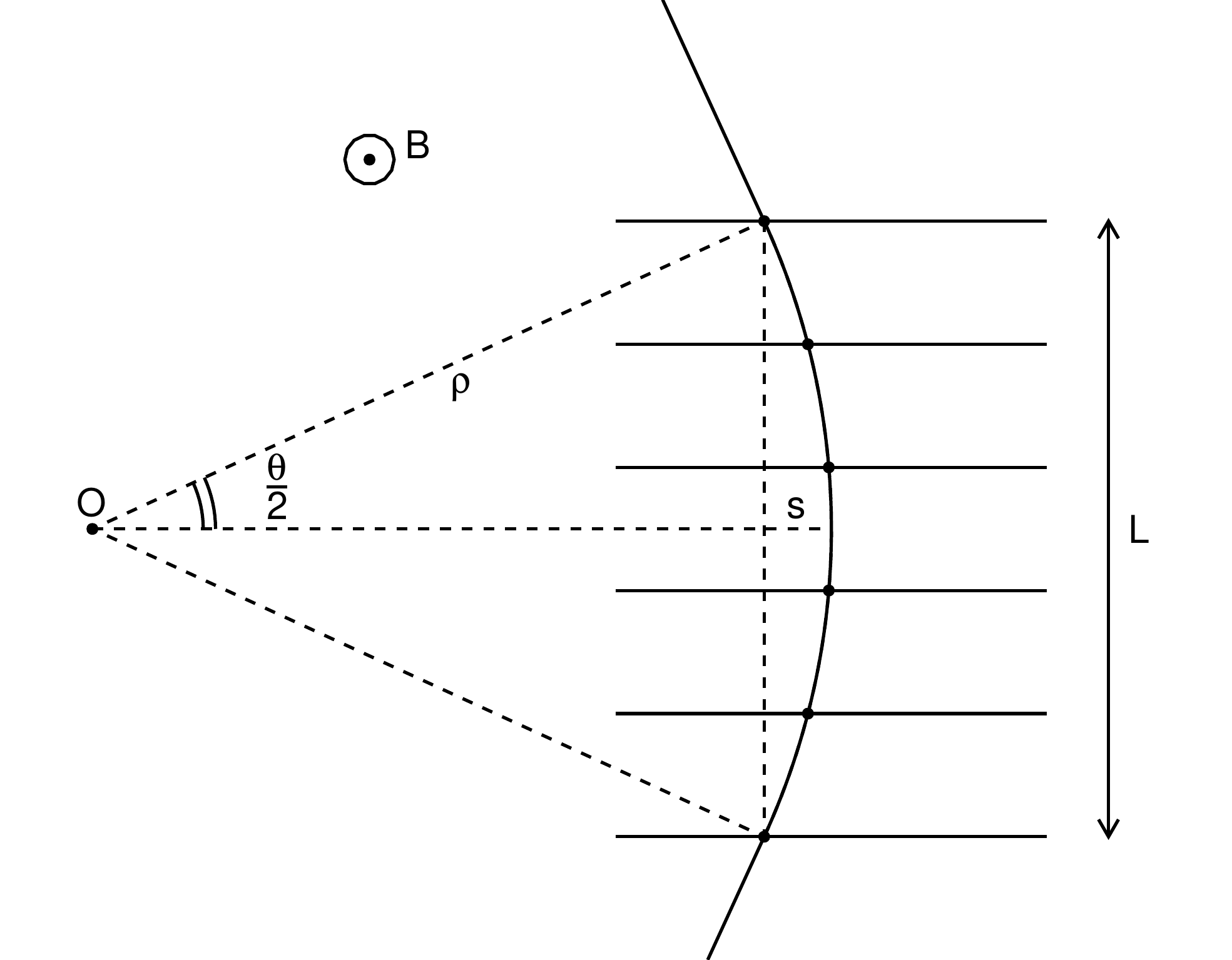}
  \caption{Sketch of the momentum measurement in a uniform magnetic field
    (in the bending view).
    Here and in the following of this section $L$ is the height of the
    region permeated by the magnetic field, $\rho$ is the radius of
    curvature of the track, $\theta_{\rm B}$ is the deflection angle, and $s$ is
    the sagitta of the track.}
  \label{fig:magnetic_spectrometer}
\end{figure}

Figure~\ref{fig:magnetic_spectrometer} illustrates the basic principle
of a magnetic spectrometer measuring the sagitta.
The curvature radius for a particle with charge $z$ (in units of the electron
charge $e$) and transverse momentum $p$ (in the following we shall neglect the
trivial drift in the longitudinal direction) in a magnetic field $B$ is
\begin{align}
  \rho = \frac{p}{zeB} = \frac{R}{cB} =
  \frac{R~[{\rm GV}]}{0.3 B~[{\rm T}]}~{\rm m}.
\end{align}
In this context the rigidity $R$ is a convenient quantity to work with, as
particles with the same rigidity behave the same way in a magnetic field,
irrespectively of their charge. If $\rho$ is measured and $B$ is known, one
can recover $R$---and, assuming that the charge of the particle is also known,
the momentum $p$.

One can see in figure~\ref{fig:magnetic_spectrometer} that $L$ and the
deflection angle $\theta$ are related
\begin{align}
  \frac{L}{2} = \rho\sin\left(\frac{\theta_{\rm B}}{2}\right) \approx
  \frac{\rho\theta_{\rm B}}{2}
\end{align}
(the latter holding in small-angle approximation). The bending angle
$\theta_{\rm B}$ therefore reads
\begin{align}\label{eq:theta_bending}
  \theta_{\rm B} \approx \frac{L}{\rho} = \frac{cBL}{R} =
  \frac{0.3 B~[{\rm T}]~L~[{\rm m}]}{R~[{\rm GV}]}~{\rm rad}.
\end{align}
The details of the measurement depend somewhat on the detector setup (e.g.,
the number of measurements and relative spacing of the tracking planes, not to
mention the pattern recognition and track-fitting algorithms) but
essentially what we measure is the sagitta $s$%
\footnote{Another commonly used exemplificative metrics is the maximum
  track displacement $\Delta_b$ in the bending plane for normal incidence,
  which can be shown to be $\Delta_b = 4s$. We prefer $s$ as its geometrical
  meaning is independent of the incidence angle (though we are mainly
  interested in the orders of magnitude here, so they are effectively
  the same thing).}
of the trajectory
\begin{align}
  s & = \rho\left[1 - \cos\left(\frac{\theta_{\rm B}}{2}\right)\right] \approx
  \frac{\rho\theta_{\rm B}^2}{8} \approx \frac{L^2}{8\rho} = \frac{cBL^2}{8R} =
  \nonumber \\
  & = \frac{37.5~B~[{\rm T}]~L^2~[{\rm m^2}]}{R~[{\rm GV}]}~{\rm mm}.
\end{align}
It goes without saying that the expression we derived is typically used in the
opposite direction, i.e., one measures $s$ to deduce $R$:
\begin{align}\label{eq:spectro_rigidity}
  R = \frac{cBL^2}{2s} = 
  \frac{37.5~B~[{\rm T}]~L^2~[{\rm m^2}]}{s~[{\rm mm}]}~{\rm GV}.
\end{align}
The quantity $BL^2$ is customarily referred to as the \emph{bending power}
and it is one of the basic figures of merit for a magnetic spectrometer,
as it determines the sagitta of the track.

\begin{figure}[htb!]
  \includegraphics[width=\linewidth]{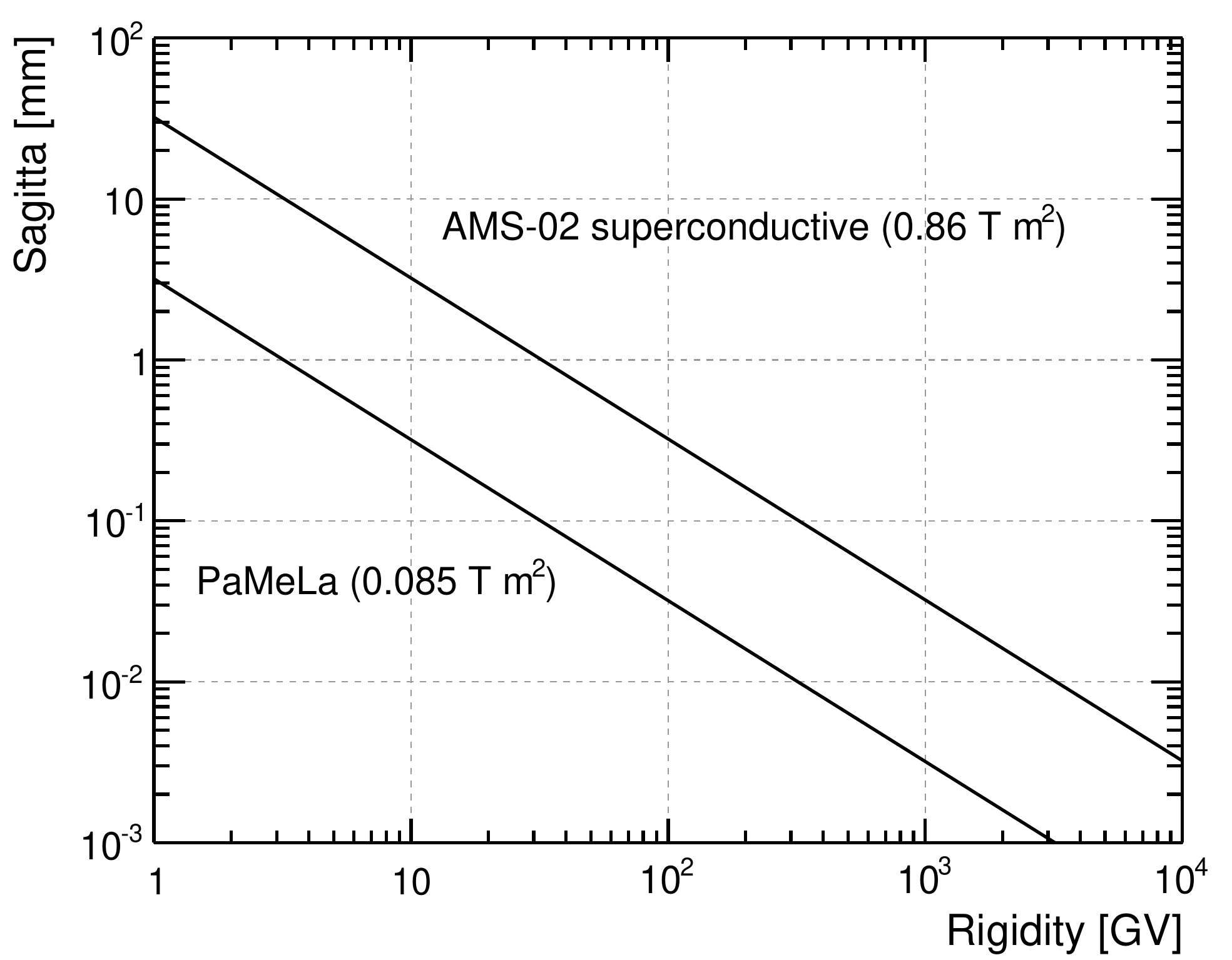}
  \caption{Trajectory sagitta as a function of rigidity
    (see figure~\ref{fig:magnetic_spectrometer}).
    The two lines correspond to two illustrative values of the bending power,
    corresponding to PaMeLa and the original design of AMS-02---with the
    superconductive magnet). We have not used the AMS-02 design
    that is operating on the ISS as the two external tracker planes do not
    quite fit in our simplified cartoon---and the original AMS-02 design
    is still representative of the current technological bleeding edge---but
    we shall come back to that in section~\ref{sec:ams_mdr}.
    Note that the plot uses the small-angle approximation that might be
    inaccurate at small rigidity.}
  \label{fig:sagitta_vs_bp}
\end{figure}

Figure~\ref{fig:sagitta_vs_bp} shows the value of the trajectory sagitta as a
function of the particle rigidity for two representative values of the bending
power. At a rigidity of $\sim 1$~TV the track displacement that one aims
measuring, with the magnetic field that can be reasonably achieved, is between
$\sim 5$ and $\sim 50~\mu$m, i.e., at the limit of a typical solid-state
tracking detector. We anticipate that this is indeed the limiting factor for
the maximum detectable rigidity, and in the high-momentum regime the relative
uncertainty on the rigidity increases linearly with the rigidity itself
\begin{align}
  \frac{\sigma_R}{R} = \frac{\sigma_s}{s} \propto{R},
\end{align}
as the uncertainty on the sagitta $\sigma_s$ is a constant dictated by the
detector and the sagitta itself $s$ goes like $1/R$. This is one of the most
notable differences between spectrometers and calorimeters (see
section~\ref{sec:calorimetry}).

\subsection{Scintillators}

Scintillators utilize the ionization produced by charged particles (or by
gamma-rays converting into an electron positron pair within the material)
into light that can be in turn converted into an electric signal by a
photodetector such as a photomultiplier tube (PMT) or a pin diode.

Scintillating materials are broadly divided into \emph{organic} (most notably,
plastic) and \emph{inorganic}---the main difference being that
the first have a relatively low density (and are therefore widely employed,
e.g., in trigger, anti-coincidence and time of flight systems), while the
latter are used in applications, such as calorimetry (see
section~\ref{sec:calorimetry}), where a high stopping power is required.

The choice of a particular material for a particular application is in general
dictated by several different properties, such as the light yield, the
duration and wavelength spectrum of the output light pulse, the value of the
radiation and/or interaction lengths. We refer the reader to~\cite{PDG} and
reference therein for a more in-depth discussion.

\subsection{Calorimetry}
\label{sec:calorimetry}

Calorimeters are instruments in which the particles to be measured are
fully or partially absorbed and their energy is transformed into a
measurable signal.

Compared to magnetic spectrometers, calorimeters have the double advantage
that the energy resolution improves as $1/\sqrt{E}$ at high energy (rather than
deteriorating linearly with the particle momentum) and the characteristic
dimensions (i.e., the shower depth) scale only logarithmically with the
particle energy. In addition, calorimeters are also sensitive to neutral
particles and provide position/direction information and particle
identification capabilities (though, strictly, speaking, the latter apply to
spectrometers, too).

\subsubsection{Electromagnetic calorimeters}
\label{sec:em_cal}

\begin{figure}[htb]
  \includegraphics[width=\linewidth]{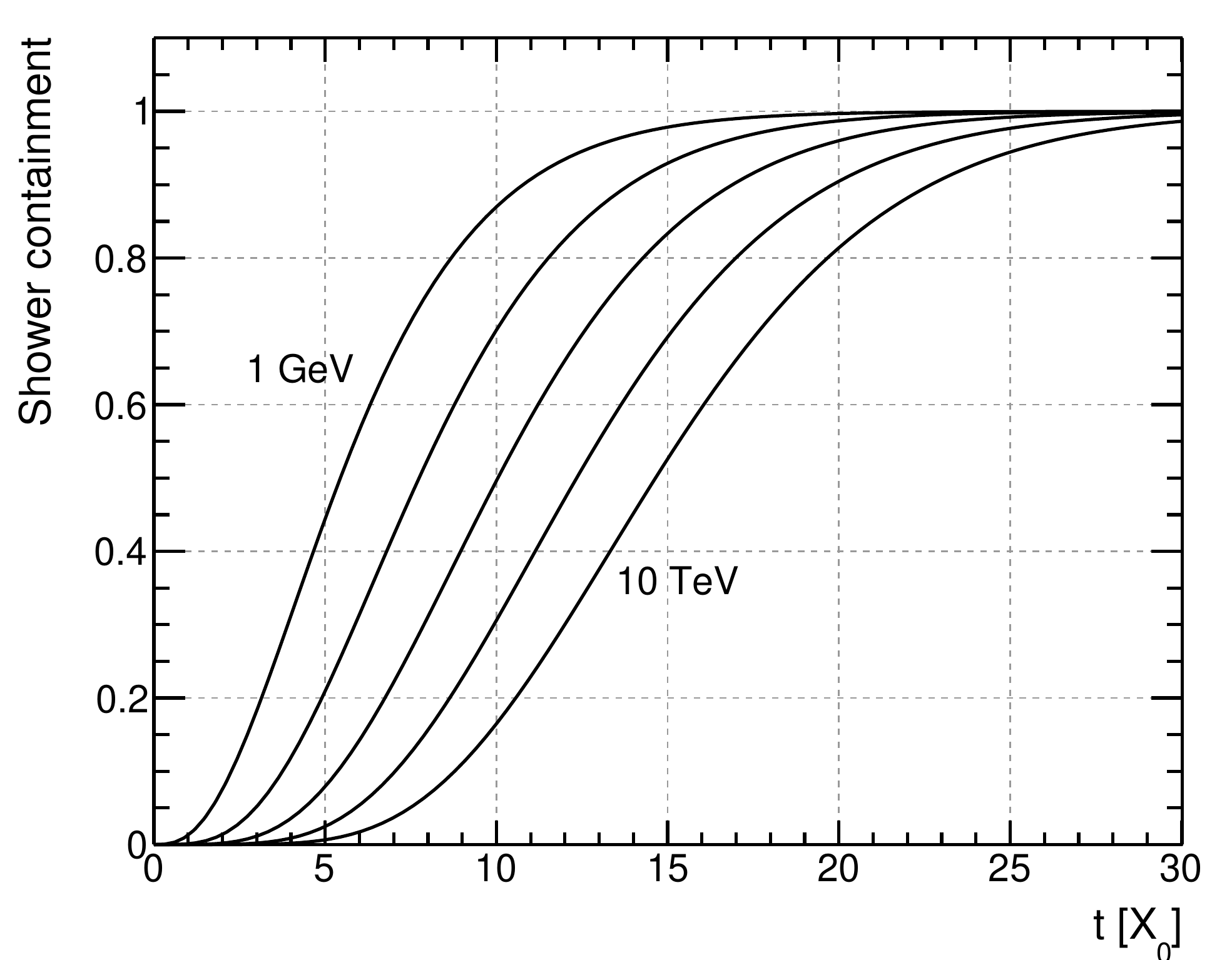}
  \caption{Average shower containment as a function of the shower
    depth (in homogeneous BGO) for electrons of different energy:
    $1$, $10$, $100$~GeV, $1$ and $10$~TeV.}
  \label{fig:shower_containment}
\end{figure}

Electromagnetic calorimeter can be homogeneous or sampling. Homogeneous
calorimeters typically feature an excellent energy resolution, but can be less
easily segmented with a potentially detrimental effect on the position
measurement and particle identification. It is fair to say that the thickness
(in terms of radiation lengths) is one of the basic figures of merit in this
context, as it determines the shower containment, as shown in
figure~\ref{fig:shower_containment}.

\begin{table}
  \begin{tabular}{p{0.25\linewidth}p{0.22\linewidth}p{0.25\linewidth}%
      p{0.2\linewidth}}
    \hline
    Experiment & Material & Depth [$X_0$] & Mass [kg]\\
    \hline
    \hline
    AMS-02 & Pb/fibers & 17 & 638\\
    ATIC & BGO & 22.6 & $\sim 500$\\
    CREAM & W/fibers & 20 & 380\\
    \Fermi-LAT & CsI(Tl) & 8.6 (10.1) & 1350\\
    PaMeLa & W/Si & 15.3 & 110\\
    \hline
  \end{tabular}
  \caption{Material and on-axis depth (in radiation lengths) for the
    calorimeters of some recent space- and balloon-borne experiments.
    The \Fermi-LAT being a pair-conversion telescope, its tracker is
    $1.5~X_0$ thick and effectively acts as a pre-shower, bringing the
    total thickness of the instrument to $10.1~X_0$ for normal incidence
    (see also the comments in section~\ref{sec:theta_instr_distr}).}
  \label{tab:exp_calorimeters}
\end{table}

The energy resolution of an electromagnetic calorimeter is usually
parameterized as
\begin{align}\label{eq:ecal_eres}
  \frac{\sigma_E}{E} = \frac{a}{\sqrt{E}} \oplus \frac{b}{E} \oplus c.
\end{align}
The stochastic term $a$ is due to the intrinsic fluctuations
related to the physical development of the shower. For homogeneous
calorimeters, assuming that the shower is entirely contained, the total
energy deposited does not fluctuate and the stochastic term is smaller
than one would expect based on counting statistics by the so called Fano
factor. In this case $a$ is typically of the order of a few \%~GeV$^{-1/2}$.
For sampling calorimeter the sampling fluctuations increase the stochastic term
and $a$ is more likely to be of the order of $5$--$20\%$~GeV$^{-1/2}$. 
The noise term $b$ is due to the electronic noise of the readout
chain and can dominate at low energy, especially when one operates at
high rate (i. e. when a large bandwidth is needed).
The constant term $c$ includes the effect of response
nonuniformities and represents the value at which the energy resolution
levels off when the stochastic term becomes negligible.

\begin{figure}[htb!]
  \includegraphics[width=\linewidth]{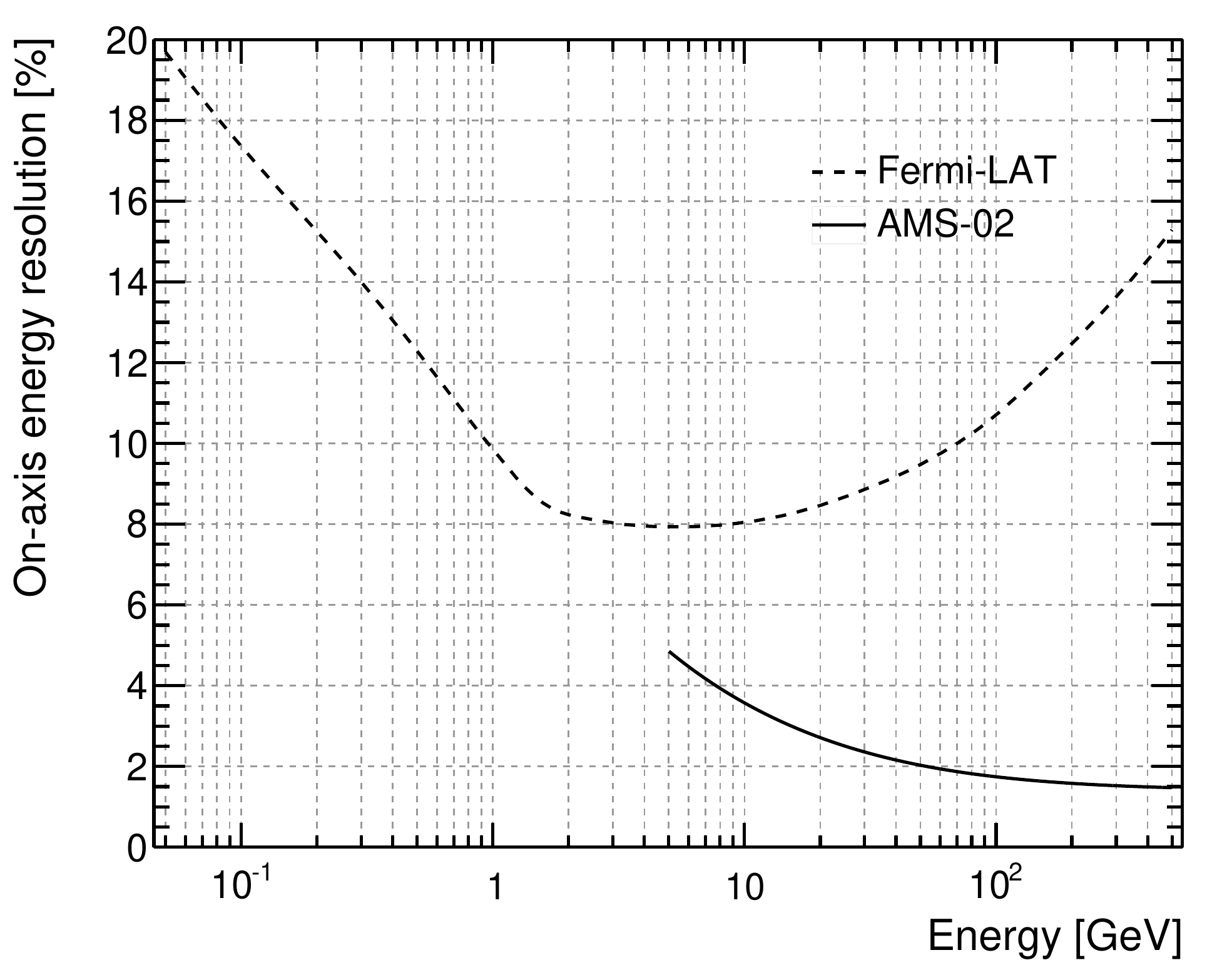}
  \caption{On-axis energy resolution, as a function of the energy, for the
    \Fermi-LAT~\cite{2012ApJS..203....4A} and
    AMS-02~\cite{2013NIMPA.714..147A}. AMS-02 features a $17~X_0$ sampling
    calorimeter whose energy resolution can be parameterized according
    to~\eqref{eq:ecal_eres} with $a = 10.4\%$ and $c = 1.4\%$.
    The \Fermi-LAT features a homogeneous calorimeter, $8.6~X_0$ thick on axis
    (and a $1.5~X_0$ tracker converter effectively acting as a pre-shower), and
    the energy resolution at high energy is dominated by the shower leakage,
    which can only be partially compensated by using the imaging capabilities
    of the detector.}
  \label{fig:calorimeter_eres}
\end{figure}

In real life there are additional contributions to the energy resolution 
of an electromagnetic calorimeter, such as the longitudinal and lateral
leakage, the upstream energy losses in other detectors and the effect of the
cracks and dead regions. Figure~\ref{fig:calorimeter_eres} shows that in
practice these effect can become dominant for relatively thin calorimeters.

\subsubsection{Hadronic calorimeters}%
\label{sec:had_cal}

From the standpoint of calorimetry, the main peculiarity of hadronic showers
is that some fraction of the energy contained in the non-electromagnetic
component does not contribute to the signal (this is what is usually called
\emph{invisible} energy). Endothermic spallation losses are responsible for
the vast majority of the invisible energy, as the nuclear binding energy of
the released protons, neutrons and heavier aggregates has to be supplied by
the shower particles that induce the reactions---and obviously does not
contribute to the signal. Furthermore nuclear recoils, neutron captures
producing delayed gamma rays, muons and neutrinos, all contribute to the
invisible energy.

In addition to that a calorimeter has a different response to
the electromagnetic and the non-electromagnetic components of a hadronic
shower, in terms of how much of the relative energy is actually translated
into \emph{visible signal}. The ratio of the efficiencies for conversion into
detector signal of the two components is customarily referred to as $h/e$, and
is characteristic of the material and detector configuration.

As a consequence of these basic facts, typical resolutions for hadronic
calorimeters are intrinsically worst than those of the best homogeneous
electromagnetic calorimeters (30--40\% being a representative figure).
It should also be noted that, being the nuclear
interaction length typically much larger than the radiation length, hadronic 
calorimeters need to be (geometrically) thick---and heavy. This is the main
reason why hadronic calorimetry, at least in the strict sense the term is used
in high-energy physics at accelerators, is not very fashionable in space.
One interesting detector concept---used, e.g., in the
ATIC~\cite{2009BRASP..73..564P} and CREAM~\cite{2011ApJ...728..122Y}
experiments---is that of exploiting a thick carbon\footnote{As shown in
figure~\ref{fig:rad_int_len}, low-$Z$ materials feature a comparatively short
nuclear interaction length, and therefore allow to realize relatively thick
(in the sense of $\lambda_I$) targets with reasonable weights.}
passive target to promote nuclear interaction of the hadrons and then
recovering the particle energy by measuring the electromagnetic component of
the shower.

\subsection{Particle identification}

There are traditionally four methods for particle identification in high-energy
physics: measurement of $dE/dx$, time of flight (TOF) systems, \cheren\
detectors and transition radiation detectors (TRD). In addition to that,
modern electromagnetic imaging calorimeters generally provide a high
discrimination power for the specific purpose of separating photon- or
electron/positron-initiated showers from hadronic showers---so we shall
briefly introduce this topic of discussion, too.

Incidentally, all of these techniques are exploited in the AMS-02 experiment
operating on the ISS, so the suite of contributions presented by the AMS-02
collaboration at the $33^{th}$ International Cosmic-Ray Conference is
potentially a good source of information on the state of the art of particle
identification in space. We refer the reader to~\cite{2012NIMPA.666..148L}
for a more accelerator-oriented review.

\subsubsection{Measurement of $dE/dx$}

The particle identification is based on the simultaneous measurement of the
momentum $p$ and the mean energy loss $dE/dx$ and relies on the fact that
the Bethe-Bloch formula is a universal function of $\beta\gamma = p/m$ for
all particle species (and therefore it's different for different masses at the
same momentum).

Particle identification systems (e.g. to discriminate protons from pions) based
on the mean energy loss have been successfully implemented up to reasonably
high energies, well in the relativistic rise (below the saturation point due
to the density effect) where typical $dE/dx$ differences are of the order of
$10\%$ or so. Since the typical width of the Landau curve for a single sample
is generally much larger than this differences, many independent measurements
are needed to achieve the required level of accuracy. Typically, due to the
long right tail of the Landau curve, some sort of truncated mean is the
simplest and most robust estimator of the average energy loss.

\begin{figure}[htb!]
  \includegraphics[width=\linewidth]{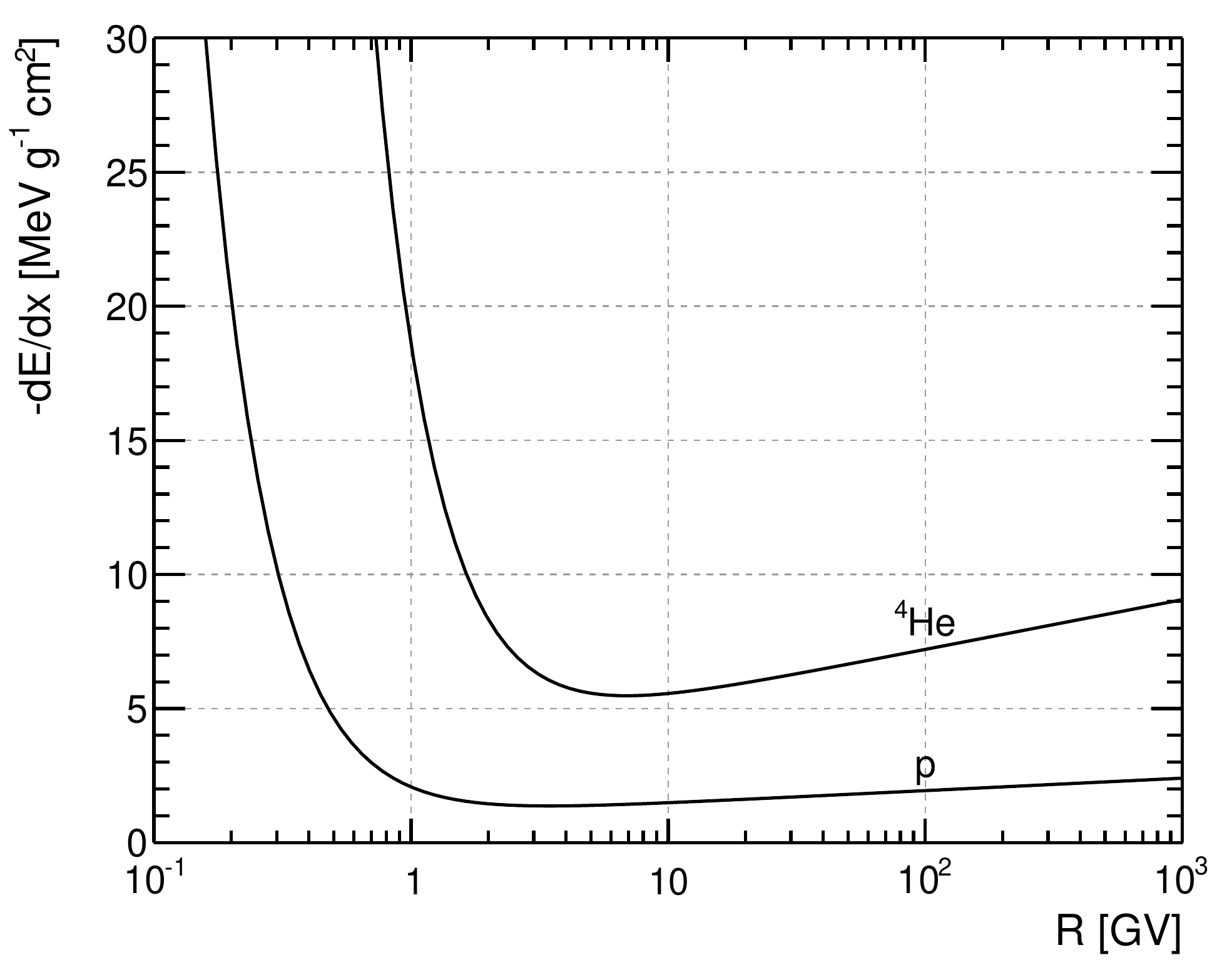}
  \caption{Ionization energy loss, as a function of the rigidity, for protons
    and He nuclei in silicon. The values are calculated according
    to~\eqref{eq:stopping_power_full} without any density correction.
    In real life the separation power of the $dE/dx$ technique is determined
    by the Landau fluctuations, determining how closely the actual events
    cluster around the average line.}
  \label{fig:dedxcharge}
\end{figure}

In the context of space-based cosmic-ray detectors the $dE/dx$ particle
identification technique is customarily used to identify the nuclear species
by measuring the atomic number $Z$ through the $Z^2$ dependence, as illustrated
in figure~\ref{fig:dedxcharge}.

\subsubsection{Time of flight}%
\label{sec:tof}

The time of flight method relies on the simultaneous measurement of the
particle momentum and velocity through the direct measurement of the time
$T$ it takes to the particles to traverse a distance $L$:
\begin{align}
  \beta c = \frac{L}{T}
\end{align}
We can invert the expression $p = m\gamma\beta c$ to obtain the mass of the
particle as a function of momentum and velocity:
\begin{align}\label{eq:tof_mass}
  m = \frac{p}{\gamma\beta c} = \frac{p\sqrt{1 - \beta^2}}{\beta c} =
  \frac{p}{c}\sqrt{\frac{1}{\beta^2} - 1}.
\end{align}
By the standard error propagation we get the relative error on the mass
measurement:
\begin{align}\label{eq:tof_mres}
  \frac{\sigma_m}{m} = \frac{\sigma_p}{p} \oplus
  \gamma^2 \frac{\sigma_\beta}{\beta}.
\end{align}
In practice $L$ is known with a good enough precision that the error on $\beta$
is essentially determined by the time resolution of the system. For
$L \sim 1.5$~m and $\sigma_T/T \sim 100$~ps (which are somewhat representative
of the AMS-02 detector), one can achieve a relative accuracy on the velocity
of the order of a few~\%. Depending on the actual momentum dependence of the
momentum resolution and the value of $\gamma$, the mass resolution can
be dominated by either one of the two terms.

\begin{figure}[htb!]
  \includegraphics[width=\linewidth]{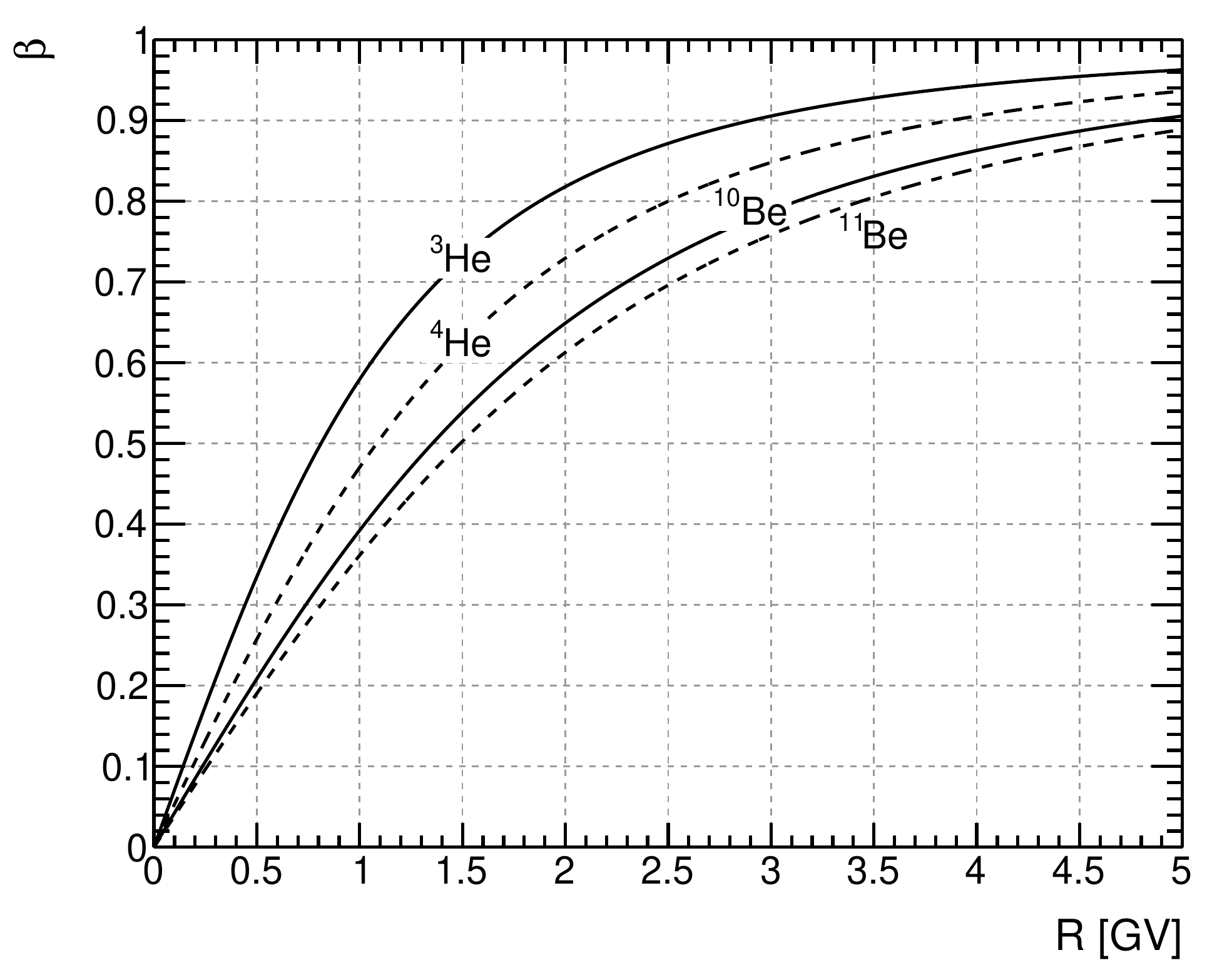}
  \caption{Relation between velocity and rigidity for different isotopes,
    according to~\eqref{eq:beta_r_isotopes}.}
  \label{fig:isotopes_rbeta}
\end{figure}

One practical application of this technique is the measurement of
cosmic ray isotopical composition (e.g. the ratio of the differential
intensities of $^3$He/$^4$He). This is usually achieved by measuring
independently the particle rigidity $R$ with a magnetic spectrometer and its
velocity $\beta$ with a TOF system (in addition to the charge $Z$, that can be
inferred by means of $dE/dx$ measurements in both the spectrometer and the TOF).
In this context \eqref{eq:tof_mass} is more conveniently written as
\begin{align}\label{eq:beta_r_isotopes}
  A = \frac{ZeR\sqrt{1 - \beta^2}}{m_p\beta c^2},
\end{align}
where $A$ is the mass number of the nucleus and $m_p$ is the mass of the
nucleon. Equivalently, for fixed $Z$, the velocity and momentum for nuclei with
different mass numbers (i.e., isotopes) will be related to each other by
\begin{align}
  \beta = \left[ 1 + \left(\frac{A m_p c^2}{ZeR}\right)^2 \right]^{-\frac{1}{2}},
\end{align}
as illustrated in figure~\ref{fig:isotopes_rbeta} for the two cases of
$^3$He vs. $^4$He and $^{10}$Be vs. $^{11}$Be.

In any practical application the finite mass resolution given
by~\eqref{eq:tof_mres} will cause events to be distributed around the
theoretical line with a finite dispersion---at the level that an event-by-event
mass separation could be impossible. Still in these case a template fitting
of the data to the separate isotopical mass distribution can in principle
allow to recover the composition, at least at intermediate rigidities, where the
curves are best separated. We refer the reader to~\cite{2011ApJ...736..105A}
for a detailed description of the light nuclei isotopical composition
measurement by the AMS-01 experiment.

\subsubsection{\cheren\ detectors}

As explained in section~\ref{sec:cherenckov_rad}, the emission of \cheren\
radiation is a threshold process taking place when a particle moves in a medium
with index of refraction $n$ with velocity $\beta > 1/n$.

\begin{figure}[htb!]
  \includegraphics[width=\linewidth]{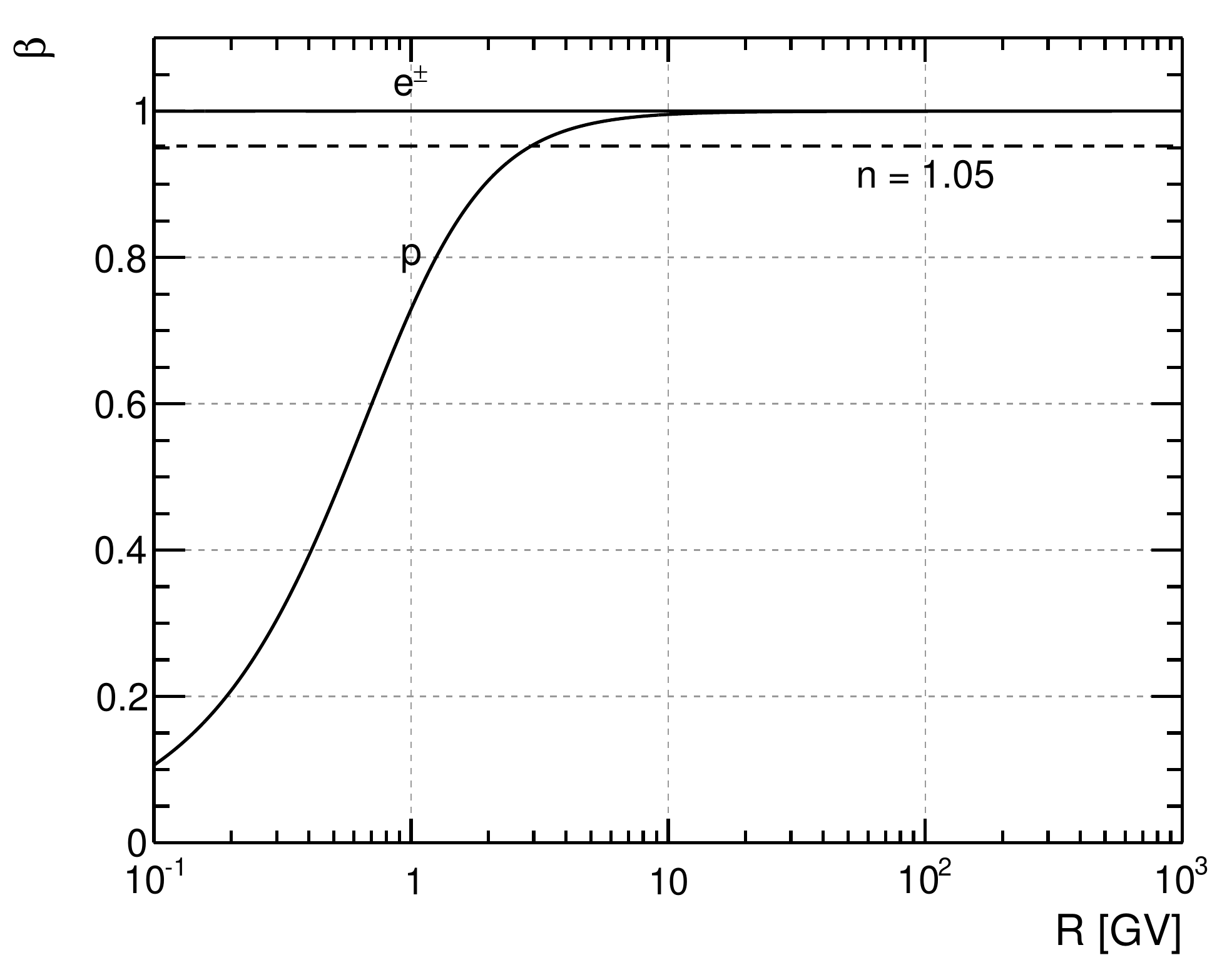}
  \caption{Relativistic $\beta$ factor, as a function of the particle rigidity,
    for electrons and protons. For illustration purposes, the threshold
    corresponding to a representative index of refraction $n = 1.05$ is shown.}
  \label{fig:beta_vs_r}
\end{figure}

In their simplest form, \cheren\ detectors exploit the existence of this
threshold acting as \emph{digital counters} that allow to distinguish, e.g.,
electrons and protons over relatively large energy (or rigidity) ranges, as
illustrated in figure~\ref{fig:beta_vs_r}. This was done in many of the early
experiments aimed at measuring the positron component of cosmic rays, e.g.,
in~\cite{1968ApJ...152..783F,1969ApJ...158..771F,1975ApJ...198..493D,
  1974PhRvL..33...34B,1975ApJ...199..669B}.
Since $\beta$ approaches $\sim 1$ for $\sim 10$~GV
protons, it is difficult to effectively exploit this technique past a few GV.

The steep slope of the intensity of the \cheren\ signal near threshold (see,
e.g., figure~\ref{fig:cherenkov_betagamma}) allows quite an accurate velocity
measurement (though in a relatively small window). This can be used to
complement TOF system in mass measurements (see~\ref{sec:tof}) and, by
using a clever arrangement of radiators with different indices of refraction,
for actual spectral measurements~\cite{2011ApJ...742...14O}.

\begin{figure}[htb!]
  \includegraphics[width=\linewidth]{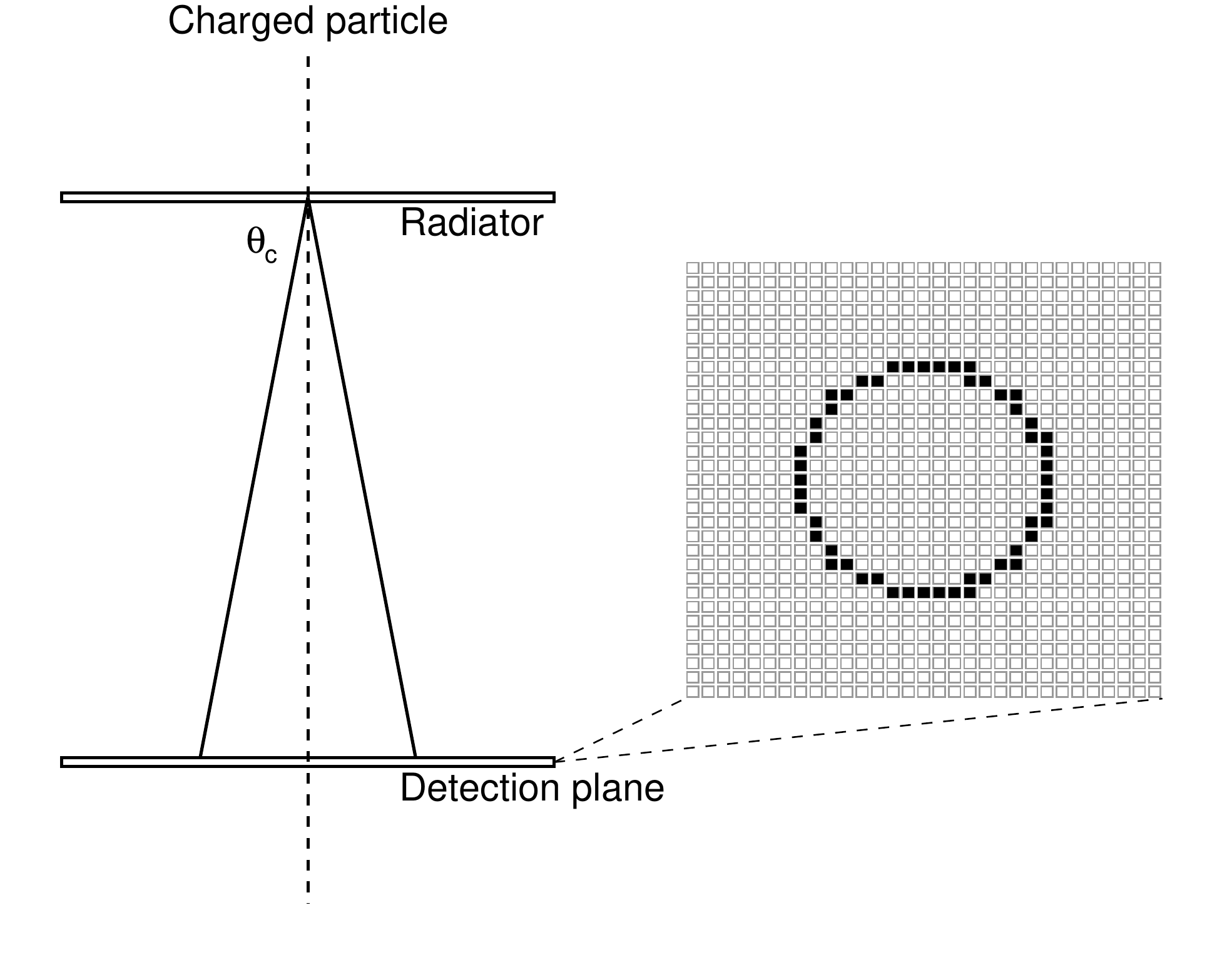}
  \caption{Functional sketch of a ring imaging \cheren\ detector. A cone
    of \cheren\ radiation produced as the charged particle traverse the
    radiator is imaged as a ring on the photon detection plane, which provides
    a measurement of both the intensity of the radiation itself and of the
    \cheren\ angle $\theta_c$.}
  \label{fig:rich_sketch}
\end{figure}

The most advanced \cheren\ detectors, such as ring-imaging \cheren\ detectors
(RICH) exploit all the properties of the \cheren\ effect, including the
intensity of the signal \emph{and} the angular aperture of the light cone
(see figure~\ref{fig:rich_sketch} for a simplified, illustrative sketch).

\subsubsection{Transition radiation detectors}

Even a brief summary of the basic considerations going into a practical
implementation of a transition radiation detector (see, e.g, \cite{PDG} and
\cite{2012NIMPA.666..130A}) are beyond the scope of this review.
Typically many layers of radiator (i.e., optical interfaces) are needed to
achieve a reasonable signal yield, interleaved with suitable radiation
detection elements---usually gas detectors. Furthermore, since the
transition radiation is emitted in the forward direction at a fairly small
angle, it effectively overlaps with the ionization due to the specific
energy loss $dE/dx$, and the latter is also an important ingredient of the mix.

\begin{figure}[htb!]
  \includegraphics[width=\linewidth]{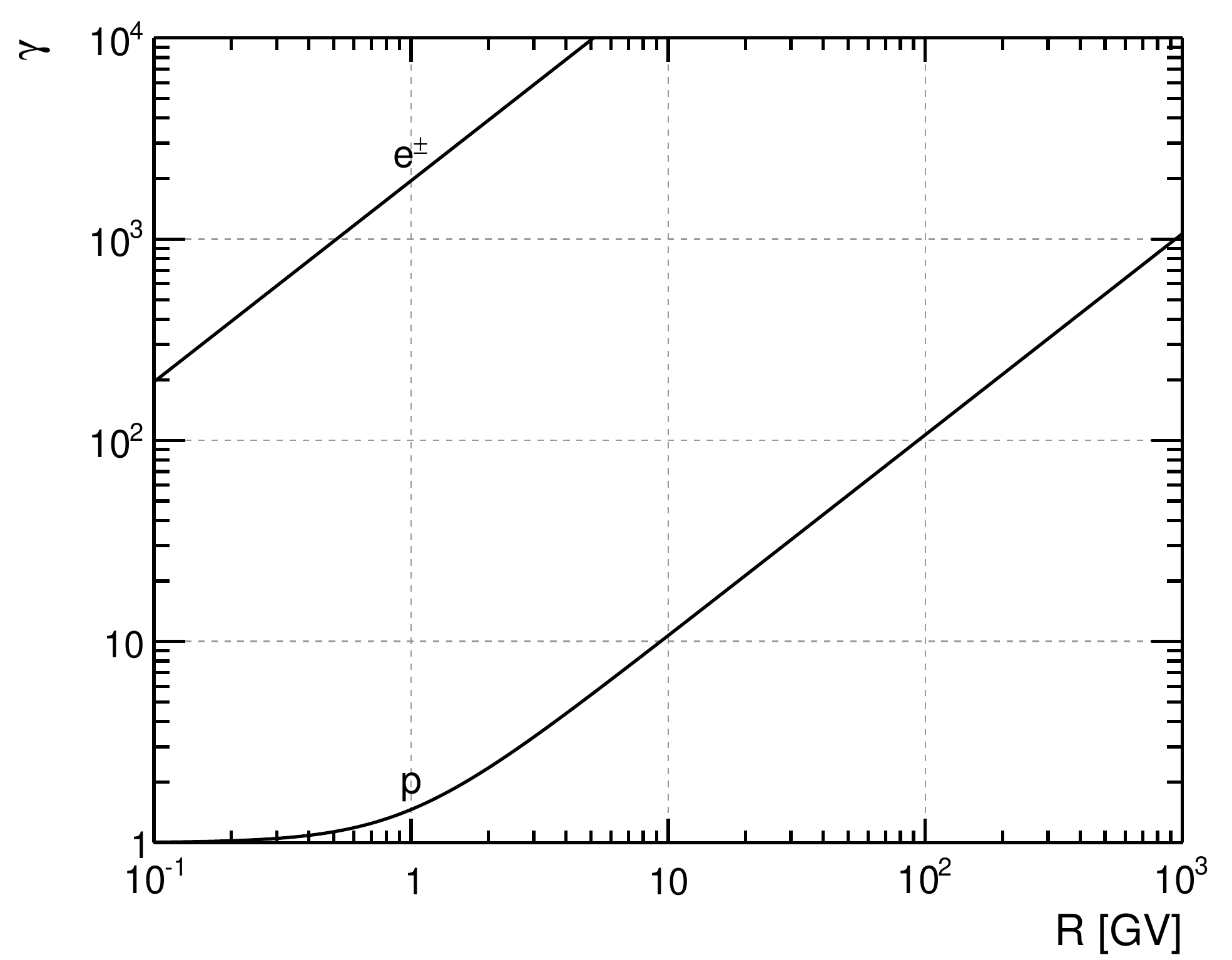}
  \caption{Relativistic $\gamma$ factor, as a function of the particle rigidity,
    for electrons and protons.}
  \label{fig:gamma_vs_r}
\end{figure}

As pointed out in section~\ref{sec:transition_rad}, a relativistic
$\gamma$ factor of the order of $10^3$ is needed for the spectrum of the
transition radiation to extend in the X-ray domain---i.e., where the emitted
photons can be effectively detected on top of the energy deposited by
ionization. Figure~\ref{fig:gamma_vs_r} shows that a relatively large rigidity
window---roughly speaking between $\sim 1$~GV and $\sim 1$~TV---exists where
the transition radiation can potentially be exploited to discriminate
electrons and positrons from protons. (Incidentally, this extends to much
higher energies than those accessible to the \cheren\ technique discussed in
the previous section).

To put things in context, the TRD of the AMS-02 experiment operating on the
ISS provides a proton rejection factor (at 90\% electron efficiency) in
excess of 100 between 1 and 600~GV, peaking around $2 \times 10^4$ at
$\sim 10$~GV~\cite{2013NuPhS.243...12T}.

\subsubsection{Electron-proton separation in imaging calorimeters}

As mentioned in section~\ref{sec:had_showers}, the differences between
electromagnetic and hadronic showers are customarily exploited to discriminate
high-energy photons, electrons and positrons from hadrons.

As shown in figure~\ref{fig:rad_int_len}, for the typical materials used in
electromagnetic calorimeters the interaction length $\lambda_I$ is 20--30 times
larger than the radiation length $X_0$ (when expressed in g~cm$^{-2}$, i.e.,
normalized to the density).
This means that the first interaction is on average at a much larger depth
for a hadronic shower than for an electromagnetic one---something like
$\sim 20$--30~cm vs. $\sim 1$~cm. Incidentally, since electromagnetic
calorimeters are typically \emph{thin} (less than one~$\lambda_I$) from the
standpoint of hadrons, a significant fraction of high-energy protons just pass
through without interacting---i.e., behaving as minimum ionizing particles.
(It goes without saying that they are easy to discriminate from photons and
electrons.)

In addition to that electromagnetic showers are typically \emph{well behaved}
in that the vast majority of the energy is contained within a few Moliere radii
from the shower axis and, fluctuations aside, the shower maximum is located at
a predictable position along the axis. In contrast hadronic showers are
generally much wider (due to secondaries from inelastic nuclear interactions
spreading out with non-negligible transverse momentum) and feature larger
fluctuations in both the longitudinal and the transverse development.

\section{Instrument Response Functions}%
\label{sec:irfs}

In broad terms the instrument response functions (IRF) are specific
parameterizations of the instrument performance allowing to convert the
count spectra registered by the detector into physically meaningful quantities
such as fluxes and spectral indices. In this section we shall try and provide a
thorough discussion of the subject, with emphasis on how the response functions
tie to the detector design and the basic interaction processes of particles and
radiation with the latter. The treatment will be somewhat simplified and aimed
at basic sensitivity studies, and the reader is referred, e.g.,
to~\cite{2012ApJS..203....4A} and references therein for a more systematic
description of how IRFs are actually used in typical gamma-ray spectral
analyses.

\subsection{The role of the event selection}%
\label{sec:event_selection}

In the context of pretty much any scientific analysis one has to deal with
the problem of separating the \emph{signal} (e.g., the particular CR species
under study) from the \emph{background} (e.g., all the other species, possibly
much more abundant, that might mimic, in a way or another, the signal we are
interested in). We shall refer to the entire process of isolating the signal
from the background as the \emph{event selection.}
While at a fist glance the reader might find awkward to start the discussion
about the instrument response functions by bringing up this seemingly unrelated
issue, we do so to stress since the beginning that
\emph{in general the instrument response functions are not intrinsic
  characteristics of the detector: they always subtend a specific event
  selection and a detector may very well have different response functions
  in the context of different analyses.}
Be wary when you read off a scientific paper or a conference presentation the
value of the acceptance for a given instrument quoted as a single plain scalar.
Appropriate as it might be for the particular context, keep in mind that there
might be hidden energy dependencies and do not forget to double check whether it
includes the effect of the selection cuts or not---the difference might be
factors to orders of magnitude!

\subsubsection{A different look at cosmic-ray spectra}

\begin{figure}[!htb]
  \includegraphics[width=\linewidth]{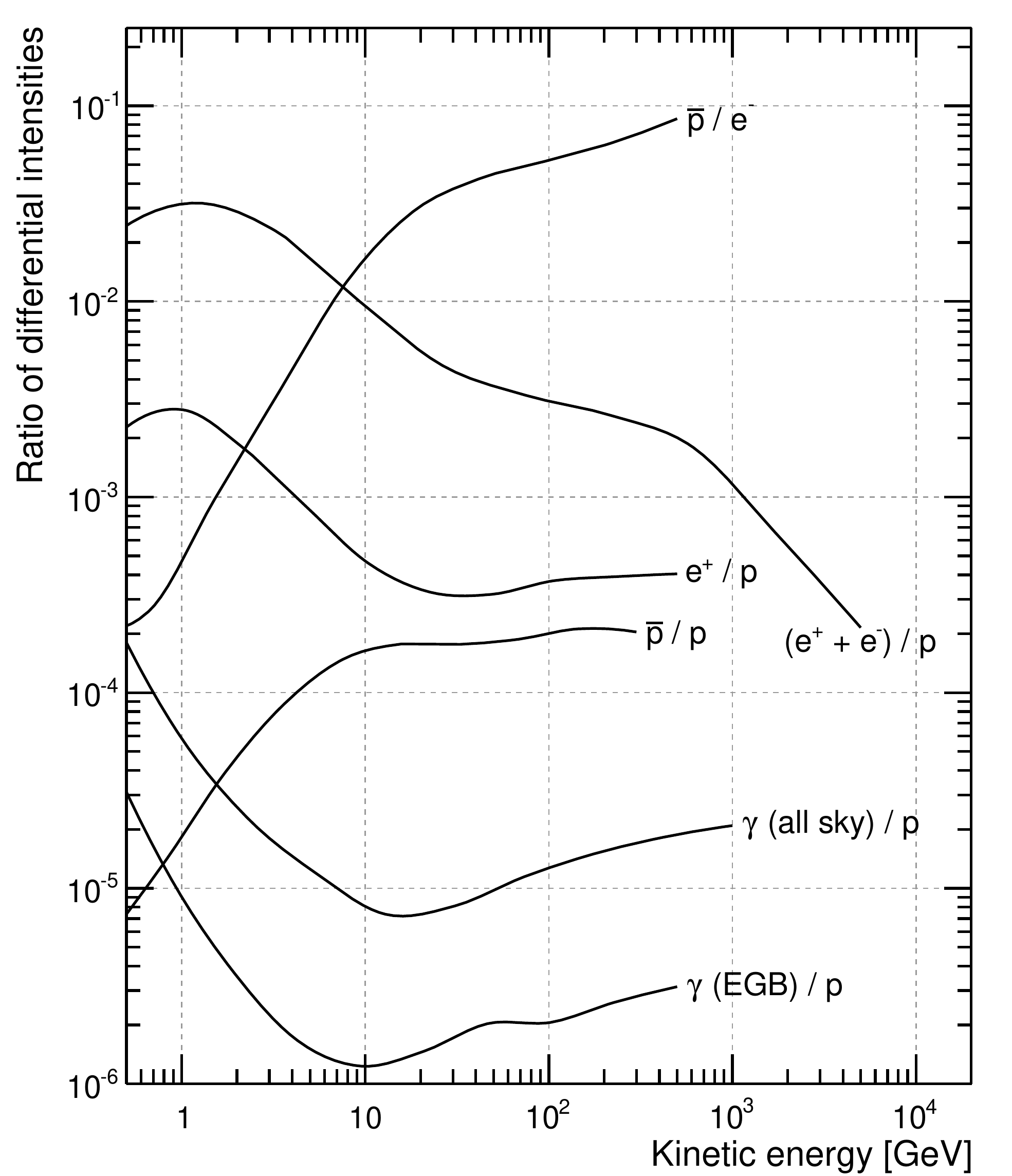}
  \caption{Ratio of the differential fluxes, as a function of the energy, for
    some specific pairs of cosmic-ray species.}
  \label{fig:cr_ratios}
\end{figure}

A comprehensive discussion of the complex (and, to many respects, subtle)
issues of event selection and background rejection is outside the scope of
this paper. We shall, however, try and put on the table some basic ideas,
starting from a somewhat different look on the cosmic-ray spectra---focusing
on the ratio between specific differential fluxes, as shown in
figure~\ref{fig:cr_ratios}.

If one is interested in measuring, say, the all-electron $(e^+ + e^-)$ spectrum
by means of a calorimetric experiment, the main challenge from the standpoint
of the background rejection, is the much larger ($10^2$--$10^4$) proton flux.
The all-electron spectrum being significantly steeper ($\Gamma_{e} \sim 3.1$)
than the proton spectrum ($\Gamma_{p} \sim 2.75$), the electron-to-proton ratio
decreases with energy---and, due to the cutoff measured by
H.E.S.S.~\cite{2008PhRvL.101z1104A}, steepens significantly past $\sim 1$~TeV.
This implies that the detector must feature a proton rejection power (see
section~\ref{sec:rejection_pwr} for more details) of at least $10^5$ in order
to have a relatively small background that can be safely subtracted in the data
analysis phase. Different how the event topologies for electrons and protons
are, this is really saying that we are only allowed to mis-tag one proton in
100,000 (while keeping a reasonable electron efficiency), which is obviously a
non trivial task.

Life is even harder for gamma rays, as they are substantially less abundant
than any of the four singly-charged CR species. Figure~\ref{fig:cr_ratios} shows
that for any celestial gamma ray there are $10^4$--$10^5$ protons of the same
energy. Photons pointing back to their sources, this does not necessarily
implies that any gamma-ray analysis is intrinsically more difficult than, say,
the measurement of the positron fraction---when studying a gamma-ray source,
restricting the sample to a small region of interest around the source itself
can allow to reduce the background by orders of magnitude.
Nonetheless, the measurement of the faint isotropic extra-galactic gamma-ray
background~\cite{2010PhRvL.104j1101A} is a good example where a proton
rejection factor of the order of $10^6$ is really required (and, to this
respect, it is fortunate that plastic scintillators can be used to assemble
efficient anti-coincidence shields to isolate neutral particles). We should
also emphasize that, while much less abundant than protons, high-energy 
electrons and positrons generally look more similar to photons in calorimetric
experiments (they all produce electromagnetic showers), and can therefore
constitute an important source of background for gamma rays.

For magnetic spectrometers most of the curves in figure~\ref{fig:cr_ratios}
are actually important. At low energy, where charge confusion is negligible,
the relevant figures are the $\bar{p}/e^-$ ratio for the measurement of the
antiproton fraction and $e^+/p$ ratio for that of the positron fraction.
While achieving a proton/electron rejection power in excess of $10^4$ is not
really a big issue with modern spectrometers such as AMS-02 or PaMeLa, this
was actually a challenge in the early days, when the particle identification
capabilities of the instruments were limited---as nicely phrased
in~\cite{1987A&A...188..145G}:
\emph{It is the opinion of the investigators that the $e^+$ observation is
substantially more difficult than the $\bar{p}$ observation [\ldots]
For negatively charged particles one has to distinguish $\bar{p}$ from a $20$
times higher flux of $e^-$ and from atmospheric mesons. In the case of $e^+$,
however, one must separate the desired particles from protons, which have the
same charge and a flux nearly $1000$ times as great.}
At high energy, on the other hand, where charge confusion is the main limiting
factor, the important figures are the $e^+/e-$ and $\bar{p}/p$ ratios---and in
this regime antiprotons are relatively harder to separate. 

In the next section we shall briefly discuss the issue of the background
rejection in a slightly more formal fashion. At this point it should be clear,
however, that different science analyses, in general, require different levels
of rejection power---and hence different event selections and different
response functions.

\subsubsection{Discriminating variables and rejection power}%
\label{sec:rejection_pwr}

The binary problem of separating signal and background in a given event
sample is very common in high-energy physics. This is generally achieved
by exploiting one or more \emph{discriminating variables}, i.e. topological
properties of the event that are different, on average, for signal and
background. The transverse size of the shower in an electromagnetic calorimeter
is a prototypical example of a discriminating variables that is useful for
separating hadrons from electrons or gamma rays. Other examples include:
the signal in a transition radiation detector or in a \cheren\ counter, 
the average $dE/dx$ in the layers of a tracking detectors, the pulse height
in a scintillator and many others. When a high level of purity is required,
it is customary to use many different variables, possibly combining them
by means of multivariate classification techniques such as Fisher discriminants,
classification trees, boosted decision trees and others.

\begin{figure}[!htb]
  \includegraphics[width=\linewidth]{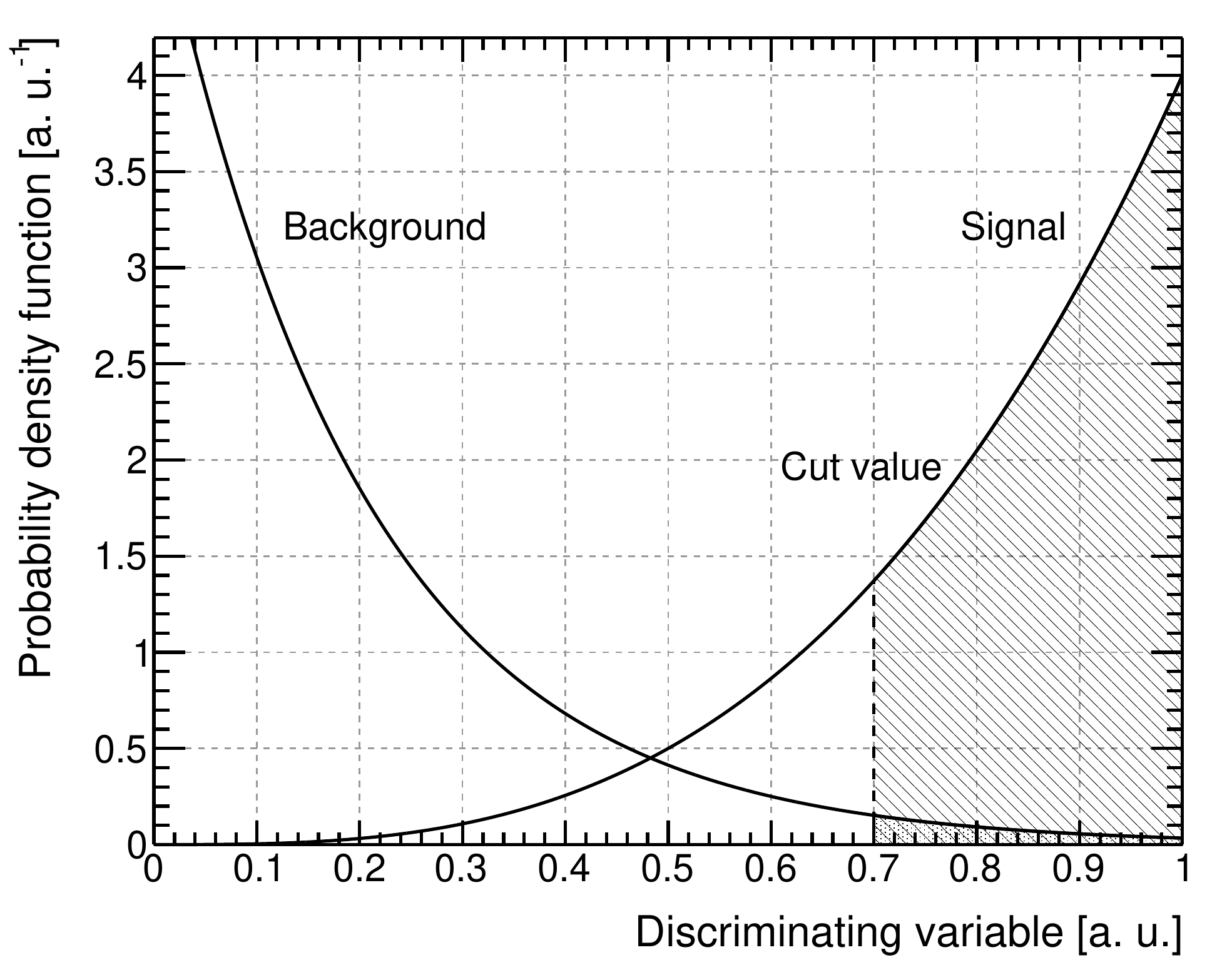}
  \caption{Illustrative probability density functions, for signal and
    background, of a generic discriminating variable (note that the shape
    of the curves, though not necessarily irrealistic, is purely fictional).}
  \label{fig:discr_variable}
\end{figure}

Figure~\ref{fig:discr_variable} shows an example of the probability density
functions, for signal and background, of a fictional discriminating variable
$x$ in the interval $[0, 1]$ that we shall use for illustrative purposes.
A \emph{cut} on the discriminating variable (e.g., the act of selecting events
for which, say, $x > x_0$) defines the values of the \emph{efficiency} for
signal and background:
\begin{align}
  \varepsilon_{\rm sig} &= \int_{x_0}^1 p_{\rm sig}(x) dx\nonumber\\
  \varepsilon_{\rm bkg} &= \int_{x_0}^1 p_{\rm bkg}(x) dx.
\end{align}
(At this point you want to be sure that
$\varepsilon_{\rm sig} > \varepsilon_{\rm bkg}$, and possibly
$\varepsilon_{\rm sig} \gg \varepsilon_{\rm bkg}$; otherwise you should probably
look for a better discriminating variable.)

\begin{figure}[!htb]
  \includegraphics[width=\linewidth]{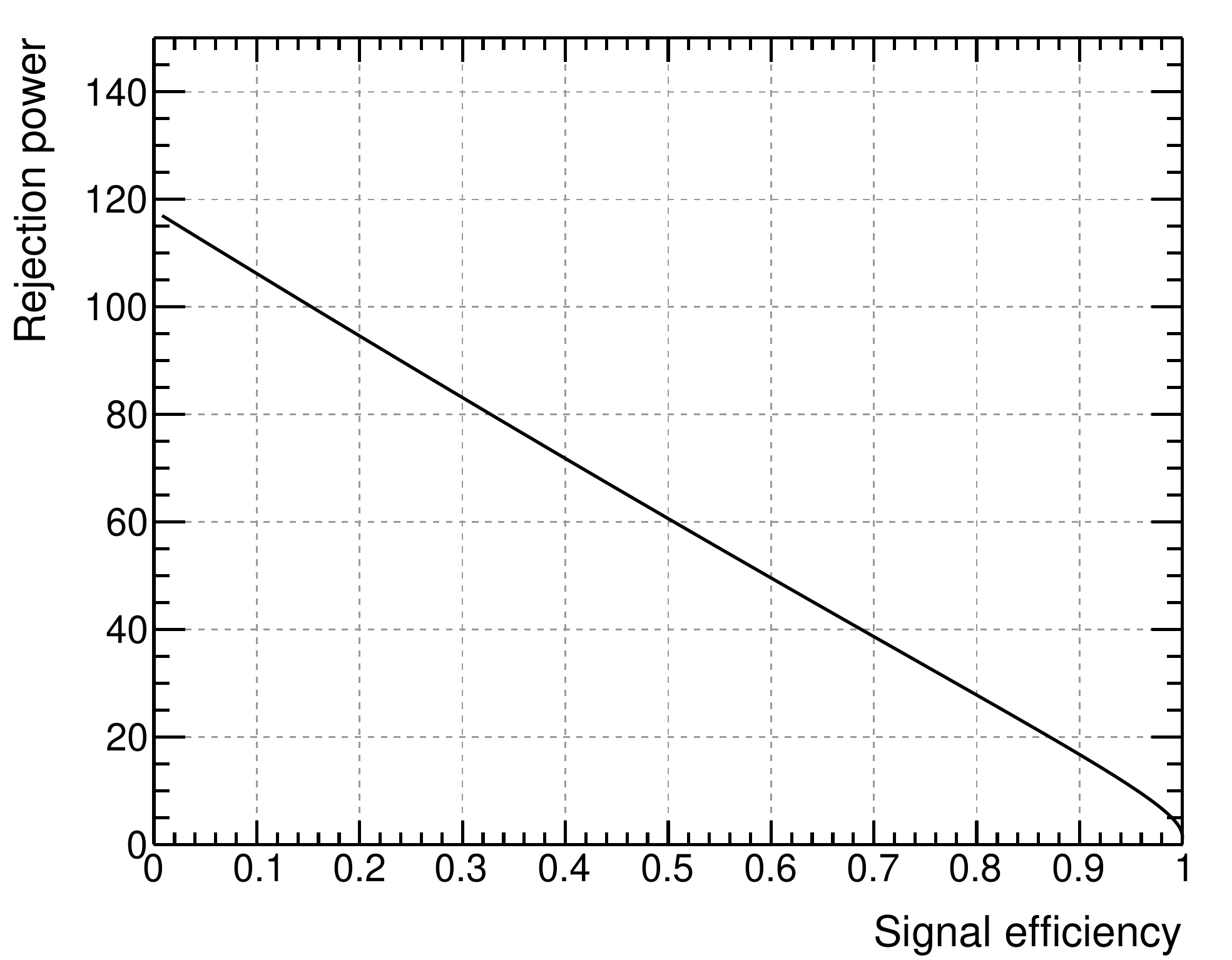}
  \caption{Illustrative trade-off between the signal efficiency and the
    rejection power. The plot refer to the probability density functions
    shown in figure~\ref{fig:discr_variable}.}
  \label{fig:rejection_power}
\end{figure}

Given a selection cut (or a set of selection cuts), the \emph{rejection power}
$\mathcal{R}$ is formally defined as
\begin{align}
  \mathcal{R} = \frac{\varepsilon_{\rm sig}}{\varepsilon_{\rm bkg}}.
\end{align}
The rejection power is a useful concept as it relates the number of
signal and background events before and after the cut
\begin{align}
  \left.\frac{n_{\rm sig}}{n_{\rm bkg}}\right|_{\rm after} =
  \mathcal{R}\left.\frac{n_{\rm sig}}{n_{\rm bkg}}\right|_{\rm before}.
\end{align}
In order to be able to do a meaningful background subtraction, one typically
wants the number of signal events to be much larger than the number of
background events in the final sample, which sets the necessary rejection
power necessary for a given analysis, once the initial ratio is known
(again, see figure~\ref{fig:cr_ratios}). It is worth stressing that
whether it is practically possible to achieve the necessary rejection
power is hostage of the shape of the probability density functions of the
discriminating variables. In real life the optimal cut value is determined
by a trade off between the signal efficiency and the rejection power, as
illustrated in figure~\ref{fig:rejection_power}.

\subsection{Effective Area, Acceptance and Field of View}

Fluxes from (gamma-ray) point sources are customarily measured in particles per
unit area, time and energy, e.g. in \fluxunits. When dealing with a (at least
approximately) isotropic flux of charged particles or photons, the intensity of
such flux is more conveniently measured in particles per unit area, time,
energy and solid angle, e.g. in \intensityunits. In the first case the
conversion factor between the differential source flux $dF/dE$ and the
differential count spectrum $dN/dE$ measured by the detector
(in s$^{-1}$~GeV$^{-1}$) is the \emph{effective area} \aeff:
\begin{align}\label{eq:aeff_flux}
  \frac{dN}{dE} = \aeff \times \frac{dF}{dE}.
\end{align}
(Strictly speaking, as we shall see in the following, this is true only
if the effects of the energy dispersion and the point-spread function are
negligible.)
In the latter case the conversion factor between the differential intensity
$dJ/dE$ and the differential count spectrum $dN/dE$ measured by the detector
(again, in s$^{-1}$~GeV$^{-1}$) is the \emph{acceptance}%
\footnote{Depending on the context, this very same quantity is also referred to
  as geometric factor, effective geometric factor, \emph{etendue}, aperture
  and geometrical aperture (there might be other synonyms the author is not
  aware of).}
\accept:
\begin{align}\label{eq:acc_flux}
  \frac{dN}{dE} = \accept \times \frac{dJ}{dE}.
\end{align}
It goes without saying that the effective area is measured in m$^2$ and the
acceptance in m$^2$~sr.

Equations~\eqref{eq:aeff_flux} and~\eqref{eq:acc_flux} are effectively
\emph{operative} definitions of the effective area and acceptance.
For the reason stated above, the concept of effective area is seldom used by
the cosmic-ray community; on the other hand the acceptance (call it with any of
the different names mentioned in the previous paragraph) is relevant for both
cosmic-ray and gamma-ray detectors.
We also note, in passing, that typically we use the equations
\eqref{eq:aeff_flux} and \eqref{eq:acc_flux} in the other direction, i.e.,
we \emph{divide} the measured count spectrum by the effective area or
acceptance to recover the actual flux or intensity. We shall see that,
when the energy redistribution due to the finite detector energy resolution is
not negligible, this has profound implications.

The effective area is in general defined for a give energy $E$ and viewing
direction. In the following of this section we shall indicate with $\theta$
and $\phi$ the polar and azimuthal angles in instrument coordinates.
In broad terms (and in a somewhat arbitrary fashion), the effective area can
be factored out in three different pieces:
\begin{align}\label{eq:aeff_factorization}
  \aeff(E, \theta, \phi) =
  A_{\rm geo}(\theta, \phi)
  \varepsilon_{\rm det}(E, \theta, \phi)
  \varepsilon_{\rm sel}(E, \theta, \phi),
\end{align}
where the $A_{\rm geo}(\theta, \phi)$ is the geometric cross-sectional area
presented by the instrument toward a given direction,
$\varepsilon_{\rm det}(E, \theta, \phi)$ is the detection efficiency at a given
energy and incidence direction (for gamma-ray detectors this includes the
conversion efficiency) and $\varepsilon_{\rm sel}(E, \theta, \phi)$ is the
efficiency of the selection cuts (e.g., for suppressing the backgrounds).
The last factor in equation~\eqref{eq:aeff_factorization} makes it clear that
the effective area, as we said, is not an intrinsic property of the detector, as
it subtends a specific event selection.

The effective area at normal incidence (or on-axis effective area)
will be used frequently in the following and deserves a dedicated notation
\begin{align}
  \aeffnorm(E) = \aeff(E, \theta = 0).
\end{align}
(Note that, for $\theta = 0$, there is a degeneracy on $\phi$ and \aeff\
only depends on $E$.)

The acceptance (or geometric factor) is formally defined as the integral
of the effective area over the solid angle:
\begin{align}
  \accept(E) = \int_{\Omega}\aeff(E, \theta, \phi) d\Omega,
\end{align}
and the field of view as the ratio between the geometric factor and the
effective area at normal incidence:
\begin{align}
  \fov(E) = \frac{\accept(E)}{\aeffnorm(E)} =
  \frac{\int_{\Omega}\aeff(E, \theta, \phi) d\Omega}{\aeffnorm(E)}.
\end{align}
(Note that when the angular dependence of the effective area is different at
different energies, the field of view does depend on energy, see, e.g.,
\cite{2012ApJS..203....4A}.)

\subsubsection{The Ideal Planar Detector}%
\label{sec:planar_detector}

In this section we shall consider the (somewhat irrealistic) scenario of a
planar (i.e. infinitely thin) detector with side $l$. We shall assume that our
detector is sensitive to all the particles coming from the upper hemisphere and
crossing it, irrespectively of the event energy. Sure enough most of the actual
detectors do not quite look like this, but as we shall see this academic
example allows to discuss the basic concepts introduced in the previous
section in a simple context.

\begin{figure}[!htb]
  \includegraphics[width=\linewidth]{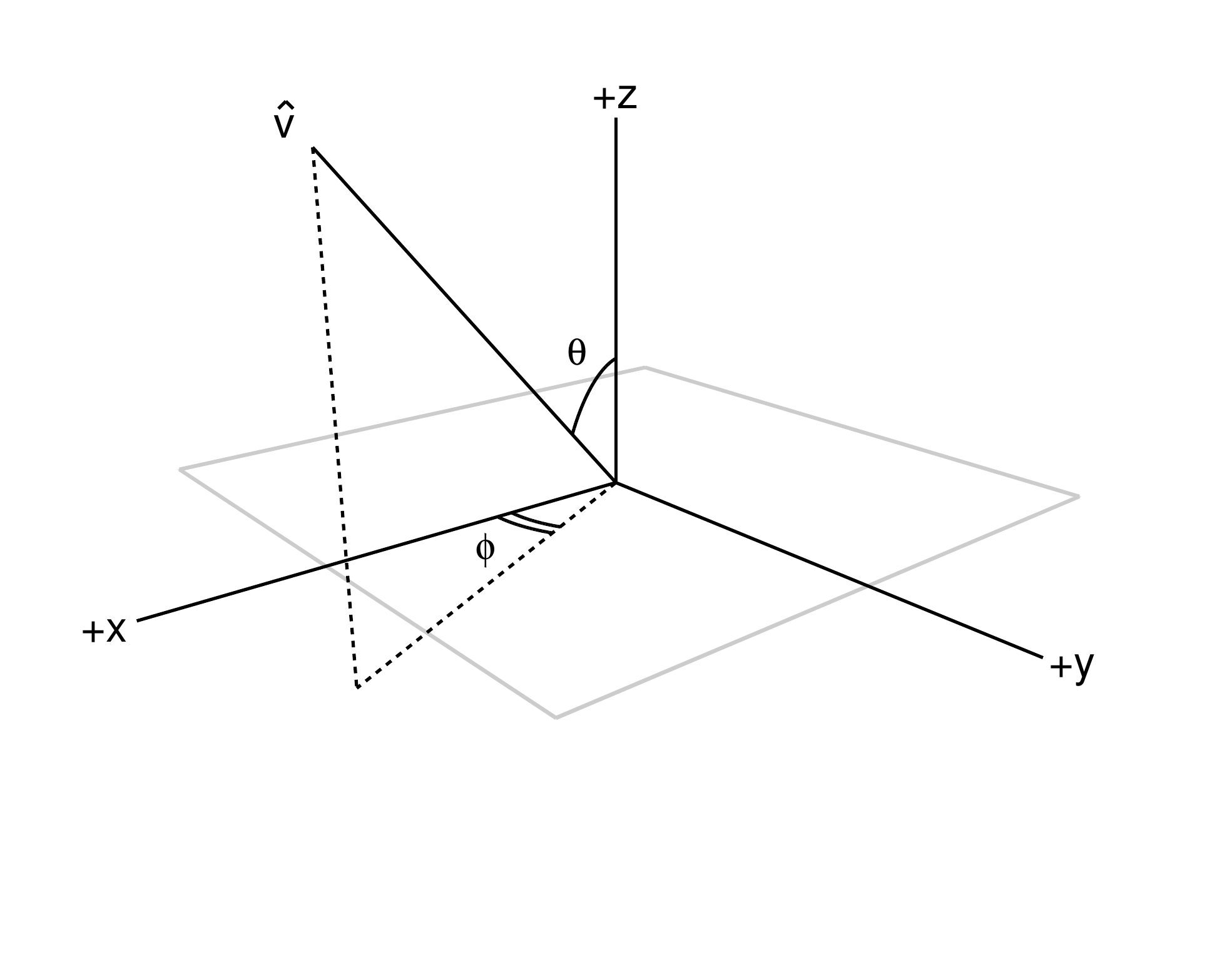}
  \caption{Sketch of an ideal planar detector. The $\theta$ and $\phi$ angles
    define the basic reference system (in instrument coordinates) used
    throughout this section to parameterize the instrument response.}
  \label{fig:plane_detector}
\end{figure}

In this case the effective area at normal incidence is simply the geometrical
area of the detector $S = l^2$. It is clear that \aeff\ only depends on the
polar angle $\theta$:
\begin{align}
  \aeff(E, \theta, \phi) = S \cos\theta,
\end{align}
and the acceptance reads:
\begin{align}
  \accept = S\int_0^{2\pi}\!d\phi
  \int_0^\pi \!\cos\theta\sin\theta d\theta = \pi S.
\end{align}
The field of view, finally, is
\begin{align}
  \fov = \frac{\accept}{S} = \pi.
\end{align}
The reader might be wondering why the field of view is not $2\pi$---after
all our detector sees the entire upper hemisphere, doesn't it? Well, that
would be if the effective area was independent of the off-axis angle. In our
case the effective area is monotonically decreasing with $\theta$, and in
fact it is zero at $\theta = 90^\circ$. This is the basic reason why the
field of view is half the value one might na\"ively expect. In turn, this
points out that the field of view \emph{is not} the solid angle subtended
by the maximum off-axis angle the detector is accepting events from. More on
that in the next section.

\subsubsection{A more realistic example}

\begin{figure}[!htb]
  \includegraphics[width=\linewidth]{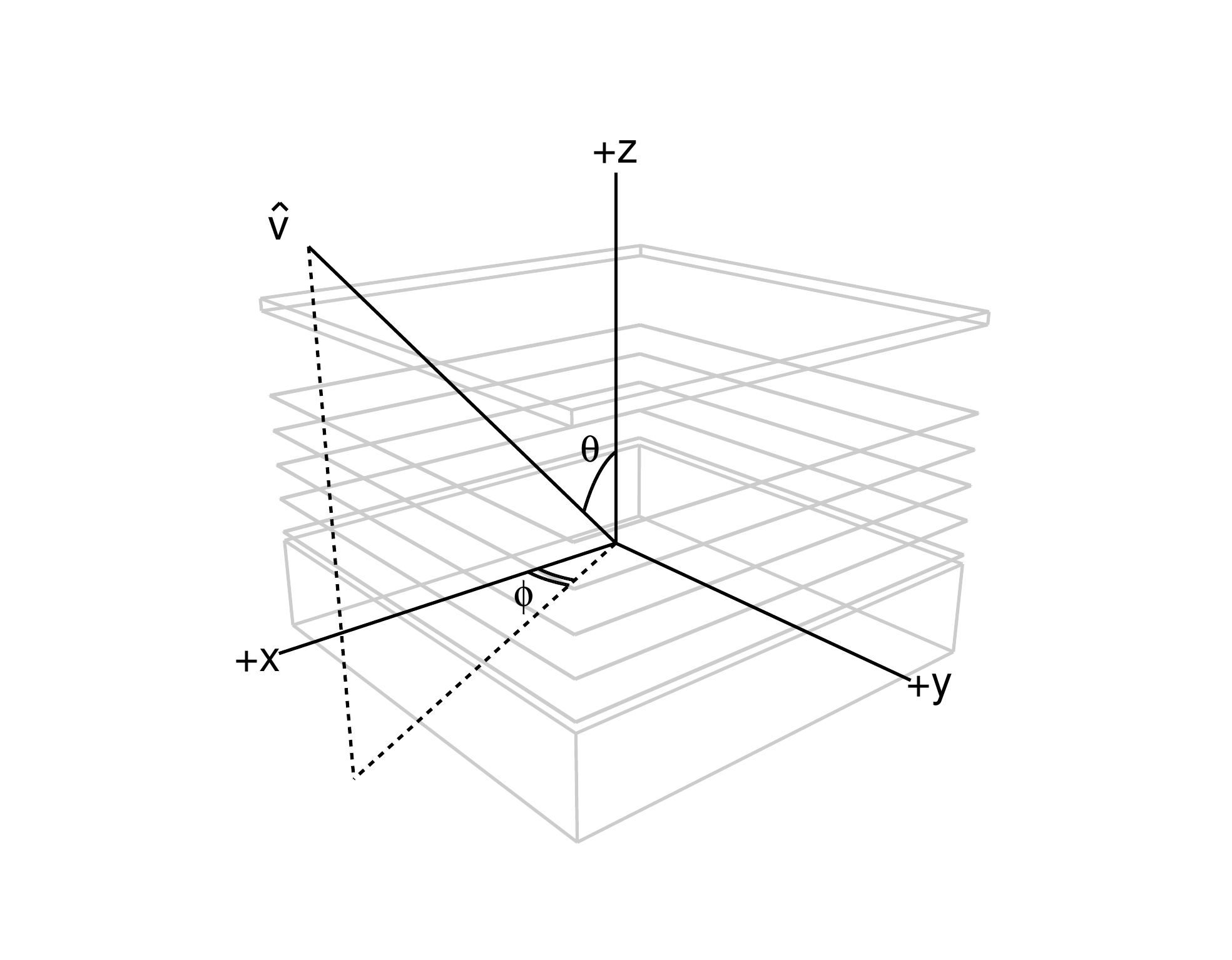}
  \caption{Sketch of a simple, semi-realistic detector. The $\theta$ and
    $\phi$ angles define the basic reference system (in instrument coordinates)
    used throughout this section to parameterize the instrument response.}
  \label{fig:simple_detector}
\end{figure}

Figure~\ref{fig:simple_detector} shows a more realistic (though still fictional)
example of cosmic-ray detectors. We have seen the sketches of some actual
detectors back in figure~\ref{fig:detectors}.

\begin{figure}[!hbt]
  \includegraphics[width=\linewidth]{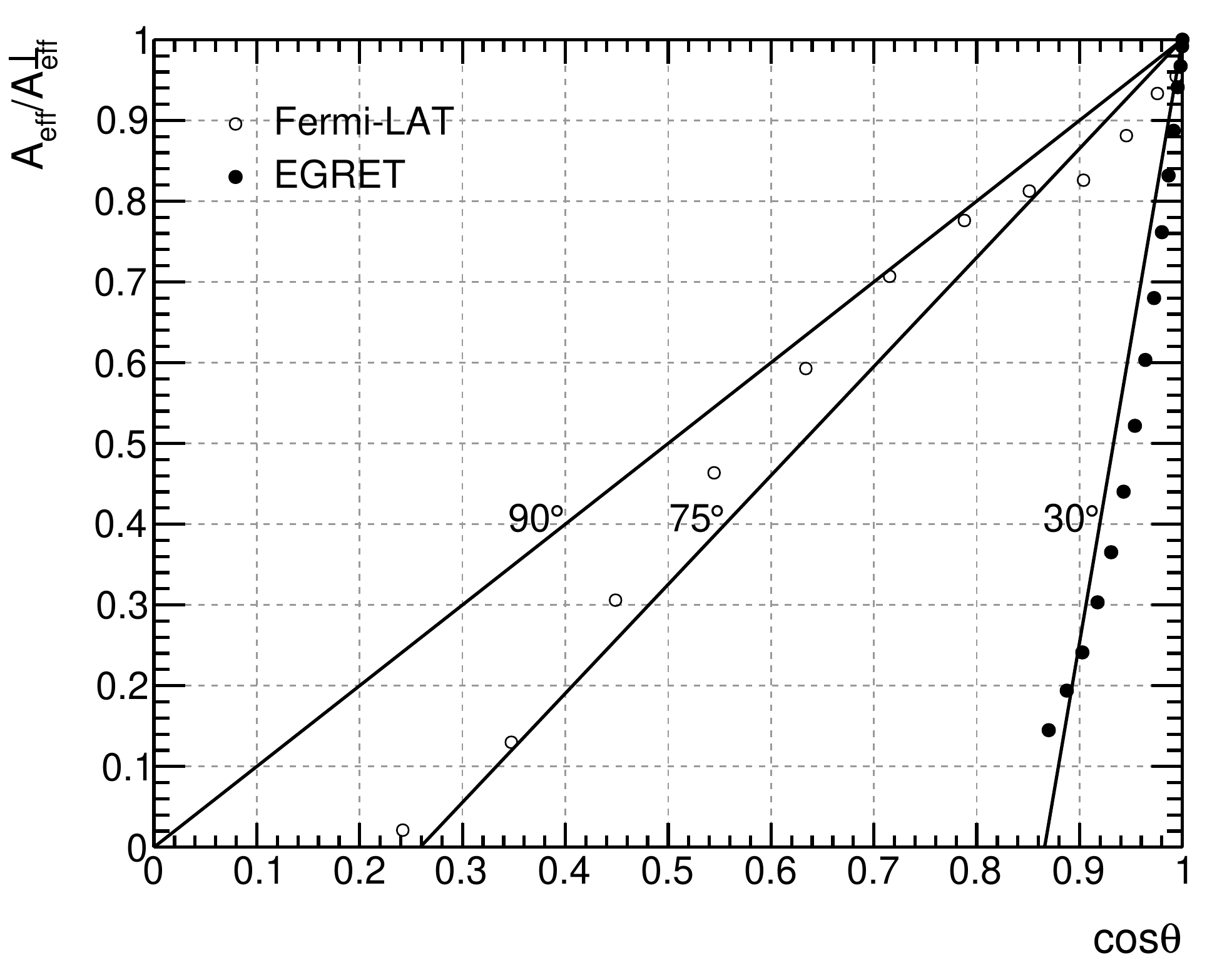}
  \caption{Graphical representation of our crude model \eqref{eq:aeff_vs_theta}
    for three illustrative values of $\theta_{\rm max}$: $90^\circ$ (the planar
    detector), $75^\circ$ and $30^\circ$. Overlaid are the actual values for
    the \Fermi-LAT and EGRET, which are reasonably represented (within $15\%$ or
    so) by the $75^\circ$ and $30^\circ$ models, respectively.}
  \label{fig:fov}
\end{figure}

The basic detector layout and the arrangement of its active elements determines
the maximum off-axis angle $\theta_{\rm max}$ at which events can be recorded
(note that $\theta_{\rm max}$ can possibly depend on the event energy, e.g., if
a longer path length in the calorimeter is required for reconstructing
high-energy events, but we shall neglect this possible dependence for the sake
of simplicity---and we shall also assume that the selection cuts do not
introduce any energy dependence).
This is in general not enough to estimate the actual field of view, as one
really needs a full parameterization of the $\theta$-dependence of the
effective area. The crudest possible approximation that one can do is to assume
that the effective area for any given theta, relative to the on-axis effective
area, decreases linearly with $\cos\theta$, with the proper boundary conditions
\begin{align}\label{eq:aeff_vs_theta}
  \aeff(\theta) = \aeffnorm
  \left(\frac{\cos\theta - \cos\theta_{\rm max}}{1 - \cos\theta_{\rm max}}\right),
\end{align}
as shown in figure~\ref{fig:fov}.

While the author would not recommend counting on this kind of models for an
actual science analysis, equation~\eqref{eq:aeff_vs_theta}
is typically adequate for back-of-the-envelope calculations. We note, in
passing, that the planar detector discussed in the previous section is a
particular case (with $\theta_{\rm max} = 90^\circ$) of this general family of
effective area curves.

The acceptance for our na\"ive detector model reads
\begin{align}
  \accept = \int_0^{2\pi}\!\!\!\!d\phi
  \int_0^{\theta_{\rm max}} \!\!\!\!\!\!\!\!\aeff(\theta)\sin\theta d\theta = 
  \pi \aeffnorm (1 - \cos\theta_{\rm max})
\end{align}
(you can explicitly calculate the integral or just look at
figure~\ref{fig:fov}: when you note that $\sin\theta d\theta = -d\cos\theta$
we are really trying to calculate the area of a triangle).

The field of view is then easily calculated as:
\begin{align}\label{eq:fov_vs_thetamax}
  \fov = \pi (1 - \cos\theta_{\rm max}),
\end{align}
which reduces to the results we obtained for the planar detector when
$\theta_{\rm max} = 90^\circ$. We note that, as in the case of the planar detector,
this is exactly half of the solid angle subtended by the cone defined by
$\theta_{\rm max}$.

\begin{figure}[!htb]
  \includegraphics[width=\linewidth]{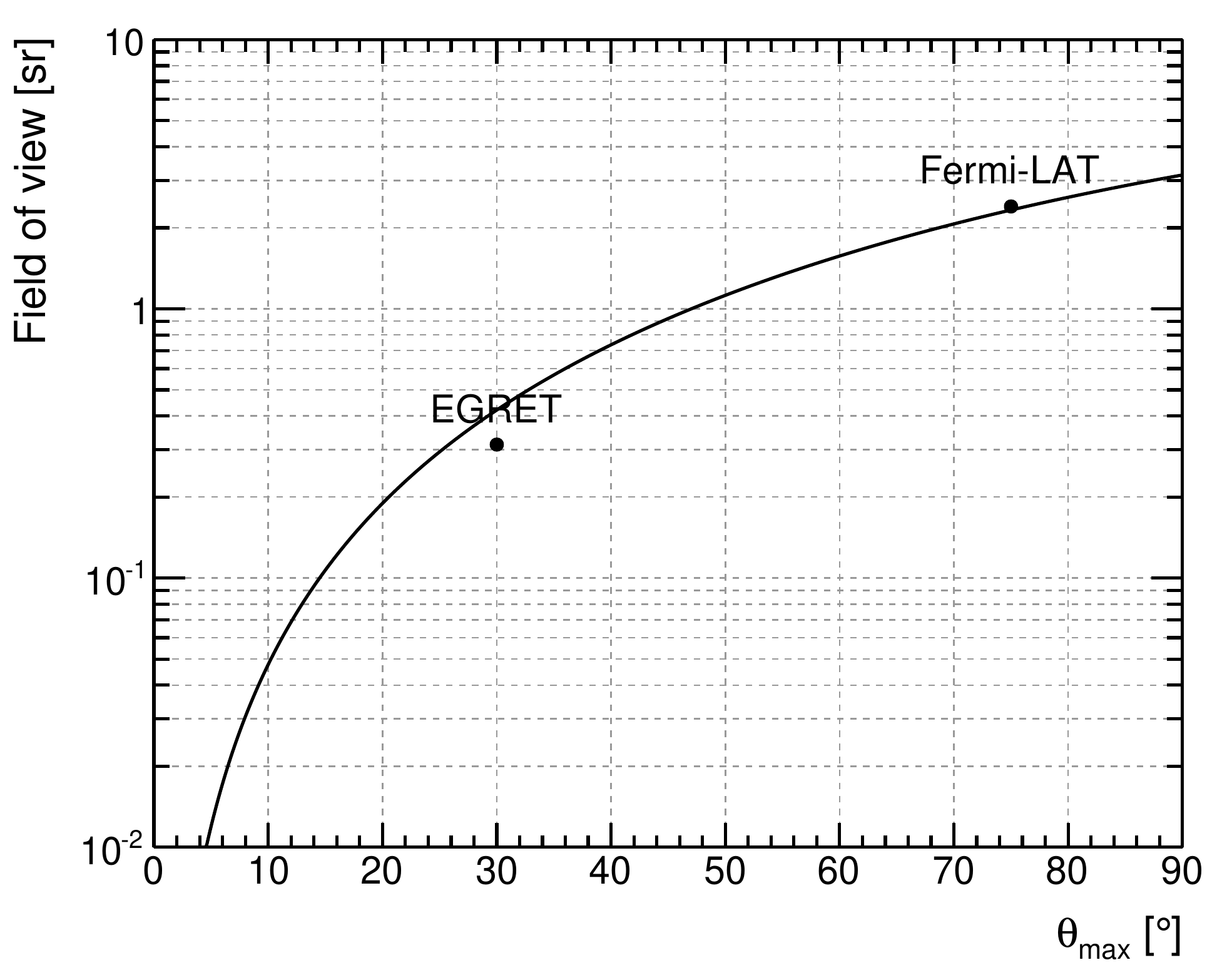}
  \caption{Field of view as a function of $\theta_{\rm max}$, as estimated
    from equation~\ref{eq:fov_vs_thetamax}. Overlaid are the actual values for
    the \Fermi-LAT and EGRET.}
  \label{fig:fov_vs_thetamax}
\end{figure}

\subsubsection{Event distributions in instrument coordinates}%
\label{sec:theta_instr_distr}

It is interesting to consider the probability density function of the event
directions in instrument coordinates---most notably that of the off-axis angle
$\theta$---for an isotropic input distribution.
Looking at equation~\eqref{eq:aeff_vs_theta} and figure~\ref{fig:fov} one might
be tempted to conclude that $dN/d\theta$ has its maximum for $\theta = 0$,
which is not quite true. Figure~\ref{fig:fov} represents $dN/d\cos\theta$, while
\begin{align}
  \frac{dN}{d\theta} = \frac{dN}{d\cos\theta} 
  \left|\frac{d\cos\theta}{d\theta}\right| = 
  \frac{dN}{d\cos\theta} \sin\theta.
\end{align}
One can easily recognize the solid angle term in the last piece.

\begin{figure}[!htb]
  \includegraphics[width=\linewidth]{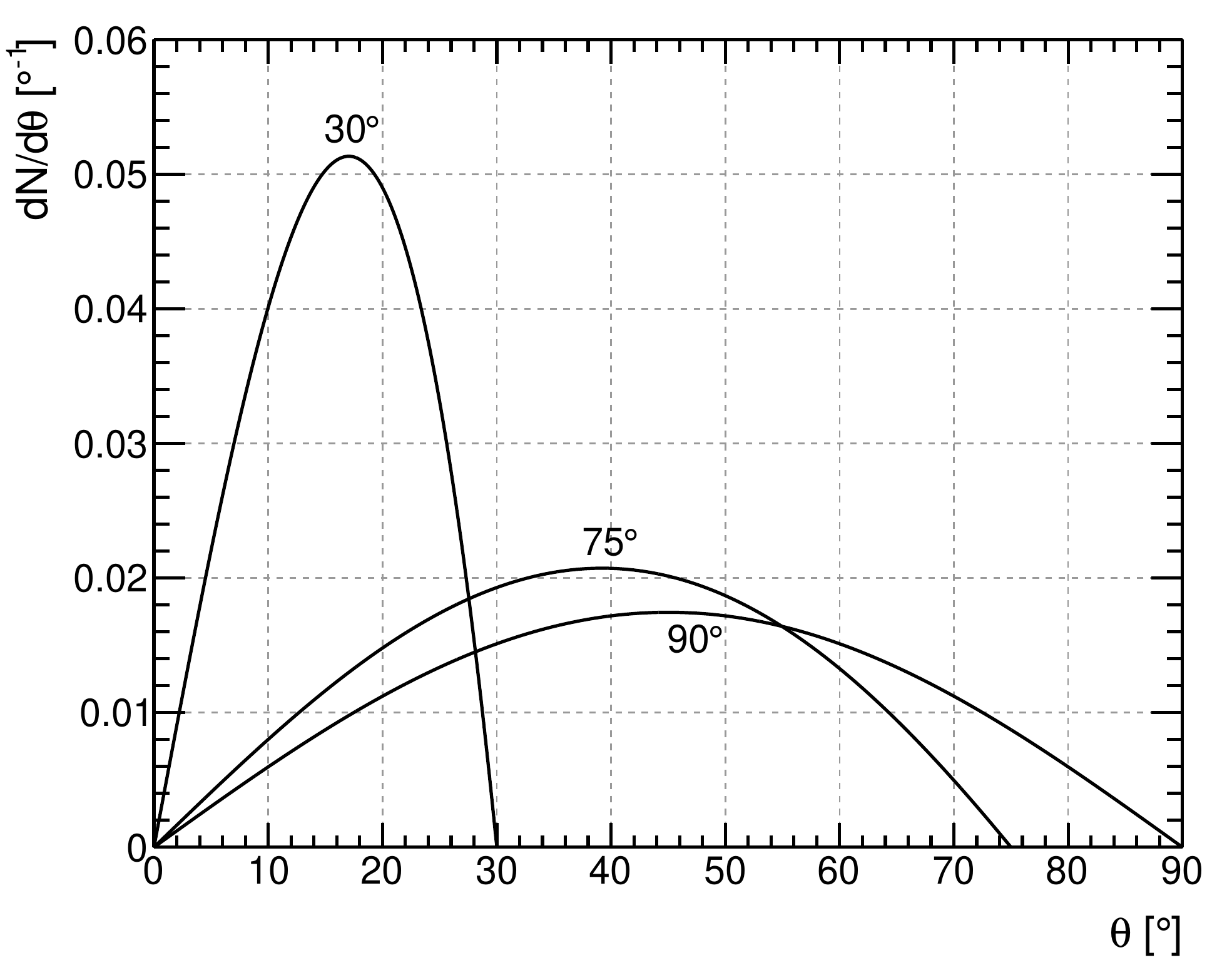}
  \caption{Illustrative $\theta$ distributions, taken from
    equation~\ref{eq:theta_dist}.}
  \label{fig:theta_dist}
\end{figure}

It is easy to calculate the normalized probability density function
\begin{align}\label{eq:theta_dist}
  \frac{dN}{d\theta} =
  \frac{(\cos\theta - \cos\theta_{\rm max})\sin\theta}%
       {(\frac{1}{2} + \frac{1}{2}\cos^2\theta_{\rm max} - \cos\theta_{\rm max})}.
\end{align}
In the limit in which our na\"ive model is a reasonable approximation of
reality, this is actually a very useful relation, as it allows to derive
a series of metrics germane to the path length of the events in the detector.
For one thing one could calculate the average value or the most probable value
of $\theta$---this is not too difficult, though somewhat lengthy.
Since the path length in the detector at a given incidence angle is given
by $1/\cos\theta$, the most important figure of merit is the expectation
value of this quantity, which physically corresponds to the average amount
of material $\left<t\right>$ traversed by an event, relative to the instrument 
depth on axis $t_0$:
\begin{align}
  \frac{\left<t\right>}{t_0} =
  \int_0^{\theta_{\rm max}} \frac{dN}{d\theta} \frac{1}{\cos\theta} d\theta.
\end{align}
The integral is not as difficult as it might seem at first glance and the
answer is:
\begin{align}\label{eq:pathlength_vs_thetamax}
  \frac{\left<t\right>}{t_0} =
  \frac{1 + \cos\theta_{\rm max} (\ln\left|\cos\theta_{\rm max}  \right| - 1)}%
       {(\frac{1}{2} + \frac{1}{2}\cos^2\theta_{\rm max} - \cos\theta_{\rm max})}.
\end{align}

\begin{figure}[!htb]
  \includegraphics[width=\linewidth]{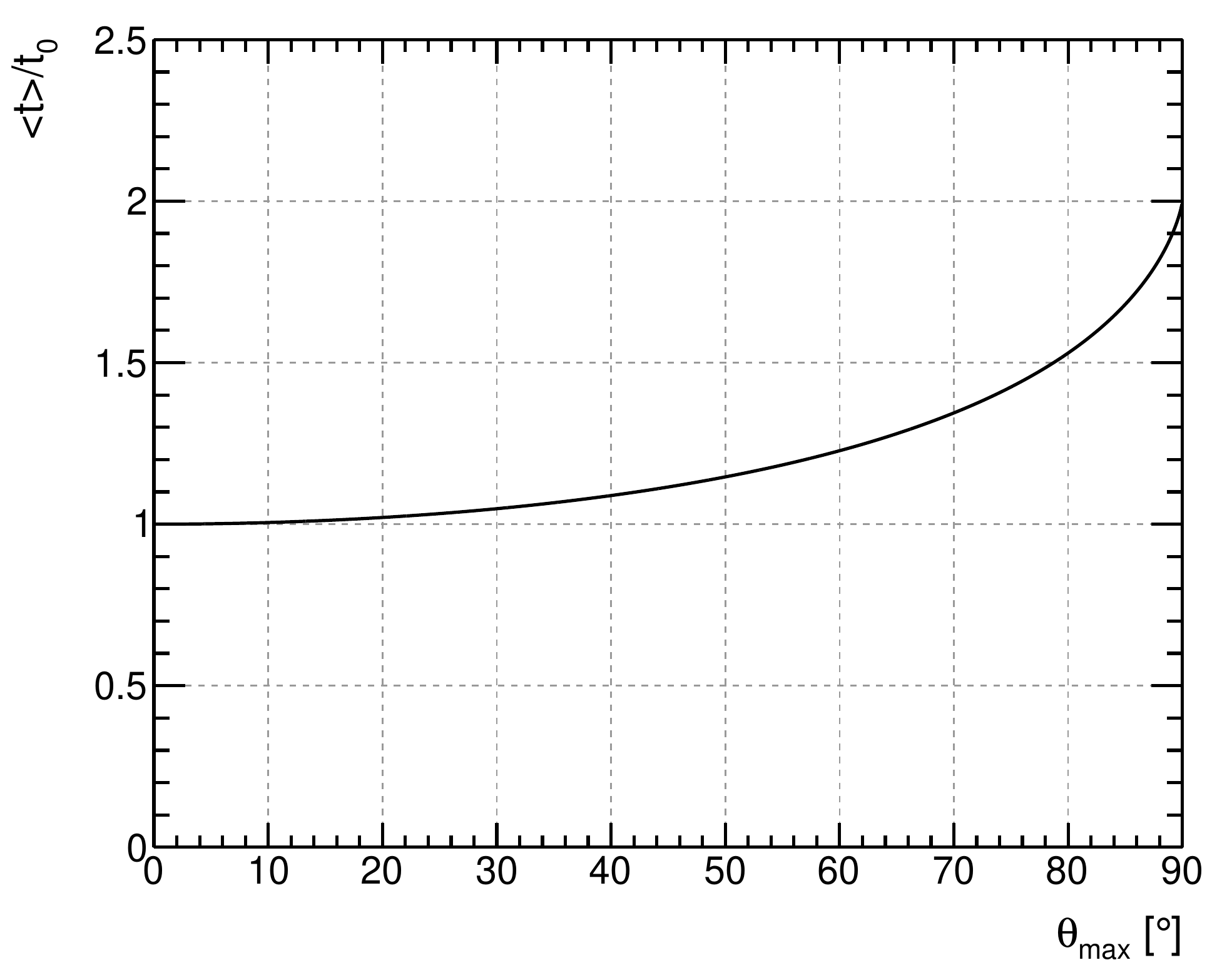}
  \caption{Average path length (relative to the depth at normal incidence)
    for an isotropic input distribution as a function of the maximum
    off-axis angle $\theta_{\rm max}$.}
  \label{fig:pathlength}
\end{figure}

This is plotted in figure~\ref{fig:pathlength}, which makes clear that the
effect can be quite noticeable for instruments with a large field of view.
To put things in context, for the \Fermi-LAT, with a total depth at normal
incidence of $10.1~X_0$ and a maximum off-axis angle of
$\theta_{\rm max} \sim 75^\circ$, equation~\eqref{eq:pathlength_vs_thetamax} gives
an average path-length of $\left<t\right> \sim 14~X_0$%
\footnote{And in fact this is is not quite right, as the correct answer
from a full Monte Carlo simulation is $12.5~X_0$ (see Figure~15 in
\cite{2010PhRvD..82i2004A}). The discrepancy is due to edge effects and the
cracks between calorimeter modules in the LAT, but we have already beaten this
subject to death and now it's time to move on.}%
.
As one can imagine, this has rather important implications for the energy
reconstruction at high energy.

\subsection{Energy Dispersion}

The energy dispersion $\edisp(E; E_{\rm true})$ is the probability density to
measure an energy $E$ for an event with (true) energy $E_{\rm true}$. In general
the energy dispersion depends on the incidence direction and impact point of
the incoming particle, but in the following of the section we shall pretend
this is not the case. Unlike the effective area, the energy dispersion is not
a scalar, but a probability density function. From an operational standpoint,
the energy dispersion is essentially the distribution of the measured energy
values for a monochromatic particle beam. It goes without saying that the
energy dispersion is in general different for different cosmic-ray
species---most notably it is typically much narrower and well-behaved for
electrons, positrons and photons than for hadrons.

The energy resolution \eresl\ is typically defined as the half width of the
smallest energy window containing 68\% of the energy dispersion, divided by the
most probable value of the energy dispersion itself. Difficult as it might seem,
this definition reduces to the ratio $\sigma/\mu$ in the gaussian case.
At any given energy the energy resolution is a scalar and it is \emph{the}
figure of merit which is customarily used to summarize the information
contained in the energy dispersion.

\subsubsection{Effect of a finite energy resolution}

\begin{figure}[!htb]
  \includegraphics[width=\linewidth]{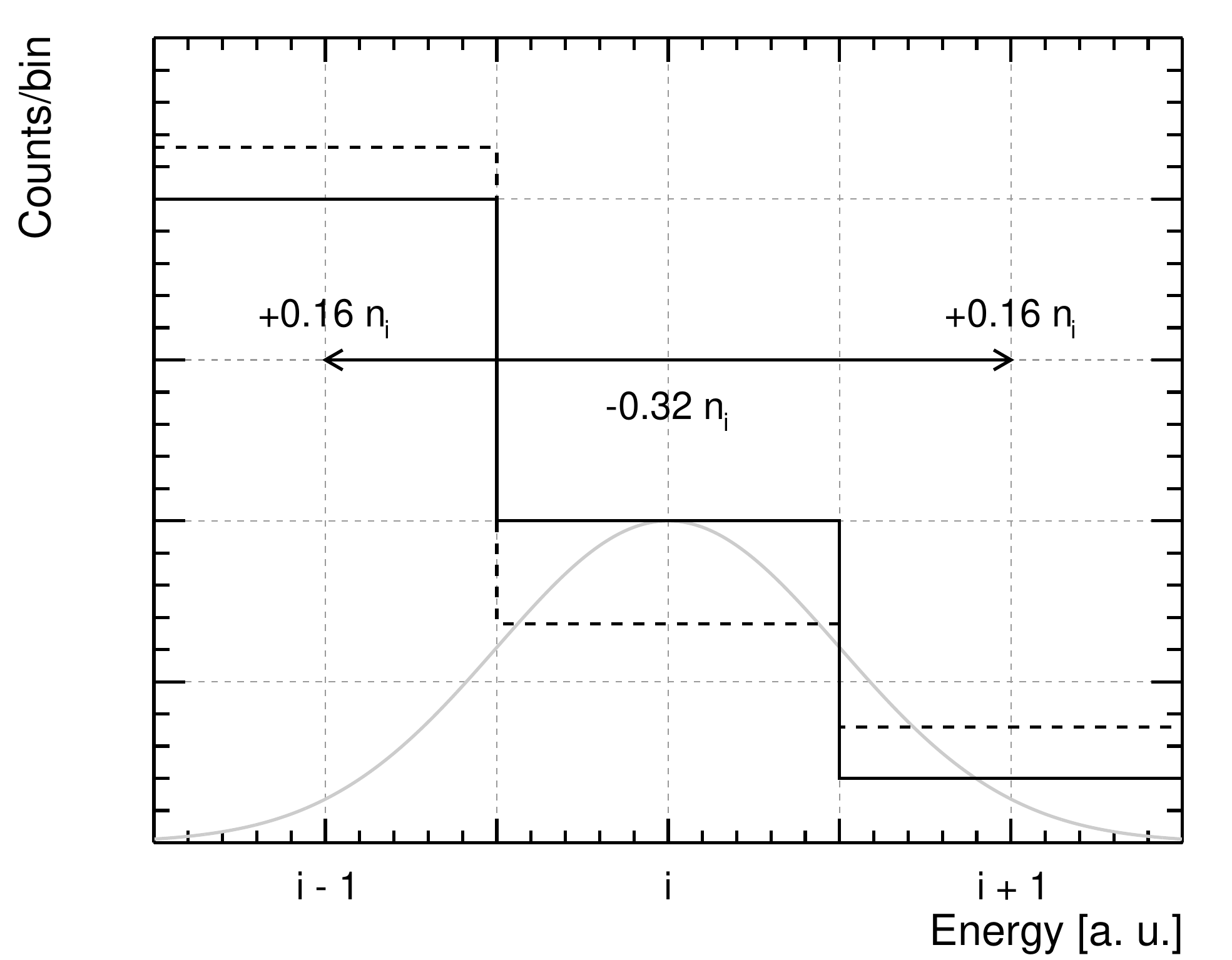}
  \caption{Graphical illustration of the energy redistribution, or bin-to-bin
    migration. In this case the bin width is equal to the energy resolution
    and the gray gaussian illustrates the energy dispersion at the
    center of the $i$-th bin.}
  \label{fig:energy_redistribution}
\end{figure}

When measuring particle spectra binned in energy, the energy dispersion
originates the phenomenon of the \emph{event migration}, or
\emph{bin-to-bin migration}---namely the fact that events that would really
belong to a given energy bin are assigned to a different (hopefully neighbor)
bin. Strictly speaking, this implies that our operational definitions of
the effective area and acceptance, equations~\eqref{eq:aeff_flux}
and~\eqref{eq:acc_flux}, no longer hold: one cannot divide the count spectrum
by the effective area or the acceptance to recover the actual source flux
or intensity. Whether the effect is negligible or not really depends on the
measurement setup, i.e., the details of the energy dispersion and the
input spectrum.
The effect is sketched in figure~\ref{fig:energy_redistribution}.

We note that, when measuring steeply falling spectra such as the typical
cosmic-ray spectra, a prominent right tail in the energy dispersion is
potentially more dangerous than a long left tail, as even a fractionally small
spillover from a low-energy bin can be significant in the next energy bins,
where the number of counts is much smaller.

\begin{figure}[!htb]
  \includegraphics[width=\linewidth]{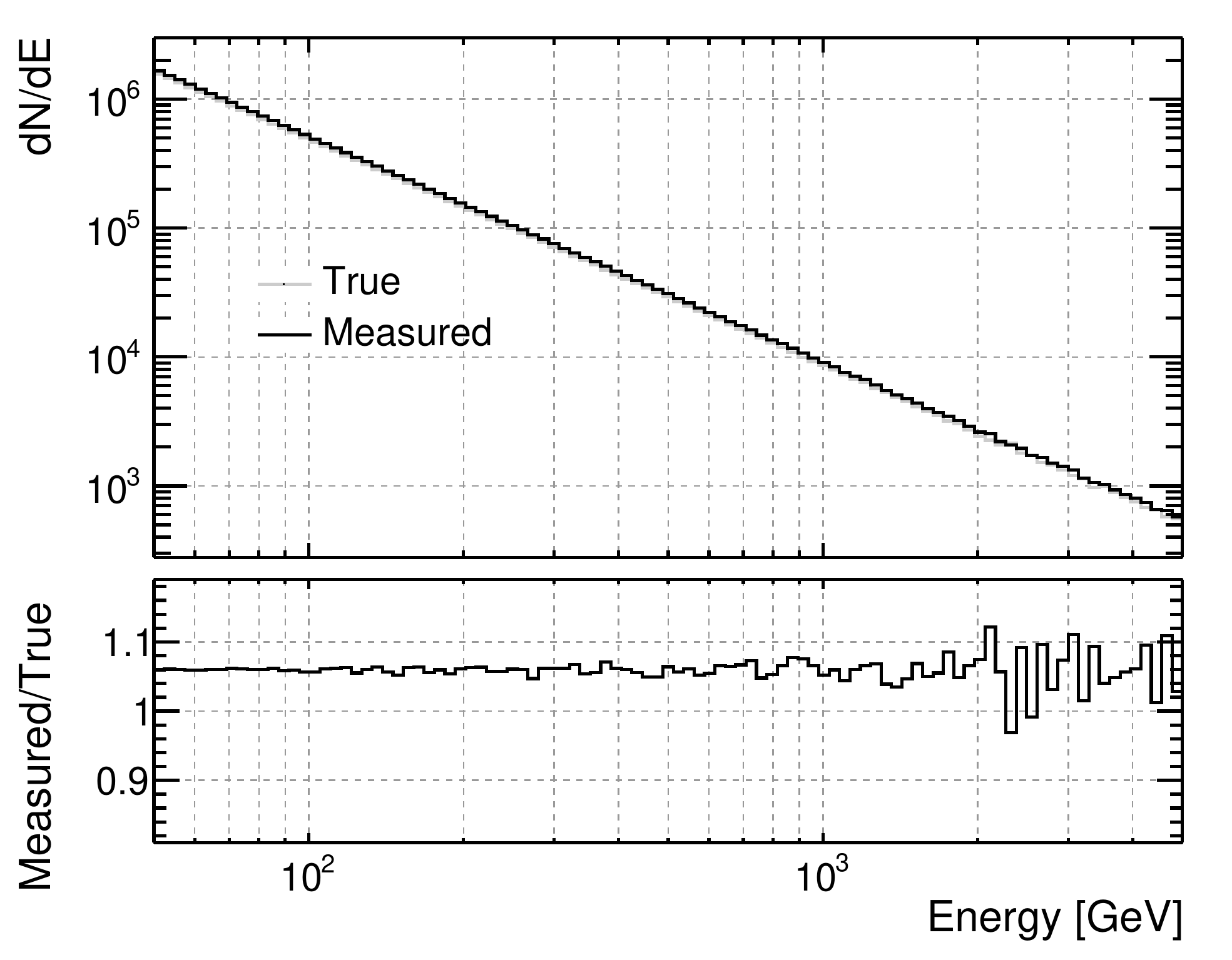}
  \caption{Toy Monte Carlo simulation of an energy spectrum proportional
    to $E^{-2.75}$, measured with an energy resolution of 30\% (constant in
    energy and with a gaussian energy dispersion).
    The setup is largely simplistic, but not totally unreasonable
    as an illustration of a typical measurement of the primary CR proton
    spectrum. The net effect, of the order of $\sim 6\%$, is something
    that must be taken into account for systematic-limited measurements.}
  \label{fig:energy_spec}
\end{figure}

Interestingly enough, a constant energy resolution \eresl\ (i.e., an energy
dispersion whose width increases linearly with the energy), when convoluted
with a power-law count spectrum with index $\Gamma$%
\footnote{Note that we are asking for an input power-law \emph{count} spectrum.
  If the actual source spectrum is a power law, this is only true in the
  limit of an acceptance flat in energy.}%
, causes an offset constant in energy%
\footnote{Technically this only works when our spectrum is binned
  logarithmically in energy, but since this is quite common in practice, we
  shall happily make this assumption.}%
, as illustrated with a toy Monte Carlo simulation in
figure~\ref{fig:energy_spec}. Formally, the convolution is written as 
\begin{align*}
  \frac{dN}{dE} = &
  \int_{-\infty}^\infty \frac{dN}{d\epsilon}\edisp(E; \epsilon)\; d\epsilon =
  \int_{-\infty}^\infty \!\!\!\!N_0\epsilon^{-\Gamma}\edisp(E; \epsilon)\; d\epsilon
\end{align*}
and, while this can be quite complicated to solve in the general case, there is
a few things than can be inferred in a model-independent fashion.
As the normalization factor of the energy dispersion is inversely proportional
to the energy resolution (the normalization factor being, e.g., $1/\sqrt{2\pi}$
in the gaussian case), we can rewrite the convolution
\begin{align*}
  \frac{dN}{dE} = &
  \int_{-\infty}^\infty \frac{N_0}{(\eresl)\epsilon}\epsilon^{-\Gamma}
  w(E; \epsilon)\; d\epsilon = \\
  & \frac{N_0}{(\eresl)} \int_{-\infty}^\infty \epsilon^{-(\Gamma + 1)}
  w(E; \epsilon)\; d\epsilon,
\end{align*}
where $w(E; \epsilon)$ is a window function which is equal to $1$ for
$E = \epsilon$ and effectively limits the integral to a region
$\sim 2k\sigma_E$ wide around $E$, $k$ being a numerical factor of the order
of $1$. If we let $s = k\eresl$, we can rewrite the convolution as
\begin{align*}
  \frac{dN}{dE} \propto & \frac{N_0}{(\eresl)}
  \int_{E(1 - s)}^{E(1 + s)} \epsilon^{-(\Gamma + 1)} \; d\epsilon =\\
  & \frac{N_0E^{-\Gamma}}{(\eresl)\Gamma}
  \left[(1 - s)^{-\Gamma} - (1 + s)^{-\Gamma}\right]
\end{align*}
The ratio $r$ between the count spectra in measured and true energy can be
written as
\begin{align*}
  r \propto \frac{1}{(\eresl)\Gamma}
  \left[(1 - s)^{-\Gamma} - (1 + s)^{-\Gamma}\right]
\end{align*}
(note that the dependence on $E$, as anticipated, disappeared).
Now, the expression in square brackets can be expanded---and you need to 
go all the way up to the third order if you want to get the first non-vanishing
correction---giving
\begin{align*}
  r - 1 \approx \frac{k^2(\Gamma - 1)(\Gamma - 2)}{3}\left(\eres\right)^2,
\end{align*}
which contains the relevant dependencies on $\Gamma$ and \eresl. Particularly,
we note that the offset scales with the \emph{square} of the energy
resolution\footnote{This is actually a good thing as, being $\eresl \ll 1$,
the effect tends to be small.}
and that it is identically $0$ for $\Gamma = 2$ (interestingly enough, this
is a typical spectral index for gamma-ray sources). We also note that the
offset is always positive for $\Gamma > 2$, which has to do with the fact that,
with a steep enough spectrum, the effect of the right tail of the energy
dispersion dominates over that of the left tail.

\begin{figure}[!htb]
  \includegraphics[width=\linewidth]{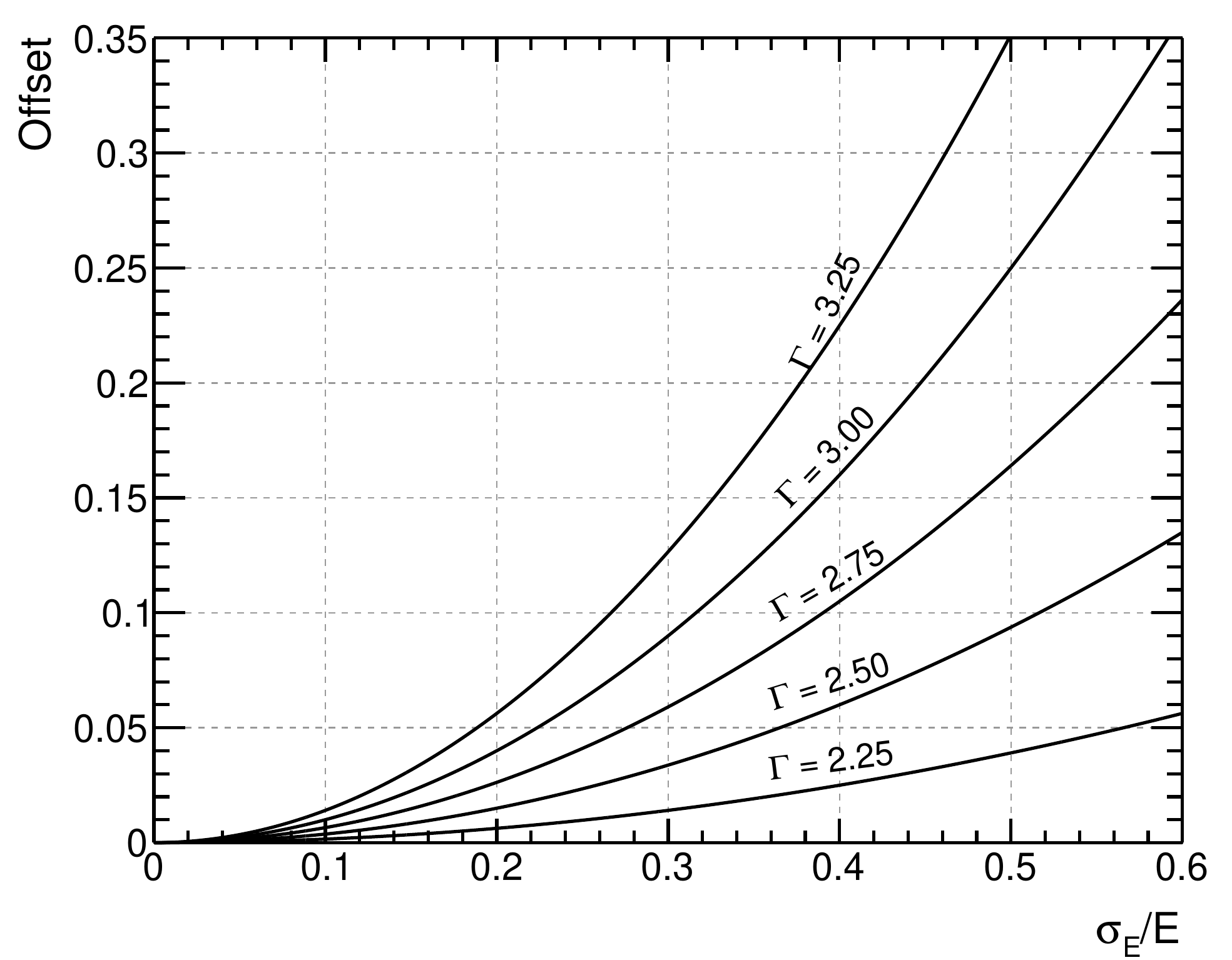}
  \caption{Offset caused by a finite energy resolution \eresl, coupled
    to a power-law count spectrum with index $\Gamma$, for different
    values of $\Gamma$, according to equation~\eqref{eq:edisp_offset}.}
  \label{fig:edisp_pl_offset}
\end{figure}

It it actually easy to explicitly work out the calculation for the
(admittedly, purely academic) example of a uniform energy dispersion and the
solution turns out to be
\begin{align}\label{eq:edisp_offset}
  r - 1 \approx \frac{(\Gamma - 1)(\Gamma - 2)}{2}\left(\eres\right)^2.
\end{align}
Since this formula actually works for the much more relevant case of a 
gaussian energy dispersion, it can be quite useful in practice.
For $\Gamma = 2.75$ and $\eresl = 30\%$, equation~\eqref{eq:edisp_offset} gives
a value of $r - 1 = 5.9\%$, in agreement with the toy
Monte Carlo in figure~\ref{fig:energy_spec}.
The general behavior of equation~\eqref{eq:edisp_offset} is illustrated in
figure~\ref{fig:edisp_pl_offset}.

We conclude noting that, when coupling an energy-dependent acceptance to the
(energy dependent) energy dispersion, one can potentially get all kinds of
spectral distortion even in the case of a plain power-law input spectrum.
In general the problem must really be studied on a case-by-case basis and
the reader is referred to~\cite{2012ApJS..203....4A} for more details.

\subsubsection{The absolute energy scale}

The uncertainty in the absolute energy scale (i.e. in the most probable value
of the energy dispersion) is another source of potential spectral deformation.
Verifying the absolute energy scale in orbit is a non trivial task, as there
isn't very many spectral features (at know position) available---see however
section~\ref{sec:calib_energy_scale} for more details. The energy scale is
typically calibrated on the ground with particle beams up to the highest
available beam energies, but then extrapolating the response at (possibly much)
higher energies is a challenge in many contexts.

It is easy enough to show that, in the case of power-law count spectrum and
a constant fractional error $s_E$ on the energy scale---e.g., the measured
energy is systematically higher or lower than the true energy by a factor
$(1 + s_E)$ on average---the net effect is again a fractional offset constant
in energy, similar to that induced by a finite energy resolution.
In this case the ratio $r$ between the count spectra $dN/dE$ in measured and
true energy is given by the combination of two effects: the numerator
$dN$ increases (or decreases, if $s_E$ is negative) by a factor
$(1 + s_E)^\Gamma$, and the denominator increases (or decreases) by a factor
$(1 + s_E)$:
\begin{align*}
  r = \frac{(1 + s_E)^\Gamma}{(1 + s_E)\phantom{^\Gamma}} =
  (1 + s_E)^{\Gamma - 1} \approx  1 + (\Gamma - 1)s_E,
\end{align*}
or
\begin{align}
  r - 1 \approx (\Gamma - 1)s_E.
\end{align}

For $\Gamma = 3$, for instance, an error of the absolute energy scale of
5\% implies a corresponding error on the flux measurement of 10\%.
As for the case of a finite energy resolution discussed in the previous
section, we conclude noting that an energy-dependent error on the absolute
energy scale (or a constant error coupled to a curved input spectrum) do
cause spectral deformations that need to be studied case by case.

\subsubsection{Correcting for the energy dispersion}%
\label{sec:correct_edisp}

As mentioned in passing before, if the effect of the energy dispersion cannot
be ignored, equations~\eqref{eq:aeff_flux} and~\eqref{eq:acc_flux} cannot
be used, as they relate the source flux to the observed counts---through the
effective area or the acceptance---\emph{all evaluated at the true event
energy.}
In general, events with true energy $E$ will contribute instead to the counts
measured at a different energy $E'$ due to the energy redistribution.
If we are doing a binned analysis, the count rate in the $i$-th (measured)
energy bin will be a sum of contributions from the source flux in several
(true) energy bins:
\begin{align}\label{eq:drm}
  n_i = C_{ij} F_j.
\end{align}
$C_{ij}$ is customarily called the detector response matrix (DRM).
If $n_i$ is measured in s$^{-1}$, or counts per second, and $F_j$ in
m$^{-2}$~s$^{-1}$, the detector response matrix has the physical dimensions of an
area---and it is in fact a generalization of the effective area.
If the detector response matrix is diagonal, we are effectively back to the
original discussion.

It goes without saying that one is typically interested in using
equation~\eqref{eq:drm} in the opposite direction, i.e., to recover the actual
source flux from the measured counts. The reader might be tempted to say
that the problem is trivially invertible by doing
\begin{align}\label{eq:unfolding_matrix_invert}
   F_i = \left( C_{ij} \right)^{-1} n_j.
\end{align}
An effective way to understand that this is actually a terrible idea is
a simple and amusing argument by Barlow~\cite{barlow_slac_lecture}. Assume that
we have two energy bins and the detector response matrix reads, in some
units:
\begin{align*}
  C =
  \begin{pmatrix}
    0.6 & 0.4\\
    0.4 & 0.6
  \end{pmatrix}.
\end{align*}
The reader can verify directly that the inverse matrix reads
\begin{align*}
    C^{-1} =
  \begin{pmatrix}
    \phantom{-}3 & -2\\
    -2 & \phantom{-}3
  \end{pmatrix}.
\end{align*}
If we measure $n = (10, 10)$, equation~\eqref{eq:unfolding_matrix_invert} gives
a flux estimate of $F = (10, 10)$---fair enough. But if we do measure
$(13, 7)$, which is within a mild statistical fluctuation from the
previous pair, we get $(25, -5)$, i.e. a nonsense. In some sense the answer
is right: $(25, -5)$, folded with our DRM, gives indeed $(13, 7)$. But it's 
an answer with little or no physical meaning.

As it turns out, \emph{unfolding} is a thoughtfully studied problem and
algorithms exist that can do a decent job under many different conditions.
If one is only interested in fitting a spectral model to a series of data
point, the \emph{forward folding} approach is typically superior.

\subsection{Rigidity Resolution}

We have worked out the basic formalism of the rigidity measurement by a
magnetic spectrometer in section~\ref{sec:magnetic_spectrometers}.
In this section we shall study the problem in some more details.

At low rigidity the track deflection is large and the the multiple scattering
is the dominant contribution to the rigidity resolution.
As both the multiple scattering angle~\eqref{eq:theta_ms} and the bending
angle~\eqref{eq:theta_bending} scale as $1/R$, the multiple scattering term
in the rigidity resolution is a constant (i.e., does not depend on $R$), at
least as long as $\beta \approx 1$:
\begin{align}\label{eq:rigidity_res_ms}
  \frac{\sigma_R^{\rm MS}}{R} = \frac{\theta_{\rm MS}^{\rm plane}}{\theta_{\rm B}}
  \sim 
  \frac{0.0136\sqrt{t} (1 + 0.038 \ln t)}{0.3~\beta~B~[{\rm T}]~L~[{\rm m}]}.
\end{align}

At high energy, in contrast, assuming that $L$ and $B$ are known with enough
precision, the momentum resolution is determined by the hit resolution
$\sigma_{\rm hit}$ of the tracking detectors:
\begin{align}\label{eq:rigidity_res_hit}
  \frac{\sigma_R^{\rm hit}}{R} = \frac{\sigma_s}{s} \sim
  \frac{8R\sigma_{\rm hit}}{eBL^2} =
  \frac{R~[{\rm GV}]~\sigma_{\rm hit}~[{\rm mm}]}%
       {37.5~B~[{\rm T}]~L^2~[{\rm m^2}~]},
\end{align}
and, as we anticipated in section~\ref{sec:magnetic_spectrometers}, degrades
linearly with the rigidity:
\begin{align}
  \frac{\sigma_R}{R} = \frac{\sigma_s}{s} \propto R.
\end{align}

It should be emphasized that we are deliberately overlooking the fact that a
magnetic spectrometer has typically several tracking and the information
from the position sensitive detectors is combined through some sort of
pattern recognition and track fitting algorithm. There are many more subtleties
involved, such as the fact that the hit resolution is different for different
particle types (e.g., for protons and alpha particles), but we'll just move
on, as we're just interested in the orders of magnitude.

We can finally sum the two contributions in quadrature to get an expression for
the rigidity resolution as a function of the rigidity:
\begin{align}\label{eq:rigidity_res}
  \frac{\sigma_R}{R} = \frac{\sigma_R^{\rm MS}}{R} \oplus
  \frac{\sigma_R^{\rm hit}}{R}.
\end{align}
As already mentioned, the multiple scattering term dominates at low energy,
while the detector resolution term, increasing linearly with $R$ dominates at
high energy.

\subsubsection{The Maximum Detectable Rigidity}%
\label{sec:mdr}

The \emph{maximum detectable rigidity} (MDR) is formally defined as the
value of the rigidity for which the rigidity (or momentum) resolution is $1$
(i.e., $100\%$). As we shall see in a second, the MDR sets the upper bound of
rigidity measurable by a given setup.
Posing the right-hand side of equation~\eqref{eq:rigidity_res_hit} equal to $1$
gives an easy expression to work with:
\begin{align}
  {\rm MDR} = \frac{37.5~B~[{\rm T}]~L^2~[{\rm m^2}]}%
  {\sigma_{\rm hit}~[{\rm mm}]}.
\end{align}
With all the caveats mentioned in the previous section, if we plug in the
relevant figures for the PaMeLa magnetic spectrometer ($B = 0.43$~T,
$L = 0.445$~m, $\sigma_{\rm hit} \approx 3~\mu$m) we end up with a MDR of
$1.05$~TV, which is close enough to the actual value of $1.2$~TV predicted by
a full Monte Carlo simulation and measured on orbit.

\begin{figure}[!htb]
  \includegraphics[width=\linewidth]{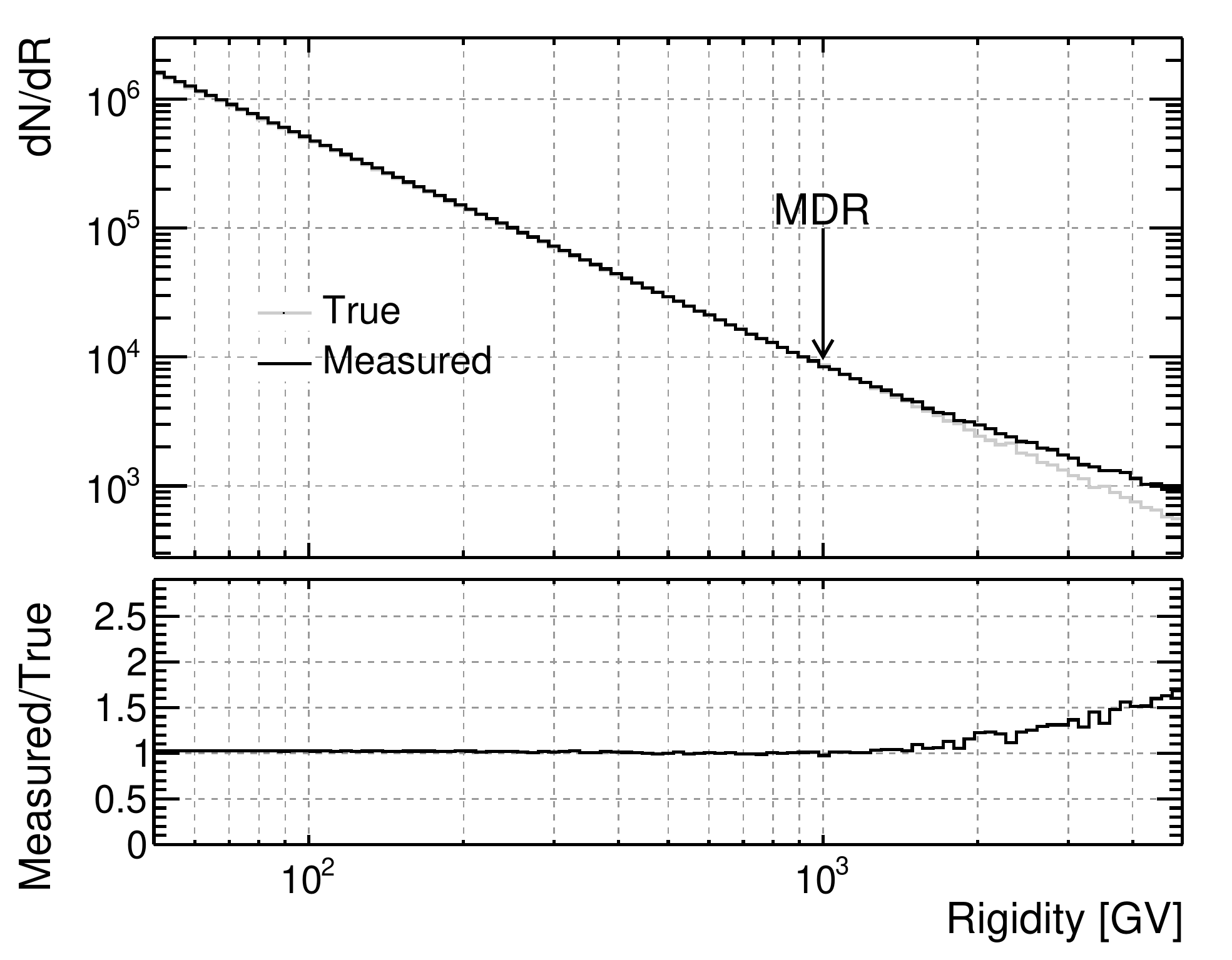}
  \caption{Effect of a finite rigidity resolution on a toy Monte Carlo
    power-law spectrum with index $2.75$, mimicking the primary cosmic-ray
    proton spectrum, for a maximum detectable rigidity of $1$~TV.
    (For completeness, the smearing used for this illustration is gaussian in
    $1/R$. Though the relative fluctuation in $R$ and $1/R$ are the same,
    the result would have been somewhat different if the smear was done in the
    rigidity space---we stress, however, that the point of the exercise
    was not to develop a realistic model of the rigidity dispersion.)}
  \label{fig:rigidity_spec}
\end{figure}

The effect of the finite rigidity resolution on a typical CR spectrum is
illustrated with a toy Monte Carlo in figure~\ref{fig:rigidity_spec}.
For rigidity values above the MDR the spectral distortions can be very
noticeable. This is germane to the discussion about the effect of the energy
dispersion in section~\ref{sec:correct_edisp}, with the noticeable difference
that when approaching the MDR correcting for the redistribution becomes quickly
difficult.

It should be also stressed that, while we presented the MDR as an intrinsic
characteristic of the detector, it is really an event-by-event quantity.
As the magnetic field is not exactly uniform in the detector and the
deflection is best measured for tracks with more hits, the quality of 
momentum reconstruction is in general different for events with the same
rigidity impinging in different points of the detector and/or from different
directions. In any real science analysis this is all properly taken into
account (e.g., it is customary to select events based on their MDR).

\subsubsection{Charge confusion}%
\label{sec:charge_confusion}

The rigidity resolution worsening at high rigidity is not the end of the
story. As tracks get more and more straight with increased rigidity,
the charge confusion (namely the fact that a track with positive curvature
can be reconstructed with negative curvature) comes into play.
The charge confusion is a combined effect with contributions from the
\emph{spillover} (i.e., errors in the sign determination due to the finite
hit resolution of the tracking detectors) and a series of (particle-dependent)
effects such as the production of delta rays and \bremss\ photons from the
primary particle and spurious hits from either electronic noise or
backsplash.

\begin{figure}[!htb]
  \includegraphics[width=\linewidth]{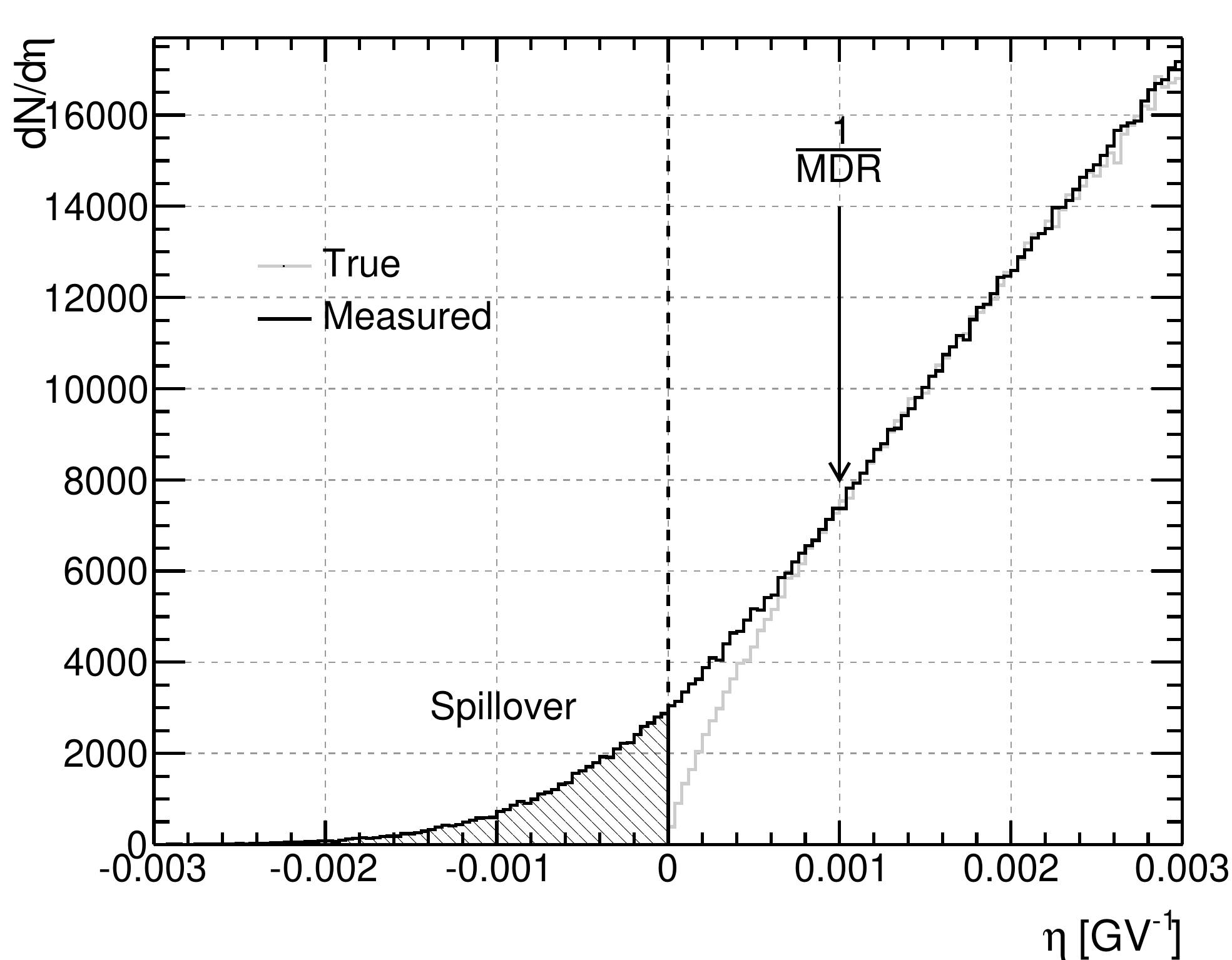}
  \caption{Illustration of the spillover effect. Here the same (true and
    measured) spectra in figure~\ref{fig:rigidity_spec} are shown in
    the $\eta$ space. We remind again that for this toy simulation the
    smearing is gaussian in $\eta$, assuming a MDR of $1$~TV.}
  \label{fig:spillover}
\end{figure}

The spillover, is better studied with the change of variable
\begin{align}
  \eta = \frac{1}{R}
\end{align}
As $R$ tends to infinity, $\eta$ tends to zero (even more importantly, keeping
its sign). As shown in figure~\ref{fig:spillover} the effect becomes noticeable
as the particle rigidity approaches the MDR. While this might be irrelevant for
the more abundant species, the spillover can be \emph{the} limiting factor for
rare species (e.g., cosmic-ray protons are so much more abundant than
antiprotons that any $\bar{p}$ spillover has virtually no effect, but the
opposite is clearly not true).

\subsubsection{A case study: the AMS-02 rigidity resolution}%
\label{sec:ams_mdr}

The AMS-02 magnetic spectrometer, in the permanent magnet configuration
currently operating on the ISS, includes an inner tracker (with 7 layers of
double-sided silicon-strip detectors) immersed in the magnetic field and two
additional layers (that we shall refer to as L1 and L9) aimed at increasing
the lever arm (and therefore the MDR). A sketch of the apparatus is shown
in figure~\ref{fig:detectors}.
Table~\ref{tab:ams02_mdr} summarizes the basic figures that enter into
modeling the rigidity resolution of the detector.

\begin{table}[htb!]
  \begin{tabular}{p{0.65\linewidth}p{0.3\linewidth}}
    \hline
    Quantity & Value\\
    \hline
    \hline
    Hit resolution ($\sigma_{\rm hit}$) & $10~\mu$m\\
    Overall tracker thickness ($t_{\rm TKR}$) & $0.5~X_0$\\
    Height of the inner tracker ($L_{\rm inner}$) & $0.82$~m\\
    Total tracker height ($L_{\rm TKR}$) & $2.95$~m\\
    Magnet height ($L$) & $1.1$~m\\
    Magnetic field ($B$) & $0.14$~T\\
    \hline
  \end{tabular}
  \caption{Basic characteristics of the AMS-02 magnetic spectrometer
    determining the rigidity resolution of the instrument
    (from~\cite{2013NuPhS.243...12T}). The total tracker thickness has been
    divided by the number of layers (9) to be plugged in
    equation~\eqref{eq:rigidity_res}.}
  \label{tab:ams02_mdr}
\end{table}

\begin{figure}[!htb]
  \includegraphics[width=\linewidth]{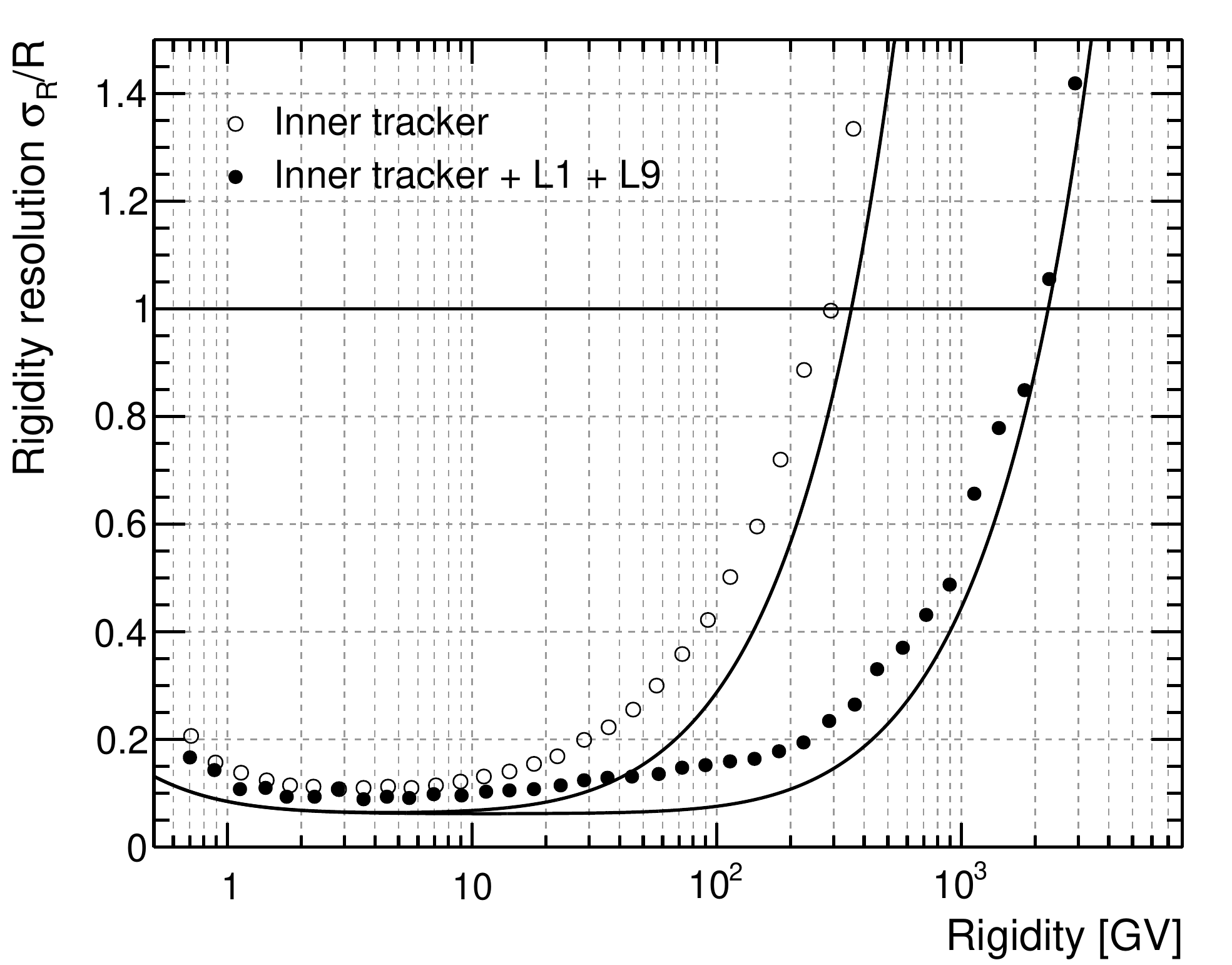}
  \caption{Na\"ive model of the proton rigidity resolution for the AMS-02
    spectrometer (lines), compared with the prediction of the full
    Monte Carlo detector simulation~\cite{ambrosi_icrc13}.
    Our simple formula~\eqref{eq:rigidity_res} is able to reproduce the general
    shape (including the raise below $\sim 1$~GeV due to the $\beta$ term in
    the multiple scattering formula) and predicts a MDR of $350$~GV and
    $2.3$~TV for the two configurations, to be compared with $250$~GV and
    $2$~TV reported in~\cite{ambrosi_icrc13}.}
  \label{fig:ams02_mdr}
\end{figure}

For the inner tracker, the numbers can be plugged directly into
equation~\eqref{eq:rigidity_res} to get a rough estimate of the rigidity
resolution as a function of rigidity. To first order the bending power for the
inner tracker $BL_{\rm inner}^2$ can be scaled to the full tracker,
including L1 and L9, by multiplying for the factor
\begin{align}
  f = \frac{L_{\rm TKR}}{L_{\rm inner}} \times \left( \frac{L}{L_{\rm inner}} \right)^2
  \approx 6.4
\end{align}
(Note that the extra tracker height in the region with no magnetic field
increases the lever arm for the measurement of the bending but not the
deflection angle, and therefore therefore the corresponding increase in the
bending power is linear and not quadratic).

In figure~\ref{fig:ams02_mdr} the prediction of our rough model are compared
with the full Monte Carlo simulation of the spectrometer, for both the
inner and the full tracker. All considered, the agreement is fair, surely
within a factor of two.

\subsection{Point-spread Function}

Similarly to the energy dispersion, the point-spread function (PSF) is the
probability density function for the (three-dimensional) space angle between
the measured and true directions. From an operational standpoint, the PSF
is the distribution of the reconstructed directions for a point source placed
at infinite distance. It goes without saying that, photons pointing back to
their sources, this is mostly a relevant metric for pair conversion
telescopes---and gamma-ray instruments in general.

As for the other instrument response functions, the point-spread function
depends, in general, on the incidence direction and impact position of the
incidence particle. Like the energy dispersion, for any point in the
instrument phase space, it is a probability density function---one notable
difference, though, is that the angular deviation is positive-definite.
In strict analogy with the energy resolution, the information contained in the
point-spread function is customarily summarized in the space angles containing
68\% and 95\% of the probability density function itself
\footnote{This so true that the 68\% containment angle of the PSF is sometimes
  (improperly) referred to as \emph{the point-spread function}.}.
The ratio between the 95\% and 68\% containment angles of the PSF (which is
exactly 2 in the gaussian case, and typically larger in real life) gives
information about the tails of the distribution.

As we shall see in the case study described in the next section, the
interplay between the multiple scattering and the hit resolution we encountered
in the determination of the rigidity resolution for a magnetic
spectrometer---with the former dominating at low energy and the latter at high
energy---has a parallel in the modelization of the point-spread function for a
pair conversion telescope.

\subsubsection{A case-study: the \Fermi-LAT PSF}%
\label{sec:lat_psf}

An accurate modeling of the PSF for a real detector is a complex task and
one has typically to resort using complex Monte Carlo simulations. Basic
physical processes aside, the angular accuracy is not only hostage of the
detector geometry (including the cracks and the inactive regions), but it
also depends on the internal of the pattern recognition in the tracking device
and of the subsequent track fitting. In this section we shall use the
\Fermi-LAT tracker as a case study and try and derive an approximate
expression for the PSF $68\%$ containment as a function of the energy---our
main goal is not to provide an accurate parameterization, but rather to
illustrate how the basic physical principles and detector characteristics
play together.

The \Fermi-LAT tracker divides into two distinct sections with noticeable
differences in terms of the point-spread function. The top section of the
tracker (\emph{front} or \emph{thin} section) is comprised of $12$ silicon
detector layers equipped with $3\%~X_0$ tungsten converters, while the
\emph{back} (or \emph{thick}) sections is comprise of $4$ layers with $18\%~X_0$
converters and $2$ additional layers with no converter. We shall derive
a simple estimate of the PSF for both sections. The basic ingredients for our
model are summarized in table~\ref{tab:lat_psf} and a sketch of the instrument
is shown in figure~\ref{fig:detectors}.

\begin{table}[htb!]
  \begin{tabular}{p{0.65\linewidth}p{0.3\linewidth}}
    \hline
    Quantity & Value\\
    \hline
    \hline
    Silicon strip pitch ($p$) & $228~\mu$m\\
    Silicon plane spacing ($d$) & $3.5$~cm\\
    Front converter thickness ($t_{\rm F}$) & $3\%~X_0$\\
    Number of front layers ($N_{\rm F}$) & $12$\\
    Back converter thickness ($t_{\rm B}$) & $18\%~X_0$\\
    Number of back layers ($N_{\rm B}$) & $4 + 2$\\
    \hline
  \end{tabular}
  \caption{Basic design metrics of the \Fermi-LAT
    tracker~\cite{2007APh....28..422A} determining the PSF of the instrument.}
  \label{tab:lat_psf}
\end{table}

At low energy the multiple Coulomb scattering (scaling as $1/E$) dominates the
angular resolution. We shall take as a representative thickness to plug
into the formula~\eqref{eq:theta_ms} that of \emph{one} converter---i.e.,
we shall pretend that the track direction is
determined by the two points close to the conversion point. In real life,
typical track-fitting routines are aware of the multiple scattering
(e.g., via Kalman filter techniques) and properly de-weight the hits as the
particle travels through the detector, progressively loosing memory of the
original direction, but this is obviously beyond the scopes of this review.
We shall also assume that the electrons and positron split equally the
photon energy $E$ and, since we are in the pair production regime
(i.e., above $10$~GeV) we shall declare $\beta = 1$ and $pc = E$.
All in all, we write the multiple-scattering term as
\begin{align}
  {\sigma_\theta}_{\rm F/B}^{\rm MS} =
  \sqrt{2}\frac{0.0136~{\rm GeV}}{E/2}\sqrt{t_{\rm F/B}} (1 + 0.038 \ln t_{\rm F/B})
\end{align}

At high energy the PSF is dictated by the strip pitch $p$ and the lever arm.
Since the detector readout is digital, the hit resolution is given by
$\sigma_{\rm hit} = p/\sqrt{12}$. In both the $x$--$z$ and $y$--$z$ views the
problem, in its simplest form, is germane to a least square fit to a straight
line with $n$ equally-spaced sampling points and constant measurement errors.
The uncertainty on the fitted slope in each projection reads
\begin{align}\label{eq:slope_least_square_errors}
  \sigma_{\rm slope} & =
  \frac{\sigma_{\rm hit}}{d\displaystyle\left[ \frac{n(n+1)(2n+1)}{6} -
      \frac{n(n+1)^2}{4} \right]^{\frac{1}{2}}} \sim \nonumber\\
  & \sim \frac{\sqrt{12}\sigma_{\rm hit}}{dn^{\frac{3}{2}}} =
  \frac{p}{dn^{\frac{3}{2}}},
\end{align}
where the simplified expression on the second row is accurate for $n \gg 1$.
Complicated as it might seems, the above expression encodes the right
dependence on the parameters of the problem, as $dn$ is the overall lever
arm for the fit (so that $p/dn$ is a proxy for the error on the slope when
only the two extreme points are considered) and the extra $\sqrt{n}$ at the
denominator accounts for the statistical improvement in the fit errors when
more measurement points are added.

Coming back to our original problem, we shall switch to the space angle by
multiplying by $\sqrt{2}$ and write
\begin{align}
  {\sigma_\theta}_{\rm F/B}^{\rm hit} =
  \frac{\sqrt{2}p}{d\left<n_{\rm F/B}\right>^{\frac{3}{2}}},
\end{align}
where $\left<n_{\rm F/B}\right>$ is the average number of hits on the 
track for events converting in the front/back sections of the tracker.
Technically we would have to properly calculate a weighted average taking into
account the conversion probabilities in the different layers and the
$3/2$ exponent in equation~\eqref{eq:slope_least_square_errors}, but since
we are neglecting a number of effects conspiring to worsen the PSF
(hard-scattering processes, delta rays, hit confusion, dead areas) we shall
just take the worst-case numbers $\left<n_{\rm F}\right> = 7$ and
$\left<n_{\rm B}\right> = 3$.

\begin{figure}[!htb]
  \includegraphics[width=\linewidth]{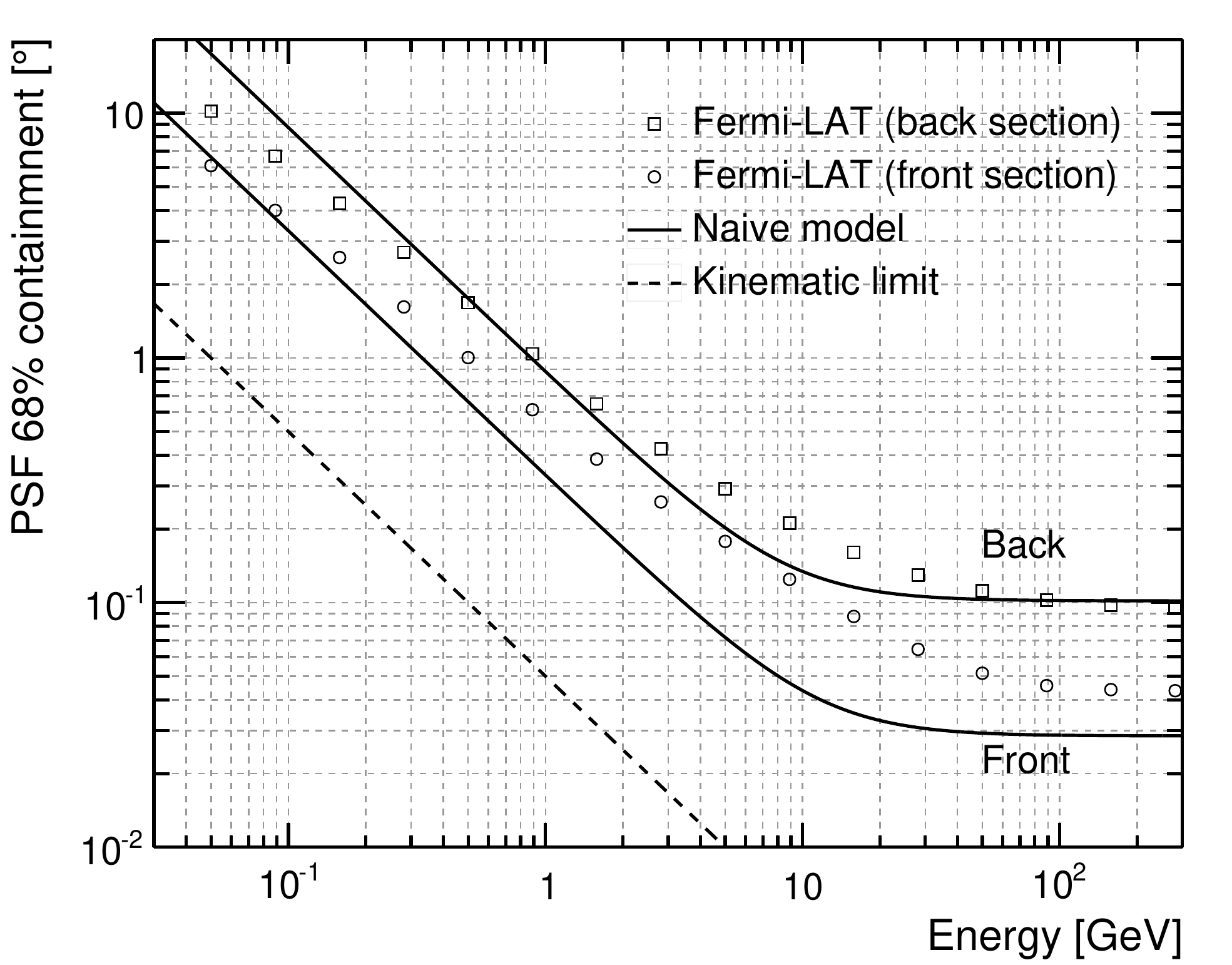}
  \caption{Comparison of our na\"ive estimates for the \Fermi-LAT PSF with the
    output of a full Monte Carlo simulation of the detector. For reference the
    kinematic limit introduced in section~\ref{sec:more_pair_production} is
    also shown.}
  \label{fig:lat_psf}
\end{figure}

That all said, our angular resolution is just the sum in quadrature of the
two terms%
\footnote{The reader might wonder why we are acting like only one member
  of the electron-positron pair existed: shouldn't we combine the two tracks?
  Well, there are many more subtleties involved, here. As we mentioned in
  section~\ref{sec:more_pair_production}, in real life there
  aren't as many clean $v$-shape events as one might expect. At low energy
  chances are high that one of the two tracks dies early in the detector and
  is not reconstructed, while at high energy the pair simply does not 
  split enough to be resolved in two tracks. Even when there are two tracks,
  combining them covariantly is not an easy task, as one would need to 
  a good estimate of the two energies. That all said, I'll be happy to give
  an extra $\sqrt{2}$, here, if the reader really insists.}%
:
\begin{align}
  {\sigma_\theta}_{\rm F/B} = {\sigma_\theta}_{\rm F/B}^{\rm MS} \oplus
  {\sigma_\theta}_{\rm F/B}^{\rm hit}
\end{align}
and this can be compared directly with the $68\%$ containment of the PSF,
as shown in figure~\ref{fig:lat_psf}. The fact that our model is within a
factor of two or so from the full detector Monte Carlo simulation over some four
orders of magnitude is remarkable, given all the things that we have (brutally)
neglected.

\section{Design principles}

Not surprisingly, the question of how one should design a scientific instrument
(be it for space or not) is a tough one---and one that requires many different
inputs. The typical logical flow is to start with a precise idea of the science
targets, i.e., \emph{what we want to measure}. The science requirements
translate more or less obviously into corresponding requirements on the
detector performance, e.g., acceptance, point-spread function and energy
resolution. How to turn them into an actual instrument implementation
that fits into all the operational constraints (e.g., the weight and dimensions
that can be accommodated by the launcher and the power budget) is often the most
difficult part of the process---and one might have to iterate the process itself
several time for it to converge.

In this section we shall briefly glance through some of the basic aspects
of instrument design. (And no, even if you do read the section all the way
to its end you should not go out and try and design your own space detector
right away.)

\subsection{Sensitivity studies}

Now that we have a basic understanding of the main figures of merit describing
the instrument performance it is time to try and tie them to the actual things
that one aims at measuring.

\subsubsection{Observing time, exposure factor and exposure}

Up to now we haven't quite talked about the observing time, but it is rather
intuitive that the total number of events collected by a given instrument
is proportional to the the exposure factor
\begin{align}\label{eq:fexposure}
  \fexposure(E) = \accept(E) \times \obstime,
\end{align}
i.e., the product of the acceptance and the exposure factor (measured, e.g., in
m$^2$~sr~year). In passing, by \emph{``observing time''}, here, we really mean
the cumulative amount of time that the instrument is collecting science data.
When defining the observing time, care should be taken to account for all
the possible effects that reduce the overall duty-cycle: instrumental dead time,
time spent for calibrations and hardware problems (and, for instruments in
low-Earth orbit, the time spent in the SAA). And, since we are only concerned
with the orders of magnitude, we shall obviously neglect all that in the
following.

That all said, the exposure factor defined in~\eqref{eq:fexposure} is mainly
relevant for charged cosmic rays, whose flux is, to an good approximation,
isotropic. On the other hand gamma-ray astronomy is by its nature
\emph{dispersive}---you see a subset of the sky at a time%
\footnote{Strictly speaking this is true for charged CRs too, but in that case
  we are essentially looking at the same thing in any direction in the sky,
  while for gamma-rays we really look at different sources in different
  directions.}.
What matters is the exposure \exposure, or the \emph{integrated effective area}
in a given direction in the sky---measured in m$^2$~year.

This is where the observation strategy comes into play: depending on how large
is the field of view, aiming at a uniform exposure (i.e., scanning the sky)
vs. pointing a particular region can give very different answers.
Phrased in a different way, different (and mutually exclusive) observing
strategies potentially provide very different sensitivities for particular
science objectives, so this is really an important piece of the puzzle.

Even in \emph{sky-survey} mode, the exposure factor are the exposure are two
very different things: assuming that the observation strategy provides an
approximately uniform sky exposure on long time scales (which is the case,
e.g., for \Fermi), the two quantities are related to each other by
\begin{align}\label{eq:exposure_acc_obstime}
  \mathcal{E}(E) \sim \frac{\fexposure(E)}{4\pi} =
  \frac{\accept(E) \times \obstime}{4\pi}.
\end{align}
It follows that \emph{an overall exposure factor of 1~m$^2$~sr~year
and an exposure of 1~m$^2$~year in a given direction in the sky are
fundamentally different things: for an instrument with an acceptance of
1~m$^2$~sr it takes 1~year of observation to achieve the former and about
$12.5$~year for the latter}%
\footnote{Incidentally, when the time spent in the SAA and
the instrumental dead-time fraction are taken into account, 1~m$^2$~year
(or $3.16 \times 10^{11}$~cm$^2$~s) is representative of the average sky
exposure at high energy (i.e., above $\sim 10$~GeV) integrated by the
\Fermi-LAT in the course of the entire mission.}.

\subsubsection{Cosmic-ray and gamma-ray intensities}

We are now ready to attack some basic sensitivity studies. The first obvious
case is calculating the expected number of cosmic rays of a particular species
(e.g., protons or electrons) for a given detector acceptance and observing time.
One obvious follow-up question would be what is the maximum energy up to which
a given detector can study a given component of the cosmic radiation before
running out of statistics.

One possible way to go about this---given a model for the differential
intensity $\intensity(E)$ under study---is to calculate the integral number of
events $N_I$ above a given energy:
\begin{align}\label{eq:int_counts}
  N_I(E) = \int_{E}^{\infty}\!\!\!\!\fexposure(E') J(E') dE'.
\end{align}
(Check for yourself that all the physical units cancel out and the result is
a pure number.) Compared to the corresponding differential quantity this has the
advantage of depending on a single energy bound.

\begin{figure}[!htb]
  \includegraphics[width=\linewidth]{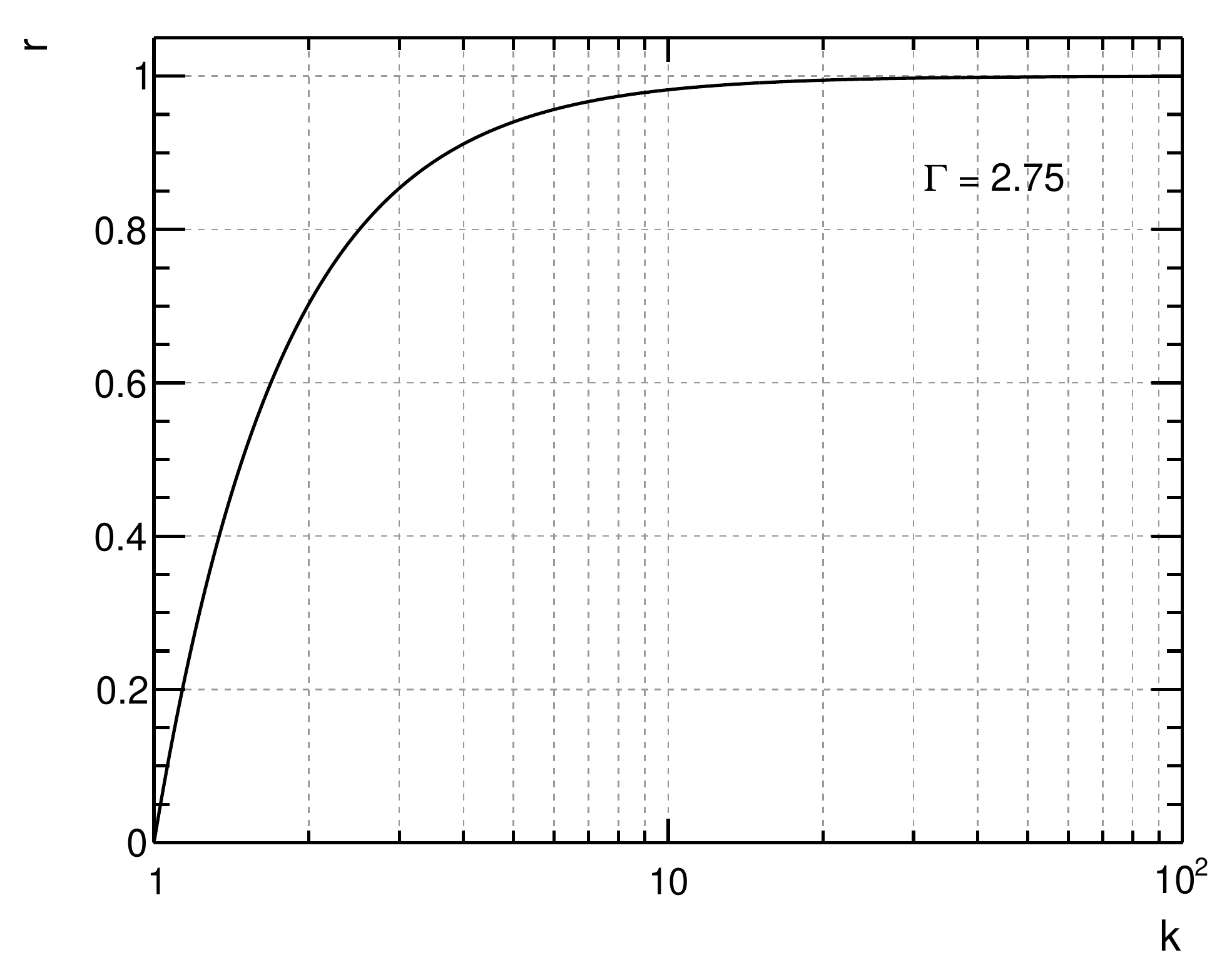}
  \caption{Accuracy of the integral spectra evaluation as a function of the
    upper extreme of integration, as given in equation~\ref{eq:int_limits}.}
  \label{fig:int_limits}
\end{figure}

\begin{figure*}[phtb]
  \includegraphics[width=0.49\linewidth]{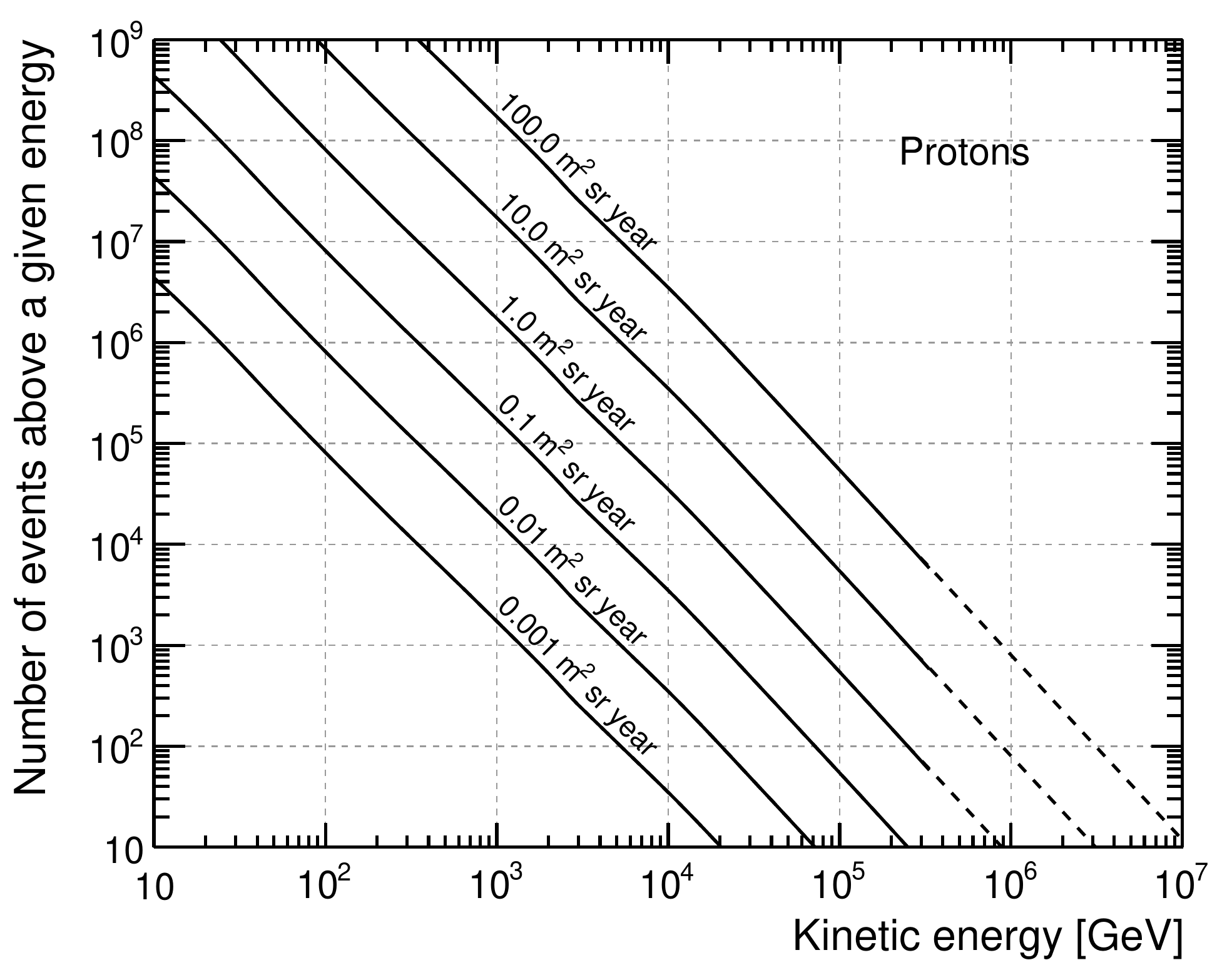}%
  \includegraphics[width=0.49\linewidth]{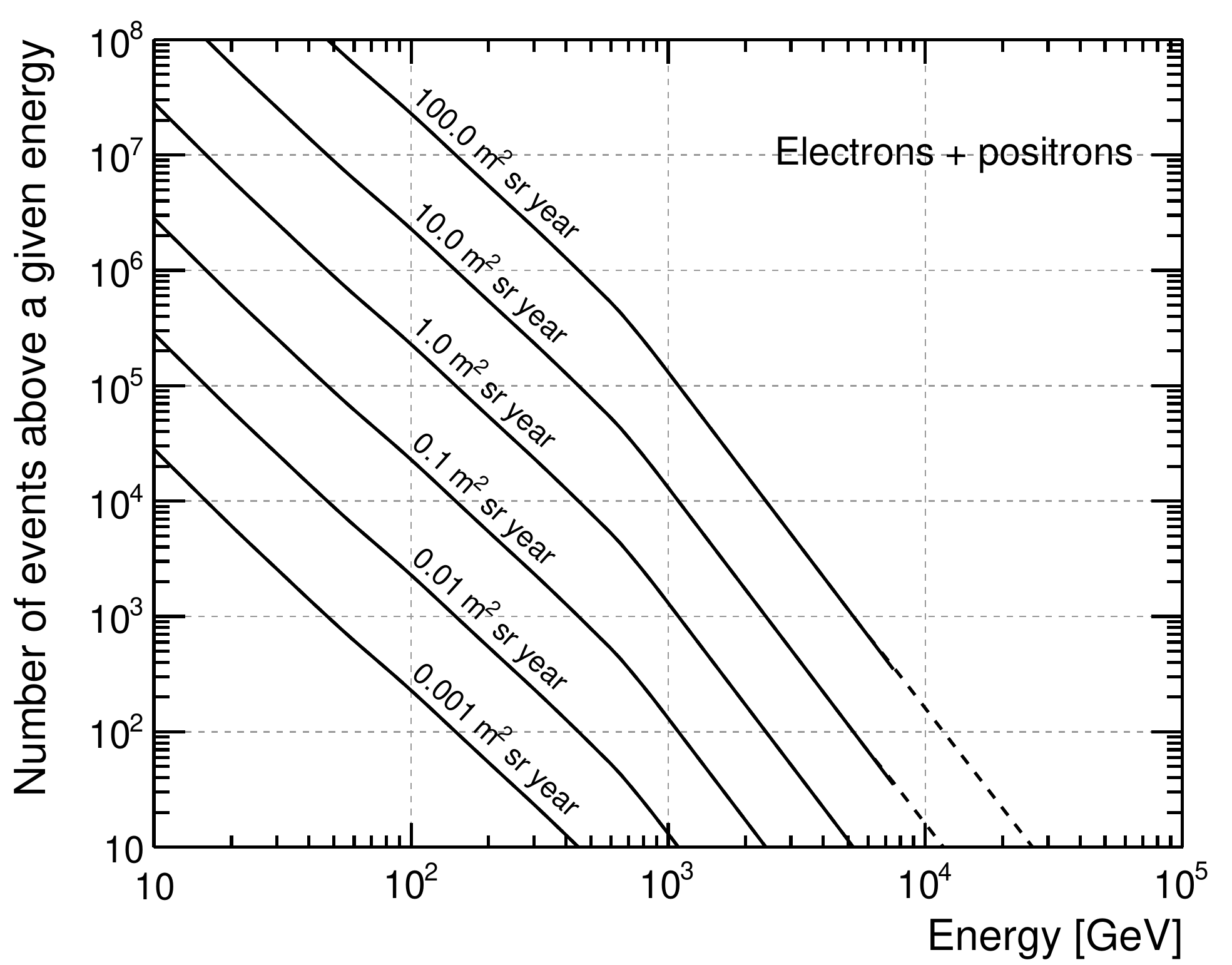}\\
  \bigskip
  \includegraphics[width=0.49\linewidth]{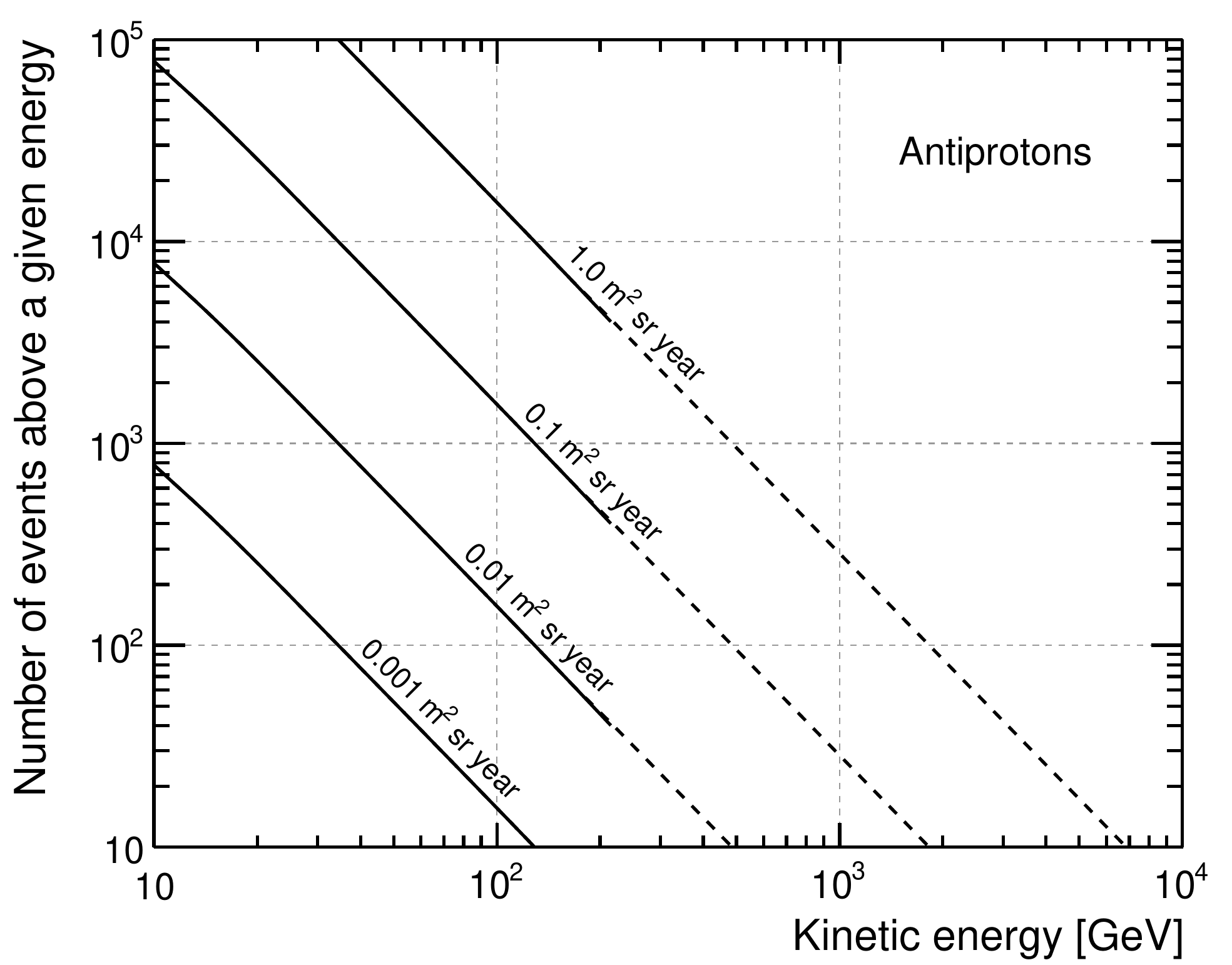}%
  \includegraphics[width=0.49\linewidth]{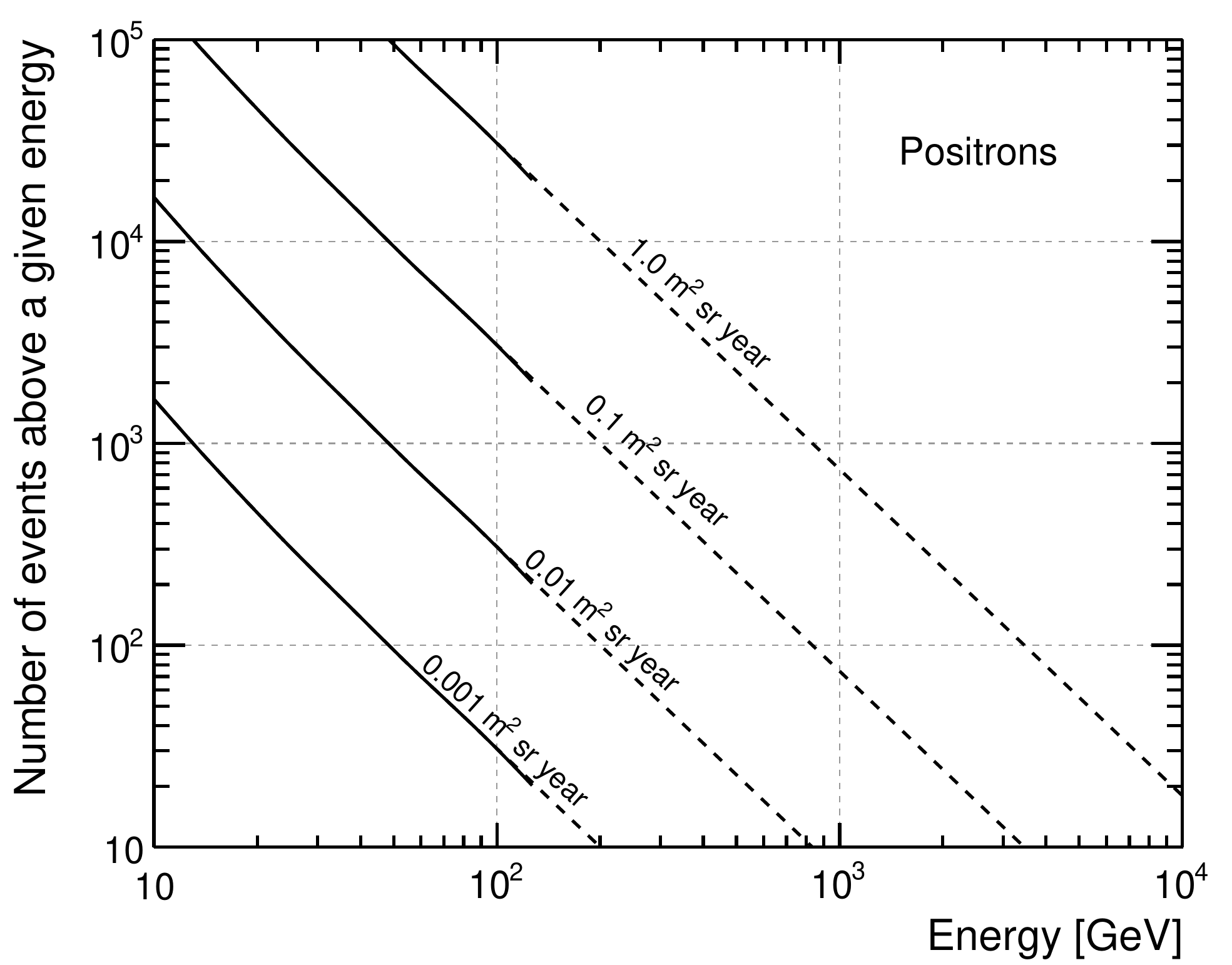}\\
  \bigskip
  \includegraphics[width=0.49\linewidth]{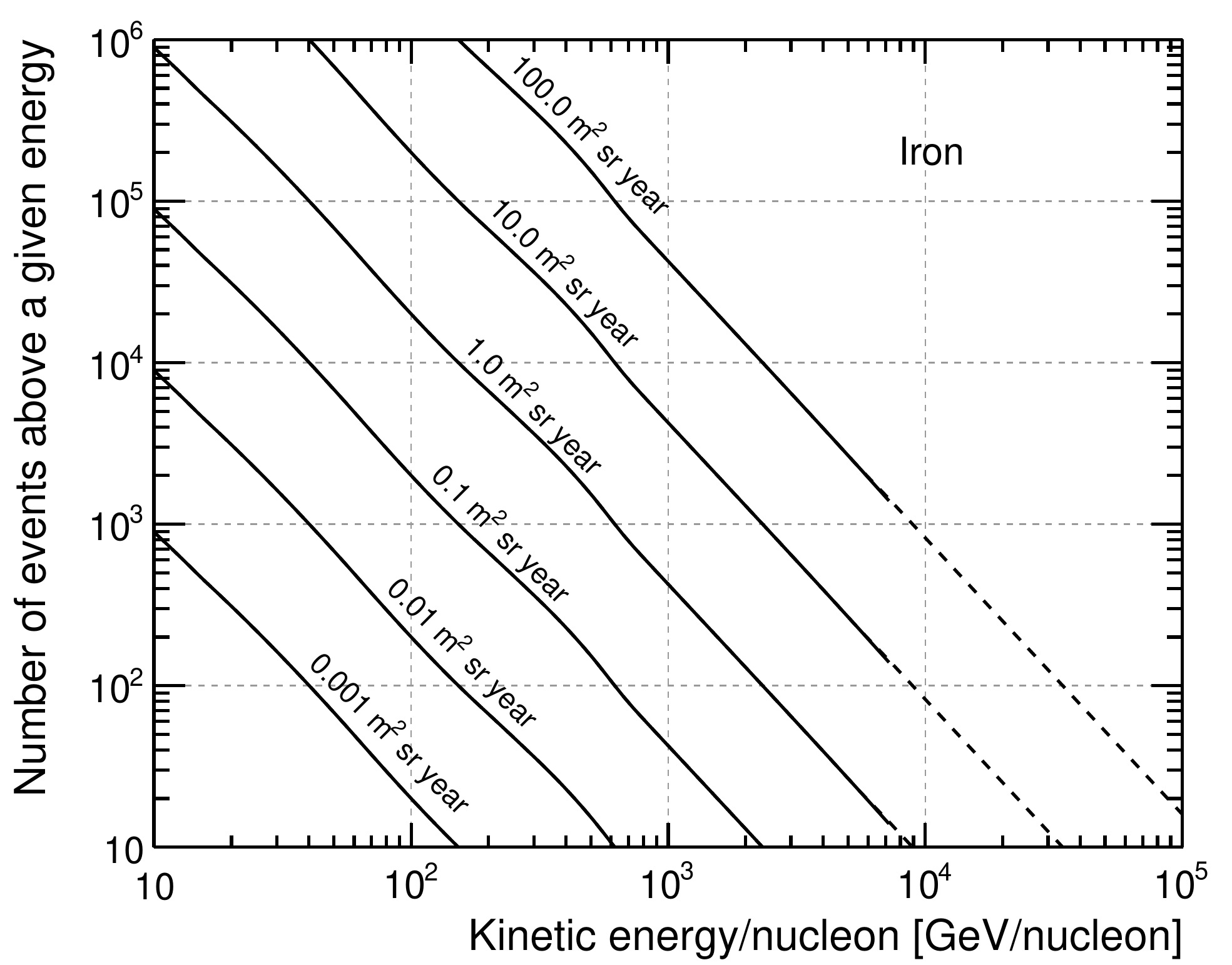}%
    \includegraphics[width=0.49\linewidth]{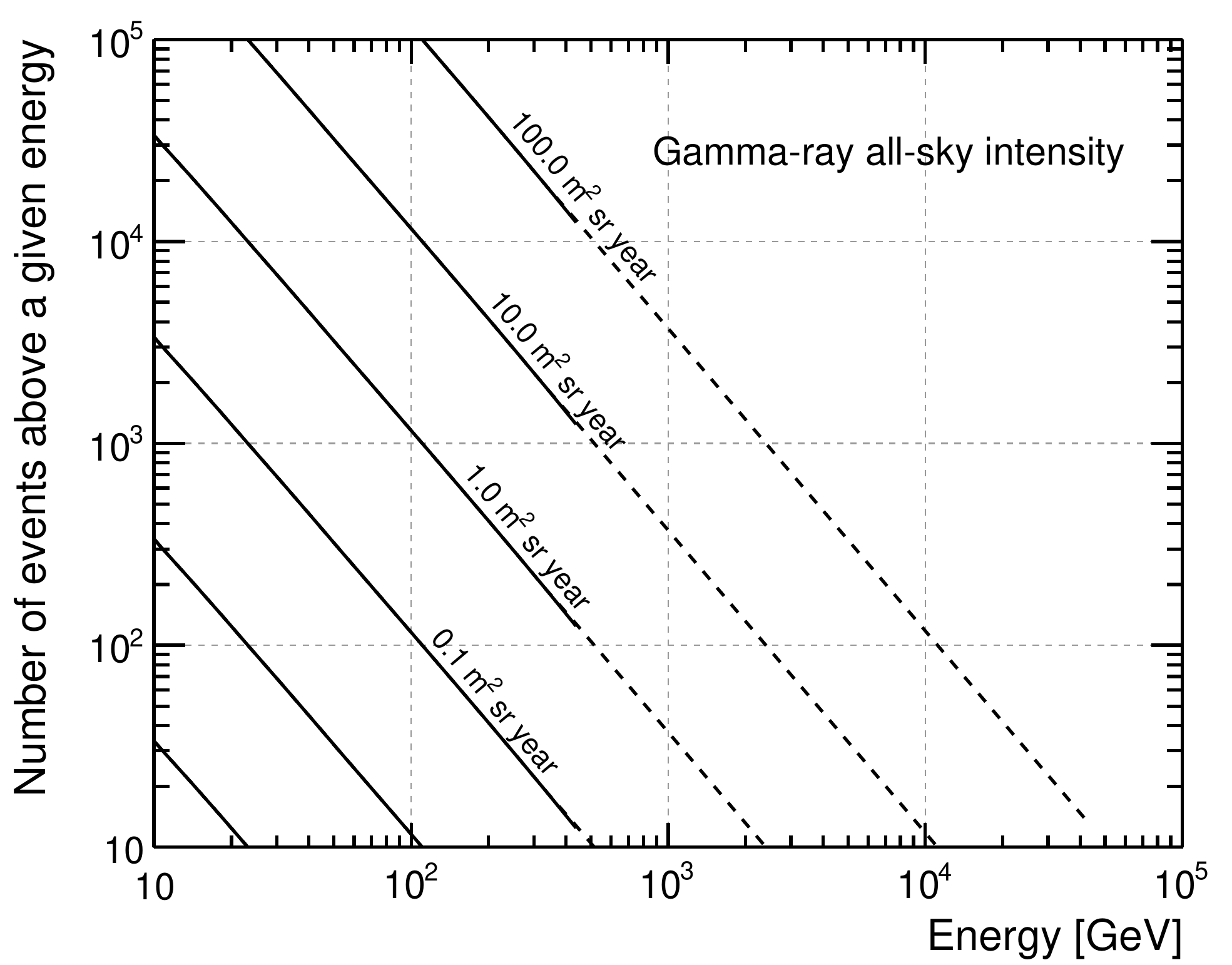}
  \caption{Integral number of events above a given energy for some interesting
    CR species, and for some representative values of the exposure factor.
    As a rule of thumb we can fairly say that, for each species, the energy for
    which a given curve drops below $\sim 10$ is the ultimate energy reach for
    an instrument integrating the corresponding exposure factor.
    Note that the axis ranges are all different. Also note that, for obvious 
    reasons, the spectrum extrapolations are quite uncertain for the
    positrons.}
  \label{fig:cr_projected_counts}
\end{figure*}

We note, in passing, that the upper extreme of integration in
equation~\eqref{eq:int_counts} is formally infinite, but given how steeply
cosmic-ray spectra are falling down, one hardly needs to get that far.
For a power law with index $\Gamma$ the relevant figure is the ratio
\begin{align}\label{eq:int_limits}
  r = \frac{\int_{E}^{kE}E'^{-\Gamma}dE'}{\int_{E}^{\infty}E'^{-\Gamma}dE'} =
  1 - \frac{1}{k^{\Gamma - 1}},
\end{align}
which is plotted in figure~\ref{fig:int_limits} for $\Gamma = 2.75$.
Integrating over a decade in energy is in practice enough to get us within a
few \% of the exact answer and a factor of $\sim 3$ in energy still provides an
answer accurate to some $\sim 15\%$. This implies that, in order to evaluate
the integral spectra above a given energy $E$, one needs a reliable estimation
of the corresponding differential intensities up to energy at least 3--4 times
higher.

If the exposure factor does not depend on energy, which is not a terrible
assumption for most high-energy instruments, at least under certain conditions,
the problem simplifies considerably
\begin{align}\label{eq:int_counts_constfexp}
  N_I(E) =  \fexposure \int_{E}^{\infty}\!\!\!\!J(E') dE',
\end{align}
as by integrating the differential intensity one effectively gets a 
\emph{universal} function that can be scaled by the exposure factor and plotted
for different values of \fexposure\ as shown in
figure~\ref{fig:cr_projected_counts}.

When looking at figure~\ref{fig:cr_projected_counts} the reader should be aware
that the we have used our na\"ive baseline models of cosmic-ray intensities,
and we have extrapolated them at high energy---which means that the numbers
you read off the plot cannot expected to be accurate to three significant
digits. That said, the plots allow to have a good idea of how many particles
one is expecting to detect in a given setup above a given energy---and, by
difference, in any given energy range.

The energy value for which the curve of interest in
figure~\ref{fig:cr_projected_counts} falls below, say, 10 events signal the
maximum energy reach that one might hope to reach for a given cosmic-ray
species with a given exposure factor. Depending on the situation this might
be realistic or no---e.g., a measurement with a magnetic spectrometer might be
dominated by the charge confusion well before reaching out that far. In other
words one should always keep in mind figure~\ref{fig:cr_ratios} when looking
at figure~\ref{fig:cr_projected_counts}.


\subsubsection{Gamma-ray source sensitivity}

As one might imagine, evaluating the source sensitivity for a given instrument
involves a fair number of ingredients---most notably the effective area and
point-spread function of the detector, the observation time toward a given
direction in the sky and the intensity of the isotropic and galactic diffuse
emission around the very same direction. In broad terms the detection
threshold (i.e., the minimum integral source flux above a given energy
$E_{\rm min}$ that is needed to detect the source itself at a given position and
in a given observation time $T_0$) is determined by the number of background
events---i.e., gamma rays from diffuse emission---into the solid angle
subtended by the (energy-dependent) \psf. It goes without saying that the
detection threshold depends on the position in the sky and is largest where
the galactic diffuse emission is most intense (e.g., on the galactic plane
and especially around the galactic center).

If we assume that the diffuse background is locally uniform, it is possible
to derive an approximate expression for the detection threshold. In the
following we shall outline a prescription described
in~\cite{2010ApJS..188..405A} in the context of the \Fermi-LAT first source
catalog.

The basic starting point is the maximum likelihood formalism customarily
used for gamma-ray source analysis---described, e.g.,
in~\cite{1996ApJ...461..396M}. The likelihood of observing $n_i$ counts
in the $i$-th bin (here the pixel index $i$ is running in space and energy)
where the model predicts $\lambda_i$ counts is given by the Poisson
probability
\begin{align}
  p_i = \frac{\lambda_i^{n_i} e^{-\lambda_i}}{n_i!}.
\end{align}
The full likelihood is the product of the (independent) probabilities for each
single pixels
\begin{align}
  L = \prod_i p_i = \prod_i \frac{\lambda_i^{n_i} e^{-\lambda_i}}{n_i!}
\end{align}
and the log-likelihood reads
\begin{align}
  \ln L = \sum_i \left( n_i\ln \lambda_i - \lambda_i - \ln n_i! \right).
\end{align}
The last term is irrelevant as it does not depend on the model, and the
quantity that we have to maximize is effectively
\begin{align}\label{eq:src_det_likelihood}
  \ln L = \sum_i \left( n_i\ln \lambda_i - \lambda_i \right).
\end{align}
Finally, the test statistics (roughly speaking, the square of the significance
of the detection) is given by (twice) the difference between the likelihood
with the source included in the model and the likelihood calculated under the
hypothesis of no source:
\begin{align}
  {\rm TS} = 2(\ln L^* - \ln L^0).
\end{align}

Assuming a locally uniform background we can rewrite the above expression 
in the form of an integral over the energy and the angular separation $\psi$
between the source position and the reconstructed direction. Let $S(E)$
be the spectrum of the source, $B(E)$ that of the background, $\psf(E, \psi)$
the spatial distribution of the events from a point source (at a given energy);
in addition, let us define the local source-to-background ratio
\begin{align}
  g(E, \psi) = \frac{S(E)\psf(E, \psi)}{B(E)}.
\end{align}
That all said, the probability density function for the number of events
detected will be
\begin{align}
  n*(E, \psi) & = T_0\aeff(E)\left( S(E)\psf(E,\psi) + B(E) \right) \nonumber\\
  & = T_0\aeff(E)B(E) (1 + g(E, \psi))
\end{align}
if the source is included and
\begin{align}
  n^0(E, \psi) = T_0\aeff(E)B(E) 
\end{align}
in case of no source (note that both have to be integrated in energy and
solid angle to get the actual number of events in the two hypotheses).
The corresponding \emph{differential} TS contribution per unit energy and
solid angle reads
\begin{align*}
  \frac{d{\rm TS}}{dEd\Omega} & =
  2 \left[ (n^*\ln n^* - n^*) - (n^* \ln n^0 - n^0) \right] = \\
  & = 2 \left[ n^*\ln \left(\frac{n^*}{n^0}\right) - (n^* - n^0) \right] = \\
  & = 2 T_0 \aeff(E)B(E) \times \\
  & \Big[(1 + g(E, \psi))  \ln(1 + g(E, \psi)) - g(E, \psi) \Big].
\end{align*}
Assuming a given spectral shape for $S(E)$, e.g., a power law with a given
index, and a model for the diffuse emission e.g., the GDE model that we have
introduced in section~\ref{sec:dge}%
\footnote{At high latitude you will also need a model for the isotropic
  background.},
one can integrate over the energy and the solid angle to get the actual TS and
calculate the value of the minimum required normalization of the source
spectrum---for a $5\sigma$ detection---by setting ${\rm TS} = 25$.

Complicated as it might seems, the previous expression simplifies considerably
in the background-dominated regime, i.e., when $g(E, \psi) \ll 1$ even for
$\psi = 0$%
\footnote{For \emph{typical} pair-conversion telescopes this is true at low
energies, where the \psf\ is comparatively poor and faint sources are always
background limited. On the other hand, at high energy the \psf\ can be narrow
enough that even for faint sources the detection threshold is determined by
the counting statistics.}%
. By expanding
\begin{align}
  (1 + x)\ln (1 + x) - x \sim (1 + x)\left(x - \frac{x^2}{2}\right) - x 
  \sim \frac{x^2}{2}
\end{align}
one gets the approximate relation
\begin{align}
  \frac{d{\rm TS}}{dEd\Omega} \sim
  \frac{T_0\aeff(E) S(E)^2 \psf(E, \psi)^2}{B(E)}.
\end{align}
Finally, by noting that
\begin{align}
  \int_0^\pi \psf(E, \psi) d\Omega \propto \frac{1}{\sigma(E)^2},
\end{align}
where $\sigma(E)$ is the angular resolution at a given energy $E$, one 
recognizes in the TS contribution per unit energy
\begin{align}
  \frac{d{\rm TS}}{dE} \propto
  \frac{T_0\aeff(E) S(E)^2}{B(E) \sigma(E)^2}
  \propto \frac{n_{\rm sig}^2}{n_{\rm bkg}^{\rm eff}}
\end{align}
the (square of) the number of signal events divided by the square root of the
number of background events within the solid angle subtended by the \psf\ of
the instrument.

We refer the reader to~\cite{2010ApJS..188..405A} for a more thorough
description of the procedure in the specific case of the \Fermi\ Large Area
Telescope.


\subsubsection{Search for spectral lines}

The detection of a high-energy spectral line on top of the continuum gamma-ray
emission (e.g., from the galactic center), is generally regarded as one of the
few instances of a potential observable phenomenon that would convincingly
point to the existence of new physics at work (e.g., two-body annihilation of
dark matter into photons). The topic has recently received a lot of attention
due to the claim~\cite{2012JCAP...07..054B,2012JCAP...08..007W} of a line-like
feature around 130~GeV in the publicly available \Fermi\ data.

From the standpoint of what we are interested in, the general subject of
the line search provides a good example to underline the interplay between
the acceptance and the energy resolution in this specific analysis.
In broad terms, the basic figure of merit $Q$ for such a line search is
\begin{align}
  Q = \frac{n_s}{\sqrt{n_b}},
\end{align}
where $n_s$ is the number of signal events and $n_b$ is the \emph{effective}
background, i.e. the number of background events integrated over the signal
p.d.f. At a fixed signal flux and detector acceptance, both the signal and
the background scale linearly with the exposure factor \fexposure, but the
width of the window over which the background is integrated is proportional to
the energy resolution, which implies that
\begin{align}\label{eq:line_sensitivity}
  Q \propto \sqrt{\frac{\fexposure}{\eres}}.
\end{align}
While a good energy resolution is surely desirable for this particular search,
equation~\eqref{eq:line_sensitivity} implies that the sensitivity increases
with a narrower energy dispersion only if one is not trading too much acceptance
for that.

\subsection{Operational constraints}

Delving into the details of the instrument design principles from the standpoint
of the operational constraints is obviously beyond the scope of the
write-up. Nonetheless we shall glance through some of the most obvious
aspects of the problem, with emphasis on space-based detectors.

\subsubsection{Weight, power and all that}

The first obvious constraints imposed by the choice of the launcher are on
the overall dimensions and weight of the instrument. The latter impacts
primarily the layout of the calorimeter, which typically accounts for a
significant fraction of the mass in most popular instrument designs.
In broad terms, once the calorimeter weight is fixed, one has the face a trade
off between the acceptance and the energy resolution---i.e., decide the
optimal compromise between the cross-sectional active surface and the depth.
It should be emphasized that for magnetic spectrometers the magnet is also
a fundamental contributor to the mass budget.

Power in space generally comes from solar panels---and is not unlimited.
Typical power budgets for space- and balloon-borne detectors are within
500--1000~W---rarely much more than that. In terms of detector implementation,
the limited power primarily impacts the tracking stage, which is typically the
one featuring the highest level of segmentation (as we mentioned in
section~\ref{sec:silicon_detectors}, silicon trackers with $10^5$--$10^6$
independent electronic channels are nowadays commonly operated in space).

The bandwidth for data down-link is expensive, too---pretty much as anything
else in space---with typical average figures ranging in the Mb/s. This has
profound implication on the overall data flow, requiring in general some
combination of zero suppression, event filtering and data compression (all
happening on board).

\subsubsection{Launch and space environment}

In contrast to ground-based detectors (e.g., experiments at accelerators),
space-based instruments must be designed, assembled and tested as to ensure
successful launch and on-orbit operation. The \emph{environmental}
verification process typically includes mechanical tests (both static and
dynamic), thermal tests (thermal and/or thermo-vacuum cycling) and
electromagnetic emission and susceptibility tests. While this general topic is
not trivial to discuss without reference to a specific instrument, we shall
briefly introduce some of the basic aspects of the problem. The reader is
referred, e.g., to~\cite{2008NIMPA.584..358B} for a thorough discussion of the 
environmental tests of the \Fermi-LAT tracker towers.

Vibration tests are aimed at verifying that the detector can sustain the
low-frequency dynamic load in the launch environment---i.e., vibrations induced
by the bearing structures (the satellite and the launcher) at the natural
frequencies of the latter (usually below 50~Hz). The \emph{transfer function}
of the instrument assembly (or any of its sub-parts) is typically measured by
means of a combination of a shaker and a series of accelerometers, that allow
to measure both the resonance frequencies and the $Q$ factors of the natural
modes (and verify that they do not change significantly in response to a
realistic simulation of the launch environment, which might indicate some kind
of structural damage).

In the space environment heat is exchanged between the detector and the ambient
essentially by \emph{radiation} (it goes without saying that convection plays
no role in vacuum, which means that you can't really use fans to dissipate
heat as you would do on the ground). Space-based detectors feature more or less
sophisticated thermal systems to ensure that the power generated by the
on-board electronics is efficiently radiated in the environment and to
mitigate the thermal gradients induced by changes of the orientation with
respect to the Sun (and Sun occultation by the Earth). Thermal and thermo-vacuum
tests serve the twofold purpose of validating the thermal model at the base
of the thermal control system and verifying that the detector is able to
operate (or survive) withing the maximum temperature excursion range expected
in orbit.

We note, in passing, that operating an instrument in vacuum is not as trivial
as one might expect, as in many designs composite structures (i.e.,
aggregates of several layers of different materials) are used in order to
maximize the mechanical stiffness within the mass budget---and the details
of manufacturing processes become critical.

There are many more considerations---connected with the environment---that
go into the design of a space detector (e.g., electromagnetic interference
effect and radiation damage) but we shall just pause here and move forward
to a different subject.

\subsection{Design trade-offs}

Trade-offs arise naturally when designing an instrument, as---even pretending
for a second that the operational constraints do not exist---it is impossible
to optimize all the performance metrics at the same time and it is necessary
to do compromises aimed at maximizing the overall sensitivity for a given
science target. In addition to that, different science targets may require
different (and often conflicting) optimizations, which makes the entire process
even more difficult.

In this section we shall elaborate on the concept of trade-off by looking
at a couple of (largely) academic examples.

\subsubsection{A case study: the tracking stage for a pair-conversion telescope}

Let's pretend for a second that your favorite space agency contacted you to
design the tracking stage for a pair-conversion telescope. The rules are: you
have a cubic space, with side $L = 1$~m, to fill up (completely) with
$n$ equally-spaced aggregates, each including a double-sided silicon-strip
detection plane with strip pitch $p$, immediately followed by a high-Z
converter with thickness $t$. (In other words, you have 1~m$^3$ available and
you get to pick your favorite choices for $n$, $p$ and $t$%
\footnote{This is a pretty bogus offer, as in real life you would presumably
  have control over the height of the tracker, as well, which in this case is
  fixed to $L$.}.) Last but not least, the agency is offering you a
non-negotiable power budget of 100~W.

There's a few performance metrics that we can calculate right away. Assuming
a 100\% detection and selection efficiency, for instance, the peak (i.e.,
high-energy) on-axis effective area can be estimated as
\begin{align}
  \aeffnorm = L^2 [1 - \exp(-nt)] = 1 - \exp(-nt)~{\rm m}^2.
\end{align}
In addition, following section~\ref{sec:lat_psf}, we can write the high-energy
68\% containment angle of the \psf\ as
\begin{align}
  \psf_{\rm HE} = \frac{\sqrt{2}p}{d(n/2)^{3/2}} = \frac{4p}{Ln^{1/2}} = 
  \frac{p~[\mu{\rm m}]}{250n^{1/2}}~{\rm mrad}
\end{align}
(note we are taking $n/2$ as the average number of planes crossed).
The corresponding \psf\ at 100~MeV reads
\begin{align}
  \psf_{100} = \sqrt{2} \frac{13.6}{50} \sqrt{t} = 385 \sqrt{t}~{\rm mrad}
\end{align}
(note we have neglected the extra logarithmic term in
equation~\eqref{eq:theta_ms}).

\begin{figure}[htb]
  \includegraphics[width=\linewidth]{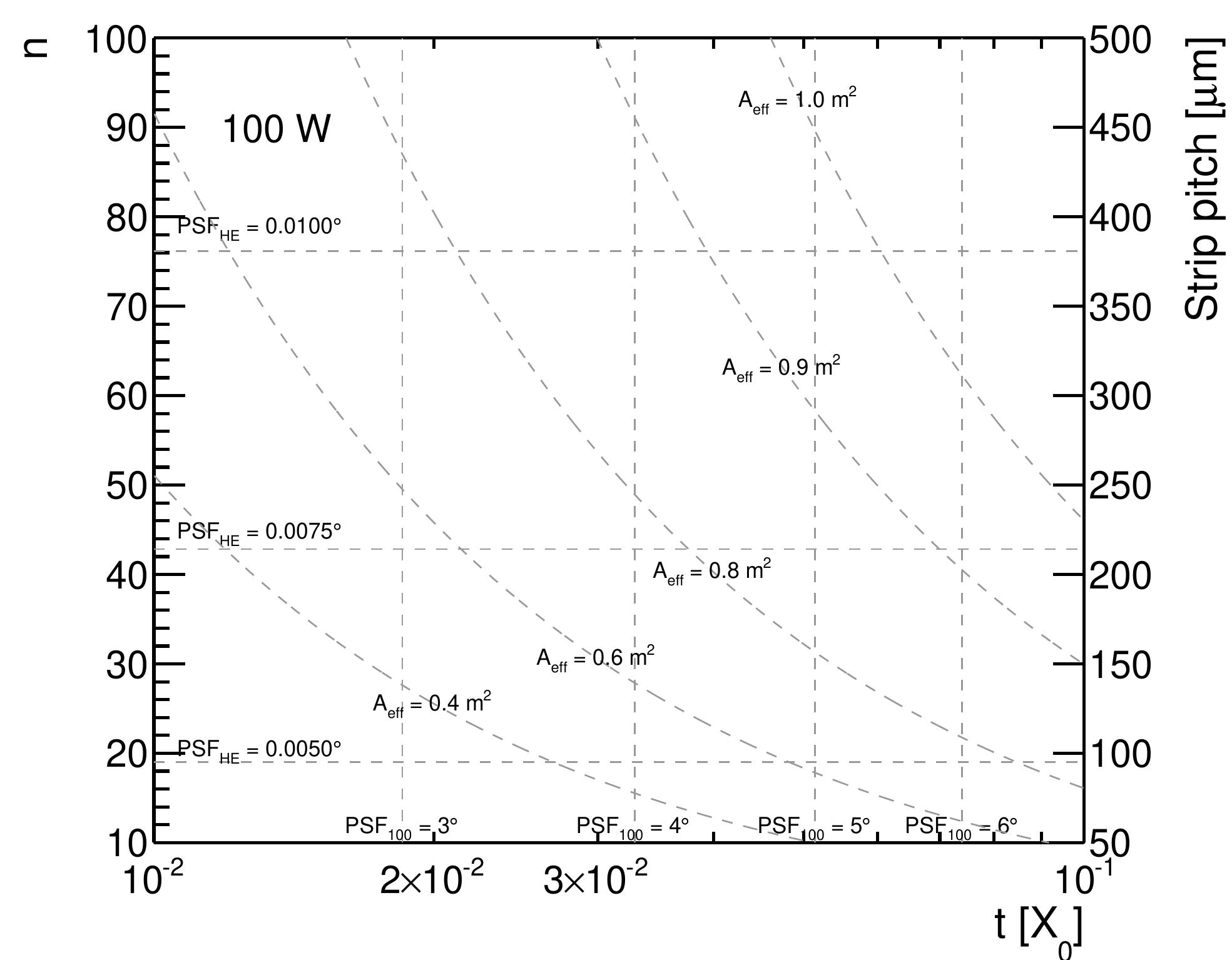}
  \caption{Iso-performance curves in the $n$--$t$ phase space for our
  benchmark pair conversion telescope. The lines at constant $\psf_{100}$ are
  vertical, while those at constant $\psf_{\rm HE}$ are horizontal. The curves
  are iso-\aeff\ lines.}
  \label{fig:telescope_optimization}
\end{figure}

That all said, with three free parameters it is not trivial to imagine how one
would go about the optimization. One thing that we can easily calculate is the
overall number of channels%
\footnote{We note, in passing, that assembling and reading out 1~m long strips
  is not trivial (in reality one would probably implement this detector concept
  as an array of at least $2\times 2$ modules), but we shall happily neglect
  this, for the time being.}:
\begin{align}
  \text{\# channels} = \frac{2nL}{p}
\end{align}
(the factor 2 at the numerator accounts for the fact that, for each plane, we
have strips in both directions---i.e., the silicon detectors are double sided).
If we assume a power consumption of 250~$\mu$W per channel, the total necessary
power $P$ reads
\begin{align}
  P = \frac{500 n}{p~[\mu{\rm m}]}~W,
\end{align}
i.e., with 100~W available you can do, say, 10 layers at 50~$\mu$m pitch or
100 layers at 500~$\mu$m pitch (see the right axis in
figure~\ref{fig:telescope_optimization}). Good: this introduces a constraint
that our design parameters must satisfy
\begin{align}
  p = 5n~\mu{\rm m},
\end{align}
and effectively reduces the dimensionality of the problem from 3 to 2.
In fact we can now rewrite the high-energy \psf\ as
\begin{align}
  \psf_{\rm HE} = \frac{n^{1/2}}{50}~{\rm mrad}
\end{align}
(if you are surprised seeing that the \psf\ is worsening as one adds more
layers, remember that the power budget is forcing us to make the strip
pitch larger).

Figure~\ref{fig:telescope_optimization} summarizes this trade-off study in
the $n$--$t$ plane. In our setup the low-energy \psf\ depends only on the
thickness of the converters and the high-energy \psf\ depends only on the
number of layers (which in turn, at fixed total power, determines the strip
pitch). Hyper-surfaces at constant on-axis high-energy effective area are
represented by the dashed curves.

There's a few interesting remarks that we can do about
figure~\ref{fig:telescope_optimization}. Obviously one would like to choose
a small value of $t$ to optimize the \psf\ at low energy. In addition, in
our setup it is advantageous to limit the number of detection planes and
take advantage of the smaller strip pitch to optimize the high-energy \psf.
By choosing the bottom-left corner of the phase space one effectively
optimizes the angular resolution over the entire energy range. Unfortunately
this is also the region where the effective area is relatively low
(no wonder: you have a small number of thin conversion layers). In contrast,
increasing the effective area requires trading off some angular resolution,
at either low or high energies---or both.

There is several things that we left off, including cost considerations and
other performance metrics such as the field of view. Nonetheless, fictional
as it is, this exercise does illustrates some of the basic aspects of
instrument design.

\section{In-flight calibration}

Space- and balloon-borne detectors are typically calibrated with particle beams
prior to launch in order to benchmark the Monte Carlo simulations used
to generate the instrument response function. While this is of primary
importance, one should recognize that that the space environment is different,
to many respects, than that in which ground tests are performed. In addition,
cosmic-ray and gamma-ray detector often aim at measuring fluxes and intensities
at much higher energies that those accessible by accelerators.

In this section we shall glance through the topic of in-flight instrument
calibration. There are two main aspects to this---namely the monitoring of the
performance stability in time and the assessment of the systematic
uncertainties associated to the instrument response functions
(e.g., effective area, point-spread function and absolute energy scale).
As we shall see in a second, the latter is generally much harder---and it goes
without saying that it is more critical for satellite (as opposed to balloon)
experiments, since, due the much larger exposure factors, many of the
measurements are systematic-limited.

\subsection{Time stability}

The stability in time of the basic instrument performance figures is usually
monitored both through dedicated calibration systems embedded in the
readout electronics (e.g., charge-injection circuits to measure the gain 
and noise of the front-end amplifiers, or laser systems to measure the 
alignment of tracking devices) and using the particle populations available in
orbit. In this latter respect minimum ionizing protons are a good example
of an abundant \emph{calibration source} with well-known properties that can be
used by large-acceptance detectors to monitor the performance stability
over relatively short time-scales.

\begin{figure}[!htb]
  \includegraphics[width=\linewidth]{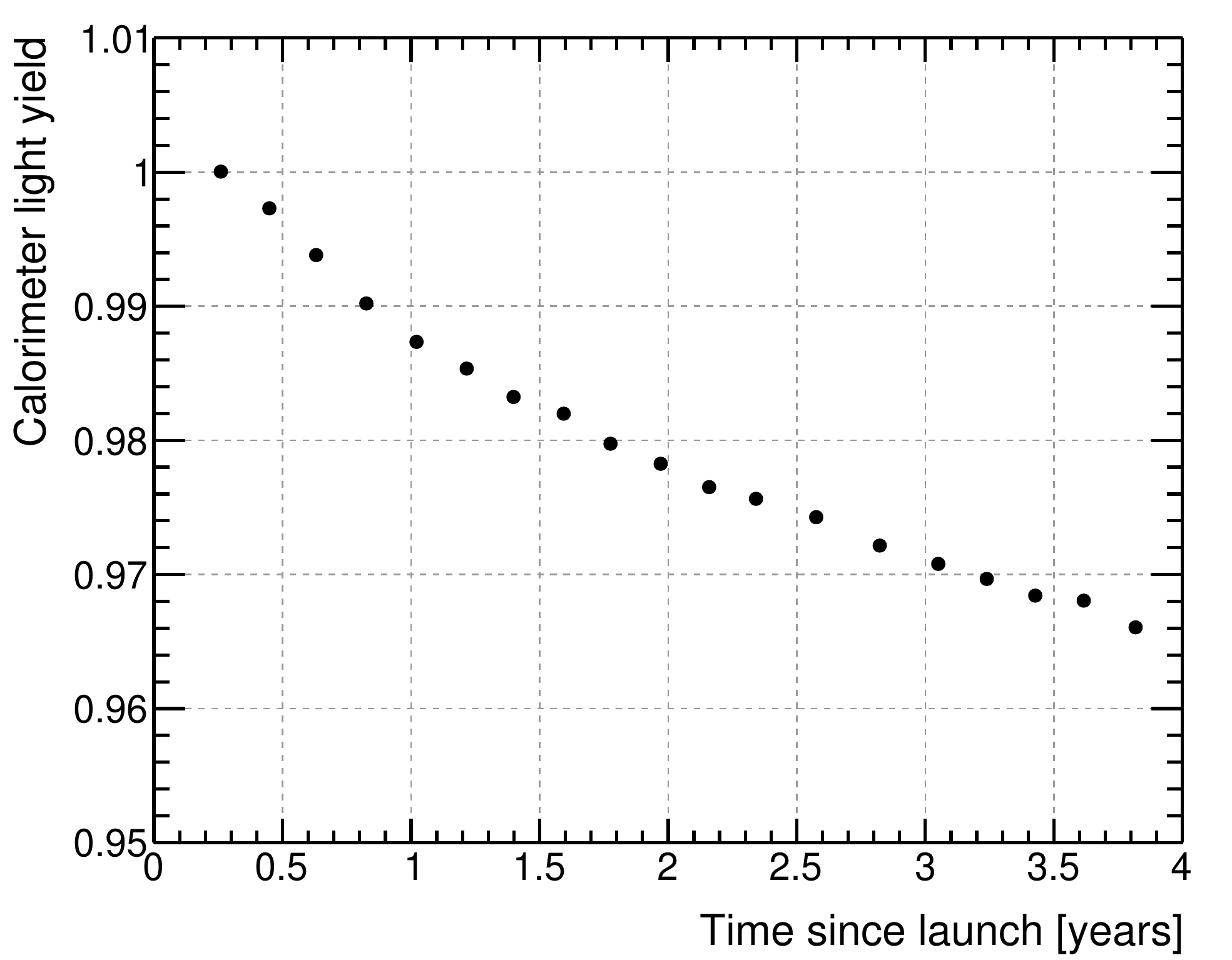}
  \caption{Relative light yield for the \Fermi-LAT calorimeter, inferred
    from the path-length-corrected energy deposition of on-orbit minimum
    ionizing protons, throughout the first four years of the mission
    (adapted from~\cite{2013arXiv1304.5456B}). The small downward trending is
    due to the (expected) radiation damage of the calorimeter crystals.}
  \label{fig:cal_light_yield}
\end{figure}

Figure~\ref{fig:cal_light_yield} shows the relative light yield of the
\Fermi-LAT calorimeter, measured through the first four years of mission
by means of the path-length-corrected energy deposition of on-orbit minimum
ionizing protons~\cite{2013arXiv1304.5456B}. The slight ($\sim 1\%$ per year)
downward trend (due to the anticipated radiation damage of the CsI crystals) is
a good example of an effect that, if not corrected through time-dependent
calibrations, might produce observable consequences in the high-level science
analysis---first and foremost a drift of the absolute energy scale.
(On a related note, it is worth noticing that it's easier to use
minimum-ionizing proton to monitor the stability of the energy scale, rather
than calibrating it directly.)

The on-orbit refinement and monitoring of the AMS-02 tracker alignment
described, e.g., in~\cite{ambrosi_icrc13_tkr_align} constitutes another
interesting topic of discussion germane to the main subject of this section.
As we know, by now, for magnetic spectrometers the tracker alignment is
crucial, as the high-energy rigidity resolution is hostage to the spatial
resolution in the bending plane. In the case of AMS-02 is impressive how the
instrument team managed to cope with the time-dependent displacements of the
outer tracker layers induced by temperature variation (up to hundreds of
$\mu$m on sub-hour time scales) and achieve a systematic error on the alignment
of the order of $\sim 3~\mu$m---significantly smaller than the hit resolution
of the position-sensitive detectors.

\subsection{Systematic uncertainties on the IRFs}

Measurements by modern, large-acceptance detectors are in many (possibly
most) cases systematic-limited---one has so many events at hand that the
statistical errors are just negligible over most of the instrumental phase
space. When that is the case the level of accuracy of our knowledge of the
detector performance figures (i.e., the systematic errors on the instrument
response functions) becomes, to some extent, more important than the absolute
values of the figures themselves. It is important, therefore, to devise methods
study those systematic errors in the environment where the instrument operates.

\subsubsection{Effective area and acceptance}

The effective area is typically studied and parameterized by means of detailed
Monte Carlo simulations and \emph{measured} with particle beams, when possible,
in discrete points of the phase space.

Given that there is no such thing as a source with a \emph{known} flux (this
is typically what one is trying to measure) it is not straightforward to study
the systematic errors on the effective area on orbit. There are ways around
this: one can select reasonably clean, signal-enriched event samples and
study the efficiency of any given selection criterium with respect to a
baseline; or try and measure the same thing with sub-samples of events---e.g.,
on-axis and off-axis events. We refer the reader to~\cite{2012ApJS..203....4A}
for a somewhat detailed discussion of such possible strategies in the context
of the analysis of \Fermi-LAT gamma-ray data, but we stress that this is a
difficult problem to discuss in abstract terms.

\subsubsection{Point-spread function}

The point-spread function is, to some extent, the easiest thing to
\emph{calibrate} in orbit (at least where it matters, i.e., in gamma rays),
in that bright gamma-ray point sources%
\footnote{In a sense, there is no such thing as a point source, but in
  practice the actual angular size for many gamma-ray sources is much
  (much) smaller than PSF one can realistically achieved.}
serve essentially the same purpose that a monochromatic particle beam would
serve for energy calibration: not only the measured photon directions are
distributed around the \emph{true} position according to the PSF (this is by
definition), but in many cases this true position is known%
\footnote{Technically, the position is measured at other wavelengths with
  a much greater accuracy than that achievable in gamma rays, which allows to
  calibrate not only the point-spread function, but also the absolute
  pointing accuracy in the sky.} 
from measurements at other wavelengths.

Bright pulsars are prototypical examples of sources that can be used for the
in-flight calibration of the PSF---in that case one can also take advantage the
phase information to select the calibration sample. One notable limitation is
due to the fact that most pulsars feature a spectral cutoff at $\sim$~GeV
energy, so that they cannot be readily used at very high energies.
Bright active galactic nuclei (and possibly a stack of them) are a viable
alternatives.
We refer the reader to~\cite{2012ApJS..203....4A,2013ApJ...765...54A} for a
thorough discussion of the in-flight calibration of the \Fermi-LAT PSF.

\subsubsection{Absolute energy scale}%
\label{sec:calib_energy_scale}

Unlike the point-spread function, the absolute energy scale is notoriously
difficult to calibrate in orbit due to the lack of sharp spectral features
\emph{at know energies} in the GeV-to-TeV range. As a matter of fact, the
discovery of such a feature (e.g., a line) would be a major one, but still
useless from the standpoint of the calibration of the energy scale.

Magnetic spectrometers with an electromagnetic calorimeter on board have the
advantage of measuring the energy \emph{and} the rigidity (which is to say,
they measure the energy twice and independently) for electrons and positrons
with a good resolution over a relatively wide energy range. The
energy-rigidity matching provides a good handle on the systematic uncertainties
on the absolute scales, but for purely calorimetric experiments life is
relatively harder.

\begin{figure}[!htb]
  \includegraphics[width=\linewidth]{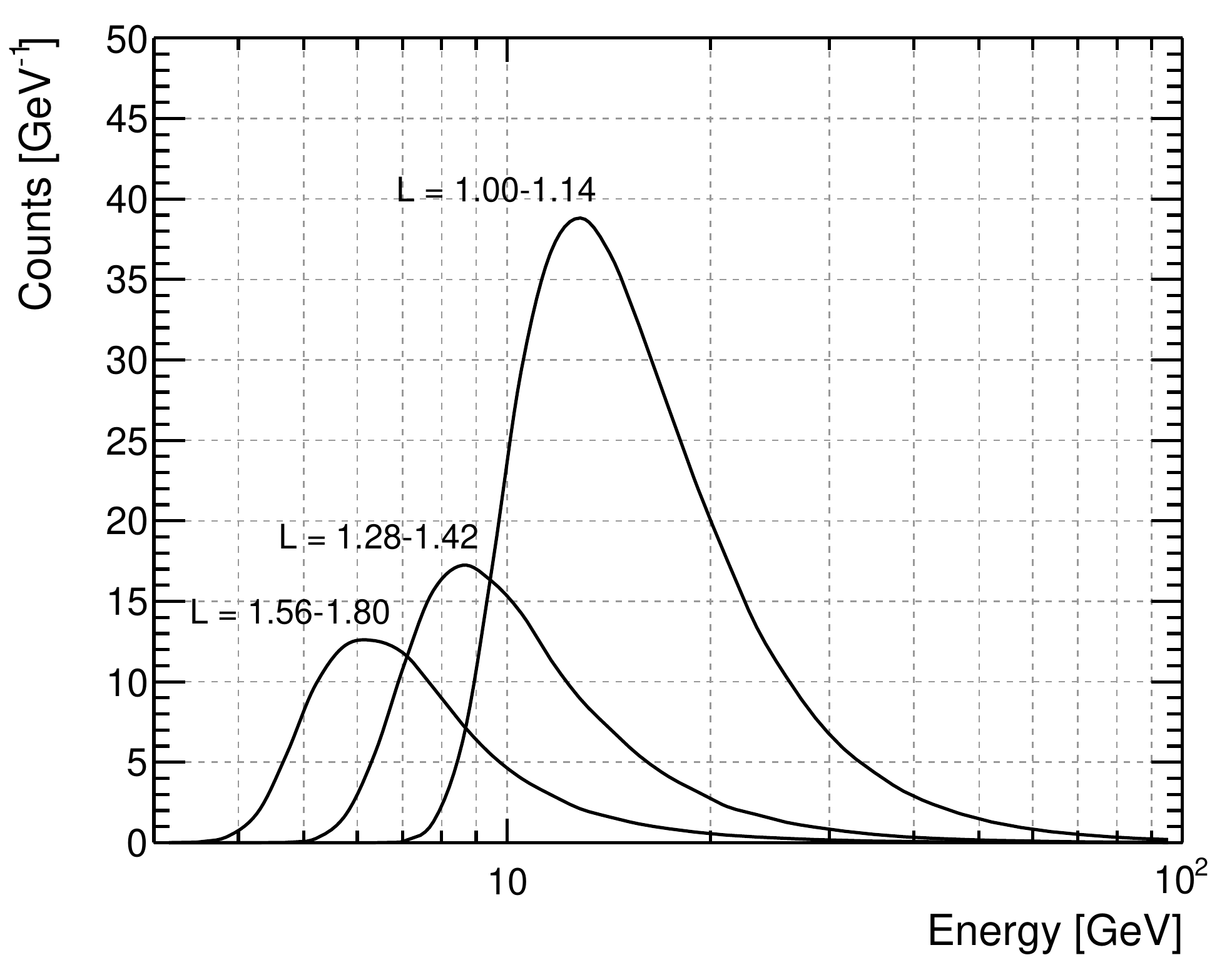}
  \caption{Examples of simulated cosmic-ray electron count spectra in bins of
    McIlwain~$L$, representative of the \Fermi\ orbit---averaged over the
    \Fermi\-LAT field of view and folded with the \Fermi\-LAT energy resolution
    (adapted from~\cite{2012APh....35..346A}). The peaked shape is the result
    of the convolution of the power-law spectrum of primary cosmic-ray electrons
    with the screening effect of the geomagnetic field. The different peaks
    at different values of McIlwain~$L$, whose position can be reliably
    predicted through ray-tracing techniques, effectively provide a series of
    calibration points for the absolute energy scale.}
  \label{fig:cre_cutoff_fermi}
\end{figure}

We end this section by briefly mentioning the verification of the absolute
energy scale performed by the \Fermi-LAT~\cite{2012APh....35..346A} using the
rigidity cutoff induced by the geomagnetic field (see
section~\ref{sec:geomagnetic_cutoff}).
While the geomagnetic cutoff is a sharp feature in any given direction, it
gets smeared when averaged over the finite field of view of an instrument, as
shown in figure~\ref{fig:cre_cutoff_fermi}. Still, by selecting events
in bins of the McIlwain~$L$ variable, the convolution between the power-law
spectrum of primary cosmic-ray electrons and the shielding effect due to the
magnetic field results in narrowly-peaked spectra whose shape can be
predicted by means of ray-tracing techniques. We refer the reader
to~\cite{2012APh....35..346A} for more details about the method and the
results.

\section*{Acknowledgments}

The author would like to thank (in strict alphabetical order)
Roberto Aloisio,
Pasquale Blasi,
Eric Charles,
Marco Cirelli, 
Stefano di Falco,
Seth Digel,
Alex Drlica-Wagner,
Fiorenza Donato,
Nicolao Fornengo,
Dario Grasso,
Eric Grove,
Marco Incagli,
Piersimone Marrocchesi,
David Maurin,
Warit Mitthumsiri,
Melissa Pesce-Rollins,
Carmelo Sgr\`o,
Luigi Tibaldo
for useful discussions.

It cannot be stressed enough what a useful service to the community the
cosmic-ray database described in~\cite{CRDB} represents.

\section*{Change log}

We are trying and log in this section the most relevant changes that are done
to the manuscript as it evolves. Cross-links are in the context of the
most recent version of the write-up where possible.

\subsubsection*{Version v2, last edited on August 4, 2014}

\begin{itemize}
\item Filled in some more numbers in table~\ref{tab:pres_fut_instruments}---most
  notably Gamma-400 and HERD were added.
\item Added $z$-axis labels to figures~\ref{fig:sea_level_mag_field},
  \ref{fig:high_alt_mc_ilwain_L} and \ref{fig:high_alt_rigidity_cutoff}.
\end{itemize}

\vfill

\bibliography{crdb,crdetectors}

\clearpage

\onecolumngrid
\appendix

\section{Cosmic-ray measurements}

\setlength\LTleft{0pt}
\setlength\LTright{0pt}
\begin{longtable}{@{\extracolsep{\fill}}llll}
  \caption[]{List of cosmic-ray measurements from~\cite{CRDB}, ordered by
    particle type and by year within each type.}
  \label{tab:cr_measurements}\\

  \hline
  \hline
  \tabrowspace
  \multicolumn{1}{l}{Experiment} &
  \multicolumn{1}{l}{Energy range} &
  \multicolumn{1}{l}{Date} &
  \multicolumn{1}{l}{Reference}\\
  \tabrowspace
  \hline 
  \endfirsthead

  \multicolumn{4}{c}{\tablename\ \thetable{} -- continued from previous page}\\
  \hline
  \tabrowspace
  \multicolumn{1}{l}{Experiment} &
  \multicolumn{1}{l}{Energy range} &
  \multicolumn{1}{l}{Date} &
  \multicolumn{1}{l}{Reference}\\
  \tabrowspace
  \hline
  \endhead

  \multicolumn{4}{r}{{Continued on next page}}\\
  \endfoot

  \hline\hline
  \endlastfoot

  \tabularnewline\hline
\tabrowspace\multicolumn{4}{c}{$^{10}$B/$^{11}$B}\\\tabrowspace
\hline
AMS01 & \makebox[0.15\textwidth]{0.20--1.4\hfill}[GeV/n] & 1998 Jun 02--1998 Jun 12 & {\cite{2011ApJ...736..105A}}\\
ISOMAX & \makebox[0.15\textwidth]{0.28--0.95\hfill}[GeV/n] & 1998 Aug 04--1998 Aug 05 & {\cite{2004ApJ...611..892H}}\\
\tabularnewline\hline
\tabrowspace\multicolumn{4}{c}{$^{10}$B/B}\\\tabrowspace
\hline
Balloon & \makebox[0.15\textwidth]{0.16--0.42\hfill}[GeV/n] & 1973 Aug 15--1973 Aug 15 & {\cite{1977ApJ...212..262H}}\\
Voyager1\&2 & \makebox[0.15\textwidth]{0.04--0.12\hfill}[GeV/n] & 1977 Jan 01--1998 Dec 31 & {\cite{1999ICRC....3...41L}}\\
\end{longtable}

\clearpage
\section{Cosmic-ray and gamma-ray experiments}

\setlength\LTleft{0pt}
\setlength\LTright{0pt}
\begin{longtable}{@{\extracolsep{\fill}}llllp{0.55\textwidth}}
  \caption[]{List of some of the cosmic-ray experiments relevant for this
    review (note that most of the balloon experiments---notably the ones
    with no name---are not included). The legend for the instrument techniques
    reads:
    CR (\cheren\ range),
    EC (emulsion chamber),
    IC (imaging calorimeter),
    MS (magnetic spectrometer),
    PCT (pair-conversion telescope),
    SC (scintillator counter).
    Note that we have been fairly liberal in the definition of ``imaging
    calorimeters'', in that most of those listed here do include some kind of
    tracking and/or charge-measurement stage.}
  \label{tab:cr_experiments}\\

  \hline
  \hline
  \tabrowspace
  \multicolumn{1}{l}{Experiment} &
  \multicolumn{1}{l}{Technique} &
  \multicolumn{1}{l}{Location} &
  \multicolumn{1}{l}{Launch date} &
  \multicolumn{1}{l}{Web site}\\
  \tabrowspace
  \hline 
  \endfirsthead

  \multicolumn{4}{c}{\tablename\ \thetable{} -- continued from previous page}\\
  \hline
  \tabrowspace
  \multicolumn{1}{l}{Experiment} &
  \multicolumn{1}{l}{Technique} &
  \multicolumn{1}{l}{Location} &
  \multicolumn{1}{l}{Launch date} &
  \multicolumn{1}{l}{Web site}\\
  \tabrowspace
  \hline
  \endhead

  \multicolumn{5}{r}{{Continued on next page}}\\
  \endfoot

  \hline\hline
  \endlastfoot

  ACE &  & Satellite & 1997 &
\smallurl{http://www.srl.caltech.edu/ACE/CRIS_SIS/index.html} \\

AESOP & MS &  Balloon & 1994 &
\smallurl{http://www.bartol.udel.edu/gp/balloon} \\

AGILE & PCT &  Satellite & 2007 &
\smallurl{http://agile.rm.iasf.cnr.it/} \\

ALICE & CR &  Balloon & 1987 &
\smallurl{http://ida1.physik.uni-siegen.de/alice.html} \\

AMS-01 & MS &  Shuttle & 1998 &
\smallurl{http://ams.cern.ch/AMS/Welcome.html} \\

AMS-02 & MS &  ISS & 2011 &
\smallurl{http://www.ams02.org} \\

ATIC & IC &  Balloon & 2000 &
\smallurl{http://atic.phys.lsu.edu/index.html} \\

BESS & MS &  Balloon & 1993 &
\smallurl{http://astrophysics.gsfc.nasa.gov/astroparticles/programs/bess/BESS.html} \\
\end{longtable}

\end{document}